\documentclass[aps, prx, twocolumn, showpacs, 10pt, floatfix, superscriptaddress, floatfix, longbibliography,nofootinbib]{revtex4-2}
\usepackage[acronym]{glossaries}
\makeglossaries
\usepackage{lipsum}  
\usepackage[english]{babel}
\usepackage[utf8]{inputenc}
\usepackage[T1]{fontenc}
\usepackage{ragged2e}
\usepackage{amsmath}
\usepackage{makecell}
\usepackage{braket}
\usepackage{amssymb}
\usepackage[normalem]{ulem}
\usepackage{dsfont}
\usepackage{bm}
\usepackage{bbm}
\usepackage[dvipsnames]{xcolor}
\usepackage{stmaryrd}
\usepackage{bbm}
\usepackage{mathtools}
\usepackage{hyperref}
\usepackage{physics}

\usepackage{graphicx}
\usepackage{dcolumn}
\usepackage{xcolor}
\usepackage{hyperref}
\hypersetup{
    colorlinks=true,
    citecolor=Green,
    linkcolor=Maroon,
    urlcolor=Blue
    }

\usepackage{cancel}
\usepackage{verbatim}
\usepackage{tikz}
\usetikzlibrary{shapes.geometric, arrows.meta, positioning}

\DeclareSymbolFont{usualmathcal}{OMS}{cmsy}{m}{n}
\DeclareSymbolFontAlphabet{\mathcal}{usualmathcal}

\newcommand{\be}{\begin{equation}}
\newcommand{\ee}{\end{equation}}
\newcommand{\f}{\frac}

\newcommand{\bea}{\begin{eqnarray}}
\newcommand{\eea}{\end{eqnarray}}
\newcommand{\ba}{\begin{align}}
\newcommand{\ea}{\end{align}}

\newcommand{\beq}{\begin{equation}}
\newcommand{\eeq}{\end{equation}}

\makeatletter
\newcommand{\thickhline}{%
\noalign {\ifnum 0=`}\fi \hrule height 1.2pt
\futurelet \reserved@a \@xhline%
}
\newcolumntype{"}{@{%
\hskip\tabcolsep\vrule width 1.2pt\hskip\tabcolsep}%
}
\makeatother

\renewcommand{\Re}{\operatorname{Re}}
\renewcommand{\Im}{\operatorname{Im}}

\begin{document}


\preprint{APS/123-QED}

\title{Bootstrapping Symmetries in Quantum Many-Body Systems \\ from the Cross Spectral Form Factor}%

\author{Chen Bai}
\thanks{These authors contributed equally to this work.}
\affiliation{Kavli Institute for Theoretical Sciences, University of Chinese Academy of Sciences, Beijing
100190, China
}

\author{Zihan Zhou}
\thanks{These authors contributed equally to this work.}
\affiliation{%
Department of Physics, Princeton University, Princeton, New Jersey 08544, USA%
}%

\author{Bastien Lapierre}
\affiliation{Philippe Meyer Institute, Physics Department, École Normale Supérieure (ENS), Université PSL, 24 rue Lhomond, F-75231 Paris, France}

\author{Shinsei Ryu}
\affiliation{%
Department of Physics, Princeton University, Princeton, New Jersey 08544, USA%
}%

\date{\today}


\begin{abstract}
Symmetries play a central role in quantum many-body physics, yet uncovering them systematically remains challenging.
We introduce a bootstrap framework designed to reconstruct the representation theory of hidden finite group symmetries of quantum many-body lattice Hamiltonians, using only a known symmetry subgroup $N$ and spectral correlations between its symmetry sectors. We introduce the cross spectral form factor (xSFF) as a subgroup-resolved spectral diagnostic of hidden symmetry, which we compute via exact diagonalization to seed the bootstrap algorithm. By applying the constraints derived from these data alongside the algebraic conditions of the fusion rules, our bootstrap procedure sharply restricts the set of candidate groups $G$. Remarkably, without any prior assumptions regarding the full symmetry group $G$, our method can systematically recover its representation-theoretic data, including the number and dimensions of the irreducible representations, their branching rules with respect to $N$, the fusion algebra, and the full character table.
This framework applies equally well to chaotic and integrable many-body systems and accommodates both unitary and anti-unitary symmetries. Through various examples, we demonstrate that the underlying group $G$ can be uniquely identified. In particular, our bootstrap independently recovers the $\mathbb{Z}_4$ symmetry at the self-dual point of the three-state quantum torus chain, detects signatures of projective representations in the effective Hamiltonian of the driven Bose-Hubbard model, and rediscovers the $\eta$-pairing $\mathrm{SO}(4)$ symmetry of the one-dimensional Fermi-Hubbard model. Leveraging the universal random-matrix ramp in chaotic systems, we demonstrate the possibility of ramp-only bootstrap alongside an example with hidden commuting symmetries. Furthermore, we establish the xSFF as a diagnostic tool for conserved order parameters in a symmetry-breaking spin-$1$ chain and introduce an explicit ancilla-assisted measurement circuit to facilitate direct experimental realization. Our framework thus establishes a practical route to identify symmetries directly from dynamical spectral observables.
\end{abstract}



\maketitle
\begingroup
\makeatletter
\let\l@subsection\@gobbletwo
\let\l@subsubsection\@gobbletwo
\tableofcontents
\endgroup

\section{Introduction\label{sec:Intro}}

Symmetry is a fundamental organizing principle in physics \cite{Wigner1959,Coleman1985,Sachdev2011}. It underlies conservation laws and constrains the form of physical laws. In quantum systems, symmetry organizes states and spectra, structuring Hilbert spaces into representations. In particle physics, states and particles are organized into irreducible representations of symmetry groups, with quantities such as spin, charge, and mass labeling these representations. Furthermore, symmetry provides a framework for understanding phases of matter and collective behavior, including phases with spontaneous symmetry breaking and symmetry-protected topological phases, which can be characterized in terms of quantum anomalies. Finally, symmetry guides effective theories and the emergence of universality. Even when microscopic details differ or are unknown, systems with the same symmetry properties often exhibit identical large-scale behavior.

Some symmetries, such as global spin rotations and crystalline symmetries, are manifest in the Hamiltonian. Others, however, are “hidden,” or remain unresolved in that they are not directly apparent. A classic example is the hydrogen atom, which possesses a hidden $\mathrm{SO}(4)$ symmetry beyond the manifest $\mathrm{SO}(3)$ rotational symmetry. In many-body quantum systems, which are the focus of the present work, another example is the $\eta$-pairing symmetry of the spin-$\tfrac{1}{2}$ Fermi–Hubbard model~\cite{Yang:1989hms}.
At half-filling, the system possesses an exact $\mathrm{SU}(2)$ pseudo-spin symmetry that remains hidden in the position-space representation of the Hamiltonian. This hidden symmetry expands the full symmetry group to $\mathrm{SO}(4)$, constraining the excited eigenstates and generating extensive spectral degeneracies~\cite{2020PhRvB.102h5140M}.
Hidden symmetries may also 
arise from unconventional conserved quantities~\cite{Abanin_2019}, Hilbert space fragmentation~\cite{Moudgalya:2021ixk}, emergent algebraic structures~\cite{YangZhang1990}, or non-local generators~\cite{Kennedy:1992ifl}. 

Existing methods for identifying symmetries in quantum many-body systems fall into two broad categories. 
On the algebraic side, commutant-based and related operator-algebraic methods provide a systematic route to characterize symmetries of (suitable families of) local Hamiltonians and, in favorable cases, to construct explicit conserved quantities~\cite{Moudgalya:2022nll,Moudgalya:2022gtj,Moudgalya:2023tmn,Moudgalya:2023yon,chen2026bridgingcommutantpolynomialmethods}. These methods are especially powerful when one has microscopic, operator-level access to the local terms defining the Hamiltonian family: their direct output is a constructive description of the commutant algebra and its sector structure, from which symmetry generators or group-theoretic labels may require further interpretation.
The second category is spectral: 
Level-spacing and gap-ratio statistics can detect hidden discrete symmetries and can even estimate the number and relative sizes of unresolved symmetry sectors~\cite{PhysRevX.12.011006,Lasek:2024ess,He2026StatisticalSignatures,Frey:2023lql,He2026StatisticalSignatures,He:2026ipo}. 

However, these diagnostics remain coarse: they can signal the presence of hidden structure without identifying the symmetry group, its irreducible representations (irreps), or how these irreps combine and decompose. 
Here, we show that hidden symmetry can be constrained and, in many cases, fully reconstructed from spectral information, provided the spectrum is resolved with respect to a known subgroup $N$ of the full symmetry group $G$. 
Inspired by bootstrap approaches in quantum field theory
\cite{Rattazzi:2010yc,Poland:2018epd},
many-body physics
\cite{PhysRevLett.125.041601,Cho:2024owx,Scheer:2024eyu},
and entanglement structures
\cite{Shi:2019mlt,Shi:2020rne},
our method may be viewed as a form of bootstrap, in which partial spectral data, together with general consistency conditions, constrain the underlying symmetry structure.

Our starting point is that a subgroup $N$ of the full symmetry group $G$ is
known together with its action on the Hilbert space. Specifically,
the available symmetry generators must permit resolution into the
irreducible representations of $N$ ($N$-irreps), while the symmetries
outside $N$ remain unresolved.
The central task is to infer the action of the full symmetry group $G$ 
on the Hilbert space from the partial information. 
More precisely, given a decomposition of the Hilbert space into $N$-irreps, we seek to determine how these $N$-irreps are grouped together into irreps of $G$.
The key unknowns are the branching structure (or branching rules), 
which describe how each irrep of $G$ decomposes upon restriction to $N$;
these data are encoded in a set of non-negative integers, known as the branching multiplicities.
In particular, they specify how many times each $N$-irrep appears inside a $G$-irrep.
Our strategy is to treat the branching structure as a fingerprint of the hidden symmetry. Once sufficiently constrained, it restricts the admissible $G$-irreps.
Furthermore,  
combined with algebraic consistency conditions on the fusion (tensor product) structure of representations, we can reconstruct much of the representation theory of $G$, including, in particular, the fusion rules and the character table. 

To constrain the branching multiplicities, we introduce the
cross spectral form factor (xSFF), a matrix of spectral
auto-correlations and cross-correlations between $N$-irrep
sectors. The xSFF generalizes the standard spectral form factor (SFF) from auto-correlations within a single symmetry sector to cross-correlations between distinct $N$-sectors. 
Intuitively, the xSFF quantifies the extent to which two $N$-irreps remain correlated by descending from the same irrep of the larger group $G$. 
In this way, it provides direct information about how the known $N$-resolved structure is embedded within the full symmetry $G$.
The crucial physical insight underlying our bootstrap framework is that the late-time averaged plateaus of the xSFF are governed by the branching rules. Unlike the intermediate-time ramp, which is sensitive to whether a system is chaotic or integrable, the plateau is universal:
it is determined by the branching structure of the hidden symmetry rather than by chaotic versus integrable dynamics. As a consequence, detailed features of the plateau, such as which entries vanish or remain finite, degeneracies among plateau heights, and their precise asymptotic values, provide rigid numerical constraints on the branching multiplicities.

Building on this observation, we formulate a bootstrap algorithm for finite group symmetries that takes as input a known subgroup $N$, xSFF plateau data extracted from exact diagonalization, and
algebraic consistency conditions on the fusion rules. Starting from these inputs, the algorithm reconstructs the admissible $G$-irreps, their branching rules under restriction to $N$, the fusion algebra,
and the character table. In the examples studied here, this information is sufficient to uniquely identify the hidden symmetry group. Our target is the representation structure of exact group symmetries, which is a kinematic property rather than a thermodynamic observable.
Moderate system sizes can therefore suffice when the relevant irreps are present and their spectral signatures are resolved,
without extrapolation to the thermodynamic limit. The complete bootstrap is formulated for finite groups, with extensions to projective representations and compact Lie groups.

For chaotic systems with a universal random matrix regime, the xSFF ramp provides unweighted branching constraints that enable plateau-independent reconstruction and the detection of hidden commuting symmetries, as shown in Appendix~\ref{app:xSFF-ramp}. Applying the xSFF to a spin-$1$ chain further allows us to test whether symmetry breaking arises from a hidden conserved order parameter in Appendix~\ref{sec:xSFF-order-parameter}. Finally, in Sec.~\ref{sec:Experiment}, we detail an ancilla-assisted randomized-measurement circuit for the experimental extraction of both diagonal and off-diagonal xSFF components.

The rest of the paper is organized as follows. In Sec.~\ref{sec:Setting}, we introduce the physical setting as well as the representation-theoretic framework underlying our bootstrap approach, including branching data and fusion rules. In Sec.~\ref{sec:Bootstrap}, we derive the bootstrap constraints and introduce the xSFF, which provides additional numerical constraints on the branching structure. In particular, the algorithmic implementation of the bootstrap procedure is provided in this section. In Secs.~\ref{sec:S3-Inv-OF-Chain}--\ref{sec:3-QTC}, we investigate various quantum many-body lattice models with finite group symmetry. The examples are ordered by increasing complexity, illustrating our method and progressively extending its scope. This includes cases with symmetry hidden by non-local transformations, higher branching multiplicities, and anti-unitary symmetries. In particular, our bootstrap independently recovers the $\mathbb{Z}_4$ symmetry at the self-dual point of the quantum torus chain. In Sec.~\ref{sec:extensions_beyond_finite}, we discuss extensions of our approach to projective representations as well as to compact Lie groups. There, the xSFF detects signatures of projective representations in the effective Hamiltonian of the driven Bose-Hubbard model, and rediscovers the $\eta$-pairing $\mathrm{SO}(4)$ symmetry of the Fermi-Hubbard model. In Sec.~\ref{sec:Experiment}, we propose a possible protocol to experimentally probe the xSFF in quantum simulators.
We conclude in Sec.~\ref{sec:Summary-Outlook},
listing some natural extensions and follow-ups.
Some technical details are deferred to the appendices. 

\section{Physical Setting and Representation-Theoretic Framework\label{sec:Setting}}
\subsection{Physical Setting}

In this work, we consider quantum many-body systems with global symmetries forming a group $G$, such that the total Hilbert space $\mathcal{H}$ decomposes into the direct sum of subspaces, $\{\mathcal{H}_{\alpha}\}$, i.e.
\be\label{eq:Isotypic-Decomposition}
    \mathcal{H}\cong\bigoplus_{\alpha\in\hat{G}}\mathcal{H}_{\alpha},
    \quad\mathcal{H}_{\alpha}\cong V_{\alpha}\otimes M_{\alpha},
\ee
where $V_{\alpha}$ is an irrep space with $\dim V_{\alpha}=d_{\alpha}\in\mathbb{N}_+$, and $M_{\alpha}$ is the corresponding multiplicity space with $\dim M_{\alpha}=m_{\alpha}$. We label the irreps of $G$ by $\alpha \in \hat{G}$, where $\hat{G}$ denotes the complete set of all inequivalent $G$-irreps. Then, the dimensions of the total Hilbert space and of the subspace $\mathcal{H}_{\alpha}$ are $\dim\mathcal{H}=D_{\mathcal{H}}=\sum_{\alpha\in\hat{G}}d_{\alpha}m_{\alpha}$ and $\dim\mathcal{H}_{\alpha}=d_{\alpha}m_{\alpha}$, respectively. Note that Eq.~\eqref{eq:Isotypic-Decomposition} is equivalent to the decomposition form of $\mathcal{H} \cong \bigoplus_{\alpha\in\hat{G}} V_{\alpha}^{\otimes m_{\alpha}}$. Specifically, if we view $M_{\alpha}$ as a diagonal matrix $M = \text{diag}(\dots,m_{\alpha},\dots)$, then the tensor product relates to the direct sum as:
\begin{equation}
    V_{\alpha}\otimes M_{\alpha}\cong V_{\alpha}^{\otimes m_{\alpha}} = \underbrace{V_{\alpha }\oplus V_{\alpha} \oplus \cdots \oplus V_{\alpha}}_{m_{\alpha}\text{times}}.
\end{equation}

\subsection{Restriction and Branching as Fingerprints of hidden symmetry}
The known subgroup $N\subset G$ can be identified from manifest symmetries of the system. Examples include the global three-state clock rotation, which generates $\mathbb Z_3$, and global $\pi$ rotations about two orthogonal axes in a spin-$1$ chain, which generate $\mathbb Z_2\times\mathbb Z_2$. The corresponding generators can be verified directly to satisfy the requisite group relations and to commute with the Hamiltonian. We label the irreducible representations of $N$ by $\lambda\in\widehat N$. We can then decompose the Hilbert space according to $N$-irreps
\begin{equation}
    \mathcal{H} \cong \bigoplus_{\lambda \in \hat{N}} V_{\lambda} \otimes M_{\lambda} ~,
\end{equation}
with $V_{\lambda}$ the $N$-irrep vector space and $M_{\lambda}$ the corresponding multiplicity space. The key data linking $G$ to the known subgroup $N$ is the branching rules, i.e. how each irrep of $G$ decomposes under restriction to $N$.
By restricting each $G$-irrep $V_{\alpha}$ to subgroup $N$, we obtain the branching decomposition
\be\label{eq:Branching-Decomposition}
    \text{Res}^G_N V_{\alpha}\cong\bigoplus_{\lambda\in\hat{N}} V_{\lambda}\otimes \mathbb{C}^{b_{\lambda,\alpha}},
    \quad 
    \dim V_{\lambda}=d_{\lambda},
\ee
where $V_{\lambda}$ is a $N$-restricted symmetry sector ($N$-irrep), $b_{\lambda,\alpha}\in\mathbb{N}$ is the branching multiplicity of $\lambda$ inside the restriction of $\alpha$,
and 
$\mathbb{C}^{b_{\lambda,\alpha}}$ means $b_{\lambda,\alpha}$-dimensional complex vector space:
$
    \mathbb{C}^{b_{\lambda,\alpha}}\cong\underbrace{\mathbb{C}\oplus\cdots\oplus \mathbb{C}}_{b_{\lambda,\alpha}\text{ times}}
$.
Then, Eq.~\eqref{eq:Isotypic-Decomposition} can be re-written as
\be\label{eq:Hilbert-Space-Decomposition-Subset}
\begin{aligned}
    \mathcal{H} & \cong \bigoplus_{\alpha\in\hat{G}} \left(\bigoplus_{\lambda\in\hat{N}} V_{\lambda}\otimes \mathbb{C}^{b_{\lambda,\alpha}}\right)\otimes M_{\alpha} \\
    & = \bigoplus_{\lambda \in \hat{N}} V_{\lambda} \otimes \Bigg(\bigoplus_{\alpha \in \hat{G}} M_{\alpha} \otimes \mathbb{C}^{b_{\lambda,\alpha}}\Bigg),
\end{aligned}
\ee
which immediately implies
\begin{equation}
    M_{\lambda} = \bigoplus_{\alpha \in \hat{G}} M_{\alpha} \otimes \mathbb{C}^{b_{\lambda,\alpha}} ~,
\end{equation}
and
\be\label{eq:Dim-Condition-1}
    d_{\alpha}=\sum_{\lambda\in\hat{N}}b_{\lambda,\alpha}d_{\lambda}.
\ee
Given that the dimensions $d_{\lambda}$ are fully specified for all $\lambda \in \hat{N}$, we can leverage this relation to constrain the permissible dimensions $d_\alpha$ of the $G$-irreps. For example, consider $G=S_3$, the symmetry group of an
equilateral triangle, and its rotation subgroup
$N\cong\mathbb{Z}_3$. Label the trivial and sign singlets by
$\alpha_0,\alpha_1$, and the two-dimensional irrep by
$\alpha_2$. The subgroup charges $\lambda=0,1,2$ correspond
to rotation eigenvalues $e^{2\pi i\lambda/3}$.
Both singlets are invariant under rotations, whereas the
doublet carries the two nontrivial eigenvalues. Hence,
\begin{equation}
\label{eq:Setting-S3-Branching}
\begin{aligned}
    \operatorname{Res}^{S_3}_{\mathbb{Z}_3}V_{\alpha_0}
    &\cong
    \operatorname{Res}^{S_3}_{\mathbb{Z}_3}V_{\alpha_1}
    \cong V_0,\\
    \operatorname{Res}^{S_3}_{\mathbb{Z}_3}V_{\alpha_2}
    &\cong V_1\oplus V_2.
\end{aligned}
\end{equation}
For an $S_3$-invariant Hamiltonian, each doublet supplies
degenerate states in the charge-$1$ and charge-$2$ sectors,
and both singlet types contribute to charge $0$.

When we restrict our attention to finite groups or compact Lie groups. The branching multiplicities evaluate to
\be\label{eq:Branching-Multiplicity}
\begin{split}
    b_{\lambda,\alpha}&=\left\langle\chi_{\lambda},\text{Res}^G _N\chi_{\alpha}\right\rangle_{N}=\left\langle\text{Ind}^G_N\chi_{\lambda},\chi_{\alpha}\right\rangle_{G},
\end{split}
\ee
where $\chi_{\alpha}$ and $\chi_{\lambda}$ are the characters of the representations $V_{\alpha}$ and $V_{\lambda}$, while $\text{Res}^G_N$ and $\text{Ind}^G_N$ denote the restriction and induction, respectively. The second equality follows from Frobenius reciprocity~\cite{serre1977linear}.
Here, $\langle \cdot, \cdot \rangle_{G}$ denotes the standard inner product of group functions on $G$ (and similarly for the subgroup $N$). For arbitrary functions $f$ and $h$ on $G$, this is defined as 
\be\label{eq:Inner-Product-Group-Function-1}
\begin{split}
    \left\langle f,h\right\rangle_{G}&=\begin{cases}
    \displaystyle
        \f{1}{|G|}\sum_{g\in G}f(g)h(g)^*~&\text{finite},\\
        \\
        \displaystyle
        \int_G {\dd  }\mu(g)f(g)h(g)^*~&\text{compact},\\
    \end{cases}
\end{split}
\ee
where $\dd\mu(g)$ denotes the normalized Haar measure on $G$. As the characters are class functions that are invariant under conjugation, we can equivalently express their inner product by integrating or summing directly over the space of conjugacy classes $\mathcal{C}_G$:
\be\label{eq:Inner-Product-Group-Function-2}
\begin{split}
    \left\langle f,h\right\rangle_{G}&=\begin{cases}
    \displaystyle
        \f{1}{|G|}\sum_{c\in\mathcal{C}_G}|c|f(c)h(c)^*~&\text{finite},\\
        \\
        \displaystyle
        \int_{\mathcal{C}_G} \dd \mu(c)f(c)h(c)^*~&\text{compact},\\
    \end{cases}
\end{split}
\ee
where $|c|$ denotes the size of the conjugacy class $c$ for finite groups, and $\dd\mu(c)$ denotes the induced measure (from $G$) on the conjugacy class space for the compact Lie group $G$.

Crucially, once the groups $N$ and $G$ are fixed up to isomorphism, the branching multiplicities are \textit{uniquely} determined (see Appendix~\ref{app:Proof-I} for a proof). In the search for hidden  symmetries where $G$ is unknown, the set $\{b_{\lambda,\alpha}\}$ forms a diagnostic fingerprint of $G$ imprinted on $N$. Accordingly, determining these branching multiplicities serves as the foundational step of our reconstruction procedure. Once these multiplicities are established, the exact Hilbert space decompositions in Eq.~\eqref{eq:Hilbert-Space-Decomposition-Subset} can be extracted for all $\alpha\in\hat{G}$, providing both the total number of $G$-irreps and their respective dimensions.

\subsection{Fusion Algebra and Character Theory}
Alongside the branching multiplicities, the fusion rules (tensor-product structure) of the $G$-irreps impose a second, strong set of constraints on the hidden symmetry. 
While
the branching data describe how irreps of $G$ decompose upon restriction to the known subgroup $N$, 
the fusion rules describe how these irreps combine within the representation theory of $G$ under tensor product. 

For any two $G$-irreps $V_{\alpha_i}$ and $V_{\alpha_j}$, their fusion is defined by the tensor product decomposition
\be\label{eq:G_fusion_rules}
    V_{\alpha_i}\otimes V_{\alpha_j}\cong \bigoplus_{\alpha_k\in\hat{G}}N_{\alpha_i\alpha_j}^{\alpha_k}V_{\alpha_k},
\ee
where the non-negative integers $N_{\alpha_i\alpha_j}^{\alpha_k}$ are the fusion coefficients. These fusion coefficients are computed directly using the character inner product over $G$, as defined in Eqs.~\eqref{eq:Inner-Product-Group-Function-1} and \eqref{eq:Inner-Product-Group-Function-2}
\be\label{eq:Group-Verlinde-Formula}
\begin{split}
    N_{\alpha_i\alpha_j}^{\alpha_k}&=\langle \chi_{\alpha_i}\chi_{\alpha_j},\chi_{\alpha_k}\rangle_G,
\end{split}
\ee
which is the group-theoretic analogue of the Verlinde formula~\cite{DiFrancesco:1997nk}. A formal derivation of this relation is provided in Appendix~\ref{app-sec:Group-Verlinde-Formula}.

The goal of our bootstrap program, developed below,  
is to obtain the full branching and fusion data constrained by the manifest subgroup $N\subset G$ and the spectral input. 
Its outputs are  
\be\label{eq:Bootstrap-Target}
\begin{split}
    \text{Outputs}=\left\{
    \{b_{\lambda,\alpha}\},\hat{G},\{d_{\alpha}\},\{N_{\alpha_i\alpha_i}^{\alpha_k}\},\{\chi_{\alpha}\},{\cal C}_G\right\}.
\end{split}
\ee
These data are not completely independent:
the $G$-irreps and their dimensions $d_{\alpha}$ emerge as direct byproducts of the branching multiplicities, while the characters $\chi_{\alpha}$ and the set of conjugacy classes ${\cal C}_G$ are extracted from the fusion rules.

In certain restricted cases, fusion data uniquely determine the abstract group. For instance, the character table of a finite group can be directly extracted from its fusion matrices (see Appendix~\ref{app:Character-F-Matrices}), and for finite simple groups, the character table alone suffices to determine the group up to isomorphism~\cite{mueger2010tensorcategoriesselectiveguided}. The same holds for compact abelian groups~\cite{McMullen1984DualObject} and connected compact Lie groups~\cite{doi:10.1142/S0129167X93000054}. It should be noted that, however, the fusion data alone do not generally determine $G$ up to isomorphism: non-isomorphic groups, such as $D_4$ and $Q_8$, can share identical fusion rules. Consequently, relying on any single mathematical structure is generally inadequate for uniquely identifying a hidden symmetry group $G$. As briefly reviewed alongside category theory notions in Appendix~\ref{app:Category-theory}, Tannakian duality~\cite{DeligneMilneOgusShih1982} states that the category $\text{Rep}(G)$ of finite-dimensional complex representations of a finite group $G$ uniquely determines the group $G$, provided it is equipped with its full structure as a symmetric monoidal fusion category. 
Namely, 
to uniquely reconstruct $G$ from its representation data, 
one also needs the $F$- and $R$-symbols, and the fiber functor~\cite{DeligneMilneOgusShih1982} (also reviewed in Appendix~\ref{app:Category-theory}), which 
satisfy the strict consistency conditions of $\text{Rep}(G)$.

The known subgroup supplies information beyond the fusion rules. For example, the dihedral group $D_4$ of order eight contains $N\cong\mathbb{Z}_2\times\mathbb{Z}_2$, generated
by a half-turn and a reflection of the square.
The quaternion group $Q_8$ has only one element of order two
and cannot contain this subgroup. Thus, knowledge of the subgroup $\mathbb{Z}_2 \times \mathbb{Z}_2$
excludes $Q_8$ despite its identical fusion rules.
More generally, restriction preserves tensor products,
linking the branching and fusion coefficients through
consistency conditions. Together with compatibility of
the subgroup embedding, these conditions restrict the
candidate groups, although uniqueness is not guaranteed
in general.

At the core of our bootstrap program are the algebraic constraints satisfied by the fusion data:
The fusion coefficients are not arbitrary non-negative integers, but are constrained by basic properties of tensor products of group representations. 
In the following, we list the constraints used in our bootstrap.

First, tensor products of two finite-dimensional representations of $G$ are again representations of $G$, which decomposes into irreps with non-negative integer multiplicities. This ensures a closed fusion algebra over $\hat{G}$. Second, the inherent associativity of the tensor product imposes a constraint on the fusion coefficients, implying
\begin{equation}\label{eq:Fusion-Associativity}
    \sum_{\alpha_k\in\hat{G}}N_{\alpha_i\alpha_j}^{\alpha_k}N_{\alpha_k\alpha_l}^{\alpha_m}
    =
    \sum_{\alpha_k\in\hat{G}}N_{\alpha_j\alpha_l}^{\alpha_k}N_{\alpha_i\alpha_k}^{\alpha_m}.
\end{equation}
Third, the trivial representation, denoted by $\alpha_0$, serves as the identity element under fusion. Its tensor product with any $G$-irrep acts trivially, such that
\be\label{eq:Fusion-Unity}
    N_{\alpha_0\alpha_j}^{\alpha_k}=N_{\alpha_j\alpha_0}^{\alpha_k}=\delta_{\alpha_j,\alpha_k},
\ee
for all $\alpha_j,\alpha_k\in\hat{G}$. Fourth, for ordinary group representation theory, the tensor product is symmetric under interchange of the two lower indices, which gives
\be\label{eq:Fusion-Symmetric}
    N_{\alpha_i\alpha_j}^{\alpha_k}=N_{\alpha_j\alpha_i}^{\alpha_k}.
\ee
Finally, every finite-dimensional $G$-irrep $V_{\alpha}$ admits a dual (or contragredient) representation $V_{\overline{\alpha}} \equiv V_{\alpha}^*$, where the star stands for the dual vector space. More precisely, the dual representation is defined via the action
\be\label{eq:Fusion-Dual}
\begin{split}
    \left(\rho_{\overline{\alpha}}(g)f\right)(v)=f\left(\rho_{\alpha}(g^{-1})v\right),
\end{split}
\ee
for all $f \in V_{\alpha}^*$ and $v \in V_{\alpha}$. $\rho_{\alpha}$ and $\rho_{\overline{\alpha}}$ denote the representation operators acting on $V_{\alpha}$ and $V_{\overline{\alpha}}$ respectively. Consequently, the character of the dual representation satisfies $\chi_{\overline{\alpha}}(g) = \chi_{\alpha}(g^{-1})$ for arbitrary $G$-element $g$. As a direct consequence of Schur's lemma~\cite{serre1977linear}, the trivial representation $\alpha_0$ is contained exactly once within the tensor products $V_{\alpha}\otimes V_{\overline{\alpha}}$ and $V_{\overline{\alpha}}\otimes V_{\alpha}$. Therefore, the corresponding fusion coefficients are strictly unity:
\be\label{eq:Fusion-Dual-Unity}
    N_{\alpha\overline{\alpha}}^{\alpha_0}=N_{\overline{\alpha}\alpha}^{\alpha_0}=1,
    \quad \forall\alpha\in\hat{G}.
\ee

These structural properties tightly constrain the fusion algebra, adding a crucial second layer of information to our reconstruction procedure beyond branching.
In particular, the fusion coefficients are inextricably related to the branching multiplicities, because the restriction of representations from $G$ to $N$ inherently commutes with the tensor product. This fundamental compatibility establishes one of the central consistency conditions in our bootstrap procedure, playing a pivotal role in restricting the space of admissible bootstrap solutions.

A particularly powerful reformulation of the fusion rules becomes manifest when we encode the fusion coefficients into a set of fusion matrices. For each $G$-irrep $V_{\alpha}$, we define the corresponding fusion matrix via its matrix elements
\be\label{eq:Fusion-Matrix}
    (N_{\alpha})_{\alpha_j\alpha_k}=N_{\alpha\alpha_j}^{\alpha_k}.
\ee
Since the tensor product of representations is both associative and commutative, these fusion matrices form a mutually commuting family and can therefore be simultaneously diagonalized. This matrix representation proves especially advantageous for finite groups, where the simultaneous eigenvectors are explicitly determined by the character table of $G$.\footnote{For compact Lie groups, the same character-multiplication identity continues to hold, but the discrete character table is replaced by a continuous family of character functions on conjugacy classes.} A detailed proof of this spectral relationship is provided in Appendix~\ref{app-eq:Group-Verlinde-Formula}. Once the character table is established, elementary representation-theoretic quantities, such as orthogonality relations and conjugacy classes, can be systematically extracted. For our reconstruction framework, the character table therefore serves not only as an auxiliary output from the fusion matrices, but as a central diagnostic tool; it complements the branching and fusion data to identify the hidden symmetry.

\subsection{Fingerprints from Fusions\label{sec:Res-Fusion}}
The branching multiplicities are the direct result of restricting representations from $G$ to $N$. However, they are not the sole imprint left by the hidden symmetry. When we apply this restriction to the fusion rules of $G$, the $G$-fusion algebra leaves an equally constrained fingerprint on the representation theory of the subgroup $N$. 

As previously established, the restriction of representations from $G$ to $N$ commutes with the tensor product of simple objects. This property allows us to explicitly derive the fingerprint that the fusion rules of $G$ leave on the subgroup $N$. By combining the branching decomposition in Eq.~\eqref{eq:Branching-Decomposition} with the $G$-fusion rules in Eq.~\eqref{eq:G_fusion_rules}, we obtain the isomorphism
\begin{equation}
    \begin{split}
        &\bigoplus_{\lambda_a,\lambda_b,\lambda_c \in \hat{N}} N_{\lambda_a\lambda_b}^{\lambda_c}V_{\lambda_c} \otimes \mathbb{C}^{b_{\lambda_a,\alpha_i}\times b_{\lambda_b,\alpha_j}} \\
        \cong &\bigoplus_{\lambda_c \in \hat{N},\alpha_k \in \hat{G}} N_{\alpha_i\alpha_j}^{\alpha_k} V_{\lambda_c}\otimes \mathbb{C}^{b_{\lambda_c,\alpha_k}},
    \end{split}
\end{equation}
where $N_{\lambda_a\lambda_b}^{\lambda_c}$ denotes the known fusion coefficients of the subgroup $N$. In the $S_3$ example, the tensor product of two doublets
contains the two singlets and one doublet:
\begin{equation}
\label{eq:Setting-S3-Fusion}
    V_{\alpha_2}\otimes V_{\alpha_2}
    \cong V_{\alpha_0}\oplus V_{\alpha_1}\oplus V_{\alpha_2}.
\end{equation}
Thus,
$N_{\alpha_2\alpha_2}^{\alpha_0}
=N_{\alpha_2\alpha_2}^{\alpha_1}
=N_{\alpha_2\alpha_2}^{\alpha_2}=1$,
with dimensions satisfying $2\times2=1+1+2$.
Matching the dimensions of the multiplicity spaces for each $N$-irrep $V_{\lambda_c}$ directly yields the fundamental algebraic constraint (i.e., monoidality constraint)
\begin{equation}\label{eq:Monoidality_Condition}
    \begin{split}
      \sum_{\lambda_a,\lambda_b\in \hat{N}} &N_{\lambda_a\lambda_b}^{\lambda_c} b_{\lambda_a,\alpha_i} b_{\lambda_b,\alpha_j} \\
      = \sum_{{\alpha_k}\in \hat{G}} &N_{\alpha_i\alpha_j}^{\alpha_k} b_{\lambda_c,{\alpha_k}}. 
    \end{split}
\end{equation}
These subgroup fusion rules inherently satisfy associativity, admit a fusion identity (unity), exhibit commutativity (symmetry under the exchange of the two lower indices), and possess dual representations. Conceptually, this monoidality constraint elegantly captures the equivalence of two distinct operations. On the left-hand side, the objects are first restricted to $N$ and then fused within the set of $N$-irreps. On the right-hand side, the two $G$-irreps $V_{\alpha_i}$ and $V_{\alpha_j}$ are first fused within the representation theory of $G$, after which the entire tensor product is restricted to the subgroup $N$. 

As a concrete illustration, consider the restriction from $S_3$ to its
subgroup $\mathbb{Z}_3$. Let $V_\lambda$, with $\lambda=0,1,2$, denote
the three one-dimensional irreducible representations of $\mathbb{Z}_3$.
Their fusion rules are
\begin{equation}
    V_{\lambda_a}\otimes V_{\lambda_b}
    \cong V_{(\lambda_a+\lambda_b)\bmod 3}.
\end{equation}
Since the $S_3$ doublet restricts as
$\operatorname{Res}^{S_3}_{\mathbb{Z}_3}V_{\alpha_2}
\cong V_1\oplus V_2$, restricting first and then taking the tensor
product gives
\begin{equation}
\begin{aligned}
    \left(\operatorname{Res}^{S_3}_{\mathbb{Z}_3}V_{\alpha_2}\right)
    \otimes
    \left(\operatorname{Res}^{S_3}_{\mathbb{Z}_3}V_{\alpha_2}\right)
    &\cong (V_1\oplus V_2)\otimes(V_1\oplus V_2) \\
    &\cong V_0^{\oplus 2}\oplus V_1\oplus V_2.
\end{aligned}
\end{equation}
On the other hand, taking the tensor product in $S_3$ first, using
Eq.~\eqref{eq:Setting-S3-Fusion}, and then restricting gives
\begin{equation}
\begin{aligned}
    \operatorname{Res}^{S_3}_{\mathbb{Z}_3}
    \left(V_{\alpha_2}\otimes V_{\alpha_2}\right)
    &\cong
    \operatorname{Res}^{S_3}_{\mathbb{Z}_3}
    \left(
        V_{\alpha_0}\oplus V_{\alpha_1}\oplus V_{\alpha_2}
    \right) \\
    &\cong V_0^{\oplus 2}\oplus V_1\oplus V_2.
\end{aligned}
\end{equation}
The two routes therefore yield the same $\mathbb{Z}_3$ representation,
as required by monoidality. In particular, the multiplicities must agree
in every subgroup charge sector; equality of the total dimensions alone
is insufficient.

Identical branching data need not determine the hidden fusion rules.
Let $N\cong\mathbb{Z}_2$ be embedded either in $\mathbb{Z}_4$ or in
$\mathbb{Z}_2\times\mathbb{Z}_2$. In both cases, two irreps restrict to
the trivial $N$-irrep $V_+$ and two to the nontrivial $N$-irrep $V_-$,
giving the same branching matrix
\begin{equation}
    (b_{\lambda,\alpha})=
    \begin{pmatrix}
        1&1&0&0\\
        0&0&1&1
    \end{pmatrix}.
\end{equation}
At the level of $N\cong \mathbb{Z}_2$, one sees only the fusion rule
$V_-\otimes V_-\cong V_+$. This is compatible with both
$V_1\otimes V_1\cong V_2\neq V_0$ in $\mathbb{Z}_4$ and
$V\otimes V\cong\mathbf{1}$ in
$\mathbb{Z}_2\times\mathbb{Z}_2$, because both $V_2$ and $\mathbf{1}$
restrict to $V_+$. Thus, the tensor product structure visible within
$N\cong\mathbb{Z}_2$ cannot distinguish whether the hidden symmetry is
$\mathbb{Z}_4$ or $\mathbb{Z}_2\times\mathbb{Z}_2$.

It is instructive to examine the rigidity of the fusion rules under subgroup restriction. Recall that every $G$-irrep $V_{\alpha}$ admits a dual representation $V_{\overline{\alpha}}$. As duality is a contravariant operation that reverses the order of the tensor product, taking the dual (denoted by an overline) of both sides of Eq.~\eqref{eq:G_fusion_rules} yields
\be\label{eq:Fusion-Rigidity-Der-1}
    \overline{V_{\alpha_i}\otimes V_{\alpha_j}}\cong V_{\overline{\alpha}_j}\otimes V_{\overline{\alpha}_i}\cong \bigoplus_{\alpha_k\in\hat{G}}N_{\alpha_i\alpha_j}^{\alpha_k}V_{\overline{\alpha}_k}.
\ee
Alternatively, evaluating the fusion of these dual irreps directly, and exploiting the symmetry of the fusion coefficients under the interchange of the lower indices, gives
\be\label{eq:Fusion-Rigidity-Der-2}
    V_{\overline{\alpha}_j}\otimes V_{\overline{\alpha}_i}\cong \bigoplus_{\overline{\alpha}_k\in\hat{G}}N_{\overline{\alpha}_j\overline{\alpha}_i}^{\overline{\alpha}_k}V_{\overline{\alpha}_k}.
\ee
Equating these two decompositions immediately yields the first rigidity condition for the fusion rules
\be\label{eq:Fusion-Rigidity-1}
    N_{\alpha_i\alpha_j}^{\alpha_k}=N_{\overline{\alpha}_j\overline{\alpha}_i}^{\overline{\alpha}_k},
\ee
for all $\alpha_i,\alpha_j,\alpha_k\in\hat{G}$. Next, let $\alpha_k = \alpha_0$ be the trivial representation, which enforces the definition of rigidity~\cite{beer2018categoriesanyonstravelogue}, requiring
\be\label{eq:Fusion-Rigidity-2}
    N_{\alpha_i\alpha_j}^{\alpha_0}=\delta_{\alpha_j,\overline{\alpha}_i},
    \quad\forall\alpha_i,\alpha_j\in\hat{G}.
\ee
Furthermore, as taking the dual commutes with the restriction operator acting on $G$-irreps, we have
\be
    \text{Res}^G_N(V_{\overline{\alpha}})\cong \overline{\text{Res}^G_N(V_{\alpha})}.
\ee
Applying this relation to Eq.~\eqref{eq:Branching-Decomposition} yields the analogous rigidity condition for the branching multiplicities
\be\label{eq:Branching-Rigidity}
    b_{\overline{\lambda},\alpha}=b_{\lambda,\overline{\alpha}},
    \quad\forall\lambda\in\hat{N},
    \alpha\in\hat{G}.
\ee
Eqs.~\eqref{eq:Fusion-Rigidity-1}, \eqref{eq:Fusion-Rigidity-2}, and \eqref{eq:Branching-Rigidity} will play a pivotal role in the next section, providing post-hoc consistency conditions for our bootstrap procedure.

\section{Symmetry Bootstrap and Cross Spectral Form Factor\label{sec:Bootstrap}}
\subsection{Bootstrap Constraints from Fusion Rules\label{sec:Bootstrap-Fusion-Condition}}
Building upon the theoretical framework established in the previous section, we now provide details on the specific constraints governing our bootstrap procedure. Crucially, the inputs of the algorithm are limited to information derivable from the known symmetry subgroup $N$. In physical settings, this prior knowledge consists solely of the many-body Hamiltonian and the set of established symmetry generators, from which the complete representation theory of $N$ can be explicitly constructed.

By analyzing the fusion rules of $G$ and their restriction to the subgroup $N$, we establish the following algebraic constraints:
\begin{itemize}
    \item \textbf{Constraint I (Monoidality)}: The monoidality condition is defined in Eq.~\eqref{eq:Monoidality_Condition}.
    \item \textbf{Constraint II (Dimensionality and Non-negativity)}: The strict preservation of dimensions. This encompasses the branching dimension sum of Eq.~\eqref{eq:Dim-Condition-1}, as well as the dimensional consistency required by the fusion rules of $\mathrm{Rep}(G)$:
    \be\label{eq:Dim-Condition-2}
        d_{\alpha_i}d_{\alpha_j}=\sum_{\alpha_k\in\hat{G}}N_{\alpha_i\alpha_j}^{\alpha_k}d_{\alpha_k},
    \ee
    where $d_{\alpha_i}$, $d_{\alpha_j}$, and $d_{\alpha_k}$ denote the dimensions of the respective $G$-irreps $V_{\alpha_i}$, $V_{\alpha_j}$, and $V_{\alpha_k}$. Furthermore, the fusion coefficients are strictly bounded to be non-negative integers:
    \be
        N_{\alpha_i\alpha_j}^{\alpha_k}\in\mathbb{N}_+,
        \quad \forall\alpha_i,\alpha_j,\alpha_k\in\hat{G}.
    \ee
    \item \textbf{Constraint III (Commutativity and Associativity)}: The fusion coefficients must satisfy the commutativity and associativity conditions defined in Eqs.~\eqref{eq:Fusion-Symmetric} and \eqref{eq:Fusion-Associativity}, respectively.
    \item \textbf{Constraint IV (Unity)}: Denoting a trivial representation by $\alpha_0 \in \hat{G}$, the fusion rules are required to satisfy the unity condition of Eq.~\eqref{eq:Fusion-Unity}. Furthermore, the branching multiplicities must obey a unity constraint: the restriction of the trivial $G$-irrep must yield precisely the trivial $N$-irrep, $\lambda_0 \in \hat{N}$. This requires
    \be\label{eq:Fusion-Unity-b}
        b_{\lambda,\alpha_0}=\delta_{\lambda,\lambda_0},
        \quad \forall\lambda\in\hat{N}.
    \ee
    \item \textbf{Constraint V (Rigidity)}: The rigidity conditions govern the fusion rules via Eqs.~\eqref{eq:Fusion-Rigidity-1} and \eqref{eq:Fusion-Rigidity-2}, and restrict the branching multiplicities according to Eq.~\eqref{eq:Branching-Rigidity}.
\end{itemize}
While the above conditions primarily constrain the fusion rules, strong constraints on the branching multiplicities are so far missing. This is a critical bottleneck, as Constraints I, IV, and V explicitly require prior knowledge of these branching rules. 
To overcome the limitations of relying only on algebraic conditions defined on the subgroup $N$, we need extra physical input from the model at hand, as we discuss below.

\subsection{Cross Spectral Form Factor}
The SFF serves as a dynamical probe of energy-level correlations and is formally defined as the Fourier transform of the spectrum
\be\label{eq:Def-SFF}
    K(t) = \left\langle \sum_{i,j}e^{i(E_i-E_j)t}\right\rangle = \left\langle\tr[U(t)]\tr[U^{\dagger}(t)] \right\rangle,
\ee
where $U(t)$ is the unitary time-evolution operator. The angle brackets denote an appropriate average, specifically, an ensemble average for disordered systems, or a time average for any individual, non-random Hamiltonian. In the study of quantum chaos~\cite{Berry1985SemiclassicalSpectralRigidity,Sieber2001CorrelationsPeriodicOrbits,Muller2004SemiclassicalFoundationUniversality,Muller2005PeriodicOrbitTheoryUniversality}, the SFF stands as a crucial diagnostic tool; it explicitly encodes the crossover from system-specific, short-time decay to the universal ramp-and-plateau signatures characteristic of Random Matrix Theory (RMT). The two characteristic times are the Thouless time (decay to ramp) $t_{\text{Th}}$ and the Heisenberg time (ramp to plateau) $t_{\text{H}}$.
It has been instrumental in multiple frontiers
of physics, such as quantum many-body chaos~\cite{Chan2018MinimalModelManyBodyChaos,Kos2018ManyBodyQuantumChaos,Chan2018SpectralStatisticsSpatiallyExtended,Bertini2018ExactSpectralFormFactor}, black hole physics~\cite{Kobrin:2020xms,Miyaji:2025yvm,Kitaev2015SimpleModelQuantumHolography,GarciaGarcia2016SpectralThermodynamicSYK,Cotler2017BlackHolesRandomMatrices,Saad2018SemiclassicalRampSYKGravity}, and many-body localization~\cite{Basko2006MetalInsulatorTransition,Oganesyan2007LocalizationInteractingFermions,Suntajs2020QuantumChaosChallengesMBL,Prakash2021UniversalSFFMBL}. Notably, while integrable systems lack the characteristic RMT ramp, their long-time averaged plateau persists, as detailed in Appendix~\ref{app:Universal-Plateau}.

Surprisingly, in systems where the full symmetry is unresolved and only a subgroup $N$ is known, we can still extract numerical constraints on the branching multiplicities by examining the universal plateau of a subgroup-restricted SFF. We formulate this via the xSFF, denoted $K^{(N)}(t)$, which can be seen as a symmetric matrix generalization of the standard SFF. As it captures critical features of the branching data, the xSFF provides the essential constraints needed to drive our bootstrap framework. Labeled by pairs of $N$-irreps, the elements of the xSFF matrix are defined as
\be\label{eq:xSFF-Elements}
    K_{\lambda_a,\lambda_b}^{(N)}(t)=\f{\text{Re}\left\langle\tr\left[P_{\lambda_a}U(t)\right]\tr\left[P_{\lambda_b}U^{\dagger}(t)\right]\right\rangle}{d_{\lambda_a}d_{\lambda_b}},
\ee
where $P_{\lambda}$ projects onto the $N$-restricted subspaces defined in Eq.~\eqref{eq:Hilbert-Space-Decomposition-Subset}. At early times, this correlator can be complex valued and is non-universal. We instead focus on the late-time plateau, obtained after averaging on timescales of the order of the Heisenberg time and beyond. In this regime, similar to the standard SFF, the oscillatory contributions with non-zero energy differences tend to dephase in both integrable and chaotic systems, ensuring that only exactly degenerate contributions survive. This ensures that the final plateau is strictly real, which justifies taking the real part in our definition of the xSFF. Moreover, semiclassical analyses~\cite{Braun2011} demonstrate that cross-correlations between distinct, non-degenerate symmetry sectors vanish in the universal regime.\footnote{In the special case $N\cong\mathbb{Z}_2$, the off-diagonal xSFF
reduces to the cross component $S_{+-}(t)$ of the
symmetry-resolved SFF introduced in
Ref.~\cite{Delacretaz:2022ojg}, up to normalization
and averaging when the same spectral window is used.
There, the even and odd spectra at fixed momentum are treated
as independent, and their cross component is used to subtract
the smooth-density slope when the normalized sector densities
approximately agree. Here, only a known subgroup $N$ of an
unknown $G$ is resolved. Distinct $N$-sectors can share a
hidden $G$-irrep and hence common multiplicity-space spectra.
We use these cross-correlations to constrain branching
multiplicities and combine them with fusion consistency
to reconstruct hidden representation-theoretic data.}

The relation to the ordinary SFF follows from the completeness
relation $\sum_{\lambda\in\hat{N}}P_{\lambda}=I$.
Equations~\eqref{eq:Def-SFF} and \eqref{eq:xSFF-Elements} give
\begin{equation}
\label{eq:SFF-xSFF-relation}
\begin{aligned}
    K(t)
    &=
    \left\langle
    \left|
    \sum_{\lambda\in\hat{N}}\tr[P_{\lambda}U(t)]
    \right|^2
    \right\rangle \\
    &=
    \sum_{\lambda_a,\lambda_b\in\hat{N}}
    d_{\lambda_a}d_{\lambda_b}
    K_{\lambda_a,\lambda_b}^{(N)}(t),
\end{aligned}
\end{equation}
which holds exactly at all times with the same averaging
prescription. The ordinary SFF therefore contains cross-sector
correlations but does not retain their assignment to specific
pairs of $N$-irreps. The xSFF preserves this information:
its diagonal entries are the SFFs of the projected subspaces
divided by $d_{\lambda}^2$, and its off-diagonal entries
resolve correlations between distinct sectors, providing
additional constraints on the branching structure.

For finite $N$, once the $N$-resolved spectra are available,
both observables can be constructed from the same traces
$\tr[P_{\lambda}U(t)]$. Summing the eigenvalue phases to obtain
all these traces requires $\mathcal{O}(D_{\mathcal{H}})$
operations per time and Hamiltonian realization. Summing the
traces before taking the modulus squared gives the ordinary SFF.
Forming their pairwise products instead gives the xSFF matrix
with $\mathcal{O}(|\hat{N}|^2)$ additional operations and no
additional diagonalization.

According to the Hilbert space decomposition Eq.~\eqref{eq:Hilbert-Space-Decomposition-Subset}, the projector $P_{\lambda}$ and the unitary time-evolution operator $U(t)$ explicitly decompose as
\be\label{eq:Ut-Decomposition}
\begin{split}
    &P_{\lambda}=\bigoplus_{\alpha\in\hat{G}}P_{\lambda}^{(\alpha)}\otimes I|_{M_{\alpha}},\\
    &U(t)=\bigoplus_{\alpha\in\hat{G}}I|_{V_{\alpha}}\otimes U_{M_{\alpha}}(t),\\
\end{split}
\ee
where $P_{\lambda}^{(\alpha)}$ is the projector onto the subspace $V_{\lambda}\otimes \mathbb{C}^{b_{\lambda,\alpha}}$, satisfying $\mathrm{tr}(P_{\lambda}^{(\alpha)}) = b_{\lambda,\alpha}d_{\lambda}$. Here, the unitary operator $U_{M_{\alpha}}(t)$ acts non-trivially only within the multiplicity space $M_{\alpha}$ associated with the $G$-irrep $V_{\alpha}$. Employing the operator decompositions from Eq.~\eqref{eq:Ut-Decomposition}, the trace of the projected time-evolution operator evaluates to
\be\label{eq:Single-Tr-U}
\begin{split}
    \tr[P_{\lambda}U(t)]&=\sum_{\alpha}
    \mathrm{tr}
    \big[P_{\lambda}^{(\alpha)}\big]
    \tr\left[U_{M_{\alpha}}(t)\right]\\
    &=d_{\lambda}\sum_{\alpha}b_{\lambda,\alpha}\tr\left[U_{M_{\alpha}}(t)\right].
\end{split}
\ee
At late times, specifically within the universal plateau regime ($t > t_{\mathrm{H}}$), we assume that the trace correlators associated with distinct, non-degenerate $G$-irreps asymptotically decouple upon averaging. This allows us to write
\be\label{eq:Ind-LS-1}
\begin{split}
    &\left\langle\tr\left[U_{M_{\alpha_i}}(t)\right]\tr\left[U_{M_{\alpha_j}}^{\dagger}(t)\right]\right\rangle_{t>t_{\text{H}}}\approx\delta_{\alpha_i,\alpha_j}K_{\alpha_i},\\
\end{split}
\ee
where the non-zero diagonal terms take the plateau value
\be\label{eq:Ind-LS-2}
    K_{\alpha}\approx K_{\alpha}(t)|_{t>t_{\text{H}}}=\left\langle\left|\tr\left[U_{M_{\alpha}}(t)\right]\right|^2\right\rangle_{t \gg t_{\text{H}}}.
\ee
Here, $K_{\alpha}(t)$ denotes the standard SFF evaluated strictly within the multiplicity space $M_{\alpha}$, and $K_{\alpha}$ is its asymptotic plateau value. We employ approximate equalities here to reflect that, in numerical calculations, the late-time SFF typically exhibits residual fluctuations around the asymptotic mean $K_{\alpha}$. Note that Eq.~\eqref{eq:Ind-LS-1} explicitly confirms our earlier physical argument: the off-diagonal components evaluate to real values at late times. Assuming no degeneracies within $M_\alpha$ beyond possible
Kramers pairs~\cite{Altland:1997zz,Haake:2010fgh}, the
exact infinite-time plateau is
\begin{equation}
\label{eq:SFF-Plateau-Malpha}
    K_\alpha=\nu m_\alpha,\quad
    \nu=
    \begin{cases}
        1,&\text{nondegenerate spectrum},\\
        2,&\text{Kramers pairs only}.
    \end{cases}
\end{equation}
Time averaging retains precisely the equal-energy terms,
so each distinct energy contributes the square of its
degeneracy. This gives $m_\alpha$ for a nondegenerate
spectrum and $4(m_\alpha/2)=2m_\alpha$ for Kramers pairs.
The result is exact at finite $m_\alpha$ and independent
of integrability. Numerical estimates remain subject to
finite averaging time and sampling accuracy.

Before discussing how the xSFF plateaus constrain the branching multiplicities, we first introduce several key algebraic structures. We define the branching matrix $B$ as
\be\label{eq:Def-Branching-Matrix}
\begin{split}
    &B_{\lambda,\alpha}=(\vec{b}_{\alpha_0},\vec{b}_{\alpha_1},\cdots)=(\vec{b}_{\lambda_0},\vec{b}_{\lambda_1},\cdots)^T,
\end{split}
\ee
whose rows are indexed by $\lambda_a \in \hat{N}$ and columns by $\alpha_i \in \hat{G}$, with $a, i \in \mathbb{N}$. The branching vectors associated with the $G$-irreps and $N$-irreps are respectively defined as $\vec{b}_{\alpha_i} = (b_{\lambda_0,\alpha_i}, b_{\lambda_1,\alpha_i}, \dots)^T$ and $\vec{b}_{\lambda_a} = (b_{\lambda_a,\alpha_0}, b_{\lambda_a,\alpha_1}, \dots)^T$. Constructing this exact branching matrix $B$ is one of the target outputs of our bootstrap. Furthermore, we define the corresponding Gram matrix as
\be\label{eq:Def-Gram-Matrix}
    G_{\lambda_a\lambda_b}=\vec{b}_{\lambda_a}^T\vec{b}_{\lambda_b}=\sum_{\alpha\in\hat{G}}b_{\lambda_a,\alpha}b_{\lambda_b,\alpha},
\ee
whose properties will serve as input constraints for our algorithm.

We now examine how the late-time plateaus of the xSFF matrix, computed using ED, translate into explicit numerical constraints on the branching multiplicities. By substituting Eqs.~\eqref{eq:Ind-LS-1} and \eqref{eq:SFF-Plateau-Malpha} into the xSFF definition [Eq.~\eqref{eq:xSFF-Elements}], the asymptotic plateau heights evaluate to
\be\label{eq:xSFF-Elements-Plateau-Height}
\begin{split}
    &K_{\lambda_a,\lambda_b}^{(N)}=\left.K_{\lambda_a,\lambda_b}^{(N)}(t)\right|_{t>t_{\text{H}}}\approx\sum_{\alpha\in\hat{G}}b_{\lambda_a,\alpha}b_{\lambda_b,\alpha}K_{\alpha}.
\end{split}
\ee
Since the individual plateau heights $K_{\alpha}$ and the branching multiplicities $b_{\lambda,\alpha}$ are strictly non-negative, these xSFF plateaus are positive if and only if the corresponding Gram matrix elements [Eq.~\eqref{eq:Def-Gram-Matrix}] are positive. This yields our first explicit bootstrap constraint: if a non-vanishing plateau ($K_{\lambda_a,\lambda_b}^{(N)} \gg  0$)\footnote{
Here, $\gg0$ indicates a positive plateau resolved above
residual finite-time and sampling fluctuations.
This is a numerical resolution requirement, not a
thermodynamic-limit assumption.
} is numerically observed for a pair of $N$-irreps $\lambda_a, \lambda_b \in \hat{N}$, the associated Gram matrix element must be strictly positive ($G_{\lambda_a\lambda_b} > 0$); otherwise, it is identically zero.

Moreover, this logic directly yields a second, stronger constraint: if a distinct pair of $N$-irreps exhibits identical diagonal and off-diagonal plateau heights (specifically, $K_{\lambda_a,\lambda_b}^{(N)} = K_{\lambda_a,\lambda_a}^{(N)} = K_{\lambda_b,\lambda_b}^{(N)} \gg 0$), then their corresponding branching vectors must be identical, i.e., $\vec{b}_{\lambda_a} = \vec{b}_{\lambda_b}$\footnote{In practice, this constraint is enforced only when a complete numerical overlap is observed, meaning that the full temporal profiles of the curves, not just their averaged asymptotic plateau heights, must agree within numerical error. Therefore, we use the strict equal sign instead of the approximate sign.}. 
This can be rigorously formulated as follows:
\begin{itemize}
    \item \textbf{Statement I: }\textit{Given $m_{\alpha} > 0$ for all $\alpha\in\hat{G}$, the condition $K_{\lambda_a,\lambda_b}^{(N)} = K_{\lambda_a,\lambda_a}^{(N)} = K_{\lambda_b,\lambda_b}^{(N)}$ implies the identity $\vec{b}_{\lambda_a} = \vec{b}_{\lambda_b}$.}
\end{itemize}
A mathematical proof of this statement is provided in Appendix~\ref{app:Proof-II}.

The diagonal elements of the xSFF matrix, $K_{\lambda,\lambda}^{(N)}$ for $\lambda\in\hat{N}$, provide a third crucial constraint. By combining Eqs.~\eqref{eq:SFF-Plateau-Malpha} and \eqref{eq:xSFF-Elements-Plateau-Height}, the asymptotic plateau height for these diagonal components evaluates to 
\be
    K_{\lambda,\lambda}^{(N)}\approx\nu\sum_{\alpha\in\hat{G}}b_{\lambda,\alpha}^2m_{\alpha}.
\ee
Using Eq.~\eqref{eq:Ut-Decomposition}, we introduce a theoretical benchmark line for each $N$-irrep $\lambda\in\hat{N}$, defined as
\be\label{eq:Def-Benchmark-Line}
    R_{\lambda}=\f{\nu}{d_{\lambda}}\tr\left[P_{\lambda}\right]=\nu\sum_{\alpha}b_{\lambda,\alpha}m_{\alpha}.
\ee
If the numerically extracted plateau height equals the benchmark line, i.e., $K_{\lambda,\lambda}^{(N)}\approx R_{\lambda}$, we conclude that $b_{\lambda,\alpha} \in \{0,1\}$ for all $\alpha$. This directly follows from equating the two above equations,
\be
    \sum_{\alpha}b_{\lambda,\alpha}^2m_{\alpha}\overset{?}{=}\sum_{\alpha}b_{\lambda,\alpha}m_{\alpha}.
\ee

Consequently, the xSFF provides a robust set of numerical constraints that explicitly complement the five purely algebraic conditions stated in Sec.~\ref{sec:Bootstrap-Fusion-Condition}. We formalize this physical input as follows:
\begin{itemize}
    \item \textbf{Constraint VI (Numerical Constraints):} By extracting the late-time xSFF plateaus via ED, we impose three structural conditions on the branching data: (1) we determine which pairs of $N$-irreps yield non-zero Gram matrix elements; (2) we isolate identical branching vectors by observing degenerate plateau heights; and (3) we deduce which $N$-irreps are multiplicity-free ($b_{\lambda,\alpha} \in \{0,1\}$) by comparing the diagonal xSFF plateau values against their theoretical benchmark lines.
\end{itemize}
Integrating all six constraints allows us to simultaneously restrict the branching structure and the fusion rules, thereby establishing the full set of constraints required to apply our bootstrap approach and yield the dataset defined in Eq.~\eqref{eq:Bootstrap-Target}.

\subsection{Implementation of Bootstrap Algorithm\label{sec:Bootstrap-Algorithm}}
We introduce the algorithmic implementation in this subsection, and provide a visual summary of the computational workflow in Fig.~\ref{fig:workflow}. 

\begin{figure*}[t]
\centering
\begin{tikzpicture}[
    node distance=0.7cm and 1.5cm,
    >={Stealth[length=5pt]},
    process/.style={rectangle, draw, thick, rounded corners=3pt, fill=blue!6,
                    minimum width=5.8cm, minimum height=0.9cm, align=center, font=\small},
    io/.style={trapezium, trapezium left angle=70, trapezium right angle=110,
               draw, thick, fill=green!8, minimum width=4.2cm, minimum height=0.9cm,
               align=center, font=\small},
    decision/.style={diamond, draw, thick, fill=orange!10, aspect=2.8,
                     minimum width=2.2cm, inner sep=1pt, align=center, font=\small},
    label/.style={font=\scriptsize\bfseries, text=gray},
    arrow/.style={->, thick},
]

\node[io] (input) {Subgroup $N$ data \& xSFF plateau matrix $K^{(N)}_{\lambda_a,\lambda_b}$};

\node[process, below=of input] (rank) {Set trial rank $r = r_{\min}$};

\node[process, below=of rank] (step1) {Deduce allowable branching vector types\\[-2pt]
{\scriptsize (equivalence classes $\to$ quotient graph $\tilde{P}$ $\to$ cliques)}};

\node[process, below=of step1] (step2) {Assemble candidate branching matrices $B_{\lambda,\alpha}$\\[-2pt]
{\scriptsize (filter by xSFF overlap and multiplicity constraints)}};

\node[process, below=of step2] (step3) {Enumerate fusion decompositions\\[-2pt]
{\scriptsize (solve Eq.~\eqref{eq:Monoidality_Condition})}};

\node[process, below=of step3] (step4) {Backtrack over fusion set\\[-2pt]
{\scriptsize (enforce rigidity, associativity, commutativity)}};

\node[decision, below=0.9cm of step4] (dec) {Solution\\found?};

\node[process, below=0.9cm of dec] (step5) {Classify surviving solutions\\[-2pt]
{\scriptsize (dimensions $d_\alpha$, characters $\chi_\alpha$, conjugacy classes $c$)}};

\node[process, below=of step5, fill=purple!8, minimum width=7cm] (step6) {%
\begin{tabular}{c@{\hskip 1.5cm}c}
\textbf{Linear rep branch} & \textbf{Corep branch} \\[-2pt]
{\scriptsize $|G|=\sum d_\alpha^2=\sum|c|$} & {\scriptsize magnetic group $G^{\text{mag.}}=N^{\text{uni.}}\sqcup KN^{\text{uni.}}$}
\end{tabular}};

\node[io, below=of step6] (output) {Outputs: $\left\{
    \{b_{\lambda,\alpha}\},\hat{G},\{d_{\alpha}\},\{N_{\alpha_i\alpha_i}^{\alpha_k}\},\{\chi_{\alpha}\},{\cal C}_G\right\}$};

\node[process, right=1.8cm of dec, fill=red!8, minimum width=2.2cm] (incr) {$r \to r+1$};

\draw[arrow] (input) -- (rank);
\draw[arrow] (rank) -- (step1);
\draw[arrow] (step1) -- (step2);
\draw[arrow] (step2) -- (step3);
\draw[arrow] (step3) -- (step4);
\draw[arrow] (step4) -- (dec);
\draw[arrow] (dec) -- node[left, font=\small] {Yes ($r=r_*$)} (step5);
\draw[arrow] (dec) -- node[above, font=\small] {No} (incr);
\draw[arrow, rounded corners=8pt] (incr) |- (step1);
\draw[arrow] (step5) -- (step6);
\draw[arrow] (step6) -- (output);

\node[label, left=3pt of step1.north west, anchor=south east] {Step 1};
\node[label, left=3pt of step2.north west, anchor=south east] {Step 2};
\node[label, left=3pt of step3.north west, anchor=south east] {Step 3};
\node[label, left=3pt of step4.north west, anchor=south east] {Step 4};
\node[label, left=3pt of step5.north west, anchor=south east] {Step 5};
\node[label, left=3pt of step6.north west, anchor=south east] {Step 6};

\end{tikzpicture}
\caption{Computational workflow of the bootstrap algorithm. At each trial rank $r$, the algorithm deduces allowable branching vector types (Step~1), assembles candidate branching matrices (Step~2), enumerates fusion decompositions (Step~3), and backtracks over the full fusion set enforcing associativity and rigidity (Step~4). If no consistent solution is found, $r$ is incremented. Otherwise, the surviving solutions at the termination rank $r=r_*$ are classified by their dimensions and character tables (Step~5) and split into linear representation or corepresentation branches (Step~6).}
\label{fig:workflow}
\end{figure*}

The independent inputs to the algorithm are the following:
\begin{itemize}
    \item The algebraic data of the manifest subgroup $N$: specifically, the projectors onto its irreps $\lambda \in \hat{N}$ and the irrep dimensions $\dim V_{\lambda} = d_{\lambda}$.
    \item The late-time xSFF plateau matrix, composed of entries $K_{\lambda_a,\lambda_b}^{(N)}$, alongside the benchmark lines $R_{\lambda}$, where $\lambda, \lambda_a, \lambda_b \in \hat{N}$.
\end{itemize}
No prior branching or fusion data for $G$ are assumed.

When no nontrivial subgroup is available for symmetry resolution,
one can formally take $N=\{e\}$, where $e$ is the identity.
Its single one-dimensional irrep $\lambda_0$ gives
\begin{equation}
    P_{\lambda_0}=I,\quad
    b_{\lambda_0,\alpha}=d_{\alpha},\quad
    K_{\lambda_0,\lambda_0}^{(\{e\})}(t)=K(t).
\end{equation}
The xSFF then reduces to the ordinary SFF, and the monoidality
constraint in Eq.~\eqref{eq:Monoidality_Condition} reduces to
dimensional consistency, Eq.~\eqref{eq:Dim-Condition-2}.
The cross-sector constraints are therefore lost, and the remaining
data do not generally determine $G$. The search is performed rank by rank, where the trial rank $r$ denotes the number of irreps in the candidate hidden group $G$. In most of the cases, we focus on the solution with the minimal rank
\be
    r_*=\text{Min}\left\{r\in\mathbb{N}_+|\exists~\text{at least one solution at rank }r\right\},
\ee
which is highly likely to be the physically relevant one. The final output is the complete set of consistent solutions at this rank.

At a fixed $r$, the algorithm proceeds step by step as follows:
\begin{enumerate}
    \item \textbf{Deduce allowable branching vector types.} The hidden symmetry irreps $\alpha \in \hat{G}$ (where $\alpha = 0, 1, \dots, r-1$) are encoded by the branching vectors $\vec{b}_{\alpha}$ defined in Eq. \eqref{eq:Def-Branching-Matrix}. The xSFF plateaus impose numerical constraints on these vectors. First, we group the $N$-irreps into equivalence classes $\mathcal{C}_1, \ldots, \mathcal{C}_m$, where $\lambda_a$ and $\lambda_b$ belong to the same class whenever $K^{(N)}_{\lambda_a,\lambda_a} = K^{(N)}_{\lambda_a,\lambda_b} = K^{(N)}_{\lambda_b,\lambda_b}$, see Fig.~\ref{fig:column-type-enum}(a) Within each class, all irreps share identical branching multiplicities: $b_{\lambda_a,\alpha} = b_{\lambda_b,\alpha}$ for all $\alpha$.
    \begin{figure*}[t]
    \centering
    \begin{tikzpicture}[
    irrep/.style={circle, draw, thick, minimum size=8mm, inner sep=0pt, font=\small},
    qnode/.style={ellipse, draw, thick, fill=blue!8, minimum width=16mm, minimum height=9mm, inner sep=1pt, font=\small},
    >=stealth
    ]

    \node[font=\small\bfseries] at (1.5, 5.2) {(a)};

    \node[irrep] (l0) at (0, 3) {$\lambda_0$};
    \node[irrep, fill=red!12] (l1) at (1.5, 4.2) {$\lambda_a$};
    \node[irrep, fill=red!12] (l2) at (1.5, 1.8) {$\lambda_b$};
    \node[irrep] (l3) at (3, 3) {$\lambda_c$};

    \draw[dashed, thick, red!50, rounded corners=4pt] (0.7, 1.1) rectangle (2.3, 4.9);
    \node[red!60, font=\scriptsize] at (3.0, 4.6) {$\vec{b}_{\lambda_a}=\vec{b}_{\lambda_b}$};

    \draw[thick, green!50!black] (l0) -- (l1) node[midway, above left=-1pt, font=\tiny] {$K_{\lambda_0\lambda_a}^{(N)}\!\gg\!0$};
    \draw[thick, green!50!black] (l0) -- (l2) node[midway, below left=-1pt, font=\tiny] {$K_{\lambda_0\lambda_b}^{(N)}\!\gg\!0$};

    \draw[->, very thick, gray] (4.0, 3) -- (5.0, 3) node[midway, above, font=\scriptsize] {merge};

    \node[font=\small\bfseries] at (7.8, 5.2) {(b)};

    \node[qnode] (C1) at (5.8, 3) {$\mathcal{C}_1$};
    \node[qnode, fill=red!12] (C2) at (7.8, 3) {$\mathcal{C}_2$};
    \node[qnode] (C3) at (9.8, 3) {$\mathcal{C}_3$};

    \node[font=\tiny, below=2pt] at (5.8, 2.35) {$\{\lambda_0\}$};
    \node[font=\tiny, below=2pt] at (7.8, 2.35) {$\{\lambda_a,\lambda_b\}$};
    \node[font=\tiny, below=2pt] at (9.8, 2.35) {$\{\lambda_c\}$};

    \draw[thick, green!50!black] (C1) -- (C2);

    \draw[->, very thick, gray] (10.8, 3) -- (11.8, 3) node[midway, above, font=\scriptsize] {cliques};

    \node[font=\small\bfseries] at (14.5, 5.2) {(c)};

    \node[anchor=north, font=\small] at (14.5, 4.7) {
    \begin{tabular}{@{}l@{\;\;$\to$\;\;}l@{}}
    \textit{Clique} & \textit{Vector type} \\[4pt]
    $\{\mathcal{C}_1\}$ & $\vec{b}_\alpha = (1,0,0,0)^T$ \\[2pt]
    $\{\mathcal{C}_2\}$ & $\vec{b}_\alpha = (0,1,1,0)^T$ \\[2pt]
    $\{\mathcal{C}_3\}$ & $\vec{b}_\alpha = (0,0,0,1)^T$ \\[2pt]
    $\{\mathcal{C}_1, \mathcal{C}_2\}$ & $\vec{b}_\alpha = (1,1,1,0)^T$
    \end{tabular}
    };

    \end{tikzpicture}
    \caption{Branching vector type deduction illustrated for a system with four $N$-irreps $\hat{N}=\{\lambda_0, \lambda_a, \lambda_b, \lambda_c\}$, assuming $b_{\max} = 1$. \textbf{(a)}~Green edges connect irreps with positive cross-correlation $K^{(N)}_{\lambda_a,\lambda_b} \gg 0$. $\lambda_a$ and $\lambda_b$ merge into an equivalence class denoted in dashed box because $K_{\lambda_a,\lambda_a}^{(N)} = K_{\lambda_b,\lambda_b}^{(N)} = K_{\lambda_a,\lambda_b}^{(N)}$. \textbf{(b)}~The quotient graph $\tilde{P}$: equivalence classes become vertices, inheriting edges from the original co-occurrences. \textbf{(c)}~Each clique (fully connected subgraph) of $\tilde{P}$ yields a branching vector type. $G$-irreps in the same equivalence class receive the same branching multiplicity.
    }
    \label{fig:column-type-enum}
    \end{figure*}
    
    We then construct the quotient graph $\tilde{P}$, see Fig.~\ref{fig:column-type-enum}(b): each equivalence class $\mathcal{C}_i$ becomes a vertex, and an edge connects $\mathcal{C}_i$ to $\mathcal{C}_j$ ($i \neq j$) whenever $K^{(N)}_{\lambda_a,\lambda_b} \gg 0$ for some $\lambda_a \in \mathcal{C}_i$, $\lambda_b \in \mathcal{C}_j$. The condition that $K^{(N)}_{\lambda_a,\lambda_b} = 0$ for $\lambda_a \neq \lambda_b$ strictly prohibits the components $b_{\lambda_a,\alpha}$ and $b_{\lambda_b,\alpha}$ from being simultaneously non-zero within the same vector. As a result, the non-zero entries of any vector $\vec{b}_\alpha$ are restricted to irreps whose equivalence classes are mutually connected in $\tilde{P}$. Hence, the support of $\vec{b}_\alpha$ must correspond to a \textit{clique} (a fully connected subgraph) of $\tilde{P}$.

    We enumerate all cliques of $\tilde{P}$ using the Bron-Kerbosch algorithm~\cite{bron1973algorithm}. For each clique $\mathcal{Q}$ and every assignment of multiplicity values $v(\mathcal{C}_i) \in \{1, \dots, b_{\max}(\mathcal{C}_i)\}$ across the vertices $\mathcal{C}_i \in \mathcal{Q}$, we construct a corresponding column type. Specifically, we set $b_{\lambda,\alpha} = v(\mathcal{C}_i)$ if $\lambda \in \mathcal{C}_i \in \mathcal{Q}$, and $b_{\lambda,\alpha} = 0$ otherwise [see Fig.~\ref{fig:column-type-enum}(c)]. The multiplicity bound $b_{\max}(\mathcal{C}_i)$ for each equivalence class is restricted to one if $K^{(N)}_{\lambda,\lambda}/R_\lambda \approx 1$ for $\lambda \in \mathcal{C}_i$; otherwise, it is set to a predetermined cutoff $b_{\max}$. The union of all such constructed vectors types constitutes the complete set of allowable branching vectors.
    \item \textbf{Assemble candidate branching matrix.} The algorithm forms candidate branching matrices $B_{\lambda,\alpha}$ by combining $r-1$ allowable vectors with the fixed trivial vector $\vec{b}_{\alpha_0}=(1,0,\cdots,0)^T$. Each candidate matrix is then filtered by configuration-level xSFF constraints. Specifically, if $K^{(N)}_{\lambda_a,\lambda_b} > 0$ for $\lambda_a \neq \lambda_b$, there must exist at least one column $\alpha$ for which both $b_{\lambda_a,\alpha} > 0$ and $b_{\lambda_b,\alpha} > 0$. Furthermore, if $K^{(N)}_{\lambda,\lambda}/R_\lambda > 1$, at least one column $\alpha$ must satisfy $b_{\lambda,\alpha} \geq 2$.
    \item \textbf{Enumerate all decompositions. }For each candidate branching matrix and each unordered pair of unresolved $G$-irreps $(\alpha_i,\alpha_j)$, the algorithm computes the fusion target vector
    \be\label{eq:Visible-Fusion-Target}
        t^{(\alpha_i,\alpha_j)}_\lambda = \sum_{\lambda_a,\lambda_b \in \hat{N}} N^{\lambda}_{\lambda_a\lambda_b}  b_{\lambda_a,\alpha_i}  b_{\lambda_b,\alpha_j},
    \ee
    as shown in Eq.~\eqref{eq:Monoidality_Condition}. This is the $N$-irrep content that the unresolved product of $G$-irreps $\alpha_i$ and $\alpha_j$ must reproduce. The fusion coefficients $N^{\lambda}_{\lambda_a\lambda_b}$ are not separate inputs; they are intrinsically determined once the manifest subgroup $N$ is specified. Then, Eq.~\eqref{eq:Monoidality_Condition} requires 
    \be\label{eq:Column-Decomposition}
        t^{(\alpha_i,\alpha_j)}_\lambda = \sum_{\alpha_k \in \hat{G}} N^{\alpha_k}_{\alpha_i\alpha_j}b_{\lambda,\alpha_k},\quad N^{\alpha_k}_{\alpha_i\alpha_j} \in \mathbb{N}.
    \ee
    We enumerate all solutions of Eq.~\eqref{eq:Column-Decomposition} by iterating over $\alpha_k = \alpha_0, \alpha_i, \ldots, \alpha_{r-1}$ and trying each value of $N^{\alpha_k}_{\alpha_i\alpha_j}$ from $0$ up to the     bound
    \be\label{eq:Fusion-Upper-Bound}
    N^{\alpha_k}_{\alpha_i\alpha_j} \leq \min_{\{\lambda : b_{\lambda,\alpha_k} > 0\}} \left\lfloor \f{t^{(\alpha_i,\alpha_j)}_\lambda}{b_{\lambda,\alpha_k}} \right\rfloor,
    \ee
    since each non-zero component $b_{\lambda,\alpha_k}$ limits the number of copies of column $\vec{b}_{\alpha_k}$ that fit into the target. If there is no valid solution for the pair $(\alpha_i,\alpha_j)$, then the entire branching matrix is discarded.
    \item \textbf{Backtrack over unresolved fusion set.} We build the complete fusion set $\{N^{\alpha_k}_{\alpha_i\alpha_j}\}$ by assigning a decomposition to one pair $(\alpha_i,\alpha_j)$ at a time. Products involving the trivial representation $V_{\alpha_0}$ are fixed by unity condition: $N^{\alpha_k}_{\alpha_0\alpha_j} = \delta_{\alpha_j,\alpha_k}$. After each tentative assignment, two local consistency checks are applied immediately. The first is a partial rigidity check: if a newly assigned pair $(\alpha_i,\alpha_j)$ yields $N^{\alpha_0}_{\alpha_i\alpha_j} > 0$ but $\alpha_i$ (or $\alpha_j$) was already assigned a different dual partner by a previous entry, the candidate is immediately rejected. The second is an incremental associativity check on all triples whose products are already defined. Only complete associative fusion sets survive.
    \item \textbf{Classify the surviving fusion sets. } For each surviving solution, the algorithm computes the dimensions of unresolved $G$-irreps $d_{\alpha}$, tests rigidity condition, and constructs the fusion matrices $(N_{\alpha_i})_{\alpha_j\alpha_k}=N^{\alpha_k}_{\alpha_i\alpha_j}$. For rigid solutions, these fusion matrices are simultaneously diagonalized to reconstruct candidate character tables and conjugacy class information. The eigenvalue equations are given by
    \be\label{eq:Eigenvalue-Eq-Character}
        N_{\alpha}v_{c_i}=\chi_{\alpha}(c_i)v_{c_i},\quad\alpha\in\hat{G},
    \ee
    where $c_i$ denotes the conjugacy classes and $v_{c_i}$ are the common eigenvectors of all fusion matrices. 
    \item \textbf{Split into the linear and corepresentation branches.} The bootstrap algorithm accommodates both linear representations and Wigner's corepresentations (involving an unresolved anti-unitary symmetry)~\cite{wigner1959group,RevModPhys.40.359,rumynin2021realrepresentationsc2gradedgroups,Mock_2016} via two parallel branches. The former imposes no restriction on the choice of subgroups, whereas the latter requires an additional constraint: the input manifest subgroup must be the unitary subgroup. The bifurcation between the two branches occurs only at the final classification stage. 
    
    In the linear representation branch, the reconstructed irrep dimensions, characters, and class sizes are tested against the consistency conditions for an ordinary finite group; namely, $|G| =\sum_{\alpha \in \hat{G}} d_{\alpha}^2=\sum_{c} |c|$, where $|c|$ denotes the size of the conjugacy class $c$. If these criteria are satisfied, the solution is interpreted as a candidate linear representation theory for $G$.

    If the solution is rigid but fails the ordinary group test, it is passed to the corepresentation branch. Now, the unresolved sectors are interpreted as candidate irreducible corepresentations (ICRs) of a magnetic group~\cite{wigner1959group,RevModPhys.40.359,rumynin2021realrepresentationsc2gradedgroups,Mock_2016}
    \be
        G^{\text{mag.}}\cong N^{\text{uni.}}\sqcup KN^{\text{uni.}},
    \ee
    where $N^{\text{uni.}}$ is the unitary normal subgroup and $K$ denotes an unresolved anti-unitary symmetry. A brief overview of Wigner's corepresentation theory for magnetic groups is provided in Appendix~\ref{app:Corepresentation}. The algorithm searches over subsets $S$ of even-dimensional, non-trivial hidden symmetry sectors—treating them as candidate type (c) ICRs\footnote{The classification of ICR types is detailed presented in Appendix~\ref{app:Corepresentation}.}—and unpacks them to determine the order of the unitary subgroup
    \be
        |N^{\text{uni.}}|=\sum_{i=0}^{r-1}d_{\alpha_i}^2-\f{1}{2}\sum_{\alpha\in S}d_{\alpha}^2.
    \ee
    A subset $S$ is retained if and only if $|N^{\text{uni.}}| > 0$ and the result is compatible with the known order of the unitary subgroup $N^{\text{uni.}}$. If optional time-reversal pairing data for the $N$-irreps are available, they are imposed at this stage as supplementary constraints on the branching vectors. Among all admissible subsets, the algorithm keeps the most parsimonious magnetic-group interpretation, namely the one with minimal $|G^{\text{mag.}}|=2|N^{\text{uni.}}|$. At present, the corepresentation search algorithm requires two restrictive prerequisites: (1) the input subgroup must constitute the full unitary normal subgroup, and (2) the system must contain only a single hidden anti-unitary operator $K$. In contrast, the linear representation branch requires no such assumptions. Extending the algorithm to accommodate more general anti-unitary cases remains a key direction for future research.
    \item \textbf{Increase rank if no solution exists.} The previous steps are executed iteratively for $r\in\{r_{\text{min}},r_{\text{min}}+1,\cdots,r_{\text{max}}\}$, where $r_{\text{min}}$ and $r_{\text{max}}$ are predetermined cut-offs. If no consistent solution is found at rank $r$, the algorithm increases $r$ by one and repeats the above steps until the first consistent solution is found at rank $r=r_*$. The algorithm then systematically enumerates all consistent solutions at $r_*$ before terminating. The final output comprises the complete set of valid solutions at $r_*$, explicitly classified as either linear representations or corepresentations. Note that, in general, two distinct termination ranks $r_*$ may exist, corresponding respectively to the linear representation and corepresentation branches.
\end{enumerate}

The xSFF uses dynamics to extract kinematic symmetry data.
For a fixed exact finite internal symmetry $G$ and subgroup
embedding $N\hookrightarrow G$, the branching coefficients
$b_{\lambda,\alpha}$ and fusion coefficients
$N_{\alpha_i\alpha_j}^{\alpha_k}$ are independent of $L$.
Equation~\eqref{eq:Single-Tr-U} separates these data from the
size-dependent multiplicity spaces and spectra.
The required size is therefore determined by representation
content and the resolution of the spectral constraints, not
by convergence to thermodynamic behavior. Moderate sizes can
already supply the reconstruction data, provided the irreps
being reconstructed occur with $m_\alpha>0$ and their
contributions are resolved by averaging. Larger systems may improve spectral sampling or expose previously absent irreps, but no $L\to\infty$ extrapolation of the branching or fusion coefficients is required.

The computational cost separates into spectral acquisition
and algebraic reconstruction. Let
$D_\lambda=\tr P_\lambda=d_\lambda\dim M_\lambda$,
with $\sum_{\lambda\in\hat N}D_\lambda=D_{\mathcal H}$.
Once the subgroup bases and the projected
$\mathcal{O}(L)$ local terms are available, block assembly,
dense diagonalization, and xSFF evaluation require
\begin{equation}
\label{eq:Bootstrap-Spectral-Cost}
\begin{split}
    \mathcal{O}\Bigl(n_{\rm ens}\Bigl[
    &\sum_{\lambda\in\hat N}
    \left(D_\lambda^3+L D_\lambda^2\right)+n_t\left(D_{\mathcal H}+|\hat N|^2\right)
    \Bigr]\Bigr)
\end{split}
\end{equation}
arithmetic operations at fixed precision, where $n_{\rm ens}$
and $n_t$ count realizations and sampled times.
The time-dependent terms arise from summing eigenvalue
phases and forming sector-pair products.
All xSFF entries reuse the same spectra, and testing
additional symmetry candidates requires no further
diagonalization or Hamiltonian eigenvectors.
For local dimension $q$ and approximately equal Abelian
sectors, the dense ED cost is
$\mathcal{O}(n_{\rm ens}q^{3L}/|\hat N|^2)$.
Subgroup resolution thus reduces the diagonalization cost,
although full-spectrum ED remains exponential in $L$.

Our implementation reuses subgroup bases and projected
local terms across realizations. Retaining these dense
arrays requires
$\mathcal{O}(D_{\mathcal H}^2+
L\sum_\lambda D_\lambda^2)$ storage, besides temporary
workspaces and sampled traces.
Dense projector diagonalization adds a one-time
$\mathcal{O}(|\hat N|D_{\mathcal H}^3)$ cost, while
projecting sparse local terms through dense bases costs
$\mathcal{O}(L D_{\mathcal H}\sum_\lambda D_\lambda^2)$.
Where the known symmetry action permits, charge- or
orbit-based bases and direct sector assembly avoid
these full-space transformations.
For non-Abelian $N$, a fully symmetry-adapted basis
also permits diagonalization on $M_\lambda$, of dimension
$D_\lambda/d_\lambda$, with the known factor $d_\lambda$
restored in the trace.

Once the numerical constraints are fixed, the algebraic
search depends on the subgroup data, trial rank $r$, and
multiplicity cutoffs, with no explicit dependence on
$D_{\mathcal H}$. If $n_{\rm col}$ denotes the number of
allowed branching column types, then
\begin{equation}
\label{eq:Bootstrap-Column-Count}
    n_{\rm col}
    \leq
    \prod_{i=1}^{m}
    \left[1+b_{\max}(\mathcal C_i)\right]-1.
\end{equation}
Each equivalence class is either absent or assigned an
allowed positive multiplicity, with the zero column
excluded. The clique condition further restricts the
supports. Fixing the trivial column leaves
$\binom{n_{\rm col}+r-2}{r-1}$ column multisets before
the configuration-level constraints.
Fusion reconstruction then uses the bounded integer
decompositions in Eq.~\eqref{eq:Column-Decomposition},
rejecting inconsistent partial tables through rigidity
and associativity.\footnote{
For a fixed branching matrix, let $p$ be the largest
number of admissible decompositions of a nontrivial
unordered product. There are at most
$p^{r(r-1)/2}$ complete fusion assignments before pruning.
A direct associativity scan costs $\mathcal{O}(r^5)$,
from the four external indices and one summed index
in Eq.~\eqref{eq:Fusion-Associativity}.}
Thus, the spectral constraints restrict the search
before complete fusion tables are constructed.
The remaining search is combinatorial, and exhaustiveness
within the specified cutoffs requires that computational
stopping caps are not reached.

This separation converts finite-system spectral data into
representation-theoretic information without constructing
the hidden symmetry generators on the many-body Hilbert
space. Error-controlled trace estimation, reuse of repeated
integer decompositions, and updates restricted to affected
associativity relations provide concrete directions for
extending the accessible system sizes and symmetry ranks.

In the following sections, we apply the bootstrap algorithm to several quantum many-body lattice models that exhibit hidden symmetries. We obtain solutions at various termination ranks $r_*$, demonstrating that the true $G$ can be unambiguously identified in these examples using the dataset defined in Eq.~\eqref{eq:Bootstrap-Target}.

\section{Example I: $S_3$ Symmetry\label{sec:S3-Inv-OF-Chain}}
\subsection{Hamiltonian and Symmetries}
We first illustrate the effectiveness of our bootstrap procedure by considering the O'Brien-Fendley model with $L$ sites and open boundary conditions (OBC), whose Hamiltonian is given by~\cite{OBrien:2019fvt}
\be\label{eq:OF-Hamiltonian}
\begin{split}
    H_{\theta}=&\cos(\theta)H_P+\sin(\theta)H_1,
    \quad \theta\in[0,2\pi).\\
\end{split}
\ee
Here, $H_P$ and $H_1$ are defined respectively as
\be
    H_P=-\sum_{j=1}^{L-1}J_j\left(Z_j^{\dagger}Z_{j+1}+\text{h.c.}\right)-\sum_{j=1}^L h_j(X_j+X_j^{\dagger}),
\ee
and
\be
\begin{split}
    H_1=&\sum_{j=1}^{L-1} J_j\left(3S_j^+S_{j+1}^--3{S_j^{+^2}}{S_{j+1}^{-^2}}+\text{h.c.}\right)\\
    +&\sum_{j=1}^L h_j (X_j+X_j^{\dagger}),
\end{split}
\ee
where ``h.c.'' denotes Hermitian conjugation. The couplings $\{J_j,h_j\}$ can be either fixed as constants or taken as random variables. Throughout this work, unless stated otherwise, we introduce randomness strictly to suppress finite-size oscillations and reveal clean late-time plateaus. However, the underlying bootstrap method applies equally well to translation invariant systems, irrespective of whether the system is integrable or not. To illustrate this, we provide an integrable example with fixed parameters in Appendix~\ref{app:non-random-FH-Model}. The local operators are defined (in the diagonal basis of $Z$ operator) by
\be\label{eq:HW-Z3-Generators}
\begin{split}
    &Z_j=\sum_{q=0}^{m-1}\omega^q\ket{q}\bra{q},
    \quad X_j=\sum_{q=0}^{m-1}\ket{q+1}\bra{q},\\
    &S_j^+=\f{1}{3}\left(2-\omega X_j-\omega^2X_j^{\dagger}\right)Z_j^{\dagger},
    \quad S_j^-=(S_j^+)^{\dagger},
\end{split}
\ee
where $\omega=e^{i\f{2\pi}{m}}$ and $m=3$ throughout this section. Locally, the operators $X_j,Z_j$ satisfy the Heisenberg-Weyl algebra, $Z_jX_j=\omega X_jZ_j$. 

This model has both spatial and internal symmetries. The system has a site-reflection $\mathbb{Z}_2$ symmetry $\mathcal{P}: j \mapsto L+1-j$, which commutes with all internal degrees of freedom. The reality of $H_{\theta}$ implies an anti-unitary time-reversal $\mathbb{Z}_2$ symmetry $\mathcal{T}$. In addition, the model has a unitary internal
$S_3\cong \mathbb{Z}_3\rtimes\mathbb{Z}_2$ symmetry generated by the global $\mathbb{Z}_3$ operator
\be\label{eq:X-Z3-Symmetry-Generator}
    X=\prod_{j=1}^LX_j,
    \quad X^3=I,
\ee
together with a charge conjugation operator $C$ satisfying $CXC=X^{-1}$ and
\begin{equation}\label{eq:S3-Charge-Conjugation}
    C=C^{\dagger}=\prod_{j=1}^LC_j,
    \quad 
    C_j=\left(\begin{array}{ccc}
        1 & 0 & 0 \\
        0 & 0 & 1 \\
        0 & 1 & 0 \\
    \end{array}\right),
    \quad C^2=1.
\end{equation}
The operators $X$ and $C$ both commute with $\mathcal{P}$ and $\mathcal{T}$, and together they generate the $S_3$ group:
\be 
    S_3\cong\langle X,C|X^3=C^2=1,~CXC=X^{-1}\rangle.
\ee

For our purposes, we focus on the $S_3$ internal symmetry. Although we have explicitly constructed all the symmetry operators above, we now proceed under the assumption that only the normal subgroup $\mathbb{Z}_3=\langle X|X^3=1\rangle$ is known. We then investigate whether the bootstrap procedure can successfully identify the larger $G\cong S_3$ group structure. As noted above, this model is a minimal example designed to demonstrate the bootstrap.

\subsection{Bootstrap Results}
Here, we apply our bootstrap algorithm to infer the branching matrix between the known subgroup $\mathbb{Z}_3$ and a hidden symmetry group $G$, as well as the fusion ring and character table of $G$.
\begin{figure*}[t]
    \centering
    \includegraphics[width=0.8\linewidth]{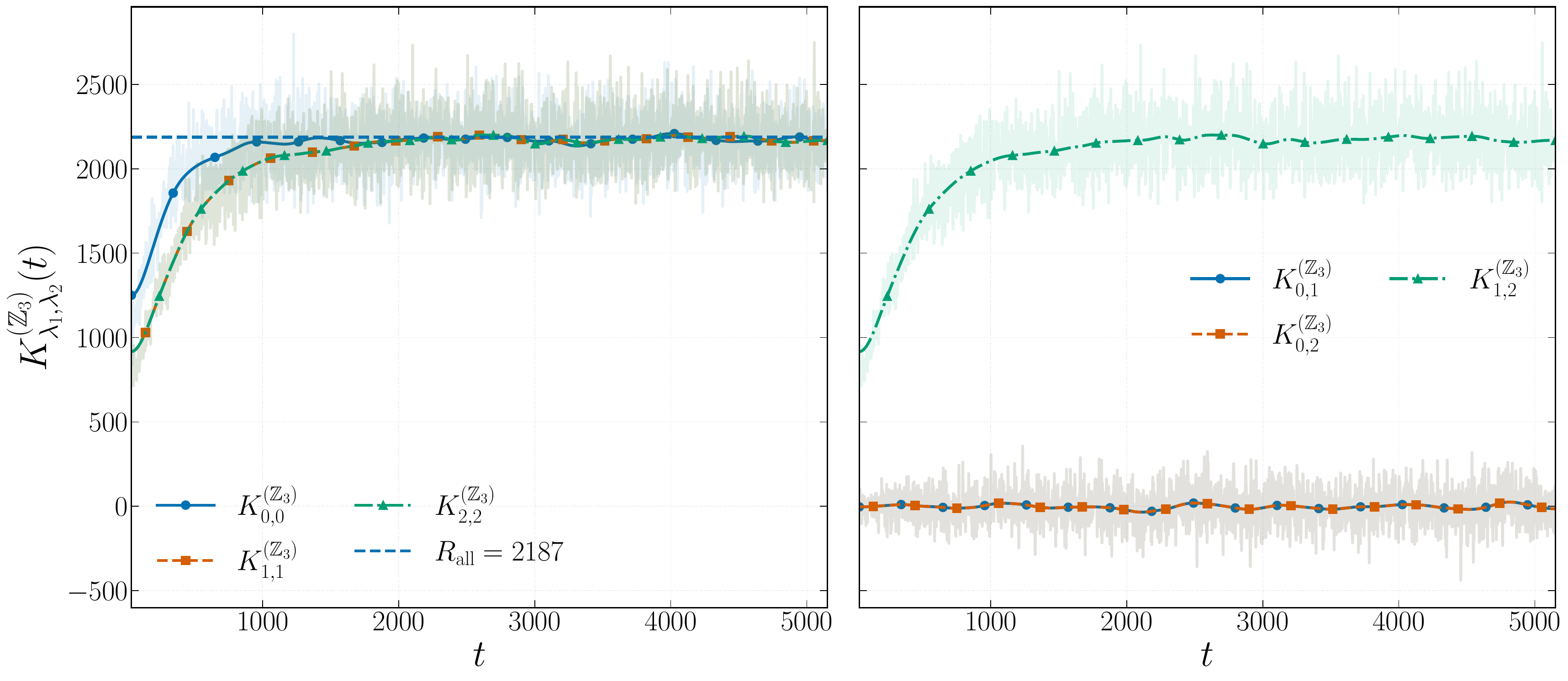}
    \caption{The xSFF for the $S_3$-invariant O'Brien-Fendley model with $L=8$ sites, projected onto the $\mathbb{Z}_3$ irreps. The y-axis plots the xSFF defined in Eq.~\eqref{eq:xSFF-Elements}. The data are averaged over 200 disorder realizations, where the bond couplings $J_j\sim\mathcal{N}(1,0.25^2)$ and the field couplings $h_j\sim\mathcal{N}(1,0.25^2)$ are independent Gaussian random variables. To clearly resolve the plateau heights, the xSFF data throughout this work are processed using auto-scale Gaussian smoothing. Solid colored lines denote the smoothed xSFF, while the underlying shaded curves represent the raw data. \textbf{Left}: The diagonal elements, whose late-time plateaus align perfectly with the corresponding benchmark lines $R_{\lambda}=R_{\text{all}}$ for all $\lambda\in\hat{\mathbb{Z}}_3$. \textbf{Right}: The off-diagonal elements of the xSFF. In particular, $K_{1,1}^{(\mathbb{Z}_3)} = K_{1,2}^{(\mathbb{Z}_3)} = K_{2,2}^{(\mathbb{Z}_3)}$, implying that the charge sectors $\mathcal{H}_1$ and $\mathcal{H}_2$ are degenerate.}
    \label{fig:S3_Chain_xSFF}
\end{figure*}

The projector onto the $\mathbb{Z}_3$-irreps labeled by $\lambda$ is explicitly given by
\be\label{eq:Z3-Projector}
    P_{\lambda}=\f{1}{3}\sum_{q=0}^2\omega^{-q\lambda}X^q,
    \quad
    \lambda\in\{0,1,2\},
\ee
and projects $\mathcal{H}$ into the charge sector $\mathcal{H}_{\lambda}$. We note that all $\mathbb{Z}_3$-irreps are one-dimensional, i.e.,  $d_{\lambda}=1$ for all three $\lambda$.

In Fig.~\ref{fig:S3_Chain_xSFF}, we show numerical results for the three diagonal elements alongside their respective benchmark lines $R_{\lambda}$, as well as the three independent off-diagonal elements \footnote{The numerical simulation data and plotting scripts are 
available at \url{https://github.com/ZihanZhou26/xSFF_data}.}. We find that the three diagonal components of the xSFF align with the benchmark values, which indicates that $b_{\lambda,\alpha} \in \{0,1\}$ for all $\lambda$ and $\alpha$. Furthermore, the numerical results in Fig.~\ref{fig:S3_Chain_xSFF} suggest the equal plateau values $K_{1,1}^{(\mathbb{Z}_3)} = K_{2,2}^{(\mathbb{Z}_3)} = K_{1,2}^{(\mathbb{Z}_3)}$. This implies that $\vec{b}_{1} = \vec{b}_{2}$. In summary, the xSFF analysis yields the following numerical constraints on the branching multiplicities:
\be\label{eq:S3_Chain_b_Condition}
\begin{split}
    &b_{\lambda,\alpha}\in\{0,1\},
    \quad\forall\lambda\in\hat{N},
     \alpha\in\hat{G},\\
    &G_{11}=G_{22}=G_{12}>0,
    \quad\vec{b}_{1}=\vec{b}_2\neq \vec{b}_0.
\end{split}
\ee

We subsequently bootstrap the complete branching matrix, the fusion rules and the character table of the hidden group $G$ by imposing the six constraints in Sec.~\ref{sec:Bootstrap}. Substituting the numerical data of the xSFF into the bootstrap algorithm at rank $r_*=3$, we identify a single candidate describing a linear representation for $G$. A separate solution found at $r_*=2$, which corresponds to a corepresentation theory including a possible unresolved anti-unitary charge conjugation, is analyzed in detail in Appendix~\ref{app:Other-S3-Corep-Solution}. However, as shown in Eq.~\eqref{eq:S3-Charge-Conjugation}, the actual charge conjugation $C$ is represented by a unitary operator. We therefore focus on the $r_*=3$ solution; an example involving a non-trivial anti-unitary symmetry will be discussed in Sec.~\ref{sec:3-QTC}. 

At $r_*=3$, the bootstrap yields the branching matrix
\be\label{eq:S3-Branching-Matrix}
    B_{\lambda,\alpha}=\begin{pmatrix}
        1 & 1 & 0\\
        0 & 0 & 1\\
        0 & 0 & 1    
    \end{pmatrix},
\ee
whose rows and columns are indexed by $\lambda\in\{\lambda_0,\lambda_1,\lambda_2\}$ and $\alpha\in\{0,1,2\}$, respectively. Expressed in the language of restrictive representations, these matrix elements are explicitly given by
\be
\begin{split}
    &\text{Res}^{G}_{\mathbb{Z}_3}V_{\alpha_0}=V_{0},\\
    &\text{Res}^{G}_{\mathbb{Z}_3}V_{\alpha_1}=V_{0},\\
    &\text{Res}^{G}_{\mathbb{Z}_3}V_{\alpha_2}=V_{1}\oplus V_{2}.
\end{split}
\ee
These restrictions clearly satisfy the necessary dimension constraints established in Eq.~\eqref{eq:Dim-Condition-1}. Furthermore, this branching structure uniquely determines the dimensions of the three $G$-irreps: we deduce two one-dimensional representations ($\dim V_{\alpha_0}=\dim V_{\alpha_1} = 1$) and one two-dimensional representation ($\dim V_{\alpha_2} = 2$).

The resulting fusion ring is obtained from the fusion coefficients
\be\label{eq:S3-Fusion-Coefficient-Alpha}
\begin{split}
    (\mathrm{a})~&N_{\alpha_0\alpha}^{\alpha}
    =N_{\alpha\alpha_0}^{\alpha}=1,
    \quad  \forall \alpha\in\hat G,\\
    (\mathrm{b})~&N_{\alpha_1\alpha_1}^{\alpha_0}=1,
    \quad  
    N_{\alpha_1\alpha_2}^{\alpha_2}
    =N_{\alpha_2\alpha_1}^{\alpha_2}=1,\\
    (\mathrm{c})~&N_{\alpha_2\alpha_2}^{\alpha_0}
    =N_{\alpha_2\alpha_2}^{\alpha_1}
    =N_{\alpha_2\alpha_2}^{\alpha_2}=1,
\end{split}
\ee
where the fusion rules are categorized as follows: (a) the unit rules, which establish that the trivial representation $V_{\alpha_0}$ acts trivially under fusion; (b) the fusion of the non-trivial one-dimensional irrep $V_{\alpha_1}$, encompassing both its self-fusion and its action on the two-dimensional irrep $\alpha_2$; and (c) the self-fusion of the two-dimensional object $V_{\alpha_2}$, which decomposes into the complete set of $G$-irreps. All other unspecified fusion coefficients are strictly zero. 

Once the fusion coefficients are determined, we can construct the fusion matrices $(N_{\alpha_i})_{\alpha_j\alpha_k} = N^{\alpha_k}_{\alpha_i\alpha_j}$. By solving the eigenvalue problems for these matrices within our bootstrap process, we obtain the characters of the group $G$ as the eigenvalues of fusion matrices, which are summarized in Table~\ref{tab:S3-Character-Table}.
\begin{table}
    \centering
    \begin{tabular}{c|ccc}
        & $c_1$ & $c_{2}$ & $c_{3}$\\
         \hline
        $\chi_{\alpha_0}$ & 1 & 1 & 1\\
        $\chi_{\alpha_1}$ & 1 & -1 & 1\\
        $\chi_{\alpha_2}$ & 2 & 0 & -1\\
        \hline
    \end{tabular}
    \caption{Character table of the hidden symmetry group for the O'Brien-Fendley model, derived using the bootstrap program. The conjugacy classes of $S_3$ are given by $c_1=\{1\}$, $c_2=\{C,XC,X^2C\}$, and $c_3=\{X,X^2\}$.}
    \label{tab:S3-Character-Table}
\end{table}
Consequently, the branching multiplicities, fusion ring, and character table shown in Eqs.~\eqref{eq:S3-Branching-Matrix}-\eqref{eq:S3-Fusion-Coefficient-Alpha} indicate that $G \cong S_3$.

\subsection{Analytical Confirmation}
In this section, as an illustrative example, we verify in detail that the candidate linear representation theory identified at $r_*=3$ precisely matches that of the group $G \cong S_3$. Analytical confirmations for all subsequent models are deferred to Appendix~\ref{app:Bootstrap-Solutions}.

Let $S_3=\left\langle X,C \right|X^3=C^2=1,CXC=X^{-1}\left. \right\rangle$.
Over $\mathbb{C}$, the normal subgroup $\mathbb{Z}_3$ has three one-dimensional irreps with characters:
\be\label{eq:Z3-Characters}
\begin{split}
    &\chi_{\lambda}(X^k)=\omega^{\lambda k}=e^{i\f{2\lambda k\pi}{3}},
    \quad \lambda=0,1,2.
\end{split}
\ee
The group $S_3$ has three conjugacy classes: $c_1=\{1\}$, $c_2=\{C,XC,X^2C\}$, and $c_3=\{X,X^2\}$. These correspond to two one-dimensional irreps ($\alpha_0, \alpha_1$) and one two-dimensional irrep ($\alpha_2$). The characters for the one-dimensional irreps are the trivial and alternating ones:
\be\label{eq:S3-Character-1-Dim}
\begin{split}
    &\alpha_0;~\chi_{\alpha_0}(c_1)=\chi_{\alpha_0}(c_2)=\chi_{\alpha_0}(c_3)=1,\\
    &\alpha_1;~\chi_{\alpha_1}(c_1)=-\chi_{\alpha_1}(c_2)=\chi_{\alpha_1}(c_3)=1.\\
\end{split}
\ee
The irrep $V_{\alpha_2}$ corresponds to the standard geometric realization of $S_3 \cong D_3$ in $O(2)$, where $X$ represents a $2\pi/3$ rotation and $C$ denotes reflection across the real axis:
\be
\begin{split}
    &\rho_{\alpha_2}(X)=\left(\begin{array}{cc}
       \cos\f{2\pi}{3}  & -\sin\f{2\pi}{3} \\
       \sin\f{2\pi}{3}  & \cos\f{2\pi}{3}
    \end{array}\right),
    \quad \rho_{\alpha_2}(C)=\left(\begin{array}{cc}
       1  & 0 \\
       0  & -1
    \end{array}\right).
\end{split}
\ee
Taking the trace yields the characters for $\alpha_2$:
\be\label{eq:S3-Character-2-Dim}
    \chi_{\alpha_2}(c_1)=2,
    \quad \chi_{\alpha_2}(c_2)=0,
    \quad \chi_{\alpha_2}(c_3)=-1.
\ee
These results confirm the $S_3$ character table presented in Table~\ref{tab:S3-Character-Table}. Applying Eq.~\eqref{eq:Branching-Multiplicity} alongside the $\mathbb{Z}_3$ characters yields the branching matrix shown in Eq.~\eqref{eq:S3-Branching-Matrix}.

Finally, we calculate the fusion coefficients to verify Eq.~\eqref{eq:S3-Fusion-Coefficient-Alpha}. These coefficients can be obtained from the inner product of the $G$-irrep characters [Eq.~\eqref{eq:Group-Verlinde-Formula}], which recover the fusion rules in Eq.~\eqref{eq:S3-Fusion-Coefficient-Alpha} and coincide with the fusion rules of $\mathrm{Rep}(S_3)$. A detailed analysis of $\mathrm{Rep}(S_3)$ can be found in~\cite{Perez-Lona:2023djo}.

\section{Example II: Non-local Hidden Symmetry\label{sec:KT-D4-MODEL}}
\subsection{Hamiltonian and Symmetries\label{sec:KT-D4-MODEL-HS}}
Having demonstrated our method in the simple illustrative case in the previous section, we now introduce a more complicated system, the Kennedy-Tasaki (KT) transformed spin-$1$ chain. 
In this example, the hidden symmetry is highly non-trivial to identify: it is encoded in a non-local unitary KT-transformation~\cite{Kennedy:1992ifl}, even though the Hamiltonian itself consists entirely of local operators.

Let us consider the following spin-$1$ Hamiltonian with OBC:
\begin{equation}
    \begin{aligned}\label{eq:KT_dual_model}
    \tilde{H} &= \sum_{i=1}^{L-1}J_i\left[- S_i^x S_{i+1}^x+S_i^y e^{i \pi S_i^z} e^{i \pi S_{i+1}^x}  S_{i+1}^y\right] \\
    & -\Delta_i S_i^z S_{i+1}^z+ \sum_{i=1}^L D_i\left(S_i^z\right)^2\\
    &+\sum_{i=1}^{L-1} 2g_i\left[A_i A_{i+1}-e^{i \pi S_i^z} B_i B_{i+1}\right]
    \end{aligned}
\end{equation}
where we defined
\begin{equation}
    A_j \equiv \left(S_j^x\right)^2-\left(S_j^y\right)^2, \quad B_j \equiv S_j^x S_j^y+S_j^y S_j^x .
\end{equation}
Here, $J_i,\Delta_i,D_i$ and $g_i$ are random couplings. The on-site operators $S^x, S^y$ and $S^z$ are in the spin-$1$ representation. In the eigenbasis of operator $S^z$, they can be written as
\begin{equation}
\begin{aligned}
    S^x & = \frac{1}{\sqrt{2}} 
    \begin{pmatrix}
        0 & 1 & 0 \\
        1 & 0 & 1 \\
        0 & 1 & 0
    \end{pmatrix}
    , \\
    S^y & = \frac{1}{\sqrt{2}}\begin{pmatrix} 0 & -i & 0 \\ i & 0 & -i \\ 0 & i & 0 \end{pmatrix}
    , \\
    S^z & =
    \begin{pmatrix}
        1 & 0 & 0 \\
        0 & 0 & 0\\
        0 & 0 & -1
    \end{pmatrix},
\end{aligned}
\end{equation}
with the algebraic relations
$[S^i, S^j]=i \epsilon^{ijk} S^k$,
where $i,j,k\in\{x,y,z\}$, and $\epsilon^{ijk}$ is the anti-symmetric tensor.

Due to its random and anisotropic couplings, the Hamiltonian lacks crystalline symmetry but remains time-reversal symmetric. The $\mathbb{Z}_2$ time-reversal symmetry is implemented by the anti-unitary operator $\mathcal{T}=U_{\pi}^yK$, where $U_{\pi}^y=\prod_{i=1}^L e^{i \pi S_i^y}$ and $K$ denotes complex conjugation. In addition, the Hamiltonian has the on-site unitary symmetries 
\begin{equation}\label{eq:KT-Dual-Model-Subgroup}
    U_{\pi}^x = \prod_{i=1}^Le^{i \pi S_i^x}, \quad U_{\pi}^z = \prod_{i=1}^Le^{i \pi S_i^z},
\end{equation} 
which commute with each other to generate the Klein four-group $V_4\cong\mathbb{Z}_2\times\mathbb{Z}_2$. This $V_4$ group also commutes with $\mathcal{T}$.

The model in Eq.~\eqref{eq:KT_dual_model} can be obtained via the non-local Kennedy-Tasaki (KT) unitary transformation~\cite{Kennedy:1992ifl}, $\tilde{H}=U_{\mathrm{KT}}HU_{\mathrm{KT}}$. Here, the unitary transformation operator is given by
\begin{equation}
    U_{\mathrm{KT}}=\prod_{1 \leq u<v \leq L} e^{i \pi S_u^z S_v^x},
    \quad U_{\mathrm{KT}}^2=1,
\end{equation}
and the pre-transformed Hamiltonian $H$ reads
\begin{equation}\label{eq:Root-Hamiltonian-KT-Model}
\begin{aligned}
    H & = \sum_{i=1}^{L-1} J_i\left( S_i^x S_{i+1}^x + S_i^y S_{i+1}^y\right) + \Delta_i S_i^z S_{i+1}^z \\
    & + \sum_{i=1}^{L} D_i (S_i^z)^2+ g_i \Big[(S_i^+)^2(S_{i+1}^+)^2 + {\rm h.c.} \Big],
\end{aligned}
\end{equation}
with the raising and lowering operators $S_j^{\pm}=S_j^x\pm iS_j^y$. The Hamiltonian in Eq.~\eqref{eq:Root-Hamiltonian-KT-Model} preserves time-reversal symmetry and possesses an on-site $D_4\cong\mathbb{Z}_4\rtimes\mathbb{Z}_2\cong V_4\rtimes\mathbb{Z}_2$ unitary symmetry. This group is generated by $U_{\pi}^x$ and $U_{\pi/2}^z=\prod_{i=1}^Le^{i \frac{\pi}{2} S_i^z}$, which satisfy the algebraic relations
\be
    (U_{\pi}^x)^2=(U_{\pi/2}^z)^4=1,
    \quad U_{\pi}^xU_{\pi/2}^z(U_{\pi}^x)^{\dagger}=(U_{\pi/2}^z)^{\dagger}.
\ee
As a direct consequence of the unitary equivalence of the commutators,
\be
    [\tilde{H},\tilde{U}_{\pi/2}^z]=U_{\mathrm{KT}}[H,U_{\pi/2}^z]U_{\mathrm{KT}}=0.
\ee
The Hamiltonian possesses a $\mathbb{Z}_4$ symmetry 
\begin{equation}
\tilde{U}^z_{\pi/2}=U_{\mathrm{KT}}U_{\pi/2}^z U_{\mathrm{KT}} = \exp\left(i \frac{\pi}{2} \sum_{j=1}^L \prod_{k<j} e^{i\pi S_k^z} S_j^z\right) ~,
\end{equation}
hidden by the KT transformation. However, note that $U_{\pi}^x$ and $U_{\pi}^z$ is invariant under the KT transformation, i.e., $U_{\mathrm{KT}}U_{\pi}^x U_{\mathrm{KT}} = U_{\pi}^x, U_{\rm KT} U_{\pi}^z U_{\rm KT} = U_{\pi}^z$, we conclude that the internal unitary symmetry group of Eq.~\eqref{eq:KT_dual_model} is fully generated by $\{U_{\pi}^x, \tilde{U}_{\pi/2}^z\}$
with the relation
\begin{equation}
    (U_{\pi}^x)^2 =1 ~, \quad (\tilde U_{\pi/2}^z)^4 = 1~, \quad \tilde U_{\pi/2}^z U_{\pi}^x (\tilde U_{\pi/2}^z)^\dagger = U_{\pi}^x U_{\pi}^z ~.
\end{equation}
This recovers the full $D_4$ symmetry structure originally present in Eq.~\eqref{eq:Root-Hamiltonian-KT-Model}.

Without prior knowledge of the KT transformation relating the local Hamiltonian in Eq.~\eqref{eq:KT_dual_model} to Eq.~\eqref{eq:Root-Hamiltonian-KT-Model}, it is highly non-trivial to identify $\tilde{U}^z_{\pi/2}$ as a symmetry and subsequently establish the full internal symmetry group as $D_4$. By employing our bootstrap procedure based on xSFF data, we bypass the need to find the explicit expression for $\tilde{U}^z_{\pi/2}$. Instead, we directly identify the full $D_4$ internal symmetry group from the obtained branching matrix, fusion rules, and character table, requiring no information beyond the known on-site symmetries $U_{\pi}^x$ and $U_{\pi}^z$.

\begin{figure*}[t]
    \centering
    \includegraphics[width=0.8\linewidth]{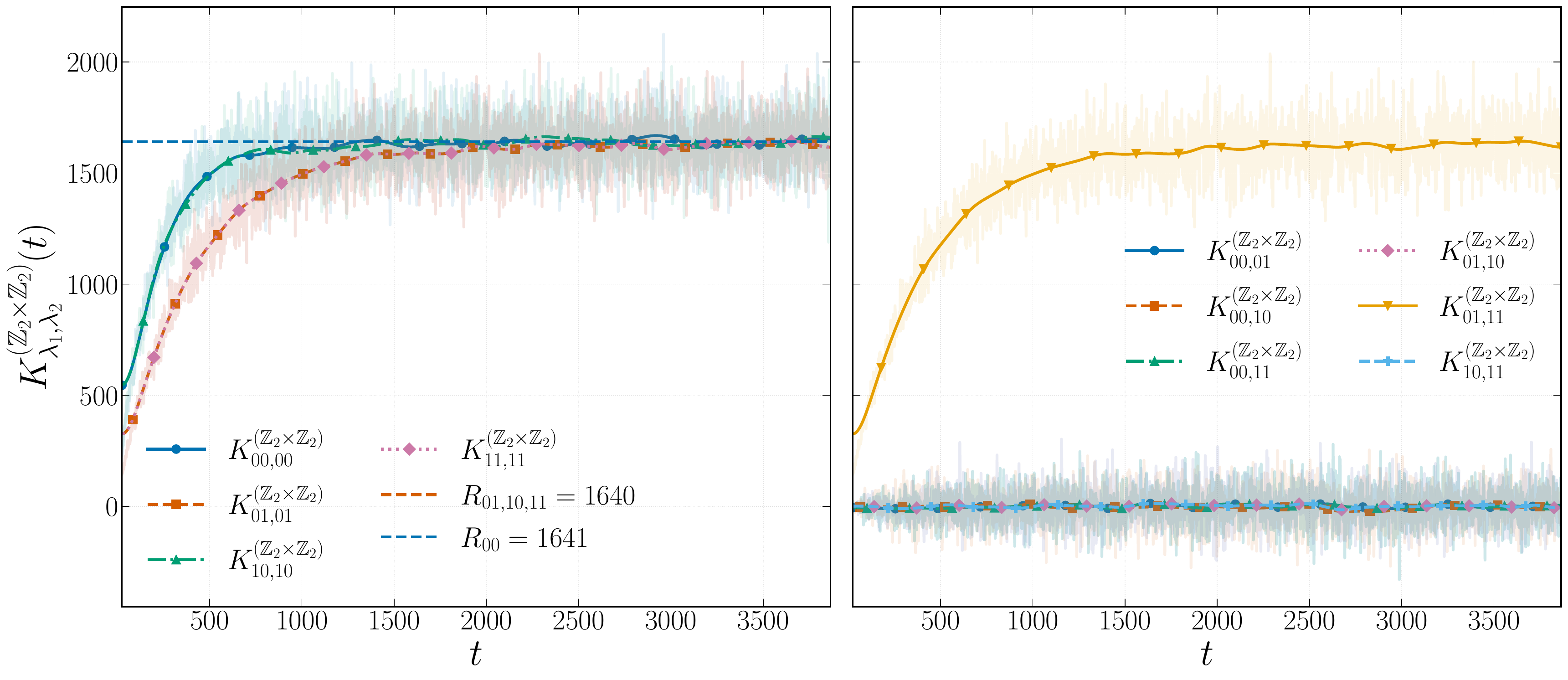}
    \caption{xSFF for the Kennedy-Tasaki transformed spin-1 model with $L=8$ sites, projected onto the $\mathbb{Z}_2 \times \mathbb{Z}_2$ irreps, averaged over 200 disorder realizations. The couplings are drawn independently as $J_j\sim\mathcal{N}(1,0.25^2)$, $\Delta_j\sim\mathcal{N}(0.7,0.25^2)$, $g_j\sim\mathcal{N}(0.2,0.25^2)$, and $D_j\sim\mathcal{N}(0.3,0.25^2)$. \textbf{Left}: The diagonal elements, whose late-time plateaus align perfectly with the corresponding benchmark lines $R_{\lambda}$. \textbf{Right}: The off-diagonal elements of the xSFF. In particular, $K_{11,11}^{(V_4)}=K_{01,01}^{(V_4)} = K_{11,01}^{(V_4)}$, implying that the charge sectors $\mathcal{H}_{01}$ and $\mathcal{H}_{11}$ are degenerate.}
    \label{fig:KT_sym}
\end{figure*}

\subsection{Bootstrap Results}
To initiate the bootstrap procedure, we first construct the projectors onto the irreps of $N$, where $N\cong V_4$ is generated by the symmetries given in Eq.~\eqref{eq:KT-Dual-Model-Subgroup}. The projectors $P_{\lambda}$ onto the $V_4$ charge sectors $\mathcal{H}_{\lambda}$ are given by
\be\label{eq:KT-Dual-Model-Subrep-Projector}
\begin{split}
    P_{\lambda}=P_{Q_x}P_{Q_z},
    \quad P_{Q_{x,z}}=\f{1+(-1)^{Q_{x,z}}U_{\pi}^{x,z}}{2},
\end{split}
\ee
with $Q_{x,z}\in\{0,1\}$ and $\lambda\in\hat{V}_4=\{00,10,01,11\}$. By substituting Eq.~\eqref{eq:KT-Dual-Model-Subrep-Projector} into Eqs.~\eqref{eq:xSFF-Elements} and \eqref{eq:Def-Benchmark-Line} and performing ED, we obtain the xSFF data presented in Fig.~\ref{fig:KT_sym}. The numerical results demonstrate that $K_{\lambda,\lambda}^{(V_4)}\approx R_{\lambda}$ for all $\lambda$, and further reveal $K_{11,11}^{(V_4)}=K_{01,01}^{(V_4)} = K_{01,11}^{(V_4)}>0$, which lead to the numerical constraints:
\begin{equation}\label{eq:Numerical-Cond-KT-Dual-Model}
\begin{aligned}
    & b_{\lambda,\alpha} \in \{0,1\},
    \quad  \forall \lambda \in \hat{N}=\hat{V_4},
      \alpha \in \hat{G}, \\
    &\vec{b}_{01}=\vec{b}_{11},
    \quad G_{01,01} =  G_{11,11} = G_{01,11} >0.
\end{aligned}
\end{equation}

Next, we apply the bootstrap procedure guided by Eq.~\eqref{eq:Numerical-Cond-KT-Dual-Model} and the algebraic constraints detailed in Sec.~\ref{sec:Bootstrap} to derive the possible branching matrices, fusion rules, and character tables at the lowest viable rank. The bootstrap algorithm yields a unique minimal solution at rank $r_*=5$. The corresponding branching matrix is given by
\begin{equation}\label{eq:KT-Dual-Model-Branching-Matrix}
    B_{\lambda,\alpha}=\begin{pmatrix}
        1 & 1 & 0 & 0 & 0 \\
        0 & 0 & 1 & 1 & 0\\
        0 & 0 & 0 & 0 & 1\\
        0 & 0 & 0 & 0 & 1
    \end{pmatrix}.
\end{equation}
Here, the rows correspond to the $V_4$-irreps $\lambda \in \{00, 10, 01, 11\}$, while the columns denote the $G$-irreps $\alpha \in \{\alpha_0, \dots, \alpha_4\}$, with $\alpha_0$ representing the trivial representation. This matrix translates directly into the following explicit branching structure
\be
\begin{split}
    &\text{Res}^G_N V_{\alpha_0}\cong V_{00},~\text{Res}^G_N V_{\alpha_1}\cong V_{00},\\
    &\text{Res}^G_N V_{\alpha_2}\cong V_{10},~\text{Res}^G_N V_{\alpha_3}\cong V_{10},\\
    &\text{Res}^G_N V_{\alpha_4}\cong V_{01}\oplus V_{11}.
\end{split}
\ee
From these restriction rules, we immediately deduce the existence of four one-dimensional $G$-irreps ($\alpha_0, \alpha_1, \alpha_2, \alpha_3$) and one two-dimensional $G$-irrep ($\alpha_4$). Summing the squares of these dimensions, we strictly determine that the candidate group $G$ must be of order $|G| = 4 \times 1^2 + 1 \times 2^2 = 8$.

We also determine the complete set of non-vanishing fusion coefficients, which are classified as follows:
\begin{equation}\label{eq:KT-Dual-Model-Fusion}
\begin{split}
    (\mathrm{a})~&N_{\alpha_0\alpha}^{\alpha}
    =N_{\alpha\alpha_0}^{\alpha}=1,
    \quad \forall \alpha\in\hat G,\\
    (\mathrm{b})~&N_{\alpha_1\alpha_1}^{\alpha_0}
    =N_{\alpha_2\alpha_2}^{\alpha_0}
    =N_{\alpha_3\alpha_3}^{\alpha_0}=1,\\
    &N_{\alpha_1\alpha_2}^{\alpha_3}
    =N_{\alpha_2\alpha_1}^{\alpha_3}
    =N_{\alpha_1\alpha_3}^{\alpha_2}
    =N_{\alpha_3\alpha_1}^{\alpha_2}
    =N_{\alpha_2\alpha_3}^{\alpha_1}\\
    &=N_{\alpha_3\alpha_2}^{\alpha_1}=1,\\
    (\mathrm{c})~&N_{\alpha_1\alpha_4}^{\alpha_4}
    =N_{\alpha_4\alpha_1}^{\alpha_4}
    =N_{\alpha_2\alpha_4}^{\alpha_4}
    =N_{\alpha_4\alpha_2}^{\alpha_4}
    =N_{\alpha_3\alpha_4}^{\alpha_4}\\  
    &=N_{\alpha_4\alpha_3}^{\alpha_4}=1,\\
    (\mathrm{d})~&N_{\alpha_4\alpha_4}^{\alpha_0}
    =N_{\alpha_4\alpha_4}^{\alpha_1}
    =N_{\alpha_4\alpha_4}^{\alpha_2}
    =N_{\alpha_4\alpha_4}^{\alpha_3}=1,
\end{split}
\end{equation}
where the classifications denote: (a) the unit rules, ensuring that fusing any irreps $\alpha\in\hat G$ with the trivial one $\alpha_0$ leaves it unchanged; (b) the fusion rules among the three non-trivial one-dimensional irreps $\alpha_1,\alpha_2,\alpha_3$, which together with $\alpha_0$ form a Klein-four subgroup under fusion; (c) the action of the one-dimensional irreps on the two-dimensional irrep $\alpha_4$, showing that fusion with any one-dimensional irrep leaves $\alpha_4$ invariant; and (d) the self-fusion of the two-dimensional irrep $\alpha_4$, which decomposes into the direct sum of all four one-dimensional irreps.
\begin{table}
    \centering
    \begin{tabular}{c|ccccc}
         & $c_1$ & $c_2$ & $c_3$ & $c_4$ & $c_5$\\
       \hline
       $\chi_{\alpha_0}$  & 1 & 1 & 1 & 1 & 1\\
       $\chi_{\alpha_1}$  & 1 & 1 & 1 & -1 & -1\\
       $\chi_{\alpha_2}$  & 1 & 1 & -1 & 1 & -1\\
       $\chi_{\alpha_3}$  & 1 & 1 & -1 & -1 & 1\\
       $\chi_{\alpha_4}$  & 2 & -2 & 0 & 0 & 0\\
       \hline
    \end{tabular}
    \caption{Character table of the hidden symmetry group for the KT transformed spin-$1$ chain \eqref{eq:KT_dual_model}, derived using the bootstrap program. There are five conjugacy classes: $c_1=\{1\}$, 
    $c_2=\{ (\tilde{U}_{\pi/2}^z)^2 \}$, $c_3=\{ \tilde{U}_{\pi/2}^z,(\tilde{U}_{\pi/2}^z)^3 \}$, $c_4=\{ U_{\pi}^x,(\tilde{U}_{\pi/2}^z)^2U_{\pi}^x \}$ and $c_5=\{ \tilde{U}_{\pi/2}^zU_{\pi}^x,
    (\tilde{U}_{\pi/2}^z)^3U_{\pi}^x \}$.}
    \label{tab:D4-Q8-Character-Table}
\end{table}

Following the approach used for the $S_3$-invariant chain, we define the commuting fusion matrices $(N_{\alpha_i})_{\alpha_j\alpha_k}=N_{\alpha_i\alpha_j}^{\alpha_k}$ with $i,j,k\in\{0,1,\cdots,4\}$. By simultaneously diagonalizing these matrices, we extract their eigenvalues to construct the character table presented in Table~\ref{tab:D4-Q8-Character-Table}. 
These results successfully recover the linear representation theory of the group $D_4$, thereby identifying $G \cong D_4$, as verified analytically in Appendix~\ref{app:Bootstrap-Solutions-II}.

\section{Example III: Higher Branching Multiplicities\label{sec:AT-Potts}}
\subsection{Hamiltonian and Symmetries}
We now consider the extended quantum Ashkin-Teller model at the four-state Potts point. Each site of an $L$-site open chain carries two Ising degrees of freedom, $\sigma$ and $\tau$, giving a local Hilbert space of dimension $4$. The Hamiltonian is
\be\label{eq:AT-Hamiltonian}
\begin{split}
    H &= \sum_{j=1}^{L-1} J_j \left(\sigma_j^z\sigma_{j+1}^z + \tau_j^z\tau_{j+1}^z + \sigma_j^z\tau_j^z\sigma_{j+1}^z\tau_{j+1}^z\right)\\
    &+ \sum_{j=1}^{L-2} J_j' \left(\sigma_j^z\sigma_{j+2}^z + \tau_j^z\tau_{j+2}^z + \sigma_j^z\tau_j^z\sigma_{j+2}^z\tau_{j+2}^z\right)\\
    &+ \sum_{j=1}^{L} h_j \left(\sigma_j^x + \tau_j^x + \sigma_j^x\tau_j^x\right),
\end{split}
\ee
where $J_j$, $J_j'$, and $h_j$ are independent random couplings. Within each term, the three sub-terms (involving $\sigma$-$\sigma$, $\tau$-$\tau$, and $\sigma\tau$-$\sigma\tau$ couplings) share the same random coefficient, which preserves the $S_4$ permutation structure of the four-state Potts model. The next-nearest-neighbor interaction is included to prevent accidental degeneracies that would obscure the hidden symmetry signal.

As the manifest symmetry, we consider $N \cong V_4$ generated by the global spin-flip operators
\be\label{eq:AT-Z2Z2-Generators}
    g_1 = \prod_{j=1}^L \sigma_j^x, \quad g_2 = \prod_{j=1}^L \tau_j^x,
\ee
which act on the local basis $\ket{s_\sigma, s_\tau}$ by flipping the $\sigma$ and $\tau$ spins, respectively. Since $N$ is abelian, its all four irreps are one-dimensional, labeled by $(a,b) \in \{0,1\}^2$, with eigenvalues $(-1)^a$ under $g_1$ and $(-1)^b$ under $g_2$, so that $d_\lambda = 1$ for all $\lambda$.

The model involves two additional hidden symmetries. The first is the global swap operator between the two Ising degrees of freedom, 
\be
\label{eq:Soperator}
\begin{split}
    &S=\prod_{j=1}^Ls_j,\quad s_j=\f{1+\sigma^x_j\tau^x_j+\sigma^y_j\tau^y_j+\sigma^z_j\tau^z_j}{2},
\end{split}
\ee
which exchanges the local basis states as $S\ket{s_{\sigma},s_{\tau}}=\ket{s_{\tau},s_{\sigma}}$. The second is the global CNOT gate, defined as
\be
\label{eq:Coperator}
\begin{split}
    &C=\prod_{j=1}^Lc_j,
    \quad c_j=\f{1+\sigma^z_j}{2}+\f{1-\sigma^z_j}{2}\tau_j^x.
\end{split}
\ee
Consequently, the unitary generators $\{g_1,g_2,S,C\}$ generate the full internal symmetry group $G\cong S_4$. We can explicitly confirm this by translating these operators into the standard conventions for permutation groups~\cite{Inui1990GroupTheory}. Specifically, their action on the four configurations $\{1,2,3,4\}$ corresponds to the following permutations (any permutation can be expressed as the product of transpositions)
\be\label{eq:S4-Standard-Group-Translation}
\begin{split}
    g_1=(13)(24)&,\quad g_2=(12)(34),\\
    S=(23)&,\quad C=(34).
\end{split}
\ee
From this basis, we already possess two adjacent transpositions, $S$ and $C$. By constructing the conjugate
\be
    g_1Cg_1^{-1}=g_1(34)g_1^{-1}=(12),
\ee
we recover the complete set of adjacent transpositions $\{(12),(23),(34)\}$, which strictly generates the symmetric group $S_4$. For our bootstrap, we take the manifest Klein four-group $N\cong V_4$, generated by $\{g_1,g_2\}$, as the only known algebraic input.
\begin{figure*}[t]
    \centering
    \includegraphics[width=0.8\linewidth]{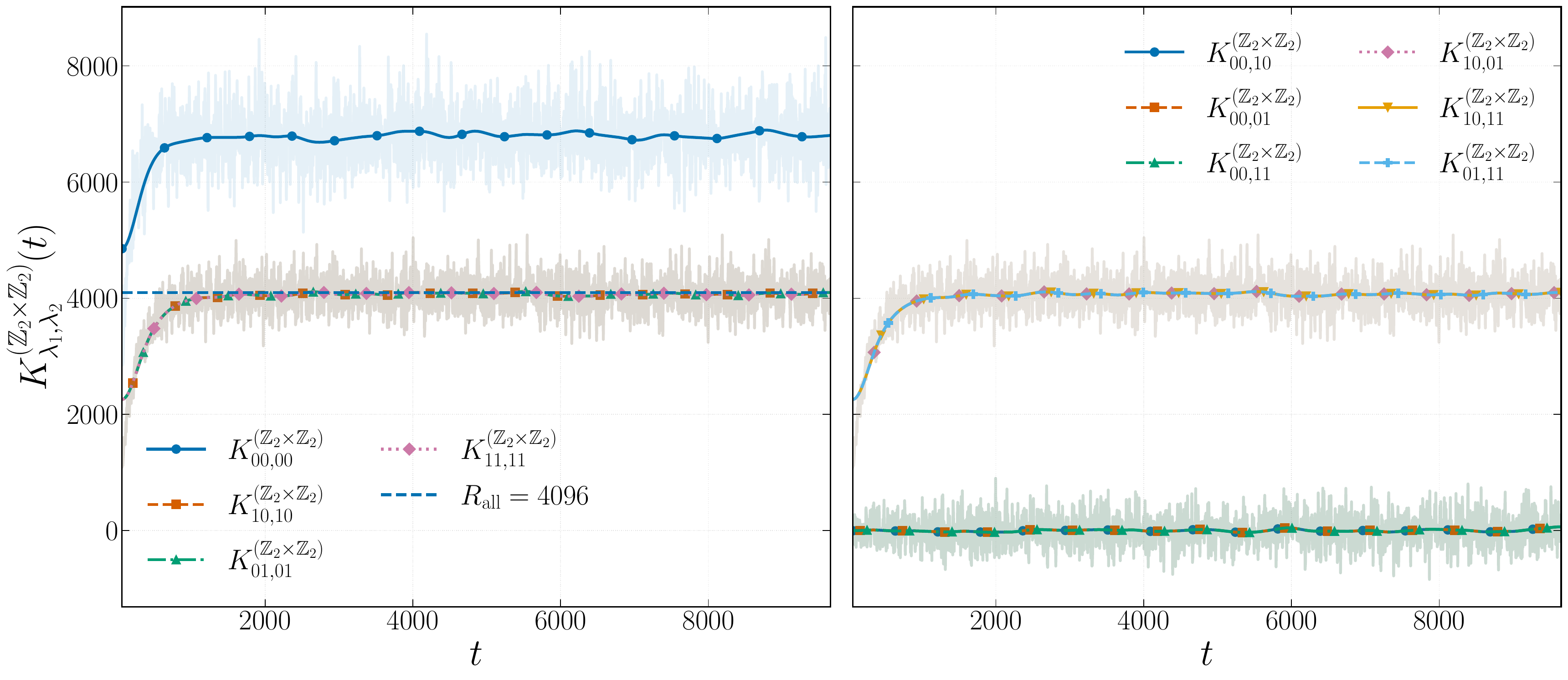}
    \caption{xSFF for the extended Ashkin-Teller chain at the Potts point with $L = 7$ sites, projected onto $N = \mathbb{Z}_2 \times \mathbb{Z}_2$ irreps, averaged over $200$ disorder realizations. The couplings are drawn independently from uniform distributions: nearest-neighbor $J_j\sim\mathrm{Unif}(0.5,1.5)$, next-nearest-neighbor $J_j\sim\mathrm{Unif}(0.5,1.5)$, and transverse field $h_j\sim\mathrm{Unif}(0.5,1.5)$. \textbf{Left}: The diagonal elements exhibit late-time plateaus that align perfectly with the corresponding benchmark lines $R_{\lambda}=R_{\text{all}}$ for all $\lambda\in\hat{V}_4$, with the exception that $K_{00,00}^{(V_4)}/R_{\text{all}}>1$. \textbf{Right}: Off-diagonal elements. The three non-trivial sectors are degenerate, with $K_{10,01}^{(V_4)} = K_{10,11}^{(V_4)} = K_{01,11}^{(V_4)} = K_{10,10}^{(V_4)}$, while the cross-correlations with the trivial sector vanish.}
    \label{fig:AT_Potts_xSFF}
\end{figure*}

\subsection{Bootstrap Results}
We calculate the xSFF by projecting onto the four $N$-irreps by the projectors
\be
\begin{split}
    &P_{(s_{\sigma},s_{\tau})}=P_{s_{\sigma}}P_{s_{\tau}}=P_{s_{\tau}}P_{s_{\sigma}},\\
    &P_{s_{\sigma}}=\f{1+(-1)^{s_{\sigma}}g_1}{2},\quad P_{s_{\tau}}=\f{1+(-1)^{s_{\tau}}g_2}{2},
\end{split}
\ee
with numerical results shown in Fig.~\ref{fig:AT_Potts_xSFF}. The diagonal components satisfy $K_{(0,0),(0,0)}^{(\mathbb{Z}_2^2)} / R_{(0,0)} \approx 1.66 > 1$, while $K_{(a,b),(a,b)}^{(\mathbb{Z}_2^2)} / R_{(a,b)} \approx 1$ for the three non-trivial $N$-irrep $(a,b) \neq (0,0)$. Furthermore, the off-diagonal elements between the trivial and non-trivial $N$-irreps vanish, $K_{(0,0),(a,b)}^{(\mathbb{Z}_2^2)} \approx 0$ for $(a,b) \neq (0,0)$, while the off-diagonal elements among the three non-trivial $N$-irreps are all equal and coincide with the corresponding diagonals:
\be
    K_{(1,0),(0,1)}^{(\mathbb{Z}_2^2)} = K_{(1,0),(1,1)}^{(\mathbb{Z}_2^2)} = K_{(0,1),(1,1)}^{(\mathbb{Z}_2^2)} = K_{(1,0),(1,0)}^{(\mathbb{Z}_2^2)}.
\ee
These observations yield the following constraints on the branching multiplicities:
\be\label{eq:AT-b-Condition}
\begin{split}
    &b_{\lambda,\alpha} \in \{0,1\}, \quad \forall (a,b) \neq (0,0),\\
    &b_{(0,0),\alpha} \in \{0,1,2,3\},\\
    &\vec{b}_{(1,0)} = \vec{b}_{(0,1)} = \vec{b}_{(1,1)},\\
    &G_{(0,0),(a,b)} = 0, \quad (a,b) \neq (0,0).
\end{split}
\ee
The first two lines follow from the diagonal ratios $K_{\lambda,\lambda}/R_\lambda$: a ratio of unity implies multiplicity-free branching (where branching multiplicities are restricted to zero or one), while a ratio exceeding unity indicates the presence of higher multiplicities, though these remain bounded by Eq.~\eqref{eq:Dim-Condition-1}. The third line follows from the equality of all xSFF elements among the three non-trivial sectors (statement I proved in Appendix~\ref{app:Proof-II}). The last line reflects the vanishing cross-correlation between the trivial and non-trivial $N$-irreps.

Applying the bootstrap algorithm (Sec.~\ref{sec:Bootstrap-Algorithm}),we find an unique rigid solution at rank $r_* = 5$. The branching matrix is
\be\label{eq:AT-Branching-Matrix}
    B_{\lambda,\alpha}=\begin{pmatrix}
        1 & 1 & 2 & 0 & 0\\
        0 & 0 & 0 & 1 & 1\\
        0 & 0 & 0 & 1 & 1\\
        0 & 0 & 0 & 1 & 1\\
    \end{pmatrix},
\ee
where the rows correspond to the $N$-irreps $((0,0),(1,0),(0,1),(1,1))$, and the columns denote the five obtained $G$-irreps, labeled by $\alpha_0$ through $\alpha_4$ from left to right. This branching matrix exactly encodes the following restrictions
\be
\begin{split}
    &\text{Res}^G_NV_{\alpha_0}\cong\text{Res}^G_NV_{\alpha_1}\cong V_{(0,0)},\\
    &\text{Res}^G_NV_{\alpha_2}\cong V_{(0,0)}\oplus V_{(0,0)},\\
    &\text{Res}^G_NV_{\alpha_3}\cong V_{(1,0)}\oplus V_{(0,1)}\oplus V_{(1,1)},\\
    &\text{Res}^G_NV_{\alpha_4}\cong V_{(1,0)}\oplus V_{(0,1)}\oplus V_{(1,1)}.\\
\end{split}
\ee
From these restrictions, we immediately deduce the dimensions of the obtained $G$-irreps: $\dim V_{\alpha_0}=\dim V_{\alpha_1}=1$, $\dim V_{\alpha_2}=2$, and $\dim V_{\alpha_3}=\dim V_{\alpha_4}=3$. Summing the squares of these dimensions yields the order of the abstract group, $|G| = 2\times 1^2+2^2+2\times 3^2 = 24$.

The non-zero fusion coefficients are given by
\begin{equation}\label{eq:AT-Fusion-Coefficients}
\begin{split}
    (\mathrm a)\quad &N_{\alpha_0\alpha}^{\alpha}
    =N_{\alpha\alpha_0}^{\alpha}=1,
    \quad \forall \alpha\in\hat G,\\
    (\mathrm b)\quad &N_{\alpha_1\alpha_1}^{\alpha_0}=1,\quad
    N_{\alpha_1\alpha_2}^{\alpha_2}
    =N_{\alpha_2\alpha_1}^{\alpha_2}=1,\\
    &N_{\alpha_1\alpha_3}^{\alpha_4}
    =N_{\alpha_3\alpha_1}^{\alpha_4}
    =N_{\alpha_1\alpha_4}^{\alpha_3}
    =N_{\alpha_4\alpha_1}^{\alpha_3}=1,\\
    (\mathrm c)\quad &N_{\alpha_2\alpha_2}^{\alpha_0}
    =N_{\alpha_2\alpha_2}^{\alpha_1}
    =N_{\alpha_2\alpha_2}^{\alpha_2}=1,\\
    &N_{\alpha_2\alpha_3}^{\alpha_3}
    =N_{\alpha_2\alpha_3}^{\alpha_4}
    =N_{\alpha_3\alpha_2}^{\alpha_3}
    =N_{\alpha_3\alpha_2}^{\alpha_4}=1,\\
    &N_{\alpha_2\alpha_4}^{\alpha_3}
    =N_{\alpha_2\alpha_4}^{\alpha_4}
    =N_{\alpha_4\alpha_2}^{\alpha_3}
    =N_{\alpha_4\alpha_2}^{\alpha_4}=1,\\
    (\mathrm d)\quad &N_{\alpha_3\alpha_3}^{\alpha_0}
    =N_{\alpha_3\alpha_3}^{\alpha_2}
    =N_{\alpha_3\alpha_3}^{\alpha_3}
    =N_{\alpha_3\alpha_3}^{\alpha_4}=1,\\
    &N_{\alpha_4\alpha_4}^{\alpha_0}
    =N_{\alpha_4\alpha_4}^{\alpha_2}
    =N_{\alpha_4\alpha_4}^{\alpha_3}
    =N_{\alpha_4\alpha_4}^{\alpha_4}=1,\\
    &N_{\alpha_3\alpha_4}^{\alpha_1}
    =N_{\alpha_3\alpha_4}^{\alpha_2}
    =N_{\alpha_3\alpha_4}^{\alpha_3}
    =N_{\alpha_3\alpha_4}^{\alpha_4}=1,
\end{split}
\end{equation}
where the classifications denote: (a) the unit rules; (b) the unique non-trivial one-dimensional irrep $\alpha_1$ squares to the identity, leaves the two-dimensional irrep $\alpha_2$ invariant, and exchanges the two inequivalent three-dimensional irreps $\alpha_3$ and $\alpha_4$; (c) the self-fusion of the two-dimensional irrep $\alpha_2$, and the fusion between the two-dimensional irrep and two three-dimensional irreps; and (d) the fusion rules among the two three-dimensional irreps, their self-fusions are identical, while the mixed fusion $V_{\alpha_3}\otimes V_{\alpha_4}$ yields the full sum $V_{\alpha_1}\oplus V_{\alpha_2}\oplus V_{\alpha_3}\oplus V_{\alpha_4}$. Subsequently, we diagonalize the fusion matrices to obtain the character table, summarized in Table~\ref{tab:AT-Character-Table}.
\begin{table}
    \centering
    \begin{tabular}{c|ccccc}
         & $c_1$ & $c_{2}$ & $c_{3}$ & $c_4$ & $c_5$\\
         \hline
        $\chi_{\alpha_0}$ & 1 & 1 & 1 & 1 & 1\\
        $\chi_{\alpha_1}$ & 1 & $1$ & $-1$ & $-1$ & 1\\
        $\chi_{\alpha_2}$ & 2 & $2$ & $0$ & 0 & $-1$\\
        $\chi_{\alpha_3}$ & 3 & $-1$ & $1$ & $-1$ & 0\\
        $\chi_{\alpha_4}$ & 3 & $-1$ & $-1$ & 1 & $0$\\
        \hline
    \end{tabular}
    \caption{Character table for the extended Ashkin-Teller chain at the Potts point, derived from the bootstrap program. The conjugacy class sizes are $|c_1| = 1$, $|c_2| = 3$, $|c_3| = 6$, $|c_4| = 6$, $|c_5| = 8$.}
    \label{tab:AT-Character-Table}
\end{table}

The above results are consistent with the interpretation that the Potts-point Hamiltonian exhibits the full permutation symmetry of its four local states ($S_4$), which is confirmed in Appendix~\ref{app:Bootstrap-Solutions-III}. The manifest $V_4$ subgroup (the Klein four-group of double transpositions) is rigorously identified as a normal subgroup of $S_4$.

\section{Example IV: Non-commuting Anti-unitary Symmetry and Non-normal Subgroup\label{sec:3-QTC}}
In the examples we have investigated so far, all the identified subgroups were normal subgroups of the hidden group $G$, and all the group generators were represented by unitary operators. To explore whether we can go beyond these two restrictions, we now study the three-state ($m=3$) quantum torus chain~\cite{PhysRevB.86.134430} as a lattice model that incorporates both generalizations.

\subsection{Hamiltonian and Symmetries}
Based on~\cite{PhysRevB.86.134430}, we define the following Hamiltonian with OBC:
\be\label{eq:Disorder-QTC-Hamiltonian}
\begin{split}
    H_{\theta}&=\sum_{j=1}^{L-1}J_j^Z(\theta)\left( Z_jZ_{j+1}^{\dagger}+\text{h.c.}\right)\\
    &+\sum_{j=1}^{L-1}J_j^X(\theta) \left(X_jX_{j+1}^{\dagger}+\text{h.c.}\right),
\end{split}
\ee
where the single-site clock and shift operators, $Z_j$ and $X_j$, are given by Eq.~\eqref{eq:HW-Z3-Generators} with $m=3$. The disordered couplings are defined as
\be
    J_j^Z = J_j\cos\theta, \quad J_j^X = J_i \sin \theta,
\ee
where $\theta\in(0,2\pi)$ and $J_i$ are independent Gaussian random variables sampled at each site. As the same random variable $J_j$ is used for both couplings on a given bond, the ratio $J_j^Z/J_j^X = \cot\theta$ remains spatially uniform across all bonds, which preserves the global symmetry structure.

The Hamiltonian in Eq.~\eqref{eq:Disorder-QTC-Hamiltonian} lacks crystalline symmetry but is manifestly time-reversal symmetric since it is purely real, with complex conjugation $K$ acting as the corresponding anti-unitary symmetry operator. Furthermore, for a generic angle $\theta$, the model possesses a $(\mathbb{Z}_3\times\mathbb{Z}_3)\rtimes \mathbb{Z}_2$ symmetry. This group is generated by the following unitary operators
\be\label{eq:Z-Z3-Symmetry-Generator}
    X = \prod_{j=1}^L X_j ~, \quad Z=\prod_{j=1}^L Z_j,
    \quad R = \prod_{j=1}^L r_j,
\ee
where $r_i$ is the local charge conjugation symmetry operator
\begin{equation}
    r_i \equiv \sum_{q=0}^2 |-q\rangle \langle q| \quad {\rm mod} \; 3 ~.
\end{equation}
These symmetries satisfy the following obvious relations
\begin{equation}
\begin{aligned}
    X^3 &= 1 ~, \quad Z^3 = 1 ~, \quad Z X = \omega^L X Z ~, \\  
    R^2 &= 1 ~, \quad R Z R = Z^{-1} ~, \quad R X R = X^{-1} ~.
\end{aligned}
\end{equation}
When $3|L$, $Z$ and $X$ form the linear representation of the subgroup $\mathbb{Z}_3 \times \mathbb{Z}_3$.  
Interestingly, the complex conjugation operator $K$ commutes with both $X$ and $R$, but it does not commute with the $Z$ operator; instead, it satisfies the relation
\be
    KZK^{-1}=Z^{-1}.
\ee
The symmetry operators $Z,~X,~R,~K$ generate a $\mathbb{Z}_2$-graded magnetic group~\cite{RevModPhys.40.359,rumynin2021realrepresentationsc2gradedgroups}
\be\label{eq:S32-Magnetic-Group}
    G^{(\text{mag.})}\cong \left(\mathbb{Z}_3^2\rtimes\mathbb{Z}_2\right)\sqcup K\left(\mathbb{Z}_3^2\rtimes\mathbb{Z}_2\right)\cong S_3^2, 
\ee
where the unitary subgroup is $N^{\text{uni.}}=\left\langle X,Z,R\right\rangle\cong\mathbb{Z}_3^2\rtimes\mathbb{Z}_2$. Consequently, analyzing this magnetic group requires Wigner's corepresentation theory rather than standard linear group representation theory~\cite{RevModPhys.40.359,wigner1959group}. 
For generic $\theta$, assuming the known subgroup is $N=N^{\text{uni.}}$, our initial objective is to investigate whether the bootstrap framework can determine minimal-rank solutions reflecting the correct Wigner's corepresentation theory associated with the full group $G\cong G^{(\text{mag.})}$ given in Eq.~\eqref{eq:S32-Magnetic-Group}.

The symmetries of the Hamiltonian in Eq.~\eqref{eq:Disorder-QTC-Hamiltonian} are further enlarged at the ``self-dual'' points, corresponding to $\theta = \frac{\pi}{4} \pmod{\pi}$, where the condition $\cos\theta = \sin\theta$ ensures equal coupling strengths. At these points, the duality transformation exchanging $X_j$ and $Z_j$ becomes an exact symmetry~\cite{PhysRevB.86.134430}. 
Our second goal in this section is to apply our bootstrap method to identify this hidden symmetry group $G$ that contains the hidden duality transformation, starting from $N\cong \mathbb{Z}_3^2\rtimes\mathbb{Z}_2$. As we will show, the subgroup $N$ in this case is not a normal subgroup of $G\cong\mathbb{Z}_3^2\rtimes \mathbb{Z}_4$, which will demonstrate the applicability of our approach beyond semi-direct product extensions.

\subsection{Bootstrap Results ($\theta\neq\frac{\pi}{4} \pmod{\pi}$)\label{eq:QTC-Bootstrap-Off-Sd}}
The subgroup $N\cong\mathbb{Z}_3^2\rtimes\mathbb{Z}_2$ features six irreps. We construct the projectors by starting from the nine irreps of its normal subgroup $\mathbb{Z}_3^2$, labeled by the charge pairs $(Q_Z\mod 3,Q_X\mod 3)\in\{0,1,2\}^2$. Since charge conjugation $R$ inverts both $X$ and $Z$, it pairs each state $(Q_Z,Q_X)$ with its conjugate charge pair, $(-Q_Z,-Q_X)$. Consequently, the zero-charge sector splits into two one-dimensional $N$-irreps, labeled by $(0,\pm)$ with $R=\pm 1$. The remaining states pair up as follows: $(1,0)$ with $(2,0)$, $(1,1)$ with $(2,2)$, $(1,2)$ with $(2,1)$, and $(0,1)$ with $(0,2)$, which altogether form four distinct two-dimensional $N$-irreps. In total, we obtain six irreps with dimensions $\{1,1,2,2,2,2\}$, labeled by
\be
    \lambda\in\hat{N}=\{(0,+), (0,-), [0,1], [1,0], [1,1], [1,2]\},
\ee
where $[Q_Z,Q_X]$ denotes the two-dimensional irrep associated with the orbit $\{(Q_Z,Q_X), (-Q_Z,-Q_X)\}$. The projectors onto irreps are defined as
\be\label{eq:Projector-QTC}
    \begin{split}
        P_{(0,\pm)}&=P_{(0,0)}P_{\pm}=P_{\pm}P_{(0,0)},\\
        P_{[Q_Z,Q_X]}&=P_{(Q_Z,Q_X)}+P_{(-Q_Z,-Q_X)},\\
        P_{(Q_Z,Q_X)}&=P_{Q_Z}P_{Q_X}=P_{Q_X}P_{Q_Z},\\
        P_{Q_{Z}}&=\f{1}{3}\sum_{q=0}^{2}e^{-i\f{2\pi q Q_{Z}}{3}}\left(Z\right)^q,\\
        P_{Q_{X}}&=\f{1}{3}\sum_{q=0}^{2}e^{-i\f{2\pi q Q_{X}}{3}}\left(X\right)^q,\\
    \end{split}
\ee
where $P_{\pm}=\f{1\pm R}{2}$. 

\begin{figure*}
    \centering
    \includegraphics[width=0.8\linewidth]{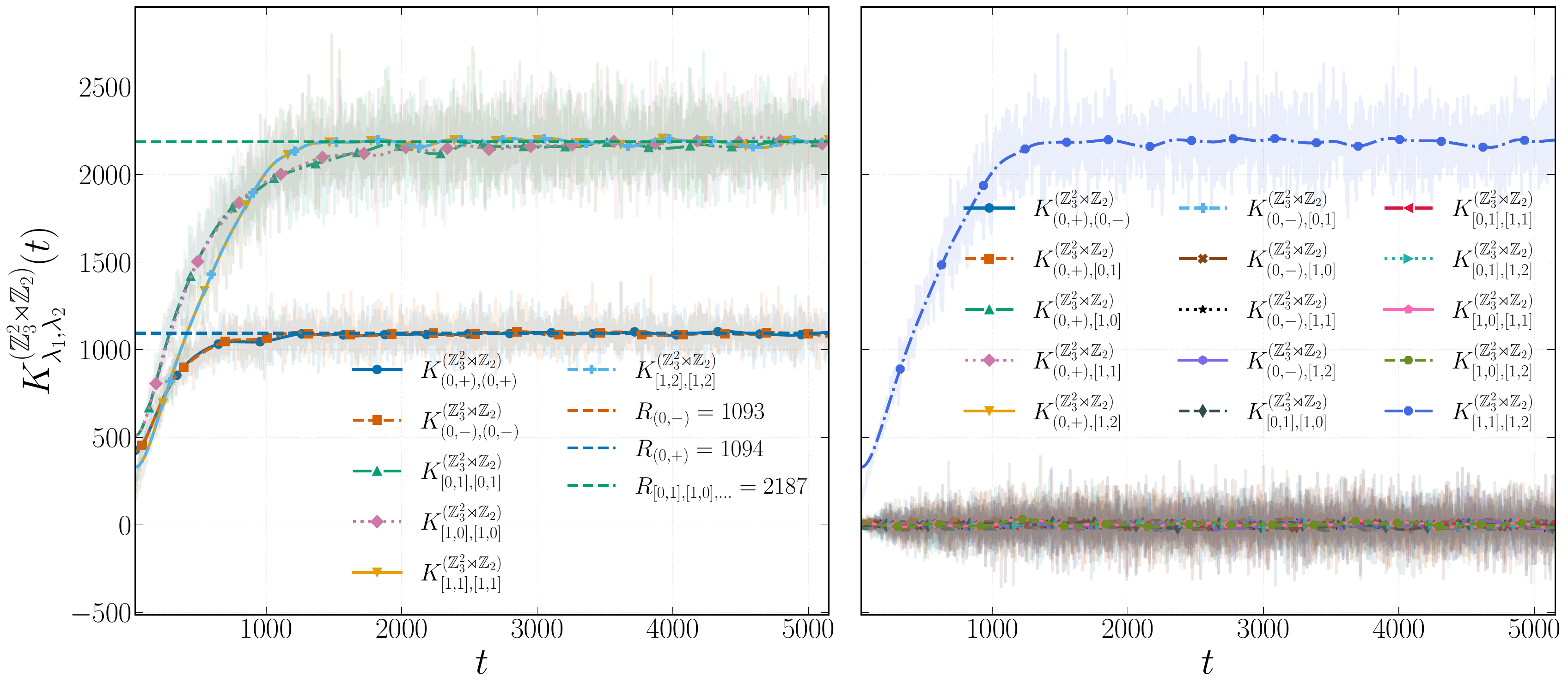}
    \caption{xSFF for the quantum torus chain at the non-self-dual point $\theta=\pi/6$, with $L = 9$ sites, projected onto $N \cong \mathbb{Z}_3^2 \rtimes \mathbb{Z}_2$ irreps, averaged over $200$ disorder realizations. The bond couplings are drawn as $J_j\sim\mathcal{N}(1,0.25^2)$, with $J_j^Z = J_j\cos\theta$ and $J_j^X = J_j\sin\theta$. \textbf{Left}: The diagonal elements exhibit late-time plateaus that align perfectly with the corresponding benchmark lines $R_{\lambda}$. \textbf{Right}: The off-diagonal elements. The paired sectors $[1,1] \leftrightarrow [1,2]$ exhibit non-vanishing xSFF values, while all other cross-block correlations vanish.}
    \label{fig:QTC-Non-Self-Dual}
\end{figure*}

Substituting these projectors into Eqs.~\eqref{eq:xSFF-Elements} and \eqref{eq:Def-Benchmark-Line} and performing ED, we obtain the xSFF data presented in Fig.~\ref{fig:QTC-Non-Self-Dual}.
The numerical data show that the diagonal ratios are $K_{\lambda,\lambda}^{(N)} / R_\lambda \approx 1$ for all $N$-irreps, indicating multiplicity-free branching, i.e., $b_{\lambda,\alpha} \in \{0,1\}$. Furthermore, the off-diagonal elements reveal a characteristic block structure: the cross-correlation within the pair $\{[1,1],[1,2]\}$ is equal to the corresponding diagonal values,
\be
\begin{split}
    &K_{[1,1],[1,2]}^{(N)} = K_{[1,1],[1,1]}^{(N)}=K_{[1,2],[1,2]}^{(N)},
\end{split}
\ee
while all other off-diagonal entries vanish. According to statement I, these observations imply the numerical constraints
\be\label{eq:QTC-b-Condition}
\begin{split}
    &b_{\lambda,\alpha}\in\{0,1\},
    \quad \forall\lambda\in\hat{N},\alpha\in\hat{G},
    \quad \vec{b}_{[1,1]}=\vec{b}_{[1,2]},\\
    &G_{[1,1],[1,2]} = G_{[1,1],[1,1]}=G_{[1,2],[1,2]}>0.\\
\end{split}
\ee
Incorporating these alongside the algebraic constraints from Sec.~\ref{sec:Bootstrap}, our bootstrap algorithm yields two distinct minimal solutions, at ranks $r_*=9$ (linear representation branch) and $r_*=5$ (corepresentation branch). 

We begin with the $r_*=9$ solution associated with the standard linear representation. This solution yields the branching matrix
\be\label{eq:QTC-Generic-S32-Lineaer-B-Matrix}
    B_{\lambda,\alpha}=\begin{pmatrix}
        1 & 0 & 0 & 1 & 0 & 0 & 0 & 0 & 0\\
        0 & 1 & 1 & 0 & 0 & 0 & 0 & 0 & 0\\
        0 & 0 & 0 & 0 & 0 & 0 & 1 & 1 & 0\\
        0 & 0 & 0 & 0 & 1 & 1 & 0 & 0 & 0\\
        0 & 0 & 0 & 0 & 0 & 0 & 0 & 0 & 1\\
        0 & 0 & 0 & 0 & 0 & 0 & 0 & 0 & 1\\
    \end{pmatrix}.
\ee
Here, the rows correspond to the $N$-irreps $(0,+), (0,-), [0,1], [1,0], [1,1]$, and $[1,2]$, while the columns represent the $G$-irreps $\alpha_0, \dots, \alpha_8$. We emphasize that only the last two rows are identical (and therefore not orthogonal), a structural feature that perfectly reproduces the numerical data shown in Fig.~\ref{fig:QTC-Non-Self-Dual}. Translating this matrix into the language of restricted representations, we obtain
\be\label{eq:S32-Group-Res}
\begin{split}
    &\text{Res}^G_N V_{\alpha_0}\cong V_{(0,+)},\quad\text{Res}^G_N V_{\alpha_1}\cong V_{(0,-)},\\
    &\text{Res}^G_N V_{\alpha_2}\cong V_{(0,-)},\quad\text{Res}^G_N V_{\alpha_3}\cong V_{(0,+)},\\
    &\text{Res}^G_N V_{\alpha_4}\cong V_{[1,0]},\quad \text{Res}^G_N V_{\alpha_5}\cong V_{[1,0]},\\
    &\text{Res}^G_N V_{\alpha_6}\cong V_{[0,1]},\quad\text{Res}^G_N V_{\alpha_7}\cong V_{[0,1]},\\
    &\text{Res}^G_N V_{\alpha_8}\cong V_{[1,1]}\oplus V_{[1,2]}.
\end{split}
\ee
With $\alpha_0$ denoting the trivial representation, these equations completely determine the dimensionality of the remaining $G$-irreps. Specifically, they yield three additional one-dimensional irreps ($\alpha_1, \alpha_2, \alpha_3$), four two-dimensional irreps ($\alpha_4, \dots, \alpha_7$), and one four-dimensional irrep ($\alpha_8$).

We obtain the following fusion rules:
\begin{equation}\label{eq:S3xS3-Fusion}
    \begin{split}
         (\mathrm{a})~&N_{\alpha_0\alpha}^{\alpha}=N_{\alpha\alpha_0}^{\alpha}=1,\quad \forall \alpha\in\hat{G},\\
         (\mathrm{b})~&N_{\alpha_1\alpha_1}^{\alpha_0}
            =N_{\alpha_2\alpha_2}^{\alpha_0}
            =N_{\alpha_3\alpha_3}^{\alpha_0}=1,\\
            &N_{\alpha_1\alpha_2}^{\alpha_3}
            =N_{\alpha_1\alpha_3}^{\alpha_2}
            =N_{\alpha_2\alpha_3}^{\alpha_1}=1,\\
        (\mathrm{c})~&N_{\alpha_1\alpha_4}^{\alpha_5}
            =N_{\alpha_1\alpha_5}^{\alpha_4}
            =N_{\alpha_1\alpha_6}^{\alpha_6}\\
            &=N_{\alpha_1\alpha_7}^{\alpha_7}=N_{\alpha_1\alpha_8}^{\alpha_8}=1,\\
            &N_{\alpha_2\alpha_l}^{\alpha_4}
            =N_{\alpha_2\alpha_5}^{\alpha_5}
            =N_{\alpha_2\alpha_6}^{\alpha_7}\\
            &=N_{\alpha_2\alpha_7}^{\alpha_6}=N_{\alpha_2\alpha_8}^{\alpha_8}=1,\\
            &N_{\alpha_3\alpha_4}^{\alpha_5}
            =N_{\alpha_3\alpha_5}^{\alpha_4}
            =N_{\alpha_3\alpha_6}^{\alpha_7}\\
            &=N_{\alpha_3\alpha_7}^{\alpha_6}=N_{\alpha_3\alpha_8}^{\alpha_8}=1,\\
        (\mathrm{d})~&N_{\alpha_4\alpha_4}^{\alpha_0}
            =N_{\alpha_4\alpha_4}^{\alpha_2}
            =N_{\alpha_4\alpha_4}^{\alpha_4}=1,\\
            &N_{\alpha_5\alpha_5}^{\alpha_0}
            =N_{\alpha_5\alpha_5}^{\alpha_2}
            =N_{\alpha_5\alpha_5}^{\alpha_4}=1,\\
            &N_{\alpha_4\alpha_5}^{\alpha_1}
            =N_{\alpha_4\alpha_5}^{\alpha_3}
            =N_{\alpha_4\alpha_5}^{\alpha_5}=1,\\
            &N_{\alpha_6\alpha_6}^{\alpha_0}
            =N_{\alpha_6\alpha_6}^{\alpha_1}
            =N_{\alpha_6\alpha_6}^{\alpha_6}=1,\\
            &N_{\alpha_7\alpha_7}^{\alpha_0}
            =N_{\alpha_7\alpha_7}^{\alpha_1}
            =N_{\alpha_7\alpha_7}^{\alpha_6}=1,\\
            &N_{\alpha_6\alpha_7}^{\alpha_2}
            =N_{\alpha_6\alpha_7}^{\alpha_3}
            =N_{\alpha_6\alpha_7}^{\alpha_7}=1,\\
        (\mathrm{e})~&N_{\alpha_4\alpha_6}^{\alpha_8}
            =N_{\alpha_4\alpha_7}^{\alpha_8}
            =N_{\alpha_5\alpha_6}^{\alpha_8}
            =N_{\alpha_5\alpha_7}^{\alpha_8}
            =1,\\
        (\mathrm{f})~&N_{\alpha_4\alpha_8}^{\alpha_6}
            =N_{\alpha_4\alpha_8}^{\alpha_7}
            =N_{\alpha_4\alpha_8}^{\alpha_8}=1,\\
            &N_{\alpha_5\alpha_8}^{\alpha_6}
            =N_{\alpha_5\alpha_8}^{\alpha_7}
            =N_{\alpha_5\alpha_8}^{\alpha_8}=1,\\
            &N_{\alpha_6\alpha_8}^{\alpha_4}
            =N_{\alpha_6\alpha_8}^{\alpha_5}
            =N_{\alpha_6\alpha_8}^{\alpha_8}=1,\\
            &N_{\alpha_7\alpha_8}^{\alpha_4}
            =N_{\alpha_7\alpha_8}^{\alpha_5}
            =N_{\alpha_7\alpha_8}^{\alpha_8}=1,\\
        (\mathrm{g})~&N_{\alpha_8\alpha_8}^{\alpha_j}=1,
        \quad j=0,1,\dots,8,        
    \end{split}       
\end{equation}
where the classifications denote: (a) the unit rules, ensuring any irrep $\alpha\in\hat{G}$ fused with the identity remains unchanged; (b) fusions between one-dimensional irreps that form a Klein-four subgroup; (c) the action of one-dimensional irreps on the two-dimensional sectors; (d) fusions between two-dimensional irreps within the respective families $(\alpha_4,\alpha_5)$ and $(\alpha_6,\alpha_7)$; (e) the mixed fusion between these two families, each yielding the unique four-dimensional irrep; (f) the fusion of the two-dimensional sectors with the four-dimensional sector; and (g) the self-fusion of the four-dimensional irrep into all sectors. By defining the commuting fusion matrices $(N_{\alpha_i})_{\alpha_j\alpha_k}=N_{\alpha_i\alpha_j}^{\alpha_k}$ from Eq.~\eqref{eq:S3xS3-Fusion}, we simultaneously diagonalize them to extract the characters and construct the character table presented in Table~\ref{tab:QTC-Character-Table-Generic}. 
\begin{table}
    \centering
    \begin{tabular}{c|ccccccccc}
        & $c_1$ & $c_2$ & $c_3$ & $c_4$ & $c_5$ & $c_6$ & $c_7$ & $c_8$ & $c_9$\\
        \hline
        $\chi_{\alpha_0}$ & 1 & 1 & 1 & 1 & 1 & 1 & 1 & 1 & 1\\
        $\chi_{\alpha_1}$ & 1 & 1 & 1 & 1 & -1 & 1 & -1 & 1 & -1\\
        $\chi_{\alpha_2}$ & 1 & 1 & 1 & 1 & 1 & -1 & 1 & -1 & -1\\
        $\chi_{\alpha_3}$ & 1 & 1 & 1 & 1 & -1 & -1 & -1 & -1 & 1\\
        $\chi_{\alpha_4}$ & 2 & 2 & -1 & -1 & 2 & 0 & -1 & 0 & 0\\
        $\chi_{\alpha_5}$ & 2 & 2 & -1 & -1 & -2 & 0 & 1 & 0 & 0\\
        $\chi_{\alpha_6}$ & 2 & -1 & 2 & -1 & 0 & 2 & 0 & -1 & 0\\
        $\chi_{\alpha_7}$ & 2 & -1 & 2 & -1 & 0 & -2 & 0 & 1 & 0\\
        $\chi_{\alpha_8}$ & 4 & -2 & -2 & 1 & 0 & 0 & 0 & 0 & 0\\
        \hline
    \end{tabular}
    \caption{Character table of the hidden symmetry group $G\cong S_3\times S_3$ for the three-site quantum torus chain with generic $\theta$. The conjugacy classes $\{c_i|i=1,\cdots,9\}$ are presented in Eq.~\eqref{eq:S32-Conjugacy-Classes} and their sizes are $\{1,2,2,4,3,3,6,6,9\}$ from left to right.}
    \label{tab:QTC-Character-Table-Generic}
\end{table}
These results are consistent with the linear representation theory of group $S_3^2$, which has been verified in Appendix~\ref{sec:QTC-Generic-Confirmation}.

Next, we consider the second minimal rank solution, namely the $r_*=5$ solution that admits a magnetic-group/corepresentation interpretation. 
To distinguish it from the ordinary hidden symmetry group above, we denote the corresponding hidden magnetic group by $G^{\text{mag.}}$, with unitary subgroup $N$. Using the same row convention as before, namely $(0,+)$, $(0,-)$, $[0,1]$, $[1,0]$, $[1,1]$, and $[1,2]$, and ordering the ICRs as $(\alpha_0,\alpha_1,\alpha_2,\alpha_3,\alpha_4)$ with sizes $(1,1,2,2,4)$, the branching matrix becomes
\be\label{eq:S32-M-Group-B-Matrix}
    B_{\lambda,\alpha}=
    \begin{pmatrix}
        1 & 0 & 0 & 0 & 0\\
        0 & 1 & 0 & 0 & 0\\
        0 & 0 & 1 & 0 & 0\\
        0 & 0 & 0 & 1 & 0\\
        0 & 0 & 0 & 0 & 1\\
        0 & 0 & 0 & 0 & 1
    \end{pmatrix}.
\ee
Here $\alpha_0$ is fixed and denotes the trivial corepresentation, $\alpha_1$ is the non-trivial one-dimensional ICR, $\alpha_2$ and $\alpha_3$ are two inequivalent two-dimensional ICRs, and $\alpha_4$ is the unique four-dimensional ICR. Their restrictions to the unitary subgroup $N$ are
\be\label{eq:S32-M-Group-B-Matrix-Res}
\begin{split}
    &\text{Res}^G_{N}V_{\alpha_0}=V_{(0,+)},\\
    &\text{Res}^G_{N}V_{\alpha_1}=V_{(0,-)},\\
    &\text{Res}^G_{N}V_{\alpha_2}=V_{[0,1]},\\
    &\text{Res}^G_{N}V_{\alpha_3}=V_{[1,0]},\\
    &\text{Res}^G_{N}V_{\alpha_4}=V_{[1,1]}\oplus V_{[1,2]}.
\end{split}
\ee
The last two rows of the branching matrix are identical—and therefore non-orthogonal, while all remaining rows are mutually orthogonal. This mathematical structure perfectly matches the numerical observations presented in Fig.~\ref{fig:QTC-Non-Self-Dual}.

The fusion rules are obtained from the bootstrap procedure:
\begin{equation}\label{eq:QTC-corep-fusion}
    \begin{split}
        (\mathrm{a})~&N_{\alpha_0\alpha}^{\alpha}
        =N_{\alpha\alpha_0}^{\alpha}=1,\quad \forall \alpha\in\hat{G}^{\text{mag.}},\\
        (\mathrm{b})~&N_{\alpha_1\alpha_1}^{\alpha_0}=1, \quad
        N_{\alpha_1\alpha_2}^{\alpha_2}
        =N_{\alpha_1\alpha_3}^{\alpha_3}
        =N_{\alpha_1\alpha_4}^{\alpha_4}=1,\\
        (\mathrm{c})~&N_{\alpha_2\alpha_2}^{\alpha_0}
        =N_{\alpha_2\alpha_2}^{\alpha_1}
        =N_{\alpha_2\alpha_2}^{\alpha_2}=1,\\
        &N_{\alpha_3\alpha_3}^{\alpha_0}
        =N_{\alpha_3\alpha_3}^{\alpha_1}
        =N_{\alpha_3\alpha_3}^{\alpha_3}=1,\\
        &N_{\alpha_2\alpha_3}^{\alpha_4}=1,\\
        (\mathrm{d})~&N_{\alpha_2\alpha_4}^{\alpha_4}=1,\quad
        N_{\alpha_2\alpha_4}^{\alpha_3}=2,\\
        &N_{\alpha_3\alpha_4}^{\alpha_4}=1,\quad
        N_{\alpha_3\alpha_4}^{\alpha_2}=2,\\
        (\mathrm{e})~&N_{\alpha_4\alpha_4}^{\alpha_4}=1,\quad 
        N_{\alpha_4\alpha_4}^{\alpha_j}=2, \quad j=0,1,2,3.
    \end{split}
\end{equation}
Here, the ICR $\alpha_1$ constitutes a $\mathbb{Z}_2$ invertible sector that acts trivially on the higher dimensional ICRs, while $\alpha_2$ and $\alpha_3$ are self-conjugate two-dimensional ICRs whose mixed fusion generates the unique Type (a) ICR $\alpha_4$ (a brief review of magnetic group corepresentation theory and detailed classifications are provided in Appendix~\ref{app:Corepresentation}).

Within our bootstrap algorithm, the characters of the ICRs can also be computed directly as the eigenvalues of the commuting fusion matrices, $(N_{\alpha_i})_{\alpha_j\alpha_k}=N_{\alpha_i\alpha_j}^{\alpha_k}$. The resulting characters are presented in Table~\ref{tab:QTC-Corep-Character-Table}. As confirmed in Appendix~\ref{sec:QTC-Generic-Confirmation}, these results precisely match Wigner's corepresentation theory for the magnetic group defined in Eq.~\eqref{eq:S32-Magnetic-Group}, whose underlying abstract group remains $S_3^2$.
\begin{table}
    \centering
    \begin{tabular}{c|ccccc}
        & $\tilde c_1$ & $\tilde c_2$ & $\tilde c_3$ & $\tilde c_4$ & $\tilde c_5$\\
        \hline
        $\chi_{\alpha_0}$ & 1 & 1 & 1 & 1 & 1\\
        $\chi_{\alpha_1}$ & 1 & 1 & 1 & 1 & -1\\
        $\chi_{\alpha_2}$ & 2 & -1 & -1 & 2 & 0\\
        $\chi_{\alpha_3}$ & 2 & -1 & 2 & -1 & 0\\
        $\chi_{\alpha_4}$ & 4 & 1 & -2 & -2 & 0\\
        \hline
    \end{tabular}
    \caption{Character table for the $r_*=5$ corepresentation solution, extracted from the fusion matrices of the quantum torus chain at generic $\theta$. Here, $\tilde{c}_i$ denote the magnetic classes, which consist exclusively of unitary elements.}
    \label{tab:QTC-Corep-Character-Table}
\end{table}

Although the solutions at $r_*=5$ and $r_*=9$ both uniquely identify the same abstract group $S_3^2$, only the $r_*=5$ solution is physically correct, as demonstrated in Appendix~\ref{sec:QTC-Generic-Confirmation}. This is because the $r_*=9$ solution effectively ignores the anti-unitary nature of the $\mathbb{Z}_2$ complex conjugation. As the bootstrap algorithm cannot intrinsically distinguish between these two solutions from the parallel branches, isolating the true physical solution requires verifying analytically whether the conjugation operator acts unitarily or anti-unitarily.

\subsection{Bootstrap Results $(\theta=\f{\pi}{4}\mod\pi)$}
\label{subsec:bootstrap_selfdual}
\begin{figure*}[t]
    \centering
    \includegraphics[width=0.8\linewidth]{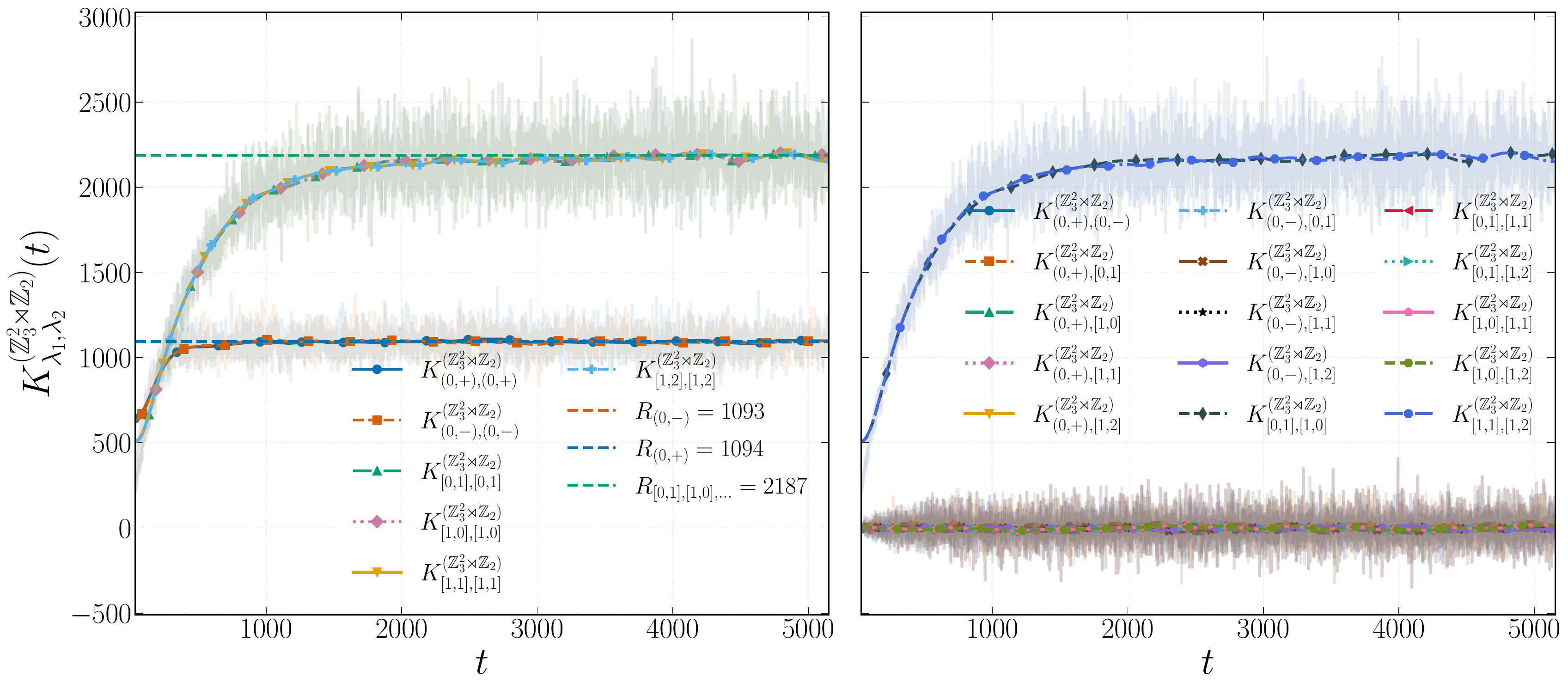}
    \caption{xSFF for the quantum torus chain at the self-dual point with $L = 9$ sites, projected onto $N \cong \mathbb{Z}_3^2 \rtimes \mathbb{Z}_2$ irreps, averaged over $200$ disorder realizations. The bond couplings are drawn as $J_j\sim\mathcal{N}(1,0.25^2)$, with $J_j^x = J_j\cos\theta$ and $J_j^y = J_j\sin\theta$ at $\theta=\pi/4$. \textbf{Left}: The diagonal elements exhibit late-time plateaus that align perfectly with the corresponding benchmark lines $R_{\lambda}$ for all $\lambda\in\hat{N}$. \textbf{Right}: The off-diagonal elements. The paired sectors $[0,1] \leftrightarrow [1,0]$ and $[1,1] \leftrightarrow [1,2]$ exhibit identical xSFF values, while all cross-block correlations vanish.}
    \label{fig:QTC_xSFF}
\end{figure*}
At the self-dual points $\theta = \pi/4 \pmod{\pi}$, the duality transformation $\theta \mapsto \pi/2 - \theta$ manifests itself as an exact symmetry of the system, enlarging the symmetry group. Since the subgroup $N \cong \mathbb{Z}_3^2 \rtimes \mathbb{Z}_2$ is preserved from the generic-$\theta$ regime, the hidden symmetry group $G$ can be conjectured to be either $G \cong \mathbb{Z}_3^2 \rtimes \mathbb{Z}_2^2$ or $G \cong N \times \mathbb{Z}_2$. 
However, the latter guess is immediately ruled out, as it does not impose the observed energy degeneracies across distinct $N$-irreps. In order to correctly resolve the hidden symmetry, we extract the xSFF data via ED and input it into our bootstrap algorithm. Strikingly, the bootstrap constraints explicitly exclude the remaining naive candidate, $\mathbb{Z}_3^2 \rtimes \mathbb{Z}_2^2$. Instead, at the lowest rank $r_*=6$, the algorithm identifies two linear representation theories. One of these yields a branching matrix, fusion rules, and a character table that uniquely pinpoint the symmetry group as $G \cong \mathbb{Z}_3^2 \rtimes \mathbb{Z}_4$. Crucially, in this case, the original subgroup $N$ is no longer a normal subgroup. The algorithm also identifies a distinct corepresentation solution at rank $r_*=4$, which we discard, because our bootstrap algorithm for corepresentations assumes that the subgroup we input is the full unitary normal subgroup. 
This assumption turns out to fail at the self-dual points, where the unitary normal subgroup is $\mathbb{Z}_3^2 \rtimes \mathbb{Z}_4$ instead of the group $\mathbb{Z}_3^2 \rtimes \mathbb{Z}_2$ used as input. Since our primary focus here is the case where $N$ is no longer a normal subgroup, the main text concentrates exclusively on the linear $G \cong \mathbb{Z}_3^2 \rtimes \mathbb{Z}_4$ solution found at $r_*=6$. Comprehensive details of an invalid alternative $r_*=6$ solution are provided in Appendix~\ref{app:Second-Solution-QTC-Self-Dual}.

In order to use ED, we note that the projectors onto $N$-irreps are defined in Eq.~\eqref{eq:Projector-QTC}, which we then substitute into the definition of the xSFF to extract the numerical constraints. By analyzing the plateaus of the xSFF elements in relation to the benchmark lines, we derive the following constraints:
\be
\begin{split}
    &b_{\lambda,\alpha}\in\{0,1\},
    \quad\forall\lambda\in\hat{N},
      \alpha\in\hat{G},\\
    &G_{[0,1][1,0]}=G_{[0,1][0,1]}=G_{[1,0][1,0]}>0,\\
    &G_{[1,1][1,2]}=G_{[1,1][1,1]}=G_{[1,2][1,2]}>0,\\
    &\vec{b}_{[0,1]}=\vec{b}_{[1,0]},
    \quad \vec{b}_{[1,1]}=\vec{b}_{[1,2]}.
\end{split}
\ee
Here, the first constraint is imposed by the observation that $K^{(N)}_{\lambda,\lambda}/R_{\lambda} \approx 1$ for all six $N$-irreps, as shown in the left panel of Fig.~\ref{fig:QTC_xSFF}. The remaining constraints are deduced from the precise overlaps between the plateaus of distinct xSFF elements in both the left and right panels of Fig.~\ref{fig:QTC_xSFF}. By integrating these numerical results with the algebraic constraints detailed in Sec.~\ref{sec:Bootstrap}, the bootstrap algorithm yields a consistent solution at rank $r_*=6$.

The resulting branching matrix is given by
\be\label{eq:QTC-Self-Dual-B-Matrix}
    B_{\lambda,\alpha}=\begin{pmatrix}
        1 & 0 & 1 & 0 & 0 & 0\\
        0 & 1 & 0 & 1 & 0 & 0\\
        0 & 0 & 0 & 0 & 0 & 1\\
        0 & 0 & 0 & 0 & 0 & 1\\
        0 & 0 & 0 & 0 & 1 & 0\\
        0 & 0 & 0 & 0 & 1 & 0\\
    \end{pmatrix}
\ee
where the rows are indexed by the set of $N$-irreps $\{(0,+), (0,-), [0,1], [1,0], [1,1], [1,2]\}$, and the columns correspond to the $G$-irreps $\{\alpha_0, \dots, \alpha_5\}$ ordered from left to right. The third and fourth rows, as well as the fifth and sixth rows, are identical and non-orthogonal. This indicates degeneracies between the $N$-irreps $[0,1]$ and $[1,0]$, and between $[1,1]$ and $[1,2]$, consistent with the spectral plateaus observed in Fig.~\ref{fig:QTC_xSFF}. From the branching matrix, we determine the restrictions of the six $G$-irreps to the subgroup $N$
\begin{equation}\label{eq:QTC-Self-Dual-Restrictions}
\begin{split}
    &\mathrm{Res}^G_N V_{\alpha_0}\cong V_{(0,+)},~
    \mathrm{Res}^G_N V_{\alpha_1}\cong V_{(0,-)},\\
    &\mathrm{Res}^G_N V_{\alpha_2}\cong V_{(0,+)},~
    \mathrm{Res}^G_N V_{\alpha_3}\cong V_{(0,-)},\\
    &\mathrm{Res}^G_N V_{\alpha_4}\cong V_{[1,1]}\oplus V_{[1,2]},\\
    &\mathrm{Res}^G_N V_{\alpha_5}\cong V_{[0,1]}\oplus V_{[1,0]}.
\end{split}
\end{equation}
These restrictions reveal four one-dimensional $G$-irreps, $\{\alpha_0, \alpha_1, \alpha_2, \alpha_3\}$, and two four-dimensional irreps, $\{\alpha_4, \alpha_5\}$. We thus deduce that the group order is $|G| = 4 \times 1^2 + 2 \times 4^2 = 36$, which is precisely the order of the group $G \cong \mathbb{Z}_3^2 \rtimes \mathbb{Z}_4$.

Then, the fusion rules are determined by the obtained non-zero fusion coefficients
\begin{equation}\label{eq:Z3sqZ4-Fusion}
\begin{split}
    (\mathrm{a})~&N_{\alpha_0\alpha}^{\alpha}
    =N_{\alpha\alpha_0}^{\alpha}=1,
\quad  \forall \alpha\in\hat G,\\
    (\mathrm{b})~&N_{\alpha_1\alpha_1}^{\alpha_2}
    =N_{\alpha_1\alpha_2}^{\alpha_3}
    =N_{\alpha_1\alpha_3}^{\alpha_0}=1,\\
    &N_{\alpha_2\alpha_2}^{\alpha_0}
    =N_{\alpha_2\alpha_3}^{\alpha_1}
    =N_{\alpha_3\alpha_3}^{\alpha_2}=1,\\
    (\mathrm{c})~&N_{\alpha_i1\alpha_4}^{\alpha_4}
    =N_{\alpha_2\alpha_4}^{\alpha_4}
    =N_{\alpha_3\alpha_4}^{\alpha_4}=1,\\
    &N_{\alpha_1\alpha_5}^{\alpha_5}
    =N_{\alpha_2\alpha_5}^{\alpha_5}
    =N_{\alpha_3\alpha_5}^{\alpha_5}=1,\\
    (\mathrm{d})~&N_{\alpha_4\alpha_4}^{\alpha_0}
    =N_{\alpha_4\alpha_4}^{\alpha_1}
    =N_{\alpha_4\alpha_4}^{\alpha_2}
    =N_{\alpha_4\alpha_4}^{\alpha_3}
    =N_{\alpha_4\alpha_4}^{\alpha_4}=1,\\
    &N_{\alpha_4\alpha_4}^{\alpha_5}=2,\\
    &N_{\alpha_4\alpha_5}^{\alpha_4}
    =N_{\alpha_4\alpha_5}^{\alpha_5}=2,\\
    &N_{\alpha_5\alpha_5}^{\alpha_0}
    =N_{\alpha_5\alpha_5}^{\alpha_1}
    =N_{\alpha_5\alpha_5}^{\alpha_2}
    =N_{\alpha_5\alpha_5}^{\alpha_3}
    =N_{\alpha_5\alpha_5}^{\alpha_5}=1,\\
    &N_{\alpha_5\alpha_5}^{\alpha_4}=2.
\end{split}
\end{equation}
Here, (a) establishes the identity rules; (b) describes the fusion among the four one-dimensional irreps. These form the invertible sector and constitute a cyclic $\mathbb{Z}_4$ subgroup generated by $\alpha_1$. Furthermore, (c) demonstrates that tensoring either four-dimensional irrep with any invertible object leaves its isomorphism class invariant; finally, (d) specifies the self-fusion and mixed fusion rules for the two four-dimensional non-Abelian sectors. From eq.~\eqref{eq:Z3sqZ4-Fusion}, we define the fusion matrices $(N_{\alpha_i})_{\alpha_j\alpha_k}=N_{\alpha_i\alpha_j}^{\alpha_k}$ with $\alpha_{i},\alpha_{j},\alpha_{k}\in\hat{G}$ and solve their eigenvalues to extract the character tables as shown in Table~\ref{tab:Z3sqZ4_character_table}. This bootstrap solution is verified to be consistent with the linear representation theory of $\mathbb{Z}_3^2\rtimes\mathbb{Z}_4$ in Appendix~\ref{sec:QTC-SD-Analytic-Confirmation}.
\begin{table}[t]
    \centering
    \begin{tabular}{c|cccccc}
         & $c_1$ & $c_2$ & $c_3$ & $c_4$ & $c_5$ & $c_6$ \\
         \hline
         $\chi_{\alpha_0}$ & 1 & 1 & 1 & 1 & 1 & 1 \\
         $\chi_{\alpha_1}$ & 1 & 1 & 1 & -1 & 0 & 0 \\
         $\chi_{\alpha_2}$ & 1 & 1 & 1 & 1 & -1 & -1 \\
         $\chi_{\alpha_3}$ & 1 & 1 & 1 & -1 & 0 & 0 \\
         $\chi_{\alpha_4}$ & 4 & 1 & -2 & 0 & 0 & 0 \\
         $\chi_{\alpha_5}$ & 4 & -2 & 1 & 0 & 0 & 0 \\
         \hline
    \end{tabular}
    \caption{Character table of the hidden symmetry group $G\cong \mathbb{Z}_3^2\rtimes\mathbb{Z}_4$ for the three-site quantum torus chain at self-dual points. The sizes of conjugacy classes $\{c_i|i=1,\cdots,6\}$ are $\{1,4,4,9,9,9\}$ from left to right.}
    \label{tab:Z3sqZ4_character_table}
\end{table}

Specifically, this bootstrap solution at rank $r_*=6$ successfully reconstructs the complete branching structure, fusion algebra, and character table of the group $G \cong \mathbb{Z}_3^2 \rtimes \mathbb{Z}_4$. This guides us to identify the explicit form of $U_H$ as a basis-changing operator, namely the product of local Hadamard gate\footnote{We are grateful to Weiguang Cao for pointing out to us that $U_H$ corresponds to the Hadamard gate, an identification previously established in Ref.~\cite{Cao:2024qjj}.}. Indeed, note that at each site, the local operators are expressed as
\begin{equation}
    Z = \sum_{q=0}^{2} e^{i \frac{2q \pi}{3}} |q\rangle \langle q| ~, \quad (X)^{-1} = \sum_0^2 e^{i \frac{2 q \pi}{3}} |\tilde q\rangle \langle \tilde q|,
\end{equation}
where $U_H$ transforms the $Z$-diagonal basis $|q\rangle$ ($q=0,1,2$) into the $X$-diagonal basis $|\tilde{q}\rangle$. The local matrix representation of $U_H$ is given by
\begin{equation}
    U_H = \frac{1}{\sqrt{3}}(|\tilde 0\rangle\langle0| + |\tilde 1\rangle\langle1| + |\tilde 2\rangle\langle 2|) = \frac{1}{\sqrt{3}}
    \begin{pmatrix}
        1 & 1 & 1 \\
        1 & \omega & \omega^2 \\
        1 & \omega^2 & \omega
    \end{pmatrix},
\end{equation}
which maps $(Z,X)$ to $\left(X^{-1},Z\right)$. Building on this symmetry, we also note that the self-dual transformation $X\leftrightarrow Z$ is implemented by the anti-unitary operator $K U_{H}^3$, since
\be
    (K U_{H}^3) Z (K U_{H}^3)^{-1} = X ,\quad(K U_{H}^3) X (K U_{H}^3)^{-1} = Z,
\ee
where $K$ is the complex conjugation operator identified in Sec.~\ref{eq:QTC-Bootstrap-Off-Sd}.

We conclude this section by discussing the alternative bootstrap solutions found at $r_*=4$ (corepresentation branch) and $r_*=6$ (linear representation branch). The $r_*=4$ solution, while mathematically consistent with Wigner's corepresentation theory for the input subgroup, fails to capture the true hidden symmetry of the system at the self-dual points. As established above, the initial subgroup $N \cong \mathbb{Z}_3^2 \rtimes \mathbb{Z}_2$ is extended by the complex conjugation $K$ and the Hadamard gate $U_H$. The full magnetic group and its unitary subgroup are generated as
\be
\begin{split}
    &G^{\text{mag.}}=\left\langle Z,X,U_H,K \right\rangle\cong N^{\text{uni.}}\sqcup KN^{\text{uni.}},\\
    &N^{\text{uni.}}=\left\langle Z,X,U_H \right\rangle\cong\mathbb{Z}_3^2\rtimes\mathbb{Z}_4.
\end{split}
\ee
From the six $N^{\text{uni.}}$-irreps, we can construct six ICRs, all of which are of Type (a). Thus, these six ICRs correspond one-to-one to the six $N^{\text{uni.}}$-irreps. This bijective relation implies that the restriction maps are trivial, yielding a $6 \times 6$ identity matrix for the branching rules. This algebraic feature is cleanly corroborated by the xSFF numerics, which exhibit a complete absence of off-diagonal plateaus when $N^{\text{uni.}}$ is taken as the known subgroup. 

Finally, we address the alternative linear representation theory found at $r_*=6$. While this spurious solution shares the same branching matrix as Eq.~\eqref{eq:QTC-Self-Dual-B-Matrix}, it deviates in the one-dimensional sector of its fusion rules. Specifically, its one-dimensional irreps fuse according to the Klein four-group $V_4 \cong \mathbb{Z}_2^2$, in contrast to the cyclic $\mathbb{Z}_4$ of the manifest subgroup $\mathbb{Z}_3^2 \rtimes \mathbb{Z}_4$. Furthermore, we can rule out the possibility that this solution describes the group $G \cong \mathbb{Z}_3^2 \rtimes V_4$. The established branching matrix leads to the absence of two-dimensional irrep, which explicitly contradicts the known representation theory of $\mathbb{Z}_3^2 \rtimes V_4$.\footnote{Moreover, as explained in Appendix~\ref{app:Second-Solution-QTC-Self-Dual}, this solution does not constitute a valid linear representation of any finite group, even at the level of category.}

\section{Extension Beyond Finite Groups}
\label{sec:extensions_beyond_finite}
In the previous sections, we bootstrapped the unitary representations of finite groups and the corepresentations of magnetic groups. We now extend our analysis to projective representations and compact Lie groups. Remarkably, projective representations leave distinct signatures on the xSFF, enabling the systematic enumeration of all admissible branching structures. However, because projective representations associated with a fixed cocycle are not closed under the tensor product, extracting their full fusion algebra remains an open algorithmic challenge. Furthermore, the primary obstacle for Lie groups lies in their infinite number of irreps, which renders the direct deduction of branching matrices computationally intractable.

\subsection{Floquet Driven Model and projective representation \label{sec:FD-Bose-Hubbard}}
To make progress with projective representations, we analyze the effective Hamiltonian of the periodically driven Bose-Hubbard model~\cite{Pieplow_2018, Pieplow_2019, 2024JSMTE2024f3104M, creffield2025reachingsachdevyekitaevphysicsshaking}. The previous study~\cite{2024JSMTE2024f3104M} has identified an emergent $\mathbb{Z}_2$ symmetry in the effective Hamiltonian in the high-frequency limit. Interestingly, our analysis not only recovers this symmetry, but also reveals that it combines with the manifest symmetries to form a projective representation.

\subsubsection{Hamiltonian and Symmetries\label{sec:KDBH-HS}}
\begin{figure*}[t]
    \centering
    \includegraphics[width=0.8\linewidth]{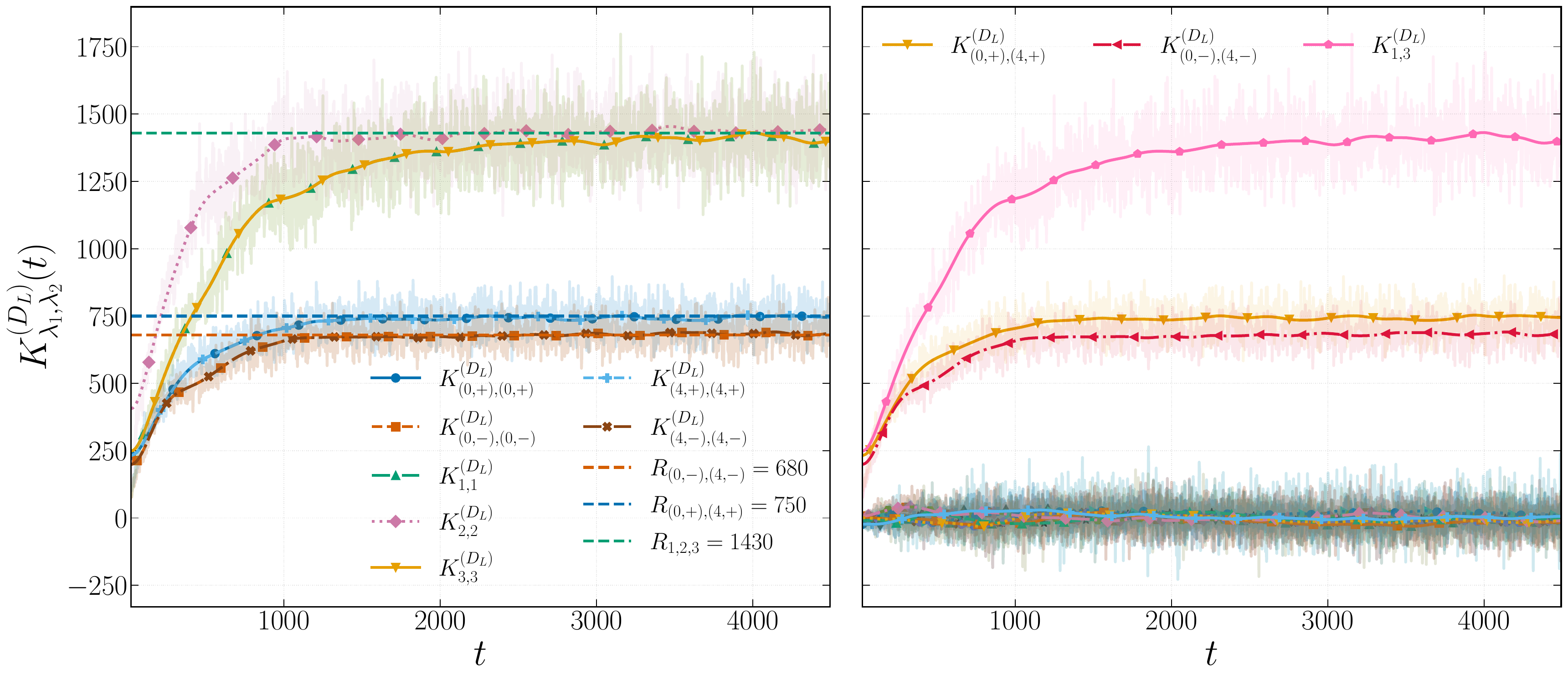}
    \caption{xSFF for the effective Hamiltonian of the driven Bose-Hubbard model with $\mathcal{N}=9$ particles and $L=8$ sites, projected onto $N = D_8$ irreps, averaged over $200$ ensembles. The Hubbard interaction is drawn as $U\sim\mathcal{N}(1,0.25^2)$. \textbf{Left}: The diagonal elements, with $K_{\lambda,\lambda}^{(D_8)}/R_{\lambda}\approx 1$ indicating branching multiplicities in $\{0,1\}$. \textbf{Right}: The off-diagonal elements. The degeneracies $K_{(0,+),(0,+)}^{(D_L)}=K_{(0,+),(4,+)}^{(D_L)}=K_{(4,+),(4,+)}^{(D_L)}$, $K_{(0,-),(0,-)}^{(D_L)}=K_{(0,-),(4,-)}^{(D_L)}=K_{(4,-),(4,-)}^{(D_L)}$, and $K_{1,1}^{(D_L)}=K_{1,3}^{(D_L)}=K_{3,3}^{(D_L)}$ constrain the branching structure. The failure of the fusion rigidity check signals that the hidden symmetry is realized as a projective representation of $D_8\times\mathbb{Z}_2$.}
    \label{fig:DBH}
\end{figure*}
We consider the kinetically driven Bose-Hubbard model~\cite{Pieplow_2018, Pieplow_2019, 2024JSMTE2024f3104M, creffield2025reachingsachdevyekitaevphysicsshaking} as an example of a time-dependent Hamiltonian with non-trivial symmetries. The Hamiltonian of the Bose-Hubbard model can be written as
\begin{equation}
    H_{\mathrm{BH}}=-J \sum_{\langle i, j\rangle}\left(a_i^{\dagger} a_j+\text{h.c.}\right)+\frac{U}{2} \sum_j n_j\left(n_j-1\right),
\end{equation}
with $J$ the hopping amplitude, $U$ is the Hubbard repulsion, which we take to be positive, and $a_i/a_i^\dagger$ are bosonic annihilation and creation operators. We periodically drive the kinetic term as
\begin{equation}
    J(t) = J \cos(\omega t) ~.
\end{equation}
In the high frequency limit, the resulting effective Floquet Hamiltonian reads
\begin{equation}
\begin{aligned}
H_{\mathrm{eff}}= & \frac{U}{2 L} \sum_{\ell, m, n, p=0}^{L-1} \delta_{k_{\ell}+k_m, k_n+k_p} \\
& \times \mathcal{J}_0\left[2 \kappa F\left(k_{\ell}, k_m, k_n, k_p\right)\right] a_{k_p}^{\dagger} a_{k_n}^{\dagger} a_{k_m} a_{k_{\ell}},
\end{aligned}
\end{equation}
where $\kappa= J/\omega < 1$, and $\mathcal{J}_0$ is the zeroth Bessel function of the first kind. In the above equation the function $F$ is given by
\begin{equation}
    F(k_\ell,k_m,k_n,k_p) = \cos(k_\ell) + \cos(k_m) - \cos(k_n) - \cos(k_p) ~.
\end{equation}

In terms of anti-unitary symmetry, the effective Hamiltonian has the time reversal symmetry which is equivalent to the complex conjugation $K a_{k} K^{-1} = a_{-k}$. In terms of unitary symmetries, the model has the obvious $\mathbb{Z}_L$ translation symmetry 
\begin{equation}
    T = \exp\Big\{ i \sum_{n=0}^L \frac{2\pi n}{L} a_{k_n}^\dagger a_{k_n}\Big\} ~,
\end{equation}
the reflection symmetry $\mathbb{Z}_2$ 
\begin{equation}
    \mathcal{P} = \prod_{i=1}^{\lfloor\frac{L-1}{2} \rfloor} S_{k_n,k_{L-n}} ~,
\end{equation}
with $S_{k_i,k_j}$ the SWAP operator between momentum $k_i$ and $k_j$
\begin{equation}
    S_{k_i,k_j} = \exp\Big\{i \frac{\pi}{2} \Big(a_{k_i}^\dagger a_{k_j} + a_{k_j}^\dagger a_{k_i} - a_{k_i}^\dagger a_{k_i} - a_{k_j}^\dagger a_{k_j}\Big)\Big\} ~.
\end{equation} 
The above crystalline symmetries form the $D_L \cong \mathbb{Z}_L \rtimes \mathbb{Z}_2$ group. In addition, the effective Hamiltonian of the driven Bose-Hubbard model also has the $\text{U}(1)$ symmetry for particle number conservation. All these symmetries form the type I magnetic group symmetry
\begin{equation}
    G^{\text{mag.}} \cong (\text{U}(1) \times D_L) \sqcup K (\text{U}(1) \times D_L) ~.
\end{equation}
These symmetries also exist in the original Bose-Hubbard model. For the effective Hamiltonian of the driven Bose-Hubbard model, the model has the well-known hidden $\pi$-reflection $\mathbb{Z}_2$ symmetry $R_{\pi}: a_k \rightarrow a_{\pi-k}$ due to the structure of $F(k_\ell,k_m,k_n,k_p)$: since $\cos(k) + \cos(\pi - k) = 0$, the function $F$ is odd under $k_j \to \pi - k_j$, and the even parity of $\mathcal{J}_0$ ensures the invariance of $H_\mathrm{eff}$ under $R_{\pi}$. We emphasize that this symmetry is emergent---it is absent in the non-driven model and arises from the structure of the high-frequency Floquet Hamiltonian.

Under $R_{\pi}$, the total momentum $Q = \sum_{j=0}^{L-1} j n_{k_j} \pmod{L}$ transforms as
\be\label{eq:KDBH-Q-transform}
    Q \;\rightarrow\; \f{\mathcal{N}L}{2} - Q \pmod{L},
\ee
where $\mathcal{N} = \sum_j n_j$ is the total particle number. The interplay between $R_{\pi}$ and the translation operator is governed by the conjugation relation
\be\label{eq:KDBH-Rpi-T-commutation}
    R_{\pi} T R_{\pi}^{-1} = (-1)^{\hat{\mathcal{N}}} T^{-1},
\ee
where $\hat{\mathcal{N}} = \sum_j n_j$ is the particle number operator. The factor $(-1)^{\hat{\mathcal{N}}}$ arises because each of the $\mathcal{N}$ particles acquires a phase $e^{i\pi} = -1$ under the momentum shift $k \to \pi - k$. Meanwhile, $R_{\pi}$ commutes with both $\mathcal{P}$ and $\text{U}(1)$.
Eq.~\eqref{eq:KDBH-Rpi-T-commutation} shows that the hidden symmetry group is \emph{not} a direct product $D_L \times \mathbb{Z}_2^{(\pi)}$ in general, since the commutation of $R_{\pi}$ with $T$ involves the $\text{U}(1)$ element $(-1)^{\hat{N}}$. The structure depends on the particle number sector:
\begin{itemize}
    \item For $\mathcal{N}$ even, $(-1)^{\mathcal{N}} = 1$ and Eq.~\eqref{eq:KDBH-Rpi-T-commutation} reduces to $R_{\pi} T R_{\pi}^{-1} = T^{-1}$, the same relation satisfied by $\mathcal{P}$. The composition $R_{\pi} \mathcal{P}$ then commutes with all elements of $D_L$, generating a central $\mathbb{Z}_2$, so that the crystalline symmetry extends to $D_L \times \mathbb{Z}_2^{(\pi)}$, with $D_L = \langle T,\mathcal{P}\rangle, \mathbb{Z}_2^{(\pi)} = \langle R_{\pi} \mathcal{P}\rangle $.
    
    \item For $\mathcal{N}$ odd, $(-1)^{\mathcal{N}} = -1$ and the relation becomes $R_{\pi} T R_{\pi}^{-1} = -T^{-1}$. We can also consider the combination of $R_{\pi} \mathcal{P}$ which has the following property $(R_{\pi} \mathcal{P}) T (R_{\pi} \mathcal{P})^{-1} = - T$. This is a projective representation of $D_L \times \mathbb{Z}_2^{(\pi)}$, and $R_{\pi}$ maps momentum $Q \to L/2 - Q$ from Eq.~\eqref{eq:KDBH-Q-transform}.
\end{itemize}
In full generality, the full symmetry group of the effective Hamiltonian is a magnetic group 
\begin{equation}
    G^{\text{mag.}}_{\rm eff} \cong (\text{U}(1) \times D_L \times \mathbb{Z}_2) \sqcup K(\text{U}(1) \times D_L \times \mathbb{Z}_2) ~.
\end{equation}
When ${\mathcal{N}}$ is even, the unitary part is realized as linear representations, while, for $\mathcal{N}$ odd, it is realized via projective representations.

\subsubsection{xSFF and constraints on branching multiplicities}
Even though the projective representation is not closed under the tensor product, we are going to show that xSFF is still powerful in terms of finding the constraints on the branching multiplicities. To make the discussion easier, we restrict ourself in the case with $\mathcal{N}=9, L =8$. We numerically evaluate the xSFF and show the results in Fig.~\ref{fig:DBH}. From the left panel, we observe that the plateau value of the diagonal piece in the of xSFF satisfies $K_{\lambda,\lambda}^{(D_8)} / R_{\lambda} \approx 1$, which means that $b_{\lambda,\alpha}\in\{0,1\}, \forall \lambda \in \hat{D_8}, \alpha \in \hat{G}$. From the right panel, we further observe that
\begin{equation}
    \begin{gathered}
K^{(D_8)}_{(0,+),(0,+)}=K^{(D_8)}_{(0,+),(4,+)}=K^{(D_8)}_{(4,+),(4,+)}, \\
K^{(D_8)}_{(0,-),(0,-)}=K_{(0,-),(4,-)}^{(D_8)}=K^{(D_8)}_{(4,-),(4,-)}, \\
K^{(D_8)}_{1,1}=K^{(D_8)}_{1,3}=K^{(D_8)}_{3,3}.
\end{gathered}
\end{equation}

Based on the bootstrap algorithm we discussed in Sec. \ref{sec:Bootstrap-Algorithm}, we find all the possible branch column vectors are
\begin{equation}
    \begin{aligned}
    \label{eq:proj_branch}
& \vec{b}_{\alpha_1}=(1,0,0,0,0,1,0)^T, \\
& \vec{b}_{\alpha_2}=(0,1,0,0,0,0,1)^T, \\
& \vec{b}_{\alpha_3}=(0,0,1,0,1,0,0)^T, \\
& \vec{b}_{\alpha_4}=(0,0,0,1,0,0,0)^T ,
\end{aligned}
\end{equation}
with the order $\{(0,+),(0,-),1,2,3,(4,+),(4,-)\}$ in $D_8$ irreps. However, none of the candidate fusion coefficients passes the rigidity check. This strongly suggests that the irreps reconstructed from the hidden symmetry do not form the fusion rule of ordinary linear representations of a finite group. We therefore interpret the hidden symmetry as a projective representation. As a consistency check, we compare the obtained branching multiplicities with the analytical construction of the projective representation of $D_8 \times \mathbb{Z}_2$ given in Appendix~\ref{app:Proj-D8Z2}. The relevant fixed-cocycle sector contains five irreducible projective represents $\{\mathcal{X}_+, \mathcal{X}_-, \mathcal{Y}_+,\mathcal{Y}_-,\mathcal{Z}\}$ with the following branching relations:
\begin{equation}
\begin{aligned}
    {\rm Res}_{N}^{G} \mathcal{X}_+ & = (0,+) \oplus (4,+), \quad {\rm Res}_N^G \mathcal{X}_- = (0,-) \oplus (4,-), \\
    {\rm Res}_N^G \mathcal{Y}_+ & = 2, \quad {\rm Res}_{N}^G \mathcal{Y}_- = 2, \quad {\rm Res}_N^G \mathcal{Z} = 1 \oplus 3.
\end{aligned}
\end{equation}
Comparing with Eq.~\eqref{eq:proj_branch}, we observe that the branching vectors $\vec{b}_{\alpha_1}, \vec{b}_{\alpha_2},$ and $\vec{b}_{\alpha_3}$ appear once, while $\vec{b}_{\alpha_4}$ appears twice, corresponding to $\mathcal{Y}_+$ and $\mathcal{Y}_-$. Since the tensor unit $\mathbf{1}$ is absent from this fixed-cocycle sector, these five projective objects do not form a fusion category by themselves.

\subsection{Extensions to Lie Groups\label{sec:Lie-Group-Case}}
We now consider a conceptually distinct case, where the hidden symmetry group $G$ is a compact Lie group. We study a specific example featuring the hidden symmetry group $G \cong [{\rm SU}(2) \times {\rm SU}(2)] / \mathbb{Z}_2 \cong {\rm SO}(4)$, the underlying group structure can still be reliably inferred directly from the numerical xSFF data.

\subsubsection{Fermi-Hubbard Model and $\eta$-paring \label{sec:Fermi-Hubbard}}
\begin{figure*}[t]
    \centering
    \includegraphics[width=0.8\linewidth]{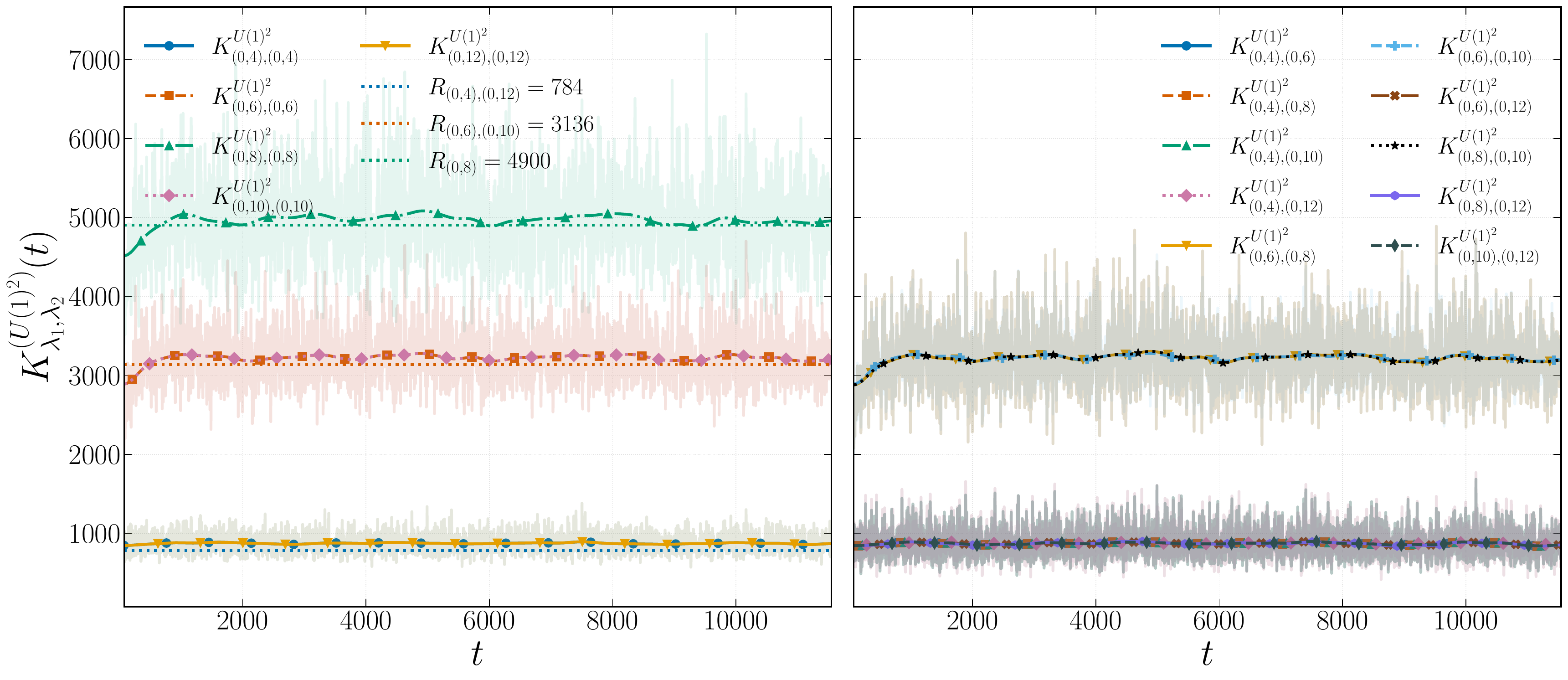}
    \caption{xSFF for the Fermi-Hubbard model with $L = 8$ sites, $U = 4$, averaged over $200$ disorder realizations with hopping $t_j \sim \mathcal{N}(3, 0.5^2)$, projected onto the $\text{U}(1)_{S^z} \times \text{U}(1)_{\mathcal{N}}$ sectors with $\mathcal{N} = 4, \ldots, 12$ (charge $Q = \mathcal{N} - L = -4, \ldots, +4$). In the figure, we only show the xSFF with $S^z=0$, i.e. $\mathcal{N}_{\uparrow} = \mathcal{N}_{\downarrow} = \mathcal{N}/2$. \textbf{Left}: Diagonal elements $K_{(0,\mathcal{N}),(0,\mathcal{N})}^{(\text{U}(1)^2)}$ for sectors near half-filling. \textbf{Right}: Off-diagonal elements $K_{(0,\mathcal{N}_1),(0,\mathcal{N}_2)}^{(\text{U}(1)^2)}$ emanating from the half-filling sector $\mathcal{N}_{\mathrm{half}} = L = 8$. Solid lines denote $|\Delta \mathcal{N}|$ even (predicted non-zero by $\mathrm{SU}(2)$ $\eta$-pairing); dashed lines denote $|\Delta \mathcal{N}|$ odd (predicted to vanish). The clear separation between even and odd $\Delta \mathcal{N}$ confirms the selection rule $\Delta Q \in 2\mathbb{Z}$ characteristic of $\mathrm{SU}(2)$.}
    \label{fig:Hubbard_xSFF}
\end{figure*}
We illustrate our approach using the one-dimensional Fermi-Hubbard model, whose Hamiltonian is
\be\label{eq:FH-Hamiltonian}
\begin{aligned}
    H & = -\sum_{j=1}^{L-1} t_j \sum_{\sigma=\uparrow,\downarrow} \left(c_{j,\sigma}^{\dagger} c_{j+1,\sigma} + \text{h.c.}\right) \\
    & \quad + U \sum_{j=1}^{L} n_{j,\uparrow} n_{j,\downarrow} - \f{U}{2}\sum_{j=1}^{L} n_j,
\end{aligned}
\ee
where $c_{j,\sigma}^{\dagger}$ and $c_{j,\sigma}$ are fermionic creation and annihilation operators at site $j$ with spin $\sigma \in \{\uparrow,\downarrow\}$, $n_{j,\sigma} = c_{j,\sigma}^{\dagger}c_{j,\sigma}$, $n_j = n_{j,\uparrow} + n_{j,\downarrow}$, and $t_j$ denotes nearest-neighbor hopping amplitude.

The model has two manifest unitary symmetries. First of all, it has the ${\rm SU}(2)_{\rm spin}$ symmetry with the symmetry generator 
\begin{equation}
    S^a=\frac{1}{2} \sum_i c_i^{\dagger} \sigma^a c_i
    \quad \text{with} \quad c_i=\binom{c_{i, \uparrow}}{c_{i, \downarrow}} ~.
\end{equation}
In particular, $S^z = (\mathcal{N}_\uparrow - \mathcal{N}_\downarrow)/2$ counts the difference between spin-up and spin-down fermion numbers. Another manifest symmetry is the U$(1)$ symmetry corresponding to total particle number conservation, generated by $\hat{\mathcal{N}} = \mathcal{N}_\uparrow + \mathcal{N}_\downarrow$. The model also has the anti-unitary time reversal symmetry $\mathcal{T} = e^{i\pi S^y} K$, where $K$ is the complex conjugation. In the Fermi-Hubbard model, $\mathcal{T}^2 = (-1)^{\mathcal{N}}$.

The hidden symmetry of the Fermi-Hubbard model is the celebrated $\eta$-pairing $\mathrm{SU}(2)_\eta$ algebra, discovered by C. N. Yang~\cite{Yang:1989hms}. The generators are
\be\label{eq:FH-eta-generators}
\begin{split}
    &\eta^{+} = \sum_{j=1}^{L} (-1)^j c_{j,\uparrow}^{\dagger} c_{j,\downarrow}^{\dagger}, \quad \eta^{-} = (\eta^{+})^{\dagger},\\
    &\eta_z = \f{\hat{\mathcal{N}} - L}{2},
\end{split}
\ee
which satisfy the $\mathfrak{su}(2)$ commutation relations $[\eta^{+}, \eta^{-}] = 2\eta_z$ and $[\eta_z, \eta^{\pm}] = \pm \eta^{\pm}$. Crucially, the operators $\eta^{\pm}$ commute with $H$ for any choice of the hopping amplitudes $t_j$, since the sublattice phase factor $(-1)^j$ ensures that the kinetic and interaction terms are separately invariant under the $\eta$-pairing algebra. This makes the $\eta$-pairing $\mathrm{SU}(2)_\eta$ a prototypical example of a hidden symmetry whose generators cannot be read off from the Hamiltonian.

\subsubsection{xSFF, branching multiplicities and bootstrap}
Now, let us apply our bootstrap algorithm to infer the branching multiplicity structure and infer the existence of the hidden ${\rm SU}(2)_\eta$ symmetry. We choose $N \cong \text{U}(1)_{S^z} \times \text{U}(1)_{\mathcal{N}}$ as a subgroup for our obvious symmetry. We evaluate the xSFF defined in Eq.~\eqref{eq:xSFF-Elements} and show the results in Fig.~\ref{fig:Hubbard_xSFF}.

We mainly focus on the $S^{z}=0$ sector. From the left panel, we observe that $K_{(0,\mathcal{N}),(0,\mathcal{N})}^{(N)}/R_{(0,\mathcal{N})}\approx 1$, which indicates that $b_{(0,\mathcal{N}),\alpha}\in \{0,1\}, \forall \alpha \in \hat{G}$. In fact, one can also calculate the xSFF for general $S^z$ charge and arrive at the conclusion 
\begin{equation}
b_{(S^z,\mathcal{N}), \alpha } \in \{0,1\}~, \quad \forall S^z, \mathcal{N}, \alpha \in \hat{G} ~.
\label{eq:Hubbard_b_constraint}
\end{equation}
From the right panel of Fig.~\ref{fig:Hubbard_xSFF}, we observe that $K_{(0,\mathcal{N}_1),(0,\mathcal{N}_2)}^{(N)} \gg 0$ once $(\mathcal{N}_1 - \mathcal{N}_2)\in 2\mathbb{Z}$. This conclusion is also true for general $S^z$ charge. 

In this particular example, we can straightforwardly bootstrap the hidden symmetry by simple mathematical arguments. Given the constraints of branching multiplicity in Eq.~\eqref{eq:Hubbard_b_constraint} and the assumption that the hidden symmetry $G$ is a connected compact Lie group whose maximal torus is
$T\cong\mathrm{U}(1)\times\mathrm{U}(1)$, i.e.\ $\mathrm{rank}(G)=2$, we are able to fix $G\cong \left[{\rm SU}(2) \times {\rm SU}(2)\right]/\mathbb{Z}_2 \cong {\rm SO}(4)$.
First of all, we can write the Lie algebra as $\mathfrak{g}=\mathfrak{g}_{\mathrm{ss}}\oplus\mathfrak{z}$,
where $\mathfrak{g}_{\mathrm{ss}}$ is semisimple and $\mathfrak{z}$ is the center,
so that $\mathrm{rank}(\mathfrak{g}_{\mathrm{ss}})+\dim(\mathfrak{z})=2$.
We claim that every simple factor of $\mathfrak{g}_{\mathrm{ss}}$ must have rank~$1$.
Indeed, consider a simple factor $\mathfrak{s}$ of rank $r_{\mathfrak{s}}\geq 2$.
The adjoint representation of $\mathfrak{s}$ has a zero-weight space equal
to the Cartan subalgebra $\mathfrak{h}\subset\mathfrak{s}$, which has
dimension~$r_{\mathfrak{h}}$.
Therefore the zero weight has multiplicity $r_{\mathfrak{h}}\geq 2$, contradicting
the assumption that all branching multiplicities are at most~$1$.
This powerful constraint rules out every rank-$2$ simple Lie algebra:
$A_2$ ($\mathfrak{su}(3)$),
$B_2\cong C_2$ ($\mathfrak{so}(5)\cong\mathfrak{sp}(4)$),
and $G_2$. We give more detailed discussions in Appendix~\ref{app:adjoint_rep}.

Since the only rank-$1$ simple compact Lie algebra is $\mathfrak{su}(2)$,
three cases remain:
(i)~$\mathfrak{g}\cong\mathfrak{su}(2)\oplus\mathfrak{su}(2)$,
(ii)~$\mathfrak{g}\cong\mathfrak{su}(2)\oplus\mathfrak{u}(1)$, and
(iii)~$\mathfrak{g}\cong\mathfrak{u}(1)\oplus\mathfrak{u}(1)$. From the non-trivial overlap between $(S^z,\mathcal{N}_1),(S^z,\mathcal{N}_2)$ shown in the right panel of Fig.~\ref{fig:Hubbard_xSFF}, we immediately rule out case (ii) and (iii).
In case~(i), the global form of $G$ depends on which subgroup of the center
$\mathbb{Z}_2\times\mathbb{Z}_2$ of the universal cover
$\mathrm{SU}(2)\times\mathrm{SU}(2)$ is modded out, yielding five possibilities:
\begin{equation}\label{eq:rank2-classification}
\begin{aligned}
    & \mathrm{SU}(2)\times\mathrm{SU}(2),\quad
    \mathrm{SO}(3)\times\mathrm{SU}(2),\quad \\
    & \mathrm{SU}(2)\times\mathrm{SO}(3),\quad
    \mathrm{SO}(3)\times\mathrm{SO}(3),\\
    &\mathrm{SO}(4)\cong \left(\mathrm{SU}(2)\times\mathrm{SU}(2)\right)/\mathbb{Z}_2.
\end{aligned}
\end{equation}
These are distinguished by which representations $(j_1,j_2)$ appear in the
spectrum:
$\mathrm{SU}(2)\times\mathrm{SU}(2)$ permits all half-integer and integer spins,
$\mathrm{SO}(4)$ requires $j_1+j_2\in\mathbb{Z}$,
and the $\mathrm{SO}(3)$ quotients restrict the corresponding factor to
integer spins only.
For the spin-$\tfrac{1}{2}$ Fermi-Hubbard model at half-filling chemical potential, both $S^z$ and $\mathcal{N}/2$ can be integer or half-integer. Therefore, we have both integer and half-integer values of $j_s$ and $j_\eta$ appear in
the spectrum. Moreover, we have the constraint $S^z + \mathcal{N}/2 \in \mathbb{Z}$ which means that $j_1 + j_2 \in \mathbb{Z}$. As a result, we uniquely select
$G\cong \left({\rm SU(2)}_{\rm spin} \times {\rm SU(2)}_\eta\right) / \mathbb{Z}_2 \cong {\rm SO}(4)$.

\section{Protocol for Probing the xSFF\label{sec:Experiment}}

In this section, we propose to probe the xSFF using state-of-the-art quantum simulation experiments. Our scheme is inspired by recent experimental realizations of the SFF~\cite{PhysRevLett.134.010402}, based on the random measurement toolbox~\cite{PhysRevX.12.011018, Elben_2022}. 
Let us first summarize this scheme for the standard SFF. In order to measure the latter quantity, one needs to evaluate the product of traces $\langle\text{tr}[U(t)]\text{tr}[U^{\dagger}(t)]\rangle$ from \textit{single-shot} experiments. To do so, one indirectly samples the trace of the unitary operator by evolving many random product states and studying their return probability~\cite{PhysRevX.12.011018}.
Concretely, one first initializes the product state $|0\rangle^{\otimes L}$, and applies random single-qubit gates on each qubit, i.e., $V= \bigotimes_{i=1}^L v_i$, with each $v_i$ drawn from local unitary 2-design Clifford gate. One then applies the time evolution operator $U(t)$ and obtains the return probability by acting with $V^{\dagger}$ and measuring the resulting bit-string in the computational basis. In other words, one prepares the state
\begin{equation}
V^{\dagger}U(t)V|0\rangle^{\otimes L},
\end{equation}
and measure it back in the computational basis to get some string $(s_1,\cdots s_L)$, with $s_i\in\{0,1\}$. This quantity needs to be averaged over realizations of the random unitary gates $V$, as well as over disorder realizations in order to estimate $\langle\text{tr}[U(t)]\text{tr}[U^{\dagger}(t)]\rangle$. In fact, define the average over $M$ shots as
\begin{equation}
\hat{K}(t) = \frac{1}{M}\sum_{r=1}^M(-2)^{-|s^{(r)}|},
\label{eq:estimator}
\end{equation}
where $|s^{(r)}|=\sum_js_j^{(r)}$. With the unnormalized trace convention in Eq.~\eqref{eq:Def-SFF}, this is an unbiased estimator of $K(t)/4^L$~\cite{PhysRevX.12.011018}; multiplying by $4^L$ recovers $K(t)$. The SFF was recently measured using this procedure for both continuous Hamiltonian evolution and Floquet circuits in superconducting quantum simulators~\cite{PhysRevLett.134.010402}. 
The moderate system sizes used there (about 5 qubits) made it possible to study the time evolution of the SFF beyond the Heisenberg time, which scales exponentially with the number of qubits, and to extract approximate plateau values.

We further develop the experimental protocol by combining linear combinations of unitaries, local random unitaries, and randomized measurements. For a finite unitary subgroup $N$, the sector projector is
\begin{equation}
P_\lambda=\frac{d_\lambda}{|N|}\sum_{n\in N}\chi_\lambda(n)^*U_n,
\label{eq:experiment-sector-projector}
\end{equation}
where $\chi_\lambda$ is the irreducible character. Defining $W_n(t)=U_nU(t)$ and
\begin{equation}
C_{nm}(t)=\left\langle\tr W_n(t)[\tr W_m(t)]^*\right\rangle,
\label{eq:experiment-trace-correlation}
\end{equation}
the character sum reconstructs the xSFF:
\begin{equation}
\begin{aligned}
&K_{\lambda_a,\lambda_b}^{(N)}(t)\\
&\quad=\frac{1}{|N|^2}\sum_{n,m\in N}
\Re\left[\chi_{\lambda_a}(n)^*\chi_{\lambda_b}(m)C_{nm}(t)\right].
\end{aligned}
\label{eq:xSFF-Exp}
\end{equation}

As seen in Eq.~\eqref{eq:xSFF-Exp}, the key quantities to measure are $\Re C_{nm}(t)$ and $\Im C_{nm}(t)$. The polarization identities express them as
\begin{equation}
\begin{aligned}
\Re C_{nm}(t)
&=\frac14\Big\langle\big|\tr[(U_n+U_m)U(t)]\big|^2\\
&\quad-\big|\tr[(U_n-U_m)U(t)]\big|^2\Big\rangle,\\
\Im C_{nm}(t)
&=\frac14\Big\langle\big|\tr[(U_n+iU_m)U(t)]\big|^2\\
&\quad-\big|\tr[(U_n-iU_m)U(t)]\big|^2\Big\rangle.
\end{aligned}
\label{eq:experiment-polarization}
\end{equation}
The standard randomized-measurement protocol~\cite{PhysRevX.12.011018} applies to unitary operators, whereas $U_n+e^{i\theta}U_m$ is generally not unitary. Motivated by the linear-combination-of-unitaries construction~\cite{Childs:2012gwh}, we introduce an ancilla and define a unitary on the enlarged Hilbert space:
\begin{equation}
\begin{aligned}
\mathcal W_\theta(t)
&=|0\rangle\langle0|_A\otimes U_nU(t)\\
&\quad+e^{i\theta}|1\rangle\langle1|_A\otimes U_mU(t)\\
&=\begin{pmatrix}
U_nU(t)&0\\
0&e^{i\theta}U_mU(t)
\end{pmatrix}.
\end{aligned}
\label{eq:experiment-ancilla-unitary}
\end{equation}
Both diagonal blocks are unitary, and its trace satisfies
\begin{equation}
\begin{aligned}
\tr_{AS}\mathcal W_\theta(t)
&=\tr[U_nU(t)]+e^{i\theta}\tr[U_mU(t)]\\
&=\tr[(U_n+e^{i\theta}U_m)U(t)].
\end{aligned}
\label{eq:experiment-trace-identity}
\end{equation}
Thus every trace square in Eq.~\eqref{eq:experiment-polarization} can be measured as the SFF of $\mathcal W_\theta(t)$.

For the manifest symmetries, we assume that $U_n$ and $U_m$ are products of local unitaries. As shown in Fig.~\ref{fig:xSFF-experimental-circuit}, $\mathcal W_\theta(t)$ is implemented by $U_nU(t)$, the ancilla phase gate $R_A(\theta)=\operatorname{diag}(1,e^{i\theta})$, and a subsequent controlled operation $U_{mn^{-1}}=U_mU_n^\dagger$. This controlled operation decomposes into controlled local gates sharing the same ancilla.

\begin{figure}[t]
\centering
\definecolor{xsffCircuitTeal}{HTML}{176980}
\definecolor{xsffCircuitGold}{HTML}{A97812}
\definecolor{xsffCircuitPlum}{HTML}{85366D}
\definecolor{xsffCircuitWire}{HTML}{45545B}
\resizebox{\linewidth}{!}{%
\begin{tikzpicture}[x=1cm,y=1cm,font=\small,line width=0.6pt,
  xsffgate/.style={draw=xsffCircuitTeal,fill=xsffCircuitTeal!9,
    minimum width=1.20cm,minimum height=0.60cm,inner sep=2pt},
  xsffwire/.style={draw=xsffCircuitWire,line width=0.7pt}]
  \draw[xsffwire] (0.20,1.05)--(8.05,1.05);
  \draw[xsffwire] (0.20,0)--(8.05,0);
  \node[anchor=east] at (0.06,1.05) {$A$};
  \node[anchor=east] at (0.06,0) {$S$};
  \node[xsffgate] at (1.00,1.05) {$v_A$};
  \node[xsffgate] at (1.00,0) {$V_r^{(S)}$};
  \node[xsffgate] at (2.70,1.05) {$R_A(\theta)$};
  \node[xsffgate,draw=xsffCircuitGold,fill=xsffCircuitGold!10]
    at (2.70,0) {$U_nU(t)$};
  \draw[draw=xsffCircuitPlum,line width=0.8pt] (4.45,1.05)--(4.45,0.30);
  \fill[xsffCircuitPlum] (4.45,1.05) circle (0.055);
  \node[xsffgate,draw=xsffCircuitPlum,fill=xsffCircuitPlum!10]
    at (4.45,0) {$U_{mn^{-1}}$};
  \node[xsffgate] at (6.15,1.05) {$v_A^\dagger$};
  \node[xsffgate] at (6.15,0) {$(V_r^{(S)})^\dagger$};
  \foreach \y in {0,1.05}{
    \draw[draw=xsffCircuitWire,fill=white]
      (7.45,\y-0.25) rectangle (8.00,\y+0.25);
    \draw[draw=xsffCircuitWire]
      (7.53,\y-0.10) arc[start angle=180,end angle=0,x radius=0.19,y radius=0.20];
    \draw[draw=xsffCircuitWire,-{Latex[length=1.2mm]}]
      (7.72,\y-0.10)--(7.86,\y+0.14);
  }
\end{tikzpicture}%
}
\caption{{Ancilla-assisted measurement of the xSFF. Independent local random gates $v_A$ and $V_r^{(S)}$, followed by their inverses, surround the joint unitary $\mathcal W_\theta(t)$. The ancilla controls $U_{mn^{-1}}=U_mU_n^\dagger$. Measuring both registers for $\theta=0,\pi,\pi/2,3\pi/2$ recovers the trace correlations through Eq.~\eqref{eq:experiment-phase-scan}.}}
\label{fig:xSFF-experimental-circuit}
\end{figure}

{We now apply the standard randomized-measurement protocol to all $L+1$ qubits. Starting from $|0\rangle_A\otimes|0\rangle_S^{\otimes L}$, we apply independent local 2-design gates, evolve with $\mathcal W_\theta(t)$, undo the random gates, and measure in the computational basis. Including the ancilla outcome in the Hamming weight, the estimator in Eq.~\eqref{eq:estimator} obeys
\begin{equation}
\begin{aligned}
\mathbb E[\hat K_\theta(t)]&=\frac{F_\theta(t)}{4^{L+1}},\quad F_\theta(t)=\left\langle\left|\tr_{AS}\mathcal W_\theta(t)\right|^2\right\rangle.
\end{aligned}
\label{eq:experiment-joint-SFF}
\end{equation}
Equations~\eqref{eq:experiment-polarization} and~\eqref{eq:experiment-trace-identity} then give
\begin{equation}
\begin{aligned}
\Re C_{nm}(t)&=\frac{F_0(t)-F_\pi(t)}{4},\\
\Im C_{nm}(t)&=\frac{F_{\pi/2}(t)-F_{3\pi/2}(t)}{4}.
\end{aligned}
\label{eq:experiment-phase-scan}
\end{equation}
Repeating these four phase settings for the required pairs $(n,m)$ supplies the correlations needed to reconstruct the xSFF through Eq.~\eqref{eq:xSFF-Exp}.}

{Although late-time decoherence can limit plateau measurements~\cite{PhysRevLett.134.010402}, an observable nonzero off-diagonal xSFF plateau would provide an experimental signature of hidden symmetry.}

\section{Summary and Outlook\label{sec:Summary-Outlook}}

\subsection{Summary}
In this work, we developed a systematic bootstrap framework for reconstructing the representation-theoretic data of finite symmetry groups for lattice models, including the number and dimensions of their irreps, the branching multiplicities compatible with known subgroups, the fusion rules, and the complete character table. 
The core idea is to combine the input from a known symmetry subgroup $N$ with algebraic consistency conditions on the branching structure and fusion algebra of the full symmetry $G$. Moreover, we complemented these algebraic constraints with numerical conditions on branching multiplicities extracted from the xSFF, as a generalization of the SFF. By integrating these algebraic and numerical inputs, our algorithm systematically outputs the total number and dimensions of the $G$-irreps, their branching multiplicities, the fusion rules, and the complete character table. Crucially, we demonstrated that this dataset unambiguously identifies $G$ in all of the examples we investigated.

We bootstrapped a sequence of finite symmetry group in various examples of quantum many-body lattice models.\footnote{Note that the bootstrap method requires only spectral data, making it applicable regardless of the system's spatial dimension. However, for simplicity, we consider only one-dimensional systems in this work.} 
Starting with an $S_3$-invariant chain featuring only a $\mathbb{Z}_3$ subgroup, the bootstrap procedure correctly reconstructed the hidden $S_3$ representation-theoretic data. We further demonstrated that the method remains efficient for models with a symmetry hidden by non-local Kennedy-Tasaki transformation, where it could correctly identify $D_4$ from a manifest subgroup {$V_4$}. We demonstrated that the bootstrap works in systems where branching multiplicities exceed one, moving beyond the multiplicity-free cases of the previous examples. 
Finally, we analyzed the three-state quantum torus chain, for which we confirmed that the bootstrap remains highly efficient even when the model involves non-commuting anti-unitary symmetry, or when the known subgroup is non-normal.  

Although a systematic bootstrap for projective representations and compact Lie groups is not yet complete, the xSFF provides powerful constraints for inferring hidden symmetry structures. For the Floquet-driven Bose-Hubbard model, the bootstrap yielded the branching vector types that pins down the emergent symmetry in the high-frequency limit.
Furthermore, the sensitivity of the xSFF to projective representations demonstrates that its diagnostic power extends well beyond ordinary linear or anti-linear representations. In the one-dimensional Fermi-Hubbard model, the xSFF imposes stringent constraints that uniquely identify the $\mathrm{SO}(4)$ symmetry underlying the paradigmatic $\eta$-pairing problem. 

{Moving beyond the plateau constraints, we demonstrate the recovery of $S_3$ branching and fusion data directly from the universal xSFF ramp without relying on plateau inputs, and we probe for additional hidden Abelian symmetries as detailed in Appendix~\ref{app:xSFF-ramp}. Furthermore, we apply the xSFF to a spin-$1$ chain to perform a finite-size consistency test for a hidden conserved order parameter mechanism of symmetry breaking, which is discussed in Appendix~\ref{sec:xSFF-order-parameter}. Finally, we introduce an explicit ancilla-assisted circuit and establish exact reconstruction identities for measuring both diagonal and off-diagonal xSFF components in Sec.~\ref{sec:Experiment}.}

\subsection{Outlook}
We close by highlighting several promising future research directions:\\

\textit{Generalized Symmetries}.
Although the present framework is formulated within Wigner's traditional paradigm, the mathematical foundation of fusion and branching is deeply rooted in category theory. Consequently, a natural future direction will consist in generalizing our bootstrap method to accommodate generalized symmetry structures in lattice models, including higher-form~\cite{Feng:2025qgg} and non-invertible symmetries~\cite{PhysRevB.111.054432, 10.21468/SciPostPhys.17.4.115}. \\

\textit{Higher-moments of xSFF}.
The higher-moments of the SFF have been studied extensively~\cite{n7rj-gwwj,Gupta:2025prv,Sohail:2025naw,Charamis:2025uly} in recent years. As the xSFF can be seen as a generalization of the SFF, its higher-moments are expected to encode more physical information.
Investigating such higher-moments could lead to more bootstrap constraints, which could potentially enable the extraction of $F$-symbols.\\

\textit{Tannakian duality and the complete reconstruction of $G$.} Our approach currently lacks sufficient constraints in order to bootstrap the \textit{complete} algebraic data of the category $\text{Rep}(G)$ and thereby uniquely reconstruct the group $G$ via Tannakian duality~\cite{DeligneMilneOgusShih1982}. Whether additional conditions can be imposed to fully reconstruct $\text{Rep}(G)$ thus remains an open question.\\

\textit{Alternative observables}. 
Although the framework presented in this paper was implemented using the xSFF, the bootstrap itself does not need, in principle, to rely on this quantity. What it fundamentally requires are observables that provide sufficiently strong constraints on the branching structure.
It would, therefore, be interesting to explore other diagnostics that can encode such information and serve as alternative inputs to the bootstrap procedure.\\ 

\textit{Hidden symmetries of states}. In this work, we have bootstrapped hidden symmetries of Hamiltonians. An important future direction is to develop a parallel bootstrap program for hidden symmetries of quantum states. In tensor-network descriptions, particularly for matrix product states \cite{Cirac:2020obd}, symmetry information is often encoded in the local tensor data and in the induced virtual action, rather than only in the microscopic Hamiltonian. This raises the prospect of extracting symmetry structure directly from wavefunction data, and using it to reconstruct projective or more general state-level symmetry patterns.\\

\section*{Acknowledgements}
We thank Bo-Ting Chen, Weiguang Cao, Frederik M. Denef, Taozhi Guo, David A. Huse, {Haruki Watanabe}, Shenghan Jiang, Biao Lian, Bowei Liu, Dachuan Lu, Yixin Ma, Weibo Mao, Yuan Miao, Sebastian Mizera, Liang-Hong Mo, Masahiro Nozaki, Shanming Ruan, Sirui Shuai, Minxuan Wang, Weijun Wu, Zhenyu Xiao, Zhenbin Yang, Wucheng Zhang, Yunqin Zheng for inspiring and helpful discussions. We acknowledge that the algorithmic implementation detailed in Sec.~\ref{sec:Bootstrap-Algorithm} was developed through an iterative human–AI workflow, in which the authors provided the physical insight and validation, while Claude supported code development, candidate enumeration, and consistency checks. C.B. gratefully acknowledges financial support from Masahiro Nozaki at the University of Chinese Academy of Sciences. Additionally, C.B. thanks Princeton University for its hospitality during the inception of this work. Z.Z. would like to especially thank Matias Zaldarriaga for organizing the AI term at the IAS and for his support in facilitating the use of Claude Code. Z.Z. thanks John Staunton for the invitation to present this work at the high energy theory group meeting at Columbia University. S.R. is supported by the National Science Foundation under Award No. DMR-2409412.

\appendix

\section{Glossary of Mathematical Notations \label{app:Glossary}}
\begin{description}
    \item[$G$] hidden symmetry group.
    \item[$|G|$] Order of $G$.
    \item[$\hat{G}$] Set of $G$-irreps, with elements $\alpha$.
    \item[$\text{Rep}(G)$] Category of representations of $G$.
    \item[$\mathcal{C}_G$] Set of conjugacy classes of $G$, with elements $c_i$.
    \item[$G^{\text{mag.}}$] Wigner's magnetic group involving unitary and anti-unitary elements.
    \item[$\hat{G}^{\text{mag.}}$] Set of irreducible corepresentations (ICRs) of $G^{\text{mag.}}$, with elements $\alpha$.
    \item[$N^{\text{uni.}}$] Unitary subgroup of magnetic group $G^{\text{mag.}}$.
    \item[$\hat{N}^{\text{uni.}}$] Set of $N^{\text{uni.}}$-irreps of $G^{\text{mag.}}$, with elements $\lambda$.
    \item[$\tilde{c}_i$] Magnetic class.
    \item[$\alpha^*$] Dual of $G$-irrep $\alpha\in \hat{G}$.
    \item[$V_{\alpha}$] Irrep space of $G$-irrep $\alpha\in \hat{G}$, with dimension $d_{\alpha}$.
    \item[$M_{\alpha}$] Multiplicity space of $V_{\alpha}$, with dimension $m_{\alpha}$.
    \item[$\chi_{\alpha}$] Character of $G$-irrep $\alpha\in\hat{G}$.
    \item[$N^{\alpha_i}_{\alpha_j \alpha_k}$] Fusion coefficients of $G$.
    \item[$N$] Known subgroup.
    \item[$|N|$] Order of $N$.
    \item[$\hat{N}$] Set of $N$-irreps, with elements $\lambda$.
    \item[$\lambda^*$] Dual of $N$-irrep $\lambda\in \hat{N}$.
    \item[$V_{\lambda}$] Irrep space of $N$-irrep $\lambda\in \hat{N}$, with dimension $d_{\lambda}$.
    \item[$\chi_{\lambda}$] Character of $N$-irrep $\lambda\in\hat{N}$.
    \item[$N^{\lambda_k}_{\lambda_i \lambda_j}$] Fusion coefficients of $N$.
    \item[$\text{Res}_N^GV_{\alpha}$] Restriction of $G$-irrep space $V_{\alpha}$ to $N$.
    \item[$\text{Ind}_N^GV_{\alpha}$] Induction of $G$ representations from $N$-irreps $V_{\lambda}$.
    \item[$\otimes$] Tensor product.
    \item[$\oplus$] Direct sum.
    \item[$\rtimes$] Semi-direct product.
    \item[$b_{\lambda, \alpha}$] Branching multiplicity with $\lambda\in\hat{N},~\alpha\in\hat{G}$. It is bounded by a cut-off $b_{\text{max}}$ in the bootstrap algorithm.
    \item[$\vec{b}_\alpha$] Branching vector for fixed $G$-irrep.
    \item[$\vec{b}_\lambda$] Branching vector for fixed $N$-irrep.
    \item[$B_{\lambda,\alpha}$] Branching matrix with rows labeled by $\lambda\in\hat{N}$ and columns labeled by $\alpha\in\hat{G}$.
    \item[$G_{\lambda_i\lambda_j}$] Gram matrix, $G_{\lambda_a\lambda_a}=\sum_{\alpha\in\hat{G}}b_{\lambda_a,\alpha}b_{\lambda_b,\alpha}$.
    \item[$\langle \cdot,\cdot \rangle_G$] $G$-inner product on class functions.
    \item[$\mathbb{Z}_n$] Cyclic group of order $n$.
    \item[$S_n$] Symmetric group of $n$ elements.
    \item[$D_n$] Dihedral group $D_n\cong \mathbb{Z}_n\rtimes\mathbb{Z}_2$.
    \item[$Q_{2^n}$] Generalized quaternion group.
    \item[$V_4$] Klein four-group, $V_4\cong\mathbb{Z}_2\times\mathbb{Z}_2$.
    \item[$G^2$] Tensor product of two identical $G$, $G^2\cong G\times G$.
    \item[$\hat{\mathcal{N}}$] Particle number operator, whose eigenvalue is $\mathcal{N}$.
    \item[$\mathcal{C}_i$] Equivalence class in the bootstrap algorithm.
    \item[$r$] Trial rank in the bootstrap algorithm: its minimal (maximal) value is $r_{\text{min}} (r_{\text{max}})$, and the target rank is denoted as $r_*$.
    \item[$\tilde{P}$] Quotient graph.
    \item[$\mathcal{Q}$] Clique, a fully connected subgraph of a quotient graph.
    \item[$v(\mathcal{C}_i)$] Assignment of branching multiplicity value within an equivalence class $\mathcal{C}_i$.
    \item[$t_{\lambda}^{(\alpha_i,\alpha_j)}$] Fusion target vector defined in the algorithm.
    \item[$\mathfrak{g}$]  Lie algebra, $\mathfrak{g}=\mathfrak{g}_{\mathrm{ss}}\oplus\mathfrak{z}$, where $\mathfrak{g}_{\mathrm{ss}}$ is semisimple and $\mathfrak{z}$ is the center.
\end{description}

\section{Mathematical Details and Proofs\label{app:Proof}}
In this appendix, we provide further mathematical derivations and proofs for several statements made in the main text.
\subsection{Group Verlinde Formula\label{app-sec:Group-Verlinde-Formula}}
Here, we review the derivation of the group Verlinde formula~\cite{DiFrancesco:1997nk}. Let $c_i\in\{\mathcal{C}_G\}$ denote the conjugacy classes of the group $G$, and let $\alpha\in\hat{G}$ label the $G$-irreps with corresponding characters $\chi_{\alpha}$; ``star'' denotes complex conjugation. We define the inner product between any two class functions $f$ and $g$ as
\be\label{app-eq:Inner-Product-Class-Function}
    \langle f,g\rangle_{G}=\f{1}{|G|}\sum_{c_i\in\mathcal{C}_G}|c_i|f(c_i)g^*(c_i),
\ee
where $|G|$ and $|c_i|$ are the sizes of the group and the conjugacy class $c_i$, respectively. This inner product is Hermitian (conjugate symmetric), satisfying
\be
    \langle f,g\rangle_{G}=\overline{\langle g,f\rangle_{G}}=\f{1}{|G|}\sum_{c_i\in\mathcal{C}_G}|c_i|g^*(c_i)f(c_i),
\ee
where the overline denotes complex conjugation. Since group characters are class functions, the orthogonality of characters with respect to this inner product implies that
\be\label{app-eq:Character-Orthogonality}
    \langle \chi_{\alpha_i},\chi_{\alpha_j}\rangle_{G}=\delta_{\alpha_i,\alpha_j},~\alpha_i,\alpha_j\in\hat{G}.
\ee

Recall that the fusion rule between two $G$-irreps $\alpha_i,\alpha_j\in\hat{G}$ is given by Eq.~\eqref{eq:G_fusion_rules}, and their corresponding characters satisfy the multiplicative property
\be
    \chi_{\alpha_i\otimes\alpha_j}=\chi_{\alpha_i}\chi_{\alpha_j}.
\ee
Consequently, the fusion coefficients can be directly computed via the inner product of characters, a relation commonly known as the group Verlinde formula~\cite{DiFrancesco:1997nk}
\be\label{app-eq:Group-Verlinde-Formula}
    N_{\alpha_i\alpha_j}^{\alpha_k}=\langle \chi_{\alpha_i\otimes\alpha_j},\chi_{\alpha_k}\rangle_{G}=\langle \chi_{\alpha_i}\chi_{\alpha_j},\chi_{\alpha_k}\rangle_{G},
\ee
where $\alpha_{i},\alpha_{j},\alpha_{k}\in\hat{G}$.

\subsection{Characters as Eigenvalues of Fusion Matrices\label{app:Character-F-Matrices}}
Here, we present the step-by-step proof showing that the fusion matrices of $\text{Rep}(G)$ commute and can be simultaneously diagonalized, with the characters emerging as their eigenvalues.

Let $G$ be a finite group and $\text{Rep}(G)$ be its category of finite-dimensional complex representations. The fusion rules describe how the tensor product of two irreps $V_{\alpha_i}$ and $V_{\alpha_j}$ is defined in Eq.~\eqref{eq:G_fusion_rules}. The fusion matrices are defined by the fusion coefficients:
\be\label{app-eq:Fusion-Matrix}
    (N_{\alpha_i})_{\alpha_j\alpha_k}=N_{\alpha_i\alpha_j}^{\alpha_k},
\ee
where $\alpha_{i,j,k}\in\hat{G}$.

First, we prove that the fusion matrices commute. We rely on the commutativity of the tensor product in $\text{Rep}(G)$: 
\be
    V_{\alpha_i}\otimes V_{\alpha_j}\cong V_{\alpha_j}\otimes V_{\alpha_i},
\ee
as well as its associativity:
\be
    (V_{\alpha_i} \otimes V_{\alpha_j}) \otimes V_{\alpha_k} \cong V_{\alpha_i} \otimes (V_{\alpha_j} \otimes V_{\alpha_k}).
\ee
Decomposing the left-hand side, we obtain:
\be
    (V_{\alpha_i} \otimes V_{\alpha_j}) \otimes V_{\alpha_m} \cong  \bigoplus_{\alpha_k, \alpha_l\in\hat{G}} N_{\alpha_i \alpha_j}^{\alpha_k} N_{\alpha_k \alpha_m}^{\alpha_l} V_{\alpha_l}.
\ee
Next, we decompose the right-hand side:
\be
    V_{\alpha_i} \otimes (V_{\alpha_j} \otimes V_{\alpha_m}) \cong  \bigoplus_{\alpha_k, \alpha_l\in\hat{G}} N_{\alpha_j \alpha_m}^{\alpha_k} N_{\alpha_i \alpha_k}^{\alpha_l} V_{\alpha_l}.
\ee
Equating the coefficients of these two decompositions due to associativity yields:
\be
    \sum_{\alpha_k\in\hat{G}} N_{\alpha_i \alpha_k}^{\alpha_l} N_{\alpha_j \alpha_m}^{\alpha_k} = \sum_{\alpha_k\in\hat{G}} N_{\alpha_i \alpha_j}^{\alpha_k} N_{\alpha_k \alpha_m}^{\alpha_l}.
\ee
By simultaneously invoking commutativity ($N_{\alpha_i \alpha_j}^{\alpha_k} = N_{\alpha_j \alpha_i}^{\alpha_k}$), we can express this relation equivalently in matrix form as:
\be
    (N_{\alpha_i} N_{\alpha_j})_{\alpha_m \alpha_l} = (N_{\alpha_j} N_{\alpha_i})_{\alpha_m \alpha_l}.
\ee
Therefore, the fusion matrices commute with each others.

As all fusion matrices commute, we can diagonalize them simultaneously to solve their eigenvalues. To show that the characters correspond to the eigenvalues, we take the character (trace) of both sides of Eq.~\eqref{eq:G_fusion_rules}. Let $\chi_{\alpha_i}(g)$ be the character of the irrep $V_{\alpha_i}$ evaluated at a group element $g \in G$. Applying this to our fusion rule $V_{\alpha_i} \otimes V_{\alpha_j} \cong \bigoplus_{\alpha_k} N_{\alpha_i \alpha_j}^{\alpha_k} V_{\alpha_k}$, we get:
\be\label{app-eq:fusion-rules-character}
    \chi_{\alpha_i}(g) \chi_{\alpha_j}(g) = \sum_{\alpha_k} N_{\alpha_i \alpha_j}^{\alpha_k} \chi_{\alpha_k}(g).
\ee

Recall that characters are class functions; they are invariant under conjugation and depend only on the conjugacy classes $c_i$ of $G$. For a fixed conjugacy class $c_i$, we define the column vector
\be
    v_{c_i}=\left(\chi_{\alpha_0}(c_i),\chi_{\alpha_i}(c_i),\cdots\right)^T,
\ee
whose entries correspond to the characters of the distinct irreps of $G$ evaluated at $c_i$. Subsequently, we apply the fusion matrix to this vector:
\be
    (N_{\alpha_i} v_{c_i})_{\alpha_j} = \sum_{\alpha_k} N_{\alpha_i \alpha_j}^{\alpha_k} \chi_{\alpha_k}(c_i).
\ee
In the end, we use the character identity $\chi_{\alpha_i\otimes\alpha_j}=\chi_{\alpha_i}\chi_{\alpha_j}$ to simplify this equation
\be
    N_{\alpha_i} v_{c_i} = \chi_{\alpha_i}(c_i) v_{c_i},
\ee
which completes the proof.

\subsection{Proof of Uniqueness of Branching Multiplicities\label{app:Proof-I}}
Here, we prove:
\begin{itemize}
    \item \textit{Once $N$ and $G$ are fixed (up to isomorphism), branching multiplicities $\{b_{\lambda,\alpha}\}$ are uniquely determined.}
\end{itemize}

Let $G$ be a group and $N\subset G$ be a non-central subgroup. Consider an irrep $V_{\alpha}$ of $G$ and its restriction on $H$:
\be
    W=\text{Res}^G_NV_{\alpha}\cong \bigoplus_{\lambda} V_{\lambda}\otimes\mathbb{C}^{b_{\lambda,\alpha}}.
\ee
Now, suppose there is another decomposition of $V_{\alpha}$ into irreps of $N$:
\be
        W\cong \bigoplus_{\lambda} V_{\lambda}\otimes\mathbb{C}^{\tilde{b}_{\lambda,\alpha}}.
\ee
To prove the statement, we just prove $\tilde{b}_{\lambda,\alpha}$ must be equal to $b_{\lambda,\alpha}$ for any $\lambda$.

For each $N$-irrep, we define the intertwiner space
\be
\begin{split}
    &\text{Hom}_N(V_{\lambda},W)=\left\{T:V_{\lambda}\overset{\text{linear}}{\to}W\right\},\\
    &T\rho_{\lambda}(n)=\rho_w(n)T,\quad \forall n\in N,
\end{split}
\ee
where $T$ are intertwiners, and $\rho_{\lambda}:N\to\text{GL}(V_{\lambda})$, $\rho_{W}:N\to\text{GL}(W)$, respectively. Furthermore, since ``$\text{Hom}$'' spaces have additivity over direct sums, we have
\be
\begin{split}
    \text{Hom}_N(V_{\lambda},W)&\cong\text{Hom}_N\left(V_{\lambda},\bigoplus_{\mu} V_{\mu}\otimes\mathbb{C}^{b_{\mu,\alpha}}\right)\\
    &\cong\bigoplus_{\mu}\text{Hom}_N\left(V_{\lambda}, V_{\mu}\otimes\mathbb{C}^{b_{\mu,\alpha}}\right)\\
    &\cong\bigoplus_{\mu}\text{Hom}_N\left(V_{\lambda}, V_{\mu}\right)\otimes\mathbb{C}^{b_{\mu,\alpha}}.\\
\end{split}
\ee
Using Schur's lemma, for a pair of $N$-irreps $V_{\lambda},V_{\mu}$, 
\be
    \text{Hom}_N\left(V_{\lambda}, V_{\mu}\right)=\delta_{\lambda,\mu}\mathbb{C}.
\ee
Therefore, the dimension of $\text{Hom}_N(V_{\lambda},W)$ is
\be
    \dim \text{Hom}_N(V_{\lambda},W)=b_{\lambda,\alpha}.
\ee
Similarly, for the second decomposition, we have $\dim \text{Hom}_N(V_{\lambda},W)=\tilde{b}_{\lambda,\alpha}$. Because $\dim \text{Hom}_N(V_{\lambda},W)$ is independent of the particular choice of decomposition of $W$, it follows that $\tilde{b}_{\lambda,\alpha}=b_{\lambda,\alpha}$ for any $\lambda$. Thus, we have proved the statement.

\subsection{Proof of statement I\label{app:Proof-II}}
We now prove the statement I:
\begin{itemize}
    \item \textit{Given $m_{\alpha} > 0$ for all $\alpha$, the condition $K_{\lambda_a,\lambda_b}^{(N)} = K_{\lambda_a,\lambda_a}^{(N)} = K_{\lambda_b,\lambda_b}^{(N)}$ implies the identity $\vec{b}_{\lambda_a} = \vec{b}_{\lambda_b}$.}
\end{itemize}

First, we define the inner product and norm as
\be
\begin{split}
    &\langle \vec{b}_{\lambda_a},\vec{b}_{\lambda_b}\rangle_{M}=\vec{b}_{\lambda_a}^TM\vec{b}_{\lambda_b}=\sum_{\alpha}b_{\lambda_a,\alpha}b_{\lambda_b,\alpha}m_{\alpha},\\
    &||\vec{b}_{\lambda}||_M^2=\langle \vec{b}_{\lambda},\vec{b}_{\lambda}\rangle_{M}=\sum_{\alpha}b_{\lambda,\alpha}^2m_{\alpha},
\end{split}
\ee
where $M=\text{diag}(\cdots,m_{\alpha},\cdots)$ with $m_{\alpha}>0$ for any $\alpha$, and $b_{\lambda,\alpha}\in\mathbb{N}$ for any $\lambda$ and $\alpha$.

Second, we consider the following weighted squared identity:
\be
\begin{split}
||\vec{b}_{\lambda_a}-\vec{b}_{\lambda_b}||^2_M &= (\vec{b}_{\lambda_a}-\vec{b}_{\lambda_b})^T M (\vec{b}_{\lambda_a}-\vec{b}_{\lambda_b}) \\
&= \vec{b}_{\lambda_a}^T M \vec{b}_{\lambda_a} + \vec{b}_{\lambda_b}^T M \vec{b}_{\lambda_b} - 2\vec{b}_{\lambda_a}^T M \vec{b}_{\lambda_b} \\
&= K_{\lambda_a,\lambda_a} + K_{\lambda_b,\lambda_b} - 2K_{\lambda_a,\lambda_b} = 0.
\end{split}
\ee
Since $M$ is positive definite—a condition naturally satisfied in our case by $m_{\alpha} > 0$ for all $\alpha$—the identity becomes:
\be
||\vec{b}_{\lambda_a}-\vec{b}_{\lambda_b}||^2_M = \sum_{\alpha} (b_{\lambda_a,\alpha} - b_{\lambda_b,\alpha})^2 m_{\alpha} = 0,
\ee
which directly implies that $\vec{b}_{\lambda_a} = \vec{b}_{\lambda_b}$.

\subsection{Long-time Averaged Plateau\label{app:Universal-Plateau}}
Here, we demonstrate that the long-time averaged plateau of the SFF is a universal feature, independent of the system's integrability. Consider an arbitrary Hermitian Hamiltonian $H$ with distinct eigenvalues $\{E_a\}_{a=1}^m$, where each $E_a$ possesses a degeneracy $\nu_a$. The long-time average of its SFF is given by
\be
\begin{split}
    K&=\lim_{T\to\infty}\sum_{a,b=1}^m\f{\nu_a\nu_b}{T}\int_0^Te^{-i(E_a-E_b)t}\\
    &=\sum_{a,b=1}^m\delta_{E_a,E_b}\nu_a\nu_b=\sum_{a=1}^m\nu_a^2,
\end{split}
\ee
where we have used the asymptotic identity for the time average
\be
    \lim_{T\to\infty}\f{1}{T}\int^Te^{-i(E_a-E_b)t}=\begin{cases}
        1,~&E_a=E_b,\\
        0,~&E_a\neq E_b.
    \end{cases}
\ee
If the spectrum is entirely non-degenerate ($\nu_a=1$ for all $a$), the plateau height evaluates simply to the dimension of the matrix. In the main text, the relevant non-degenerate subspaces are denoted by $M_{\alpha}$. Consequently, the plateaus of $K_{\alpha}(t)$, universally converge to $m_{\alpha}$, irrespective of whether the underlying dynamics are chaotic or integrable. Furthermore, if Kramers degeneracy is present within $M_{\alpha}$ (such that $\nu_a=2$ for all $a$), the plateau value becomes $\sum_{a=1}^{\frac{m_{\alpha}}{2}} 2^2 = 2m_{\alpha}$. This doubling naturally accounts for the overall factor of $\nu$ introduced in Eq.~\eqref{eq:SFF-Plateau-Malpha}.

\subsection{Brief Introduction to Wigner's Corepresentation Theory\label{app:Corepresentation}}
Here, we briefly introduce Wigner's corepresentation theory~\cite{wigner1959group,RevModPhys.40.359,rumynin2021realrepresentationsc2gradedgroups,Mock_2016}, which describes magnetic groups generated by both unitary and anti-unitary operators.

Let $G^{\text{mag.}}$ denote a finite magnetic group containing both unitary and anti-unitary operators. We define $N^{\text{uni.}}\subset G^{\text{mag.}}$ as its unitary normal subgroup. A corepresentation of $G^{\text{mag.}}$ on a vector space $V$ is a map \cite{Moore2014QuantumSymmetry}
\be\label{app-eq:AU-h-Corep-1}
    \rho:~G^{\text{mag.}}\to\text{Aut}_{\mathbb{R}}(V),
\ee
where $\text{Aut}_{\mathbb{R}}(V)$ denotes the group of invertible, real-linear transformations on $V$. The restriction to real-linear maps is necessary because anti-unitary operators are anti-linear over the complex numbers $\mathbb{C}$, but they remain strictly linear over the real numbers $\mathbb{R}$. More precisely, the transformation $\rho(g)$ is complex-linear for $g\in N^{\text{uni.}}$ and anti-linear for $g\in G^{\text{mag.}}\setminus N^{\text{uni.}}$, satisfying the standard composition rule
\be\label{app-eq:AU-h-Corep-2}
    \rho(g_1g_2)=\rho(g_1)\rho(g_2),\quad \forall g_1,g_2\in G^{\text{mag.}}.
\ee
By restricting this map to the unitary subgroup, one immediately recovers the standard linear representation theory. A convenient way to describe an anti-unitary element $h\in G^{\text{mag.}}\setminus N^{\text{uni.}}$ is
\be\label{app-eq:AU-h-Corep-3}
    \rho(h)=U_hK,\quad U_h^{-1}=U_h^{\dagger},
\ee
where $K$ denotes coefficient-wise complex conjugation.

The main idea of Wigner's corepresentation theory is simple. One first classifies the ordinary irreps of the unitary subgroup $N^{\text{uni.}}$, and then asks how an anti-unitary element acts on those irreps. Let $\Delta$ be an ordinary irrep of $N^{\text{uni.}}$ and $h \in G^{\text{mag.}} \setminus N^{\text{uni.}}$ be an anti-unitary element. Under the corepresentation $\rho$, the conjugation of a unitary element $n \in N^{\text{uni.}}$ by $h$ yields
\be\label{app-eq:AU-h-Conjugation-Irrep-1}
    \rho(h)\Delta(n)\rho(h)^{-1}=U_h\Delta(n)^*U_h^{\dagger}=\Delta(hnh^{-1}),
\ee
where $U_h$ is the unitary matrix associated with the anti-unitary operator $\rho(h)$. As $N^{\text{uni.}}$ is a normal subgroup of $G^{\text{mag.}}$, the conjugate element $hnh^{-1}$ remains in $N^{\text{uni.}}$. Thus, $h$ sends $\Delta$ to its conjugate irrep:
\be\label{app-eq:AU-h-Conjugation-Irrep-2}
    \Delta_h(n)=\Delta(hnh^{-1})^*.
\ee

Now, we arrive at the central problem in the Wigner's corepresentation theory, i.e., the classification of corepresentations. This classification is governed by Eqs.~\eqref{app-eq:AU-h-Conjugation-Irrep-1} and \eqref{app-eq:AU-h-Conjugation-Irrep-2}. Depending on whether the conjugate representation $\Delta_h$ is equivalent to the original irrep $\Delta$\footnote{Here, the symbol $\simeq$ denotes equivalence, indicating the existence of an invertible change-of-basis matrix (the intertwiner) that transforms one representation into the other.}, we distinguish three distinct types of corepresentations~\cite{rumynin2021realrepresentationsc2gradedgroups,Mock_2016}:
\begin{itemize}
    \item \textbf{Type (a) $(\Delta_h \simeq \Delta)$}:  There exists an invertible matrix (the intertwiner) $\mathcal{I}_{\Delta}$ such that
    \be\label{app-eq:Corep-Classification-Intertwiner}
        \mathcal{I}_{\Delta}\Delta(h^{-1}nh)^*\mathcal{I}_{\Delta}^{-1}=\Delta(n),
    \ee
    or all $n\in N^{\text{uni.}}$ and $h\in G^{\text{mag.}}\setminus N^{\text{uni.}}$. Furthermore, for a Type (a) corepresentation, the intertwiner obeys the constraint
    \be\label{app-eq:Corep-Classification-Intertwiner-a}
        \mathcal{I}_{\Delta}\mathcal{I}_{\Delta}^*=\Delta(h^2),
    \ee
    in which the anti-unitary symmetry can be implemented within the original irrep space. From a physical perspective, $\mathcal{I}_{\Delta}$ simply changes the basis to map the anti-unitary partner $\Delta_h$ directly back to $\Delta$. This indicates that the $N^{\text{uni.}}$-irrep is strictly self-conjugate under the action of the anti-unitary symmetry.
    \item \textbf{Type (b) $(\Delta_h \simeq \Delta)$}: This class is mathematically identical to Type (a), differing only in the intertwiner constraint, which acquires a relative minus sign:
    \be\label{app-eq:Corep-Classification-Intertwiner-b}
        \mathcal{I}_{\Delta}\mathcal{I}_{\Delta}^*=-\Delta(h^2).
    \ee
    In this case, while the unitary irrep remains self-conjugate under the anti-unitary symmetry, the dimension of the resulting irreducible corepresentation is necessarily doubled.
    \item \textbf{Type (c)} ($\Delta_h \not\simeq \Delta$): The anti-unitary element relates two inequivalent $N^{\text{uni.}}$-irreps. These irreps cannot exist as independent symmetry sectors of the full group and must instead combine into a single irreducible corepresentation with twice the original dimension. 
\end{itemize}
So far the classification has been formulated directly in terms of representations and intertwiners. In practice, however, one would like a more efficient criterion for classification. This is precisely where character theory of corepresentation enters.

On the unitary subgroup $N^{\text{uni.}}$, the character of an irrep $\Delta$ is
\be
    \chi_{\Delta}(n)=\tr\Delta(n),
    \quad\forall n\in N^{\text{uni.}}.
\ee
These characters obey the standard orthogonality relations and completely characterize the ordinary irreps of the unitary subgroup. Based on this definition, the linear representations can be classified using the Frobenius-Schur indicator, given by
\be
    \nu(\Delta)=\f{1}{|N^{\text{uni.}}|}\sum_{n\in N^{\text{uni.}}}\chi_{\Delta}(n^2).
\ee
Depending on whether $\nu(\Delta)$ evaluates to $1$, $0$, or $-1$, the irrep $\Delta$ is classified as real, complex, or quaternionic, respectively. Although the Frobenius-Schur indicator applies to finite groups as well as compact Lie groups, our current analysis considers only the finite group case.

When lifting the unitary subgroup to the full magnetic group $G^{\text{mag.}}$, we encounter a fundamental difficulty: as anti-unitary operators are anti-linear, the standard trace of the corepresentation matrix $\rho$ is no longer a basis-independent invariant. For this reason, the character theory of a magnetic group is still most naturally formulated by restricting entirely to its unitary subgroup $N^{\text{uni.}}$. The character of corepresentation $\rho$ is a map:
\be\label{app-eq:Def-Character-Corep}
    \chi_{\rho}:~N^{\text{uni.}}\to \mathbb{C},
\ee
and for $n\in N^{\text{uni.}}$ we have $\chi_{\rho}(n)=\tr\rho(n)$.

Then, one can define the Wigner-Schur indicator as~\cite{newmarch1983character,Mock_2016}
\be
    \eta(\Delta)=\f{1}{|N^{\text{uni.}}|}\sum_{n\in N^{\text{uni.}}}\chi_{\Delta}\left((hn)^2\right),
\ee
where $h\in G^{\text{mag.}}\backslash N^{\text{uni.}}$ is an anti-unitary element. Similar to the Frobenius-Schur indicator, the Wigner-Schur indicator takes only $1,0,-1$, with the interpretation:
\be
    \eta(\Delta)=\begin{cases}
        1,~&\text{Type (a)},\\
        -1,~&\text{Type (b)},\\
        0,~&\text{Type (c)}.\\
    \end{cases}
\ee
By using the Wigner-Schur indicator, we can determine the corepresentation type directly instead of constructing the intertwiner matrices.

Having established the characters and the Wigner-Schur indicator, we can now formulate the character theory in terms of the conjugacy classes of the magnetic group, referred to as magnetic classes in~\cite{newmarch1983character}. A magnetic class $\mathcal{C}$ of $G^{\text{mag.}}$ is defined as an equivalence class of unitary elements in $N^{\text{uni.}}$: $n_1,n_2\in \mathcal{C}$ if there exists either $n\in N^{\text{uni.}}$ with $nn_1n^{-1}=n_2$ or $h\in G^{\text{mag.}}\backslash N^{\text{uni.}}$ with $hn_1h^{-1}=n_2^{-1}$ (or both). Crucially, unlike the standard conjugacy classes of an ordinary group, a magnetic class exclusively contains elements from the unitary subgroup. The characters of corepresentations are invariant functions on the magnetic classes. Note that for ordinary finite groups, the order of a conjugacy class equals to the order of the group divided by the order of centralizer of any element of the class. A similar statement holds for magnetic groups, once the centralizer has been defined. Specifically, the magnetic centralizer $\mathbb{C}[L]$ of a set of linear operators $L$ in $G^{\text{mag.}}$~\cite{newmarch1983character}:
\be
    \mathbb{C}[L]=
    \left\{n,h\in G^{\text{mag.}}: nl=ln, hl=l^{-1}h, \forall l\in L\right\}.
\ee

The fusion rules for the category $\text{Corep}(G^{\text{mag.}})$ are given by
\begin{equation}\label{eq:Gmag_fusion_rules}
        V_{\alpha_i} \otimes V_{\alpha_j} \cong \bigoplus_{\alpha_k \in \hat{G}^{\text{mag.}}} N^{\alpha_k}_{\alpha_i\alpha_j} V_{\alpha_k},
    \end{equation}
where $\hat{G}^{\text{mag.}}$ denotes the complete set of irreducible corepresentations (ICRs) of the magnetic group, with $\alpha_i, \alpha_j, \alpha_k \in \hat{G}^{\text{mag.}}$. Similarly, the fusion coefficients can be computed via the character inner product on the unitary subgroup:
\be\label{app-eq:Group-Verlinde-Formula-Corep}
    N_{\alpha_i\alpha_j}^{\alpha_k}=\f{\langle \chi_{\alpha_i\otimes\alpha_j},\chi_{\alpha_k}\rangle_{N^{\text{uni.}}}}{\mu_{\alpha_k}}=\f{\langle \chi_{\alpha_i}\chi_{\alpha_j},\chi_{\alpha_k}\rangle_{N^{\text{uni.}}}}{\mu_{\alpha_k}},
\ee
where $\mu_{\alpha_k}$ is determined by Wigner's classification of the corepresentation
\be
    \mu_{\alpha}=\begin{cases}
        1,~&\alpha~\text{is Type (a)},\\
        4,~&\alpha~\text{is Type (b)},\\
        2,~&\alpha~\text{is Type (c)}.\\
    \end{cases}
\ee 
This can be derived using the same method in Appendix~\ref{app-sec:Group-Verlinde-Formula} with only one difference:
\be
    \langle \chi_{\alpha},\chi_{\alpha}\rangle_{N^{\text{uni.}}}=\mu_{\alpha}.
\ee
By retracing the earlier proof, i.e., substituting ordinary irreps and conjugacy classes with their ICRs and magnetic class counterparts, one can demonstrate that the characters defined in Eq.~\eqref{app-eq:Def-Character-Corep} are precisely the eigenvalues of the commuting fusion matrices.

\section{Analytical Confirmations for the Bootstrap Solutions\label{app:Bootstrap-Solutions}}
In this appendix, we provide analytical verification that the bootstrap solutions presented in the main text successfully recover the representation theory of the hidden symmetry group $G$ for Examples II–IV.

\subsection{Analytical Confirmation for Example II\label{app:Bootstrap-Solutions-II}}
Here, we analytically verify that the lowest-rank bootstrap solution uniquely corresponds to the representations of $G\cong D_4\cong\mathbb{Z}_4\rtimes\mathbb{Z}_2$. Although the dihedral group $D_4$ and the quaternion group $Q_8$ share identical fusion rings and character tables, $Q_8$ lacks a $V_4$ subgroup. Consequently, it can be safely ruled out as a candidate for the hidden symmetry.

For even integers $n$, the dihedral group $D_n=\langle x,y|x^n=y^2=e,yxy=x^{-1} \rangle$ possesses exactly four one-dimensional irreps and $\left(\frac{n}{2}-1\right)$ two-dimensional irreps. The sum of the squares of their dimensions confirms the total group order: $|D_n| = 4 \times 1^2 + \left(\frac{n}{2}-1\right) \times 2^2 = 2n$~\cite{serre1977linear}. The four one-dimensional irreps are constructed by mapping the generators $x$ and $y$ to $\pm 1$ in all possible combinations. Their corresponding characters, denoted by $\alpha_0, \dots, \alpha_k$, are summarized in Table~\ref{tab:Dn-Character-1D-Table}.
\begin{table}
    \centering
    \begin{tabular}{c|cc}
         & $x^k$ & $yx^k$\\
         \hline
       $\alpha_0$  & 1 & 1\\
       $\alpha_i$  & 1 & -1\\
       $\alpha_j$  & $(-1)^k$ & $(-1)^k$\\
       $\alpha_k$  & $(-1)^k$ & $(-1)^{k+1}$\\
         \hline
    \end{tabular}
    \caption{Character table for the one-dimensional $D_n$-irreps for even $n$, with $k \in \mathbb{Z}_n$.}
    \label{tab:Dn-Character-1D-Table}
\end{table}
Next, we consider the two-dimensional irreps. Let $\omega_n = e^{\frac{2\pi i}{n}}$, and let $0< h< \f{n}{2}$ be an arbitrary integer. We define a two-dimensional representation of $D_n$, denoted by $\rho^h$, with the following matrix representations~\cite{serre1977linear}:
\be\label{eq:2D-D2n-Irrep-Matrix}
\begin{split}
    &\rho^h(x^k)=\begin{pmatrix}
        \omega_n^{hk} & 0\\
        0 & \omega_n^{hk}
    \end{pmatrix},
    \quad\rho^h(yx^k)=\begin{pmatrix}
        0 & \omega_n^{hk}\\
        \omega_n^{hk} & 0
    \end{pmatrix}.
\end{split}
\ee
From this, the characters for the two-dimensional irrep $\rho^h$ are given by:
\be\label{eq:2D-D2n-Irrep-Matrix-Character}
\begin{split}
    &\chi_{\rho_h}(x^k)=\tr\left[\rho^h(x^k)\right]=2\cos\left(\f{2\pi hk}{n}\right),\\
    &\chi_{\rho_h}(yx^k)=\tr\left[\rho^h(yx^k)\right]=0.
\end{split}
\ee

For the $D_4$ symmetry relevant to our model, we specialize to $n=4$ and set the index $h=1$. Evaluating Eq.~\eqref{eq:2D-D2n-Irrep-Matrix-Character} yields the non-vanishing characters for the two-dimensional irrep $\alpha_l$:
\be\label{eq:D4-Irrep-Alpha4-Character}
\begin{split}
    &\chi_{\alpha_4}(c_1)=\chi_{\alpha_4}(e)=2,\\
    &\chi_{\alpha_4}(c_2)=\chi_{\alpha_4}(x^2)=-2,\\
\end{split}
\ee
where $e$, $x$, and $y$ correspond to the identity, $\tilde{U}_{\pi/2}^z$, and $U_{\pi}^x$ operators, respectively. Together with the one-dimensional characters in Table~\ref{tab:Dn-Character-1D-Table}, these results confirm that Table~\ref{tab:D4-Q8-Character-Table} is the complete character table of $G \cong D_4$.

To compute the branching matrix, we consider the restriction of $D_4$ to its subgroup $V_4 = \langle x^2, y \rangle$. This subgroup consists of four conjugacy classes: $\{1\}$, $\{x^2\}$, $\{y\}$, and $\{yx^2\}$. Accordingly, for an arbitrary irrep $\alpha\in\hat{D}_4$, the restricted character is given by the vector:
\be
    \text{Res}^{D_4}_{V_4}\chi_{\alpha}=\left(\chi_{\alpha}(e),\chi_{\alpha}(x^2),\chi_{\alpha}(y),\chi_{\alpha}(yx^2)\right).
\ee
Given the relations $c_1 = \{1\}$, $c_2 = \{x^2\}$, $\{y\} \subset c_3$, and $\{yx^2\} \subset c_4$, applying Eq.~\eqref{eq:Branching-Multiplicity} yields a branching matrix identical to that in Eq.~\eqref{eq:KT-Dual-Model-Branching-Matrix}. Furthermore, by employing Eq.~\eqref{app-eq:Group-Verlinde-Formula} and the characters in Table~\ref{tab:D4-Q8-Character-Table}, we verify that the fusion coefficients in Eqs.~\eqref{eq:KT-Dual-Model-Fusion} are identical to those of $D_4$-irreps (Rep$(D_4)$). Similarly to the previous $S_3$ example, a detailed analysis of $\mathrm{Rep}(D_4)$ can be found in~\cite{Perez-Lona:2023djo}.

\subsection{Analytical Confirmation for Example III\label{app:Bootstrap-Solutions-III}}
Here, we analytically confirm that the $r_*=4$ bootstrap solution found previously corresponds to the representation-theoretic data of $S_4$.

As the irreps and character table of permutation groups are standard topics in group theory textbooks, e.g.,~\cite{serre1977linear}, we skip the details and directly note that the character table of $S_4$ is given by Table~\ref{tab:AT-Character-Table}. Following standard conventions, the permutation group $S_4$ comprises five conjugacy classes corresponding to distinct cycle types: the identity $\{1=(1)(2)(3)(4)\}$, transpositions $\{(ab)\}$, double transpositions $\{(ab)(cd)\}$, three-cycles $\{(abc)\}$, and four-cycles $\{(abcd)\}$, for distinct $a,b,c,d\in\{1,2,3,4\}$. It is sufficient to express all six transpositions explicitly in terms of the generators $g_1, g_2, S,$ and $C$ introduced in Eqs.~\eqref{eq:AT-Z2Z2-Generators},~\eqref{eq:Soperator} and~\eqref{eq:Coperator} as transpositions generate the entire permutation group. Using Eq.~\eqref{eq:S4-Standard-Group-Translation}, we find
\be
\begin{split}
    (12)=g_1Cg_1^{-1},\quad (13)&=g_2(SCS^{-1})g_2^{-1},\quad (14)=g_1Sg_1^{-1},\\
    (23)=S,\quad (24)&=SCS^{-1},\quad (34)=C.\\
\end{split}
\ee

The normal subgroup $V_4$ comprises four elements that fall into the conjugacy classes $c_1$ and $c_2$ as follows: 
\be
\begin{split}
    &\{1\}\subset c_1,\quad\{g_1\}\subset c_2,\quad\{g_2\}\subset c_2,\quad\{g_1g_2\}\subset c_2.
\end{split}
\ee
The characters of its four one-dimensional irreps are given by
\be\label{eq:V4-Character}
    \chi_{(Q_1,Q_2)}(g_1^ag_2^b)=(-1)^{aQ_1+bQ_2},
\ee
where $a,b\in\mathbb{N}$ and $(Q_1,Q_2)\in\{0,1\}^2$. By restricting the $S_4$-irreps to $V_4$ and using the Table~\ref{tab:AT-Character-Table}, we obtain the restricted characters 
\be\label{eq:S4-V4-Restricted-Character}
\begin{split}
    &\text{Res}^{S_4}_{V_4}\chi_{\alpha_0}=(1,1,1,1),\quad \text{Res}^{S_4}_{V_4}\chi_{\alpha_1}=(1,1,1,1),\\
    &\text{Res}^{S_4}_{V_4}\chi_{\alpha_2}=(2,2,2,2),\quad \text{Res}^{S_4}_{V_4}\chi_{\alpha_3}=(3,-1,-1,-1),\\
    &\text{Res}^{S_4}_{V_4}\chi_{\alpha_3}=(3,-1,-1,-1),
\end{split}
\ee
where the character values correspond to the ordered elements $(1,g_1,g_2,g_1g_2)$. Substituting Eqs.~\eqref{eq:V4-Character} and \eqref{eq:S4-V4-Restricted-Character} into Eq.~\eqref{eq:Branching-Multiplicity} yields the branching multiplicities, which are fully consistent with the branching matrix presented in Eq.~\eqref{eq:AT-Branching-Matrix}. Furthermore, the direct application of Eq.~\eqref{eq:Group-Verlinde-Formula} and Table~\ref{tab:AT-Character-Table} recovers the non-zero fusion coefficients shown in Eq.~\eqref{eq:AT-Fusion-Coefficients}. 

The above results confirm that the bootstrap solution at $r_*=5$ successfully recovers the representation theory of $S_4$, without any prior knowledge beyond its normal subgroup $V_4$. Crucially, this example demonstrates that even when numerical constraints are partially relaxed, such as by allowing branching multiplicities greater than one, the bootstrap algorithm can still reach the correct solution.

\subsection{Analytical Confirmation for Example IV $(\theta\neq \f{\pi}{4}\mod\pi)$\label{sec:QTC-Generic-Confirmation}}
First, we demonstrate that a redefinition of the magnetic group's generators reveals an explicit isomorphism to the abstract group $S_3^2$. Leveraging the representation theory of $S_3^2$, we then show that at $r_*=9$, our bootstrap algorithm yields the exact branching matrix, fusion rules, and character table for its linear representations. Furthermore, we analytically verify that the $r_*=5$ solution successfully reproduces the analogous algebraic data—namely, the branching matrix, fusion rules, and character table—for the corepresentation theory of the magnetic group.  While both solutions provide mathematically valid representations of the abstract group $S_3^2$, the $r_*=9$ solution fundamentally neglects the anti-unitary nature of complex conjugation. Consequently, although the $r_*=5$ solution alone captures the physically correct representation theory, both solutions unambiguously identify the same underlying abstract group $S_3^2$.

The magnetic group is generated by the set $\{Z, X, R, K\}$, whose elements satisfy the algebraic relations:
\be\label{eq:M-Group-Algebra-S32}
\begin{split}
    &Z^3=X^3=R^2=K^2=1,\\
    &[Z,X]=[K,X]=[K,R]=0,\\
    &RZR^{-1}=Z^{-1}, \quad RXR^{-1}=X^{-1},\\
    &KZK^{-1}=Z^{-1}.
\end{split}
\ee
By introducing the redefined charge conjugation $\tilde{R}=RK$, we can decouple these relations into:
\be
\begin{split}
    &Z^3=X^3=\tilde{R}^2=K^2=1,\\
    &[Z,X]=[K,X]=[K,\tilde{R}]=[\tilde{R},Z]=0,\\
    &KZK^{-1}=Z^{-1},
    \quad\tilde{R}X\tilde{R}^{-1}=X^{-1}.
\end{split}
\ee
At the level of the abstract group, this decoupled structure makes it manifest that the magnetic group is isomorphic to the direct product $S_3^2$.

\subsubsection{Linear Representation Theory of $S_3^2$}
For the moment, we temporarily disregard the anti-unitary nature of $K$ and focus exclusively on the linear representations of the abstract group $S_3^2$. Let the two $S_3$ subgroups be generated by
\be
\begin{split}
    S_3^{(Z)}&=\left\langle Z,K|Z^3=K^2=1,~KZK^{-1}=Z^{-1} \right\rangle,\\
    S_3^{(X)}&=\left\langle X,\tilde{R}|X^3=\tilde{R}^2=1,~\tilde{R}X\tilde{R}^{-1}=X^{-1} \right\rangle.\\
\end{split}
\ee
Every element of $S_3^2$ can be uniquely written as $g=(g_1,g_2)$, where
\be
    g_1=Z^{a_1}K^{b_1}, \quad g_2=X^{a_2}\tilde{R}^{b_2},
\ee
with $a_1, a_2 \in \{0,1,2\}$ and $b_1, b_2 \in \{0,1\}$. Since the irreps of a direct product group are precisely the tensor products of the irreps of its constituent subgroups, the nine irreps of $S_3^2$ derive directly from the three $S_3$ irreps. A single $S_3$ group has the following three irreps:
\be
\begin{split}
   &\rho_{\mathbf{1}}(Z)=\rho_{\mathbf{1}}(K)=1,\\
   &\rho_{\mathrm{sgn}}(Z)=-\rho_{\mathrm{sgn}}(K)=1,\\
   &\rho_{\mathbf{2}}(Z)=\begin{pmatrix}
       \omega & 0\\
       0 & \omega^2
   \end{pmatrix},
   \quad \rho_{\mathbf{2}}(K)=\begin{pmatrix}
       0 & 1\\
       1 & 0
   \end{pmatrix},
\end{split}
\ee
where $\rho_{\mathbf{1}}$ is the trivial representation, $\rho_{\mathrm{sgn}}$ is the non-trivial one-dimensional representation, $\rho_{\mathbf{2}}$ is the two-dimensional representation, and $\omega=e^{\frac{2\pi i}{3}}$. Since the $S_3^{(Z)}$ and $S_3^{(X)}$ subgroups mutually commute, we construct the nine $S_3^2$-irreps simply by taking the tensor products of their individual irreps
\begin{equation}
\begin{split}
    &\rho_{\alpha_0}=\rho_{\mathbf{1}}\otimes\rho_{\mathbf{1}},\quad
    \rho_{\alpha_i}=\rho_{\mathbf{1}}\otimes\rho_{\mathrm{sgn}},\quad
    \rho_{\alpha_j}=\rho_{\mathrm{sgn}}\otimes\rho_{\mathbf{1}},\\
    &\rho_{\alpha_k}=\rho_{\mathrm{sgn}}\otimes\rho_{\mathrm{sgn}}, \quad\rho_{\alpha_l}=\rho_{\mathbf{2}}\otimes\rho_{\mathbf{1}}, \quad
    \rho_{\alpha_m}=\rho_{\mathbf{2}}\otimes\rho_{\mathrm{sgn}},\\    
    &\rho_{\alpha_6}=\rho_{\mathbf{1}}\otimes\rho_{\mathbf{2}},\quad
    \rho_{\alpha_7}=\rho_{\mathrm{sgn}}\otimes\rho_{\mathbf{2}},\quad
    \rho_{\alpha_8}=\rho_{\mathbf{2}}\otimes\rho_{\mathbf{2}}.
\end{split}
\end{equation}
The dimensions of these representations are $1, 1, 1, 1, 2, 2, 2, 2$, and $4$. This is in exact agreement with our bootstrap findings at $r_*=9$, which yielded four one-dimensional, four two-dimensional, and one four-dimensional irreps. To verify that no representations are missing, we compare the sum of their squared dimensions with the group order:
\begin{equation}
    |S_3\times S_3|=36=4\times 1^2+4\times 2^2+1\times 4^2.
\end{equation}
The exact match proves that this set of nine irreps is complete for $G \cong S_3^2$. 

Next, we systematically construct the nine conjugacy classes of $S_3^2$, denoted as $\{c_1, c_2, \dots, c_9\}$. Recall that a single $S_3 = \langle x,y \mid x^3=y^2=1, ~ yxy=x^{-1} \rangle$ possesses exactly three conjugacy classes
\be
\begin{split}
    &c_1^{(S_3)}=\{1\},\quad c_2^{(S_3)}=\{x,x^2\},
    \quad
    c_2^{(S_3)}=\{y,yx,yx^2\}.
\end{split}
\ee
As the conjugacy classes of a direct product group are simply the Cartesian products of the classes of its constituent subgroups, the nine classes of $S_3^2$ are given by
\be\label{eq:S32-Conjugacy-Classes}
\begin{split}
    &c_1=c_1^{(S_3)}\times c_1^{(S_3)}, \quad c_2=c_1^{(S_3)}\times c_2^{(S_3)},
    \quad  c_3=c_2^{(S_3)}\times  c_1^{(S_3)},\\
    &c_4=c_2^{(S_3)}\times  c_2^{(S_3)},
    \quad c_5=c_1^{(S_3)}\times  c_3^{(S_3)},
    \quad c_6=c_3^{(S_3)}\times  c_1^{(S_3)},\\
    &c_7=c_2^{(S_3)}\times  c_3^{(S_3)},
    \quad c_8=c_3^{(S_3)}\times  c_2^{(S_3)},
    \quad c_9=c_3^{(S_3)}\times  c_3^{(S_3)}.
\end{split}
\ee
The size of each class is the product of the sizes of its constituent $S_3$ classes, yielding
\be
    \{|c_1|,\cdots,|c_9|\}=\{1,2,2,4,3,3,6,6,9\},
\ee
which perfectly agrees with Table~\ref{tab:QTC-Character-Table-Generic}. Summing these sizes returns the full order of the group: $|G|=1+2\times 2+4+2\times 3+2\times 6+9=36$. 

To further derive the branching matrix, recall that for a tensor-product group $S_3^2$, group composition acts component-wise 
\be
    (h_Z,h_X)(g_Z,g_X)=\left(h_Zg_Z,h_Xg_X\right)\in S_3^2,
\ee
where $h_Z,g_Z\in S_3^{(Z)}$ and $h_X,g_X\in S_3^{(X)}$. We then consider the subgroup $N\cong \mathbb{Z}_3^2\rtimes \mathbb{Z}_2$ generated by $\{z=(Z,1), x=(1,X), y=(K,\tilde{R})\}$, which satisfies the relations
\be
\begin{split}
    &x^3=z^3=y^2=1, \quad[z,x]=0,\\
    &yxy=x^{-1}, \quad yzy=z^{-1}.
\end{split}
\ee
$\mathbb{Z}_3^2$ being an Abelian normal subgroup of $N$, conjugation by $y$ acts via inversion, sending $x^az^b \mapsto x^{-a}z^{-b}$. Consequently, the conjugacy classes within $\mathbb{Z}_3^2$ are exactly the inversion orbits: a single trivial class $\{1\}$ and four classes of size two, namely $\{x,x^2\}$, $\{z,z^2\}$, $\{xz,x^2z^2\}$, and $\{xz^2,x^2z\}$. 
For the remaining coset $\mathbb{Z}_3^2 y$, conjugating any element $h_2 y$ by $h_1 \in \mathbb{Z}_3^2$ yields $h_1(h_2 y)h_1^{-1} = h_1^2 h_2 y$. Since the squaring map $h_1 \mapsto h_1^2$ is a bijection on $\mathbb{Z}_3^2$, any element of this coset is conjugate to $y$. Therefore, the entire coset forms a single conjugacy class of size nine, given by $\mathbb{Z}_3^2 y=\{x^az^by \mid a,b \in \{0,1,2\}\}$. We omit the explicit construction of the irreps of $N$~\cite{serre1977linear} and instead directly provide their characters evaluated on the conjugacy classes:
\be
\begin{split}
    &\chi_{(0,+)}=(1,1,1,1,1,1),\\
    &\chi_{(0,-)}=(1,1,1,1,1,-1),\\
    &\chi_{[0,1]}=(2,2,-1,-1,-1,0),\\
    &\chi_{[1,0]}=(2,-1,2,-1,-1,0),\\
    &\chi_{[1,1]}=(2,-1,-1,-1,2,0),\\
    &\chi_{[1,2]}=(2,-1,-1,2,-1,0).\\
\end{split}
\ee
Here, the vector components correspond to the conjugacy classes $\{1\}$, $\{x,x^2\}$, $\{z,z^2\}$, $\{xz,x^2z^2\}$, $\{xz^2,x^2z\}$, and $\mathbb{Z}_3^2 y$, ordered from left to right. According to Eq.~\eqref{eq:S32-Conjugacy-Classes}, the six conjugacy classes of the subgroup $N$ are embedded within the classes of $S_3^2$ as follows
\be
\begin{split}
    &\{1\}\subset c_1, \quad \{x,x^2\}\subset c_3, \quad \{z,z^2\}\subset c_2,\\
    &\{xz,x^2z^2\}\subset c_4,\quad \{xz^2,x^2z\}\subset c_4,\quad \mathbb{Z}_3^2 y\subset c_9.
\end{split}
\ee
Notably, the $S_3^2$ class $c_4$ splits into two distinct conjugacy classes within $N$: $\{xz,x^2z^2\}$ and $\{xz^2,x^2z\}$. By applying these subset relations to restrict the representations of $S_3^2$ down to the subgroup $N \cong \mathbb{Z}_3^2\rtimes \mathbb{Z}_2$, we obtain the branching rules from Eq.~\eqref{eq:S32-Group-Res}. These decompositions can be equivalently expressed as the branching matrix presented in Eq.~\eqref{eq:QTC-Generic-S32-Lineaer-B-Matrix}. Finally, substituting the branching matrix into Eq.~\eqref{app-eq:Group-Verlinde-Formula} and performing the requisite summations reveals that the fusion coefficients in Eq.~\eqref{eq:S3xS3-Fusion} precisely reproduce the linear representation theory of $S_3^2$.

In conclusion, while the $r_*=9$ bootstrap solution accurately reflects the linear representation theory of the abstract group $S_3^2$ associated with Eq.~\eqref{eq:Disorder-QTC-Hamiltonian}, it neglects anti-unitarity. Incorporating the anti-unitary operator requires the use of Wigner's corepresentation theory, as captured by our $r_*=5$ solution.

\subsubsection{Corepresentation Theory of $G^{\text{mag.}}$}
We now turn to Wigner's corepresentation theory for the magnetic group $G^{\text{mag.}}=\langle Z,X,R,K \rangle$, which explicitly accounts for the anti-unitary nature of $K$. First, we construct the magnetic classes (detailed in Appendix~\ref{app:Corepresentation}) and identify the characters that remain invariant within each class. This requires us to enumerate the six conjugacy classes of the unitary subgroup $N\cong N^{\text{uni.}}=\langle Z,X,R\rangle$
\be
\begin{split}
    &C_1=\{1\},\quad C_Z=\{Z,Z^2\}, \quad C_X=\{X,X^2\},\\
    &C_{XZ}=\{XZ,Z^2Z^2\},\quad C_{XZ^2}=\{XZ^2,X^2Z\},\\
    &C_R=\{X^aZ^bR|a,b\in\{0,1\}\},
\end{split}
\ee
whose respective sizes are $|C_1|=1$, $|C_Z|=|C_X|=|C_{XZ}|=|C_{XZ^2}|=2$, and $|C_R|=9$. We then apply the anti-unitary operator $K$ to these sets via the conjugation $n \mapsto KnK^{-1}$ for all $n \in N^{\text{uni.}}$. According to the algebraic relations in Eq.~\eqref{eq:M-Group-Algebra-S32}, this operation leaves the classes $C_1$, $C_Z$, $C_X$, and $C_R$ invariant, while mutually exchanging $C_{XZ}$ and $C_{XZ^2}$. Recall that a magnetic class $\tilde{c}_i$ of $G^{\text{mag.}}$ is defined as an equivalence class restricted to the unitary subgroup $N^{\text{uni.}}$. Specifically, two elements $n_1, n_2 \in N^{\text{uni.}}$ belong to the same magnetic class if they are related either by a unitary conjugation, $n n_1 n^{-1} = n_2$ for some $n \in N^{\text{uni.}}$, or by an anti-unitary conjugation that yields the inverse, $h n_1 h^{-1} = n_2^{-1}$ for some $h \in G^{\text{mag.}} \setminus N^{\text{uni.}}$. Thus, we can deduce that there are exactly five magnetic classes
\be\label{eq:QTC-Generic-Magnetic-Classes}
\begin{split}
    &\tilde{c}_1=C_1,\quad \tilde{c}_2=C_Z, \quad \tilde{c}_3=C_X,\\
    &\tilde{c}_4=C_{XZ}\cup C_{XZ^2},\quad\tilde{c}_5=C_R.
\end{split}
\ee
Their respective sizes are $|\tilde{c}_1|=1$, $|\tilde{c}_2|=|\tilde{c}_3|=2$, $|\tilde{c}_4|=4$, and $|\tilde{c}_5|=9$. This is the class structure relevant for corepresentation theory, i.e., two ordinary conjugacy classes merge into one magnetic class, so the number of relevant classes drops from six to five.

To complete the character table, we must additionally identify the ICRs of $G^{\text{mag.}}$. These are obtained by applying the anti-unitary operator to the $N$-irreps. The six $N$-irreps are represented by the matrices
\be
\begin{split}
    &\rho_{(0,\pm)}(Z)=\rho_{(0,\pm)}(X)=1,\quad \rho_{(0,\pm)}(R)=\pm 1,\\
    &\rho_{[0,1]}(Z)=\begin{pmatrix}
        \omega & 0\\
        0 & \omega^2
    \end{pmatrix},
    \quad \rho_{[0,1]}(X)=\begin{pmatrix}
        1 & 0\\
        0 & 1
    \end{pmatrix},\\
    &\rho_{[0,1]}(R)=\begin{pmatrix}
        0 & 1\\
        1 & 0
    \end{pmatrix},\quad \rho_{[1,0]}(Z)=\begin{pmatrix}
        1 & 0\\
        0 & 1
    \end{pmatrix},\\
    &\rho_{[1,0]}(X)=\begin{pmatrix}
        \omega & 0\\
        0 & \omega^2
    \end{pmatrix},\quad \rho_{[1,0]}(R)=\begin{pmatrix}
        0 & 1\\
        1 & 0
    \end{pmatrix}.\\
\end{split}
\ee
It is convenient to choose the remaining representatives such that
\be
\begin{split}
    &\rho_{[1,1]}(Z)=\rho_{[1,1]}(X)=\begin{pmatrix}
        \omega & 0\\
        0 & \omega^2
    \end{pmatrix},\\
    &\rho_{[1,2]}(X)=\rho_{[1,2]}(X)^{-1}=\begin{pmatrix}
        \omega^2 & 0\\
        0 & \omega
    \end{pmatrix},\\
    &\rho_{[1,1]}(R)=\rho_{[1,2]}(R)=\begin{pmatrix}
        0 & 1\\
        1 & 0
    \end{pmatrix}.\\
\end{split}
\ee
Next, we adjoin the anti-unitary operator $K$ using Eq.~\eqref{eq:M-Group-Algebra-S32} and define the $K$-twisted $N$-irreps as
\be\label{eq:K-Twisted-N-Irrep}
    \rho_{\lambda}^{(K)}(n)=\rho_{\lambda}(KnK^{-1})^*,
    \quad \forall n\in N,
\ee
where $\lambda\in\hat{N}$. From this definition, we obtain the relations
\be
\begin{split}
    &\rho_{(0,\pm)}^{(K)}\cong\rho_{(0,\pm)},
    \quad \rho_{[0,1]}^{(K)}\cong\rho_{[0,1]},
    \quad \rho_{[1,0]}^{(K)}\cong\rho_{[1,0]},\\
    &\rho_{[1,1]}^{(K)}\cong\rho_{[1,2]},
    \quad \rho_{[1,2]}^{(K)}\cong\rho_{[1,1]},
\end{split}
\ee
where the $N$-irreps partition into four size-one $K$-orbits leaving $(0,\pm),[0,1],[1,0]$ invariant, and one size-two $K$-orbit formed by the exchange of $[1,1]$ and $[1,2]$. Consequently, four size-one $K$-orbits lead to two one-dimensional and two-dimensional ICRs and the single size-two orbit corresponds to a four-dimensional ICR. We begin with the one-dimensional ICRs:
\be
\begin{split}
    &\rho_{\alpha_0}(Z)=\rho_{\alpha_0}(X)=\rho_{\alpha_0}(R)=1,
    \quad \rho_{\alpha_0}(K)=\rho_K,\\
    &\rho_{\alpha_1}(Z)=\rho_{\alpha_1}(X)=-\rho_{\alpha_0}(R)=1,\quad \rho_{\alpha_1}(K)=\rho_K,\\
\end{split}
\ee
where $\rho_K$ denotes complex conjugation. Since $\rho_K^2=1$, these two one-dimensional ICRs are of Type (a), i.e., $\rho_{\lambda}^{(K)}\cong \rho_{\lambda}$ and $\rho_K^2=1$ (see Appendix~\ref{app:Corepresentation} for details). Consequently, $\alpha_0$ and $\alpha_i$ act on the same one-dimensional space, extending the $(0,+)$ and $(0,-)$ representations, respectively. Similarly, the two-dimensional ICR $\alpha_j$ is obtained by extending $[0,1]$ with the assignment $\rho_{\alpha_j}(K)=\rho_K$, making it a Type (a) corepresentation as well. However, for $\alpha_k$, which is constructed from $[1,0]$, simply defining $\rho_{\alpha_k}(K)=\rho_K$ is insufficient because
\be
    \rho_K\rho_{[1,0]}(X)\rho_K=\rho_{[1,0]}(X)^{-1},
\ee
which violates Eq.~\eqref{eq:S32-Magnetic-Group}. Instead, we must compose the conjugation with the swap matrix to define
\be
    \rho_{\alpha_3}(K)=\begin{pmatrix}
        0 & 1\\
        1 & 0
    \end{pmatrix}\rho_K,
    \quad \rho_{\alpha_3}(K)^2=1.
\ee
This assignment satisfies Eq.~\eqref{eq:S32-Magnetic-Group} and is realized on the same two-dimensional representation space as $[1,0]$, thereby forming another Type (a) corepresentation. Finally, we consider the four-dimensional ICR constructed from the non-self-conjugate orbit $\{[1,1],[1,2]\}$. According to the general theory of corepresentations~\cite{wigner1959group,rumynin2021realrepresentationsc2gradedgroups,Mock_2016}, this orbit does not yield two separate ICRs. Instead, it induces a single Type (c) ICR, whose restriction to $N$ is given by
\be
    \text{Res}^{G^{\text{mag.}}}_{N}V_{\alpha_4}\cong V_{[1,1]}\oplus V_{[1,2]}.
\ee
Here, the Type (c) classification arises precisely because of the inequivalence relation $\rho_{\lambda}^{(K)}\not\cong\rho_{\lambda}$. We induce the corepresentation from the $N$-irreps $[1,1]$ and $[1,2]$, defined for all $n \in N$ as
\be
\begin{split}
    \rho_{\alpha_4}(n)=\begin{pmatrix}
        \rho_{[1,1]}(n) & 0\\
        0 & \rho_{[1,2]}(n)
    \end{pmatrix}.
\end{split}
\ee
By construction, this representation is irreducible. To satisfy the algebraic relations of Eq.~\eqref{eq:S32-Magnetic-Group}, the representation of the anti-unitary element is given by
\be
    \rho_{\alpha_4}(K)=\begin{pmatrix}
        0 & 1_{2\times 2}\\
        1_{2\times 2} & 0
    \end{pmatrix}\rho_K,\quad \rho_{\alpha_4}(K)^2=1_{4\times 4},
\ee
where $1_{n\times n}$ denotes the $n\times n$ identity matrix. Consequently, this induced corepresentation is indeed a valid ICR of the magnetic group described in Eq.~\eqref{eq:S32-Magnetic-Group}. Since all ICRs are extended directly from the $N$-irreps by adjoining $K$, their restriction to the subgroup $N$ is straightforward. This restriction is explicitly given by Eq.~\eqref{eq:S32-M-Group-B-Matrix-Res}, thereby confirming the branching matrix presented in Eq.~\eqref{eq:S32-M-Group-B-Matrix}.

The characters of ICRs are defined on the magnetic classes $\{\tilde{c}_i\}$:
\be
    \chi_{\alpha}(\tilde{c}_i)=\tr\rho_{\alpha}(n),
    \quad n\in\tilde{c}_i.
\ee
By substituting the representation matrices $\rho_{\alpha}$ and the magnetic classes from Eq.~\eqref{eq:QTC-Generic-Magnetic-Classes} into this definition, we obtain the character table presented in Table~\ref{tab:QTC-Corep-Character-Table}.

With the branching matrix and character table at hand, we now proceed to evaluate the fusion coefficients. In the case of magnetic groups, the orthogonality relations are defined relative to the unitary subgroup $N$. Specifically, the squared norm of an ICR character $\chi_{\alpha}$ satisfies~\cite{rumynin2021realrepresentationsc2gradedgroups,newmarch1983character}:
\be
    \langle \chi_{\alpha},\chi_{\alpha}\rangle_{N^{\text{uni.}}}=\mu_{\alpha},
\ee
where $\mu_{\alpha}$ is called the ``intertwiner number'' and determined by the Wigner classification of the corepresentation
\be
    \mu_{\alpha}=\begin{cases}
        1,~&\alpha~\text{is Type (a)},\\
        4,~&\alpha~\text{is Type (b)},\\
        2,~&\alpha~\text{is Type (c)}.\\
    \end{cases}
\ee
The standard group Verlinde formula is thus modified to account for this normalization factor:
\be\label{eq:FC-Character-Magnetic}
\begin{split}
    N_{\alpha_i\alpha_j}^{\alpha_k}&=\f{\langle \chi_{\alpha_i}\chi_{\alpha_j},\chi_{\alpha_k}\rangle_{N^{\text{uni.}}}}{\mu_{\alpha_k}}\\
    &=\f{\sum_{\tilde{c}}|\tilde{c}|\chi_{\alpha_i}(\tilde{c})\chi_{\alpha_j}(\tilde{c})\chi_{\alpha_k}(\tilde{c})^*}{\mu_{\alpha_k}|N^{\text{uni.}}|},
\end{split}
\ee
where $\alpha_{i,j,k}$ label the ICRs of $G^{\text{mag.}}$ and the summation runs over all magnetic classes $\tilde{c}$. Applying this formalism in conjunction with the branching matrix and character table, we confirm that Eq.~\eqref{eq:QTC-corep-fusion} correctly describes the fusion rules of the corepresentation theory of $G^{\text{mag.}}$.

In summary, the $r_*=5$ bootstrap solution accurately reproduces the predictions of corepresentation theory for symmetries governed by the magnetic group $(\mathbb{Z}_3^2\rtimes\mathbb{Z}_2)\sqcup K(\mathbb{Z}_3^2\rtimes\mathbb{Z}_2)$.
Remarkably, this is achieved without any prior knowledge of the underlying anti-unitary operator $K$, or any other information beyond the unitary subgroup $\mathbb{Z}_3^2\rtimes\mathbb{Z}_2$.

\subsection{Analytical Confirmation for Example IV $(\theta= \f{\pi}{4}\mod\pi)$\label{sec:QTC-SD-Analytic-Confirmation}}
Here, starting from the analytical representation theory of $G\cong\mathbb{Z}_3^2\rtimes \mathbb{Z}_4$, we explicitly verify that it perfectly reproduces the branching matrix in Eq.~\eqref{eq:QTC-Self-Dual-B-Matrix}, the fusion rules in Eq.~\eqref{eq:Z3sqZ4-Fusion}, and the character table in Table~\ref{tab:Z3sqZ4_character_table}. Crucially, this demonstrates that the applicability of our bootstrap method extends well beyond the paradigm of normal subgroups.

Let $G=\langle Z,X,U_H\rangle\cong\mathbb{Z}_3^2\rtimes\mathbb{Z}_4$, with the algebraic relations
\be\label{eq:Z32-Z4-Algebra}
\begin{split}
    &Z^3=X^3=U_H^4=1,\quad [Z,X]=0,\\
    &U_HZU_H^{-1}=X^{-1},\quad U_HXU_H^{-1}=Z.
\end{split}
\ee
The conjugacy classes of $G$ can be systematically constructed by partitioning elements based on the normal subgroup structure. Within the normal subgroup $\mathbb{Z}_3^2 = \langle Z, X \rangle$, the conjugation action yields three distinct orbits:
\be\label{eq:Z32Z4-Conjugacy-Class-1}
\begin{split}
    &c_1=\{1\},
    \quad c_2=\{X,Z,X^2,Z^2\},\\
    &c_3=\{XZ,X^2Z^2,X^2Z,XZ^2\}.
\end{split}
\ee
The remaining conjugacy classes arise from the non-trivial cosets of $\mathbb{Z}_3^2$ in $G$. To determine these, we examine the conjugation of a generic coset element $n_1 U_H^a$ (where $n_1 \in \mathbb{Z}_3^2$ and $a \in \{1, 2, 3\}$) by an element $n_2 \in \mathbb{Z}_3^2$
\be\label{eq:Z32-Z4-Coset-Conjugation}
    n_2\left(n_1U_H^{a}\right)n_2^{-1}=\left[n_1+(1-A^a)n_2\right]U_H^a.
\ee
Here, we employ additive vector notation by identifying the group element $X^x Z^y \in \mathbb{Z}_3^2$ with the exponent column vector $\mathbf{v} = (x,y)^T \in \mathbb{F}_3^2$. The matrix $A \in \text{GL}(2, \mathbb{F}_3)$ encodes the conjugation action of $U_H$ on this vector space
\be
    U_HX^aZ^bU_H^{-1}=X^{a'}Z^{b'},
    \quad (a',b')^T=A(a,b)^T.
\ee
In this basis, the matrix $A$ is given by
\be
    A=\begin{pmatrix}
        0 & -1\\
        1 & 0
    \end{pmatrix}.
\ee
Since $(I - A^a)$ is invertible for $a \in \{1, 2, 3\}$, the equation $n_1 + (I - A^a)n_2 = 0$ always admits a solution for $n_2$ for any $n_1$. Consequently, every element $n_1 U_H^a$ is conjugate to $U_H^a$, and each coset $\mathbb{Z}_3^2 U_H^a$ collapses into a single conjugacy class. This yields three additional classes, each of size nine
\be\label{eq:Z32Z4-Conjugacy-Class-2}
\begin{split}
    &c_4=\{nU_H^2|n\in\mathbb{Z}_3^2\},\\
    &c_5=\{nU_H|n\in\mathbb{Z}_3^2\},\\
    &c_6=\{nU_H^3|n\in\mathbb{Z}_3^2\}.\\
\end{split}
\ee
Combining these with the three classes from the normal subgroup $\mathbb{Z}_3^2$, we obtain a total of six conjugacy classes. Their respective sizes are $|c_1|=1$, $|c_2|=|c_3|=4$, and $|c_4|=|c_5|=|c_6|=9$. As expected, the sum of these sizes correctly yields $1 + 4 + 4 + 9 + 9 + 9 = 36$, which matches the order of the group $G \cong \mathbb{Z}_3^2 \rtimes \mathbb{Z}_4$.

Following the standard construction for semi-direct products~\cite{serre1977linear}, the four one-dimensional $G$-irreps are trivially realized on the normal subgroup by setting $\rho_{\alpha_i}(X) = \rho_{\alpha_i}(Z) = 1$ for all $i \in \{0,1,2,3\}$. Their dependence on $U_H$ is given by:
\be
\begin{split}
    &\rho_{\alpha_0}(U_H)=-\rho_{\alpha_2}(U_H)=1,\\
    &\rho_{\alpha_1}(U_H)=-\rho_{\alpha_3}(U_H)=i.\\
\end{split}
\ee
To construct the two four-dimensional irreps, we first define the one-dimensional characters of the normal subgroup $\mathbb{Z}_3^2$ as
\be
    \chi_{(p,q)}(X^aZ^b)=\omega^{pa+qb},
    \quad \omega=e^{\f{2\pi i}{3}},
\ee
where both pairs $(p,q)^T$ and $(a,b)^T$ are treated as vectors in $\mathbb{F}_3^2$. The conjugation action of $U_H$ on $\mathbb{Z}_3^2$ naturally induces a dual action on this character group. Consequently, the eight non-trivial characters split into two disjoint $U_H$-orbits, each of length four: $\{(0,1),(2,0),(0,2),(1,0)\}$ and $\{(1,1),(2,2),(1,2),(2,1)\}.$ Characters within the same $U_H$-orbit induce equivalent representations of $G$, whereas those from different orbits yield inequivalent ones. Thus, each orbit uniquely determines a distinct induced representation. Furthermore, as each orbit contains exactly four characters, the resulting induced representations are four-dimensional. We can therefore construct the two inequivalent four-dimensional $G$-irreps by selecting one representative character from each orbit, such as $\chi_{(1,1)}$ and $\chi_{(0,1)}$, and inducing them from the normal subgroup $N = \mathbb{Z}_3^2$ to $G$, 
\be
    \rho_{\alpha_4}\cong \text{Ind}^{G}_N\chi_{(1,1)},
    \quad \rho_{\alpha_5}\cong \text{Ind}^{G}_N\chi_{(0,1)}.
\ee

Next, recall that the coset representatives of the quotient group $G/\mathbb{Z}_3^2 \cong \mathbb{Z}_4$ are given by $\{1, U_H, U_H^2, U_H^3\}$. Let the induced representation span a basis $\{e_0, e_1, e_2, e_3\}$. In this basis, the generator $U_H$ acts via a cyclic shift, namely $\rho_{\alpha_i}(U_H)e_k = e_{k+1 \pmod 4}$ for $i \in \{4, 5\}$. Consequently, for both $\alpha_4$ and $\alpha_5$, the representation matrix for $U_H$ takes the form:
\be
    \rho_{\alpha_i}(U_H)=\begin{pmatrix}
         0 & 0 & 0 & 1\\
         1 & 0 & 0 & 0\\
         0 & 1 & 0 & 0\\
         0 & 0 & 1 & 0\\
    \end{pmatrix},~i=4,5.
\ee
Applying the standard formula for induced representations~\cite{serre1977linear}, we determine the action of the normal subgroup generators $Z$ and $X$
\be
\begin{split}
    &\rho_{\alpha_4}(Z)=\begin{pmatrix}
        \omega & 0 & 0 & 0\\
        0 & \omega & 0 & 0\\
        0 & 0 & \omega^2 & 0\\
        0 & 0 & 0 & \omega^2\\
    \end{pmatrix},~\rho_{\alpha_4}(X)=\begin{pmatrix}
        \omega & 0 & 0 & 0\\
        0 & \omega^2 & 0 & 0\\
        0 & 0 & \omega^2 & 0\\
        0 & 0 & 0 & \omega\\
    \end{pmatrix},\\
    &\rho_{\alpha_5}(Z)=\begin{pmatrix}
        \omega & 0 & 0 & 0\\
        0 & 1 & 0 & 0\\
        0 & 0 & \omega^2 & 0\\
        0 & 0 & 0 & 1\\
    \end{pmatrix},~\rho_{\alpha_5}(X)=\begin{pmatrix}
        1 & 0 & 0 & 0\\
        0 & \omega^2 & 0 & 0\\
        0 & 0 & 1 & 0\\
        0 & 0 & 0 & \omega\\
    \end{pmatrix},
\end{split}
\ee
which can be explicitly verified to satisfy the algebraic relations in Eq.~\eqref{eq:Z32-Z4-Algebra}. With the full set of representation matrices for $G$ established, the characters are readily obtained by taking the trace of each matrix. This procedure yields the character table of $\mathbb{Z}_3^2 \rtimes \mathbb{Z}_4$, which is in perfect agreement with Table~\ref{tab:Z3sqZ4_character_table}.

To recover the branching matrix in Eq.~\eqref{eq:QTC-Self-Dual-B-Matrix}, we restrict the group $G$ to the subgroup $N = \langle Z, X, R=U_H^2 \rangle$. Under this restriction, the operator $R=U_H^2$ leaves the two orthogonal subspaces spanned by $\{e_0, e_2\}$ and $\{e_1, e_3\}$ invariant. By decomposing these invariant subspaces in terms of the $N$-irreps established in Sec.~\ref{sec:QTC-Generic-Confirmation}, we obtain the explicit restriction mappings in Eq.~\eqref{eq:QTC-Self-Dual-Restrictions}. This successfully provides an analytical confirmation of the branching matrix presented in Eq.~\eqref{eq:QTC-Self-Dual-B-Matrix}. Furthermore, the explicit fusion rules Eq.~\eqref{eq:Z3sqZ4-Fusion} are analytically corroborated by applying Eq.~\eqref{app-eq:Group-Verlinde-Formula}.

\section{Analytical Study of the Projective Representation of $D_8\times \mathbb{Z}_2$\label{app:Proj-D8Z2}}
In this appendix, we analytically construct the projective irreducible representations relevant to the odd-particle-number sector of the driven Bose-Hubbard model discussed in Sec.~\ref{sec:FD-Bose-Hubbard}. Let the manifest subgroup be
\be
    N=D_8=\langle T,\mathcal{P}\mid T^8=\mathcal{P}^2=1,\quad \mathcal{P}T\mathcal{P}=T^{-1}\rangle,
\ee
and let the additional generator of the abstract factor $\mathbb{Z}_2^{(\pi)}$ be denoted by
\be
    U_{\pi}=R_{\pi}\mathcal{P}.
\ee
For odd particle number $\mathcal{N}$, Eq.~\eqref{eq:KDBH-Rpi-T-commutation} implies the projective relations
\be\label{eq:Proj-D8Z2-Relations}
\begin{split}
    &U_{\pi}TU_{\pi}^{-1}=-T,\quad [U_{\pi},\mathcal{P}]=0,\quad U_{\pi}^2=1,
\end{split}
\ee
where the scalar factor $-1$ is the value of $(-1)^{\hat{\mathcal{N}}}$ in the fixed odd-$\mathcal{N}$ sector. Equivalently, one may view this projective representation as an ordinary representation of the central extension
\be
\begin{split}
    \widetilde G=\langle T,\mathcal{P},U_{\pi},z\mid &T^8=\mathcal{P}^2=U_{\pi}^2=z^2=1,\\
    &z~\text{central},\quad \mathcal{P}T\mathcal{P}=T^{-1},\\
    &[U_{\pi},\mathcal{P}]=1,\quad U_{\pi}TU_{\pi}^{-1}=zT\rangle,
\end{split}
\ee
followed by restricting to the sector in which $z$ acts as $-I$.

We first recall the ordinary irreps of $D_8$. The four one-dimensional irreps are
\be
\begin{split}
    &\rho_{(0,\pm)}(T)=1,\quad \rho_{(0,\pm)}(\mathcal{P})=\pm 1,\\
    &\rho_{(4,\pm)}(T)=-1,\quad \rho_{(4,\pm)}(\mathcal{P})=\pm 1,
\end{split}
\ee
while the three two-dimensional irreps are
\be\label{eq:D8-2D-Irreps}
\begin{split}
    \rho_{m}(T)=\begin{pmatrix}
        \omega^m & 0\\
        0 & \omega^{-m}
    \end{pmatrix}&,
    \quad
    \rho_{m}(\mathcal{P})=\begin{pmatrix}
        0 & 1\\
        1 & 0
    \end{pmatrix},\\
    \omega=e^{i\pi/4}&,\quad m=1,2,3.
\end{split}
\ee
The projective action of $U_{\pi}$ induces a twist on $D_8$-representations:
\be\label{eq:Proj-D8Z2-Twist}
    \rho^{(\pi)}(T)=-\rho(T),\quad \rho^{(\pi)}(\mathcal{P})=\rho(\mathcal{P}).
\ee
Applying this twist to the seven irreps of $D_8$ yields four orbits:
\be\label{eq:Proj-D8Z2-Orbits}
\begin{aligned}
     (0,+)\leftrightarrow(4,+)&,\quad
    (0,-)\leftrightarrow(4,-),\quad \\
     1\leftrightarrow 3&,\quad
    2\mapsto 2.
\end{aligned}
\ee
These four orbits generate the five irreducible projective objects relevant to the odd-$\mathcal{N}$ sector.

The first two projective irreps arise from the paired one-dimensional orbits in Eq.~\eqref{eq:Proj-D8Z2-Orbits}. They are two-dimensional and can be written as
\be\label{eq:Proj-D8Z2-Xpm}
\begin{split}
    \rho_{\mathcal{X}_{+}}(T)& =\rho_{\mathcal{X}_{-}}(T)=\begin{pmatrix}
        1 & 0\\
        0 & -1
    \end{pmatrix},\\
    \rho_{\mathcal{X}_{+}}(\mathcal{P})&=-\rho_{\mathcal{X}_{-}}(\mathcal{P})=1_{2\times 2} ~, \\
    \rho_{\mathcal{X}_{+}}(U_{\pi})& =  \rho_{\mathcal{X}_{-}}(U_{\pi}) =\begin{pmatrix}
        0 & 1\\
        1 & 0
    \end{pmatrix}
\end{split}
\ee
These matrices satisfy Eq.~\eqref{eq:Proj-D8Z2-Relations} and, when restricted to $N\cong D_8$, they decompose respectively into the pairs $(0,+)\oplus(4,+)$ and $(0,-)\oplus(4,-)$.

The self-twisted orbit $2\mapsto 2$ produces two inequivalent two-dimensional projective irreps:
\be\label{eq:Proj-D8Z2-Ypm}
\begin{split}
    \rho_{\mathcal{Y}_{\pm}}(T)& =\begin{pmatrix}
        i & 0\\
        0 & -i
    \end{pmatrix},
    \quad
    \rho_{\mathcal{Y}_{\pm}}(\mathcal{P})=\begin{pmatrix}
        0 & 1\\
        1 & 0
    \end{pmatrix}, \\
    \rho_{\mathcal{Y}_{\pm}}(U_{\pi})& =\pm\begin{pmatrix}
        0 & 1\\
        1 & 0
    \end{pmatrix}.
\end{split}
\ee
They are projective because $\rho_{\mathcal{Y}_{\pm}}(\mathcal{P})\rho_{\mathcal{Y}_{\pm}}(T)\rho_{\mathcal{Y}_{\pm}}(\mathcal{P})^{-1}=\rho_{\mathcal{Y}_{\pm}}(T)^{-1}=-\rho_{\mathcal{Y}_{\pm}}(T)$. The two signs lead to inequivalent projective irreps: by Schur's lemma, any intertwiner commuting with the irreducible $D_8$-action $\rho_2$ is proportional to the identity, and therefore cannot flip the sign of $U_{\pi}$.

Finally, the orbit $1\leftrightarrow 3$ yields a four-dimensional projective irrep. Using the explicit matrices of Eq.~\eqref{eq:D8-2D-Irreps}, we may choose
\be\label{eq:Proj-D8Z2-Z}
\begin{split}
    &\rho_{\mathcal{Z}}(T)=
    \begin{pmatrix}
        \omega & 0 & 0 & 0\\
        0 & \omega^{-1} & 0 & 0\\
        0 & 0 & \omega^3 & 0\\
        0 & 0 & 0 & \omega^{-3}
    \end{pmatrix},\\
    &\rho_{\mathcal{Z}}(\mathcal{P})=
    \begin{pmatrix}
        0 & 1 & 0 & 0\\
        1 & 0 & 0 & 0\\
        0 & 0 & 0 & 1\\
        0 & 0 & 1 & 0
    \end{pmatrix},
    \quad
    \rho_{\mathcal{Z}}(U_{\pi})=
    \begin{pmatrix}
        0 & 0 & 0 & 1\\
        0 & 0 & 1 & 0\\
        0 & 1 & 0 & 0\\
        1 & 0 & 0 & 0
    \end{pmatrix}.
\end{split}
\ee
The operator $\rho_{\mathcal{Z}}(U_{\pi})$ exchanges the two $D_8$ sectors and enforces $\rho_{\mathcal{Z}}(U_{\pi})\rho_{\mathcal{Z}}(T)\rho_{\mathcal{Z}}(U_{\pi})^{-1}=-\rho_{\mathcal{Z}}(T)$, exactly as required by Eq.~\eqref{eq:Proj-D8Z2-Relations}.

We can now read off the restriction of each projective irrep to the manifest subgroup $N=D_8$:
\be\label{eq:Proj-D8Z2-Restrictions}
\begin{split}
    &{\rm Res}_{N}^{G}\mathcal{X}_{+}=(0,+)\oplus(4,+),\quad
    {\rm Res}_{N}^{G}\mathcal{X}_{-}=(0,-)\oplus(4,-),\\
    &{\rm Res}_{N}^{G}\mathcal{Y}_{+}=2,\quad
    {\rm Res}_{N}^{G}\mathcal{Y}_{-}=2,\quad
    {\rm Res}_{N}^{G}\mathcal{Z}=1\oplus 3.
\end{split}
\ee
Using the ordered basis $\{(0,+),(0,-),1,2,3,(4,+),(4,-)\}$ for the $D_8$ irreps, these restrictions correspond to the branching vectors
\be\label{eq:Proj-D8Z2-Branching-Vectors}
\begin{split}
    &\vec{b}_{\mathcal{X}_{+}}=(1,0,0,0,0,1,0)^T,\quad
    \vec{b}_{\mathcal{X}_{-}}=(0,1,0,0,0,0,1)^T,\\
    &\vec{b}_{\mathcal{Z}}=(0,0,1,0,1,0,0)^T,\quad
    \vec{b}_{\mathcal{Y}_{\pm}}=(0,0,0,1,0,0,0)^T.
\end{split}
\ee
This exactly reproduces the xSFF-guided column types given in Eq.~\eqref{eq:proj_branch}: the columns $\vec{b}_{\alpha_1}, \vec{b}_{\alpha_2},$ and $\vec{b}_{\alpha_3}$ occur once, while $\vec{b}_{\alpha_4}$ occurs twice.

The projective dimensions are
\be
    \dim\mathcal{X}_{\pm}=\dim\mathcal{Y}_{\pm}=2,\quad \dim\mathcal{Z}=4,
\ee
and their squared dimensions sum to
\be
    2^2+2^2+2^2+2^2+4^2=32=|D_8\times\mathbb{Z}_2|.
\ee
Nevertheless, these five objects do not form a fusion category within a fixed cocycle sector: the tensor unit belongs to the untwisted linear sector, not to $\{\mathcal{X}_{\pm},\mathcal{Y}_{\pm},\mathcal{Z}\}$. Consequently, the tensor product of two objects in this projective sector necessarily leaves the sector. This analytically explains why the xSFF correctly reconstructs the branching data in Sec.~\ref{sec:FD-Bose-Hubbard}, while the bootstrap finds no rigid fusion ring compatible with ordinary linear representation theory.

\section{Adjoint Representations for $\mathfrak{su}(3), \mathfrak{so}(5)$ and $G_2$}
\label{app:adjoint_rep}

In this appendix, we discuss the adjoint representations for $\mathfrak{su}(3), \mathfrak{so}(5)$ and $G_2$ and its branching structure to $\mathfrak{u}(1) \oplus \mathfrak{u}(1)$. 

The $\mathfrak{su}(3)$ algebra is formed by all the complex anti-Hermitian traceless matrices. We choose the following basis
\begin{equation}
    H_1 = 
    \begin{pmatrix}
        1 & 0& 0\\
        0 & -1 & 0\\
        0 &0 &0
    \end{pmatrix}
    ~, \quad 
    H_2 = 
    \begin{pmatrix}
        0 & 0 &0 \\
        0 & 1 & 0 \\
        0 & 0 & -1
    \end{pmatrix}
    ~,
\end{equation}
with $E_{12},E_{13},E_{21},E_{23},E_{31},E_{32}$, where $E_{ij}$ is the matrix unit with $1$ in position $(i,j)$ and zeros elsewhere. $H_1,H_2$ form the Cartan sub-algebra of $\mathfrak{su}(3)$. The other generators can be constructed by considering $E_{ij}+E_{ji}, -i E_{ij} + i E_{ji}$. The adjoint representation denotes the mapping of the algebra generator $X$ to its adjoint action on algebra itself ${\rm ad}_X\equiv[X,\cdot]$. For the $\mathfrak{su}(3)$ case, once restricting this representation onto $\mathfrak{u}(1) \oplus \mathfrak{u}(1)$, we only need to the adjoint operations ${\rm ad}_{H_1}$ and ${\rm ad}_{H_2}$. It is important to notice that these two operators commute and we can therefore simultaneously diagonalize them. The eigenvalues and eigenvectors are given by
\begin{equation}
\begin{aligned}
    [H_1,E_{12}] &= 2 E_{12} ~, \quad [H_1, E_{13}] = E_{13} ~, \\
    [H_1,E_{21}] & -2 E_{21} ~, \quad [H_1, E_{23}] = - E_{23} ~, \\
    [H_1, E_{31}] & = - E_{31} ~, [H_1, E_{32}] = E_{32} ~,
\end{aligned}
\end{equation}
and 
\begin{equation}
\begin{aligned}
    [H_2, E_{12}] & = - E_{12} ~, \quad [H_2, E_{13}] = E_{13} ~, \\
    [H_2, E_{21}] & = E_{21} ~,\quad [H_2, E_{23}]  =2 E_{23} ~, \\
    [H_2, E_{31}] & = - E_{31} ~, \quad [H_2, E_{32}] = -2 E_{32} ~.
\end{aligned}
\end{equation}
The eigenvalues correspond to the weights of the generators. For example, $H_1, H_2$ both have weight $(0,0)$, $E_{12}$ has weight $(2,-1)$, and so on. Now, we can decompose the adjoint representation of $\mathfrak{su(3)}$ into irreps of $\mathfrak{u}(1) \oplus \mathfrak{u}(1)$
\begin{equation}
\begin{aligned}
    {\rm Res}_{\mathfrak{u}(1) \oplus \mathfrak{u}(1)}^{\mathfrak{su}(3)}V_{\rm adj} & = 2 V_{(0,0)} \oplus V_{(2,-1)} \oplus V_{(1,1)} \oplus V_{(-2,1)} \\
    & \quad \oplus V_{(-1,2)} \oplus V_{(-1,-1)} \oplus V_{(1,-2)} ~.
\end{aligned}
\end{equation}
The factor ``2'' appears precisely because $H_1, H_2$ both have weights $(0,0)$. This immediately shows that the branching multiplicity $b_{(0,0),{\rm adj}} > 1$, thereby excluding ${\rm SU}(3)$ as a candidate hidden symmetry from the xSFF data in Fig.~\ref{fig:Hubbard_xSFF}.

The same conclusion applies to $\mathfrak{so}(5)$ and $G_2$. The $\mathfrak{so}(5)$ algebra consists of the ten antisymmetric $5\times 5$ real matrices, spanned by $M_{ij}=E_{ij}-E_{ji}$ for $1\leq i<j \leq 5$. We adopt the Hermitian generators $L_{ij}=iM_{ij}$, which satisfy the compact-form commutation relations
\begin{equation}\label{eq:so5-commutators}
    [L_{ij},L_{kl}] = i\left(\delta_{jk} L_{il} - \delta_{jl} L_{ik} - \delta_{ik} L_{jl} + \delta_{il} L_{jk}\right) ~.
\end{equation}
Since the rotations in the $(1,2)$- and $(3,4)$-planes act on disjoint coordinate pairs, the operators
\begin{equation}
    H_1 = L_{12} ~, \quad H_2 = L_{34}
\end{equation}
mutually commute and furnish a Cartan subalgebra. The remaining eight generators are $L_{13},L_{14},L_{23},L_{24},L_{15},L_{25},L_{35},L_{45}$, and organize into four pairs of raising/lowering operators. Introduce the linear combinations
\begin{equation}
\begin{aligned}
    F^{\pm}_{1} &:= L_{15}\mp i L_{25} ~, \quad  F^{\pm}_{2} := L_{35}\mp i L_{45} ~, \\
    G^{\pm,+} &:= (L_{13}-L_{24}) \mp i (L_{14}+L_{23}) ~, \\
    G^{\pm,-} &:= (L_{13}+L_{24}) \mp i (L_{14}-L_{23}) ~,
\end{aligned}
\end{equation}
so that the nonzero commutators with $(H_1,H_2)$ take the form
\begin{equation}
\begin{aligned}
    [H_1,F^{\pm}_{1}] &= \pm F^{\pm}_{1} ~, & [H_2,F^{\pm}_{1}] &= 0 ~, \\
    [H_1,F^{\pm}_{2}] &= 0 ~, & [H_2,F^{\pm}_{2}] &= \pm F^{\pm}_{2} ~, \\
    [H_1,G^{\pm,s}] &= \pm G^{\pm,s} ~, & [H_2,G^{\pm,s}] &= sG^{\pm,s} ~,
\end{aligned}
\end{equation}
with $s \in \{+1,-1\}$.
In other words, $F^{\pm}_{1}$ has weight $(\pm 1,0)$, $F^{\pm}_{2}$ has weight $(0,\pm 1)$, and $G^{\pm,s}$ has weight $(\pm 1, s)$. Together with the two Cartan generators at weight $(0,0)$, the ten roots recover the $B_2$ root system. The adjoint representation therefore decomposes under $\mathfrak{u}(1)\oplus\mathfrak{u}(1)$ as
\begin{equation}
\begin{aligned}
    {\rm Res}^{\mathfrak{so}(5)}_{\mathfrak{u}(1)\oplus\mathfrak{u}(1)} V_{\rm adj} &= 2V_{(0,0)} \oplus V_{(1,0)} \oplus V_{(-1,0)} \\
    & \quad \oplus V_{(0,1)} \oplus V_{(0,-1)}  \oplus V_{(1,1)} \oplus V_{(-1,-1)} \\
    & \quad \oplus V_{(1,-1)} \oplus V_{(-1,1)} ~.
\end{aligned}
\end{equation}
As in the $\mathfrak{su}(3)$ case, the zero-weight space has multiplicity $2 = \mathrm{rank}(\mathfrak{so}(5))$, hence $b_{(0,0),{\rm adj}}>1$ and $\mathrm{SO}(5)$ is excluded from the xSFF data of Fig.~\ref{fig:Hubbard_xSFF}.

The exceptional Lie algebra $G_2$ is $14$-dimensional with rank two. An analogous analysis yields the $\mathfrak{u}(1)\oplus\mathfrak{u}(1)$ decomposition of its adjoint representation,
\begin{equation}
\begin{aligned}
    {\rm Res}^{G_2}_{\mathfrak{u}(1)\oplus\mathfrak{u}(1)} V_{\rm adj}
    &= 2 V_{(0,0)} \oplus V_{(1,0)} \oplus V_{(-1,0)} \\
    & \quad \oplus V_{(1/2,\sqrt{3}/2)} \oplus V_{(-1/2,-\sqrt{3}/2)} \\
    & \quad \oplus V_{(-1/2,\sqrt{3}/2)} \oplus V_{(1/2,-\sqrt{3}/2)} \\
    & \quad \oplus V_{(0,\sqrt{3})} \oplus V_{(0,-\sqrt{3})} \\
    & \quad \oplus V_{(3/2,\sqrt{3}/2)} \oplus V_{(-3/2,-\sqrt{3}/2)} \\
    & \quad \oplus V_{(-3/2,\sqrt{3}/2)} \oplus V_{(3/2,-\sqrt{3}/2)} ~,
\end{aligned}
\end{equation}
with the zero weight carrying multiplicity $b_{(0,0),{\rm adj}} \geq 2 $, so $G_2$ is again ruled out by the xSFF data.

\section{xSFF for an Integrable Model \label{app:non-random-FH-Model}}
\begin{figure*}[t]
    \centering
    \includegraphics[width=0.8\linewidth]{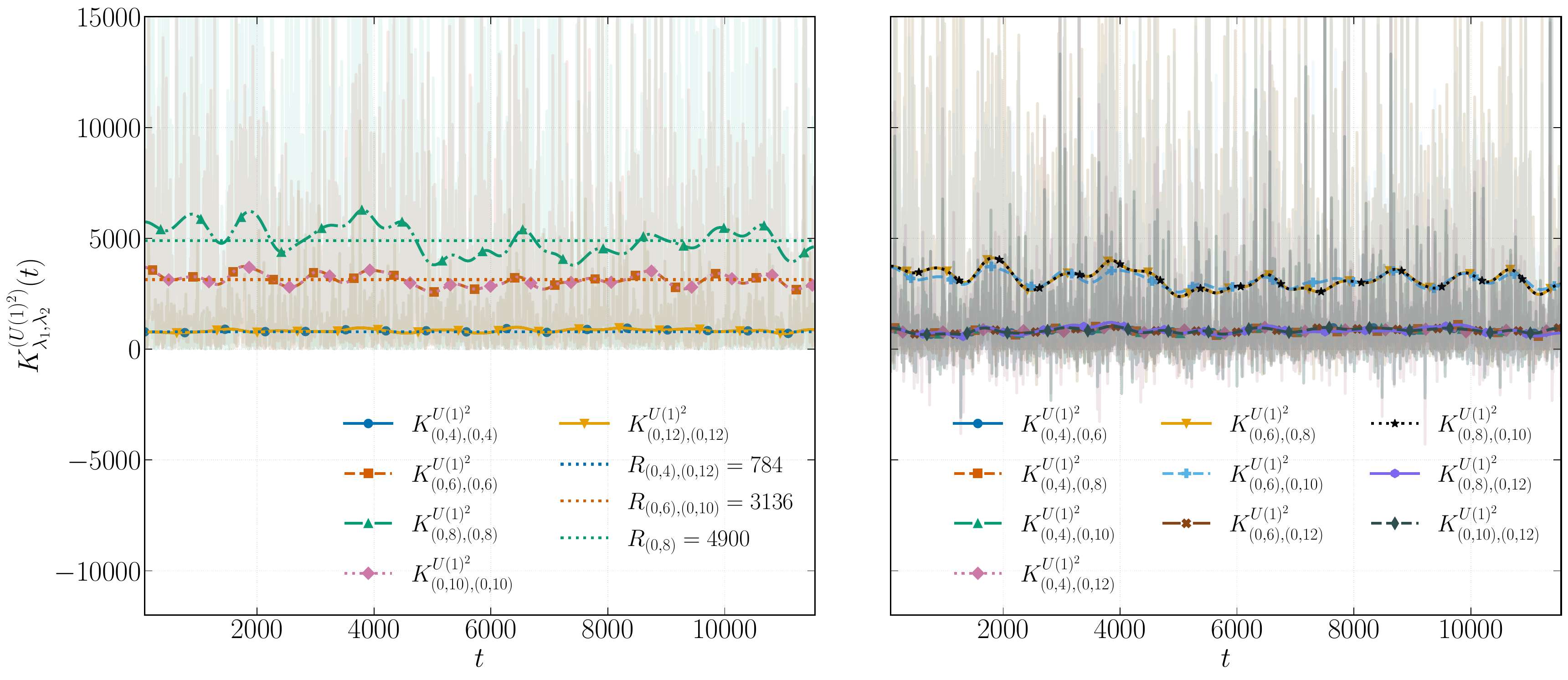}
    \caption{The xSFF for the Fermi-Hubbard model with $L = 8$ sites, constant interaction strength $U = 4$, and uniform hopping $t_{j} = t_c = 3$ for all $j$. The data are smoothed using a Gaussian moving average and projected onto the $\text{U}(1)_{S^z} \times \text{U}(1)_{\mathcal{N}}$ sectors with $\mathcal{N} = 4, \ldots, 12$ (charge $Q = \mathcal{N} - L = -4, \ldots, +4$). As in Fig.~\ref{fig:Hubbard_xSFF}, only the $S^z = 0$ sectors are shown. \textbf{Left}: Diagonal elements $K_{(0,\mathcal{N}),(0,\mathcal{N})}^{(\text{U}(1)^2)}$ for sectors near half-filling. \textbf{Right}: Off-diagonal elements $K_{(0,\mathcal{N}_1),(0,\mathcal{N}_2)}^{(\text{U}(1)^2)}$ emanating from the half-filling sector $\mathcal{N}_{\mathrm{half}} = L = 8$. Solid lines denote even $|\Delta \mathcal{N}|$ (predicted to be non-zero by $\mathrm{SU}(2)$ $\eta$-pairing), while dashed lines denote odd $|\Delta \mathcal{N}|$ (predicted to vanish). Despite larger fluctuations compared to the disordered case, the separation between even and odd $\Delta \mathcal{N}$ clearly confirms the $\Delta Q \in 2\mathbb{Z}$ selection rule.}
    \label{fig:Fixed_FH}
\end{figure*}
To ensure the plateaus remain clean, we employ random couplings throughout the main body of this paper. Since introducing randomness explicitly breaks the integrability of the model, a natural concern arises regarding whether these results generalize to Hamiltonians with fixed parameters, particularly those of integrable models. 

In this appendix, we examine the one-dimensional Fermi-Hubbard model defined in Eq.~\eqref{eq:FH-Hamiltonian}, restricting our focus to a constant interaction strength $U$ and uniform hopping parameters $t_j = t_c$ for all sites. The resulting xSFF plateaus are presented in Fig.~\ref{fig:Fixed_FH}. Although these plateaus exhibit more fluctuations than those generated with random couplings, performing a Gaussian moving average over temporal windows successfully recovers the characteristic features observed in Fig.~\ref{fig:Hubbard_xSFF}. As a result, the analyses presented in Sec.~\ref{sec:Fermi-Hubbard} are fully applicable, yielding the identical symmetry group classification $G\cong \text{SO}(4)$. Ultimately, this demonstrates that our approach is robust and effectively generalizes to deterministic Hamiltonians, irrespective of integrability. 

\section{Alternative Bootstrap Solutions\label{app:Other-Bootstrap-Solutions}}
In this appendix, we provide further details on several alternative bootstrap solutions at lowest rank, which were not discussed in the main text.

\subsection{Corepresentation Solution for $S_3$ Invariant Chain\label{app:Other-S3-Corep-Solution}}
In addition to the linear representation identified at rank $r_*=3$ in Sec.~\ref{sec:S3-Inv-OF-Chain}, the corepresentation search within our bootstrap algorithm yields an alternative solution at rank $r_*=2$.

The branching matrix is
\be
    B_{\lambda,\alpha}=\begin{pmatrix}
        1 & 0\\
        0 & 1\\
        0 & 1
    \end{pmatrix},
\ee
where the rows correspond to the $\mathbb{Z}_3$-irreps $\lambda=0, 1, 2$ from top to bottom, and the columns are indexed by the two ICRs: the one-dimensional $\alpha_0$ and the two-dimensional $\alpha_i$.

Then, the corresponding fusion rules are given by
\be
\begin{split}
    (\mathrm{a})~& N_{\alpha_0\alpha_0}^{\alpha_0}=N_{\alpha_0\alpha_1}^{\alpha_1}=1,\\
    (\mathrm{b})~&N_{\alpha_1\alpha_1}^{\alpha_0}=2,\quad N_{\alpha_1\alpha_1}^{\alpha_1}=1,
\end{split}
\ee
where (a) denotes the unit rules; and (b) describes the self-fusion of the two-dimensional irreps. Furthermore, by defining fusion matrices and solving their common eigenvalues, we obtain the character table shown in Table~\ref{tab:S3-Corep-Solution}.
\begin{table}
    \centering
    \begin{tabular}{c|cc}
         & $\tilde{c}_1$ & $\tilde{c}_2$ \\
         \hline
        $\chi_{\alpha_0}$ & 1 & 1\\
        $\chi_{\alpha_i}$ & 2 & -1\\
         \hline
    \end{tabular}
    \caption{Character table of corepresentation solution for $S_3$ invariant chain at rank $r_*=2$. The magnetic classes have sizes $|\tilde{c}_1|=1$ and $|\tilde{c}_2|=2$.}
    \label{tab:S3-Corep-Solution}
\end{table}

Applying the corepresentation theory detailed in Appendix~\ref{app:Corepresentation}, we find that this solution perfectly matches the magnetic group
\be
    G^{\text{mag.}}\cong \mathbb{Z}_3\sqcup C\mathbb{Z}_3,
\ee
where $C$ is treated as an anti-unitary operator. 

In summary, this solution could be a potential candidate for $G$ if the exact form of $C$ is not known a priori. However, at the algebraic level, both this solution and the one introduced in Sec.~\ref{sec:S3-Inv-OF-Chain} indicate that the hidden abstract group is $S_3$. Therefore, there must exist an operator $C$ such that its local components $C_j$ satisfy
\be
    C_jX_jC_j=X_j^{-1},
\ee
yielding the unitary matrix presented in Eq.~\eqref{eq:S3-Charge-Conjugation}. By direct verification, the $r_*=3$ linear representation solution introduced in Sec.~\ref{sec:S3-Inv-OF-Chain} ultimately provides the correct representation theory for the system.

\subsection{Second Linear Representation Solution for Three-State Quantum Torus Chain At Self-Dual Points\label{app:Second-Solution-QTC-Self-Dual}}
As discussed in Sec. \ref{subsec:bootstrap_selfdual}, for the quantum torus chain at the self-dual point, the algorithm identifies \textit{two} distinct linear representation solutions at rank $r_*=6$. While the first solution corresponds to the physically correct representation theory of $\mathbb{Z}_3^2\rtimes\mathbb{Z}_4$, we now investigate the second solution.

The branching matrix is given by Eq.~\eqref{eq:QTC-Self-Dual-B-Matrix}, while the fusion rules are instead given by
\begin{equation}\label{eq:QTC-SD2-Solution2-Fusion}
\begin{split}
    (\mathrm{a})~&N_{\alpha_0\alpha}^{\alpha}
    =N_{\alpha\alpha_0}^{\alpha}=1,
\quad  \forall \alpha\in\hat G,\\
    (\mathrm{b})~&N_{\alpha_1\alpha_1}^{\alpha_0}
    =N_{\alpha_2\alpha_2}^{\alpha_0}
    =N_{\alpha_3\alpha_3}^{\alpha_0}=1,\\
    &N_{\alpha_1\alpha_2}^{\alpha_3}
    =N_{\alpha_2\alpha_3}^{\alpha_1}
    =N_{\alpha_3\alpha_1}^{\alpha_2}=1,\\
    (\mathrm{c})~&N_{\alpha_1\alpha_4}^{\alpha_4}
    =N_{\alpha_2\alpha_4}^{\alpha_4}
    =N_{\alpha_3\alpha_4}^{\alpha_4}=1,\\
    &N_{\alpha_1\alpha_5}^{\alpha_5}
    =N_{\alpha_2\alpha_5}^{\alpha_5}
    =N_{\alpha_3\alpha_5}^{\alpha_5}=1,\\
    (\mathrm{d})~&N_{\alpha_4\alpha_4}^{\alpha_0}
    =N_{\alpha_4\alpha_4}^{\alpha_1}
    =N_{\alpha_4\alpha_4}^{\alpha_2}
    =N_{\alpha_4\alpha_4}^{\alpha_3}
    =N_{\alpha_4\alpha_4}^{\alpha_4}=1,\\
    &N_{\alpha_4\alpha_4}^{\alpha_5}=2,\\
    &N_{\alpha_4\alpha_5}^{\alpha_4}
    =N_{\alpha_4\alpha_5}^{\alpha_5}=2,\\
    &N_{\alpha_5\alpha_5}^{\alpha_0}
    =N_{\alpha_5\alpha_5}^{\alpha_1}
    =N_{\alpha_5\alpha_5}^{\alpha_2}
    =N_{\alpha_5\alpha_5}^{\alpha_3}
    =N_{\alpha_5\alpha_5}^{\alpha_5}=1,\\
    &N_{\alpha_5\alpha_5}^{\alpha_4}=2.
\end{split}
\end{equation}
These fusion rules are almost identical to those in Eq.~\eqref{eq:Z3sqZ4-Fusion}, with the crucial distinction that the invertible sector $\{\alpha_0, \alpha_1, \alpha_2, \alpha_3\}$ forms a Klein four-group $V_4 \cong \mathbb{Z}_2^2$. Furthermore, by extracting the eigenvalues of the associated fusion matrices, we derive the complete character table presented in Table~\ref{tab:QTC-SD2-Solution2-Character}. 
\begin{table}[t]
\centering
\begin{tabular}{c|cccccc}
 & $c_1$ & $c_2$ & $c_3$ & $c_4$ & $c_5$ & $c_6$\\
\hline
$\chi_{\alpha_0}$ & 1 & 1 & 1 & 1 & 1 & 1\\
$\chi_{\alpha_i}$ & 1 & 1 & 1 & 1 & -1 & -1\\
$\chi_{\alpha_j}$ & 1 & 1 & 1 & -1 & -1 & 1\\
$\chi_{\alpha_k}$ & 1 & 1 & 1 & -1 & 1 & -1\\
$\chi_{\alpha_l}$ & 4 & 1 & -2 & 0 & 0 & 0\\
$\chi_{\alpha_m}$ & 4 & -2 & 1 & 0 & 0 & 0\\
\hline
\end{tabular}
\caption{Character table for the second solution of the three-state quantum torus chain at self-dual points. Here, the conjugacy classes have sizes $(|c_1|,|c_2|,|c_3|,|c_4|,|c_5|,|c_6|)=(1,4,4,9,9,9)$.}
\label{tab:QTC-SD2-Solution2-Character}
\end{table}

As briefly noted in Appendix.~\ref{sec:QTC-SD-Analytic-Confirmation}, this solution does not correspond to the representation theory of any finite group. In fact, let us suppose that it describes $\text{Rep}(G)$ for some group $G$. Then, the sum of the squared dimensions of the irreps would require the group order to be $|G|=36$. In addition, the existence of exactly four one-dimensional irreps indicates that the Abelianization is $G/[G,G] \cong V_4$, which implies that the commutator subgroup $[G,G]$ has order nine. Since four and nine are coprime, the Schur–Zassenhaus theorem~\cite{serre1977linear} guarantees that $G$ decomposes as a semi-direct product, which we identify as $G \cong \mathbb{Z}_3^2 \rtimes V_4$. However, as established previously, the representation theory of this group requires two-dimensional irreps, which are entirely absent from the calculated branching matrix.
Thus, a group-theoretical origin has to be ruled out. Moreover, these fusion rules cannot describe any near-group category, such as a Tambara-Yamagami category~\cite{Siehler2003NearGroup,TambaraYamagami1998}. Indeed, near-group fusion rings are defined with exactly one single non-invertible simple object. In contrast, our solution contains two distinct non-invertible objects, $\alpha_l$ and $\alpha_m$, whose tensor products mix both sectors. In conclusion, this solution must correspond to an exotic, non-group rigid fusion ring, which we can safely rule out.

\section{Branching, Fusion within Categorical Description and Tannakian Duality\label{app:Category-theory}}
In this appendix, we adopt a categorical formulation, for two main reasons. First, such language has become standard in modern discussions of generalized, and particularly non-invertible symmetries~\cite{shao2024whatsundonetasilectures,Luo:2023ive}, providing the natural mathematical framework for future generalizations of our bootstrap approach. Second, this framework is naturally suited to finite groups, whose representation theories canonically organize into rigid symmetric monoidal fusion categories~\cite{beer2018categoriesanyonstravelogue}, and compact Lie groups, whose representation theories form rigid symmetric monoidal (Tannakian) categories. Due to Tannakian duality, once the identified $G$-irreps are endowed with the complete structure of a symmetric monoidal category $\text{Rep}(G)$, the underlying group $G$ is uniquely determined.

A category is defined by its objects and the morphisms mapping between them. We therefore operate within the representation categories $\text{Rep}(G)$ and $\text{Rep}(N)$, where the objects are the finite-dimensional representations of $G$ and its subgroup $N\subset G$, and the morphisms are the corresponding intertwining linear maps. For finite groups and compact Lie groups, both categories are semisimple, i.e., every object admits a direct-sum decomposition into simple (irreducible) objects. Within this categorical language, the sets $\hat{G}$ and $\hat{N}$ naturally index the isomorphism classes of the simple objects in $\text{Rep}(G)$ and $\text{Rep}(N)$, respectively.

The operations of group restriction and induction give rise to two functors
\be\label{eq:Res-Ind-Functor}
\begin{split}
    &\text{Res}^G_N:~\text{Rep}(G)\to\text{Rep}(N),\\
    &\text{Ind}^G_N:~\text{Rep}(N)\to\text{Rep}(G),
\end{split}
\ee
where a functor acts as a structure-preserving map between categories. Let $V_{\alpha}\in\text{Rep}(G)$ and $V_{\lambda}\in\text{Rep}(N)$ be simple objects in two categories. Exploiting semisimplicity, the restriction of $V_{\alpha}$ admits the direct-sum decomposition
\be
    \text{Res}^G_NV_{\alpha}\cong\bigoplus_{\lambda\in\hat{N}}b_{\lambda,\alpha}V_{\lambda}.
\ee
The non-negative integers $b_{\lambda,\alpha}$ define the branching multiplicities. In this formalism, $b_{\lambda,\alpha}$ is given by the dimension of the morphism space (the space of intertwiners) between $V_{\lambda}$ and the restricted object $\text{Res}^G_N V_{\alpha}$ in $\text{Rep}(N)$:
\be
\begin{split}
    b_{\lambda,\alpha}&=\dim\text{Hom}_{\text{Rep}(N)}\left(V_{\lambda},\text{Res}^G_NV_{\alpha}\right)\\
    &=\dim\text{Hom}_{\text{Rep}(G)}\left(\text{Ind}^G_NV_{\lambda},V_{\alpha}\right),
\end{split}
\ee
where the second equality is a direct consequence of Frobenius reciprocity.

In the language of category theory, the fusion of irreps corresponds to the tensor product of simple objects $V_{\alpha_i}$ and $V_{\alpha_j}$ within the category $\mathrm{Rep}(G)$, as described by Eq.~\eqref{eq:G_fusion_rules}. The fusion coefficients $N_{\alpha_i\alpha_j}^{\alpha_k}$ are non-negative integers given by the dimension of the corresponding hom-space:
\be\label{eq:Def-Category-Fusion-Coefficients}
\begin{split}
    &N_{\alpha_i\alpha_j}^{\alpha_k}=\dim\text{Hom}_{\text{Rep}(G)}\left(V_{\alpha_k},V_{\alpha_i}\otimes V_{\alpha_j}\right).
\end{split}
\ee
Furthermore, $\mathrm{Rep}(G)$ forms a rigid symmetric monoidal category. In this context, ``monoidal'' indicates the existence of the tensor product $\otimes$; ``rigid'' ensures that every object $V_{\alpha}$ admits a dual (or contragredient) object $V_{\overline{\alpha}} = V_{\alpha}^{\vee}$ ($V_{\alpha}^{\vee}$ is the same as dual representation space); and ``symmetric'' denotes the canonical isomorphism $V_{\alpha_i} \otimes V_{\alpha_j} \cong V_{\alpha_j} \otimes V_{\alpha_i}$. Collectively, these fundamental categorical structures strictly enforce the standard constraints on the fusion coefficients: associativity in Eq.~\eqref{eq:Fusion-Associativity}, the existence of a fusion identity in Eq.~\eqref{eq:Fusion-Unity}, and the contragredient properties of the dual objects.
\begin{table}
    \centering
    \begin{tabular}{|c|c|}
    \hline
       \textbf{Representations of $G$}  & \textbf{Category $\text{Rep}(G)$}\\
    \hline
       Representation  &  Object\\
       Irreducible  & Simple\\
       One-dimensional irrep  & Invertible object\\
       Higher-dimensional irrep  & Non-invertible object\\
       Trivial Representation & Tensor Unit \\
       Intertwining linear map & Morphism\\
       Operation on representations & Functor between categories\\
    \hline
    \end{tabular}
    \caption{Dictionary between equivalent concepts mapping standard representation-theoretic languages for $G$ to their formal categorical counterparts in $\mathrm{Rep}(G)$.}
    \label{tab:equivalent-concepts-rep-category}
\end{table}

In this categorical setting, the role of the character $\chi_{\alpha}$ is formally abstracted by the isomorphism class $[V_{\alpha}]$, which denotes the equivalence class of the object $V_{\alpha}$ up to isomorphism. To linearize Eq.~\eqref{eq:G_fusion_rules}, it is convenient to construct the free abelian group generated by the isomorphism classes of these simple objects, where multiplication is canonically induced by the tensor product. This algebraic structure is the well-known Grothendieck ring~\cite{serre1977linear}. Choosing $\{[V_{\alpha}] \mid \alpha\in\hat{G}\}$ as the basis of simple objects, the fusion rules are recast as
\be
\begin{split}
    [V_{\alpha_i}] [V_{\alpha_j}]= \sum_{\alpha_k\in\hat{G}}N_{\alpha_i\alpha_j}^{\alpha_k}[V_{\alpha_k}],
\end{split}
\ee
where the multiplication by $[V_{\alpha}]$ corresponds to the action of the fusion matrix defined in Eq.~\eqref{eq:Fusion-Matrix}. The associativity and symmetry of the tensor product ensure that these fusion matrices commute and can be simultaneously diagonalized. For finite groups, the common eigenvectors are precisely the columns of the ordinary character table, granting direct access to the full character data of $G$. For compact Lie groups, the underlying character-multiplication identity remains valid, with the discrete character table generalizing to a continuous family of character functions defined on conjugacy classes. By unifying branching, fusion, and character data into a single algebraic structure, the Grothendieck ring provides the rigid mathematical constraints necessary to bootstrap the hidden symmetry. To clarify the terminology used throughout this work, Table~\ref{tab:equivalent-concepts-rep-category} provides a dictionary of equivalent concepts, explicitly mapping standard representation-theoretic terms for $G$ to their formal categorical counterparts in $\mathrm{Rep}(G)$. This correspondence allows the reader to seamlessly translate between the two mathematical frameworks as these concepts arise in the main text.

Finally, we briefly review Tannakian duality established in Ref.~\cite{DeligneMilneOgusShih1982} and outline the complete algebraic data required to uniquely reconstruct a finite group $G$ from its representation category $\text{Rep}(G)$. The Tannakian duality for finite groups can be summarized as
\begin{itemize}
    \item Let $\mathcal{C}$ be a $\mathbb{C}$-linear, abelian, rigid symmetric monoidal category equipped with a faithful, exact, $\mathbb{C}$-linear, and symmetric monoidal fiber functor
    \be
        \Omega:~\mathcal{C}\to\text{Vec}_{\mathbb{C}}.
    \ee
    Here, $\text{Vec}_{\mathbb{C}}$ denotes the category of finite-dimensional complex vector spaces, with $\mathbb{C}$-linear maps as its morphisms. Then, the dual finite group $G$ is uniquely determined as
    \be
        G\cong \text{Aut}^{\otimes}(\Omega),
    \ee
    where $\text{Aut}^{\otimes}(\Omega)$ denotes the tensor automorphisms of the fiber functor $\Omega$. An arbitrary element $g\in \text{Aut}^{\otimes}(\Omega)$ is defined by the set of linear isomorphisms
    \be
        g_{V_{\alpha}}: \Omega(V_{\alpha})\to\Omega(V_{\alpha}),\quad\forall V_{\alpha}\in\mathcal{C}.
    \ee
    In the finite semisimple case $G$ is an ordinary finite group, and one has an equivalence of symmetric monoidal categories, $\mathcal{C}\simeq \text{Rep}(G)$. As a result, the Tannakian duality leads to the reconstruction theorem
    \be\label{app-eq:Tannakian-Reconstruction}
        (\mathcal{C},\Omega)\Leftrightarrow G.
    \ee  
\end{itemize}
To specify the complete algebraic data on the left-hand side of Eq.~\eqref{app-eq:Tannakian-Reconstruction} such that the right-hand side can be uniquely reconstructed, one requires the following dataset:
\be
    \left\{\hat{G},V_{\alpha_0},\overline{\alpha},N_{\alpha_i\alpha_j}^{\alpha_k},F,R,\Omega\right\}.
\ee
While the components $\hat{G}$, $V_{\alpha_0}$, $\overline{\alpha}$, and $N_{\alpha_i\alpha_j}^{\alpha_k}$ are successfully identified through our current bootstrap algorithm, the rest of categorical structures remain undetermined. Specifically, the current framework lacks the constraints necessary to explicitly reconstruct the associator ($F$-symbols), defined by the isomorphism
\be
    F^{\alpha_i\alpha_j\alpha_k}_{\alpha_l}: \sum_{\alpha_m\in\hat{G}}W_{\alpha_i\alpha_j}^{\alpha_m}\otimes W_{\alpha_m\alpha_k}^{\alpha_l}\to\sum_{\alpha_n\in\hat{G}}W_{\alpha_i\alpha_n}^{\alpha_l}\otimes W_{\alpha_j\alpha_k}^{\alpha_n},
\ee
where the multiplicity spaces are given by
\be
    W_{\alpha_i\alpha_j}^{\alpha_k}=\text{Hom}_{\text{Rep}(G)}\left(V_{\alpha_k},V_{\alpha_i}\otimes V_{\alpha_j}\right);
\ee
the symmetric braiding ($R$-symbols), given by
\be
    R_{\alpha_i\alpha_j}^{\alpha_k}:~W_{\alpha_i\alpha_j}^{\alpha_k}\to W_{\alpha_j\alpha_i}^{\alpha_k},
\ee
and the fiber functor $\Omega$. $N_{\alpha_i \alpha_j}^{\alpha_k}$ is the dimension of the multiplicity space according to Eq.~\eqref{eq:Def-Category-Fusion-Coefficients}.

As a concrete illustration, consider $G\cong\text{SU}(2)$, whose irreps are labeled by spin $j=0,\tfrac{1}{2},1,\tfrac{3}{2},\ldots$ The tensor product decomposition $V_{j_1}\otimes V_{j_2}\cong\bigoplus_{j_3=|j_1-j_2|}^{j_1+j_2}V_{j_3}$ is multiplicity-free, i.e., $N_{j_1 j_2}^{j_3}\in\{0,1\}$. Consequently, each multiplicity space is at most one-dimensional,
\be
    W_{j_1 j_2}^{j_3}\cong\begin{cases}\mathbb{C},&\text{if }|j_1-j_2|\leq j_3\leq j_1+j_2,\\0,&\text{otherwise},\end{cases}
\ee
and the $F$-symbol reduces to the well-known Wigner $6j$-symbols. To see this explicitly, consider coupling three angular momenta $j_1,j_2,j_3$ to a total spin $J$ with projection $M$. Since each $W$ space is one-dimensional, we may label basis vectors by the intermediate spin: let $|((j_1j_2)j_{12},j_3);JM\rangle$ denote the state obtained by first coupling $j_1\otimes j_2\to j_{12}$, then $j_{12}\otimes j_3\to J$, and $|(j_1,(j_2j_3)j_{23});JM\rangle$ the state from first coupling $j_2\otimes j_3\to j_{23}$, then $j_1\otimes j_{23}\to J$. The $F$-symbol acts as the change of basis between these two coupling schemes:
\begin{widetext}
\be
\begin{aligned}
     |((j_1j_2)j_{12},j_3);JM\rangle  =\sum_{j_{23}}(-1)^{j_1+j_2+j_3+J} \sqrt{(2j_{12}+1)(2j_{23}+1)}\begin{Bmatrix}j_1&j_2&j_{12}\\j_3&J&j_{23}\end{Bmatrix} |(j_1,(j_2j_3)j_{23});JM\rangle,
\end{aligned}
\ee
\end{widetext}
where the quantity in curly braces is the Wigner $6j$-symbol. Similarly, because each $W_{j_1 j_2}^{j_3}$ is one-dimensional, the $R$-symbol acts as multiplication by a scalar phase, $R_{j_1 j_2}^{j_3}=(-1)^{j_1+j_2-j_3}$, the familiar sign acquired when exchanging two angular momenta.

Note that the associativity in Eq.~\eqref{eq:Fusion-Associativity} and commutativity in Eq.~\eqref{eq:Fusion-Symmetric} of the fusion rules serve as macroscopic fingerprints of the underlying $F$- and $R$-symbols, respectively.  These relations can be readily derived from Eq.~\eqref{eq:Def-Category-Fusion-Coefficients}.

The fiber functor $\Omega$ is essential for the Tannakian reconstruction: it is the structure that allows the abstract category $\mathcal{C}$ to be identified with $\text{Rep}(G)$ for some group $G$. To illustrate its importance, consider the Fibonacci category, a unitary modular tensor category arising in $\text{SU}(2)_3$ Chern-Simons theory~\cite{beer2018categoriesanyonstravelogue}. It has two simple objects, the vacuum $\mathbf{1}$ and a non-abelian anyon $\tau$, with fusion rule $\tau\otimes\tau\cong\mathbf{1}\oplus\tau$. A fiber functor $\Omega:\mathcal{C}\to\text{Vec}_{\mathbb{C}}$ must map each simple object to a vector space and preserve tensor products, so the dimension $d_{\tau}=\dim\Omega(\tau)$ would have to satisfy $d_{\tau}^2=1+d_{\tau}$, giving the golden ratio $d_{\tau}=(1+\sqrt{5})/2$. Since vector space dimensions are positive integers, no fiber functor exists. More generally, a unitary modular tensor category possesses a non-degenerate braiding that is not symmetric, whereas a fiber functor to $\text{Vec}_{\mathbb{C}}$ necessarily maps the braiding to the symmetric swap of vector spaces. Thus, modular tensor categories, which describe anyonic excitations in $(2{+}1)$-dimensional topological phases of matter, lie outside the Tannakian framework, and no finite group $G$ can be reconstructed from them.

\section{Ramp of xSFF}
\label{app:xSFF-ramp}
\subsection{xSFF Ramp and Two Complementary Bootstrap Routes}
\label{sec:xSFF-ramp-bootstrap}

{The late-time xSFF plateau introduced above provides robust constraints on the branching structure for both chaotic and integrable systems. For non-integrable systems, the intermediate time RMT ramp contains complementary information. In particular, the plateau in Eq.~\eqref{eq:xSFF-Elements-Plateau-Height} takes the weighted form
\begin{equation}
K_{\lambda_a,\lambda_b}^{(N)}
\approx
\sum_{\alpha\in\hat{G}}
b_{\lambda_a,\alpha}b_{\lambda_b,\alpha}K_\alpha,
\end{equation}
where $K_\alpha\approx\nu m_\alpha$ depends on the multiplicity dimension $m_\alpha$. By contrast, the leading universal ramp coefficient is independent of $m_\alpha$ and directly probes the Gram matrix $G_{\lambda_a\lambda_b}$ defined in Eq.~\eqref{eq:Def-Gram-Matrix}. This difference leads to two complementary uses of the xSFF ramp. In the first route, the plateau bootstrap has already reconstructed a candidate hidden symmetry, and the ramp tests whether additional unresolved symmetry sectors remain. In the second route, the ramp itself provides numerical information about the branching Gram matrix and serves as the input to a limited bootstrap without using the plateau.}

\subsubsection{Connected xSFF and the branching Gram matrix}

{To isolate the universal two-level correlations associated with the ramp, we introduce an energy-filtered connected xSFF. We use the Gaussian filter
\begin{equation}
f(E)=\exp\left[-\frac{(E-E_c)^2}{2\sigma_E^2}\right],
\end{equation}
where $E_c$ and $\sigma_E$ denote the center and width of the selected spectral window, respectively. Following the normalization convention of Eq.~\eqref{eq:xSFF-Elements}, we define
\begin{equation}
\label{eq:connected-xSFF-ramp}
\begin{aligned}
K_{\lambda_a,\lambda_b}^{(N),{\rm c}}&(t)\\
=&\frac{1}{d_{\lambda_a}d_{\lambda_b}}\operatorname{Re}\Bigg[
\left\langle
\tr\left[P_{\lambda_a}f(H)U(t)\right]
\tr\left[P_{\lambda_b}f(H)U^\dagger(t)\right]
\right\rangle\\
&-\left\langle
\tr\left[P_{\lambda_a}f(H)U(t)\right]
\right\rangle
\left\langle
\tr\left[P_{\lambda_b}f(H)U^\dagger(t)\right]
\right\rangle
\Bigg].
\end{aligned}
\end{equation}
The connected subtraction removes the product of one-point spectral contributions and isolates the spectral correlations responsible for the universal ramp.}

{Using the Hilbert-space decomposition in Eq.~\eqref{eq:Hilbert-Space-Decomposition-Subset} and the operator decomposition in Eq.~\eqref{eq:Ut-Decomposition}, the filtered projected trace satisfies the exact relation
\begin{equation}
\label{eq:filtered-trace-branching}
\tr\left[P_\lambda f(H)U(t)\right]=
d_\lambda\sum_{\alpha\in\hat{G}}
b_{\lambda,\alpha}
\tr\left[\left.f(H)\right|_{M_\alpha}U_{M_\alpha}(t)\right].
\end{equation}
Equation~\eqref{eq:filtered-trace-branching} follows only from the branching decomposition and does not require chaotic dynamics.}

{We now specialize to a chaotic spectral window. We assume that the active multiplicity spaces $M_\alpha$ share the same RMT universality class, that connected spectral correlations between distinct non-degenerate $G$-irreps vanish after averaging, and that a common universal time regime exists between the Thouless and Heisenberg scales. For the GOE case considered below, let $\varrho_\alpha(E)$ denote the smooth density of states in the multiplicity space $M_\alpha$. The standard GOE spectral form factor with unit plateau is
\begin{equation}
\label{eq:GOE-form-factor-ramp}
k_{\rm GOE}(\tau)=
\begin{cases}
2\tau-\tau\log(1+2\tau),&0\leq\tau\leq1,\\
2-\tau\log\left(\dfrac{2\tau+1}{2\tau-1}\right),&\tau\geq1.
\end{cases}
\end{equation}
Within the universal RMT regime, the connected xSFF therefore takes the form
\begin{equation}
\label{eq:xSFF-ramp-full-RMT}
\begin{split}
    K_{\lambda_a,\lambda_b}^{(N),{\rm c}}(t) \approx& \sum_{\alpha\in\hat{G}}b_{\lambda_a,\alpha}b_{\lambda_b,\alpha}\int \dd Ef(E)^2\varrho_\alpha(E)\\
    &\times k_{\rm GOE}\left(\frac{t}{2\pi\varrho_\alpha(E)}\right).
\end{split}
\end{equation}
Equation~\eqref{eq:xSFF-ramp-full-RMT} is an RMT approximation, whereas Eq.~\eqref{eq:filtered-trace-branching} is an exact representation-theoretic identity.}

{The small-$\tau$ expansion of the GOE form factor is
\begin{equation}
k_{\rm GOE}(\tau)=2\tau-2\tau^2+\mathcal{O}(\tau^3).
\end{equation}
Consequently, the leading contribution within the universal ramp is
\begin{equation}
\label{eq:xSFF-ramp-Gram}
K_{\lambda_a,\lambda_b}^{(N),{\rm c}}(t)
\approx
\frac{t}{\pi}
\left[\int \dd Ef(E)^2\right]
G_{\lambda_a\lambda_b},
\end{equation}
where
\begin{equation}
\label{eq:ramp-Gram-definition}
G_{\lambda_a\lambda_b}=
\sum_{\alpha\in\hat{G}}
b_{\lambda_a,\alpha}b_{\lambda_b,\alpha}=
(BB^T)_{\lambda_a\lambda_b}.
\end{equation}
The local density $\varrho_{\alpha}(E)$ cancels from the leading ramp coefficient. Therefore, unlike the plateau, the leading ramp coefficient is not weighted by the multiplicity dimensions $m_\alpha$. It directly counts the overlap of the branching vectors and hence probes the unweighted Gram matrix defined in Eq.~\eqref{eq:Def-Gram-Matrix}. In finite-size numerical calculations, we fit the full GOE expression in Eq.~\eqref{eq:GOE-form-factor-ramp}, rather than fitting the linear approximation in Eq.~\eqref{eq:xSFF-ramp-Gram}.}

\subsubsection{Route I: ramp as a symmetry-completeness test}

{The first route begins after the plateau bootstrap has already reconstructed a candidate branching matrix $B_{\lambda,\alpha}$ and the associated representation data of the hidden symmetry $G$. The reconstructed branching matrix gives a definite prediction for $G_{\lambda_a\lambda_b}$ through Eq.~\eqref{eq:ramp-Gram-definition}. The ramp then provides an independent measurement of the same Gram matrix and can test whether the reconstructed representation theory accounts for all hidden symmetry sectors.}

{A particularly relevant ambiguity is an additional finite Abelian factor $G_{\rm Ab}$. Consider a candidate extension $G\times G_{\rm Ab}$, with the known subgroup $N$ embedded in the $G$ factor. Since every irrep $V_\beta$ of a finite Abelian group is one-dimensional, its restriction to $N$ satisfies
\begin{equation}
\label{eq:Abelian-extension-restriction}
\operatorname{Res}^{G\times G_{\rm Ab}}_N
\left(V_\alpha\otimes V_\beta\right)
\cong
\operatorname{Res}^{G}_N V_\alpha,
\quad
\beta\in\widehat{G_{\rm Ab}}.
\end{equation}
Thus every branching column associated with a $G$-irrep $V_\alpha$ is repeated for each $\beta\in\widehat{G_{\rm Ab}}$. Since a finite Abelian group satisfies $|\widehat{G_{\rm Ab}}|=|G_{\rm Ab}|$, the corresponding Gram matrix becomes
\begin{equation}
\label{eq:Abelian-extension-Gram}
\left.G_{\lambda_a\lambda_b}\right|_{G\times G{\rm Ab}}=
|G_{\rm Ab}|\times
\left.G_{\lambda_a\lambda_b}\right|_G.
\end{equation}
This relation assumes that the additional product irreps are realized as active hidden sectors in the spectral window under consideration.}

{The plateau-derived candidate therefore predicts $\left.G_{\lambda_a\lambda_b}\right|_G$, whereas an additional Abelian factor predicts an integer enhancement by $|G_{\rm Ab}|$. Comparing these predictions with the Gram matrix extracted from the ramp provides a direct symmetry-completeness test. Agreement with $\left.G_{\lambda_a\lambda_b}\right|_G$ supports a single copy of the reconstructed branching structure, while disagreement with $|G_{\rm Ab}|\times\left.G_{\lambda_a\lambda_b}\right|_G$ excludes the corresponding Abelian extension. In this route, the plateau reconstructs the candidate representation theory, and the ramp tests whether further unresolved sectors are required.}

\subsubsection{Route II: ramp-only limited bootstrap}

{The second route does not use plateau information. Its numerical input is the connected xSFF ramp resolved with respect to the known subgroup $N$. Fitting the universal RMT ramp yields numerical estimates of the entries $G_{\lambda_a\lambda_b}$.}

{Since the branching multiplicities satisfy $b_{\lambda,\alpha}\in\mathbb{N}$, the exact Gram matrix must satisfy
\begin{equation}
\label{eq:ramp-Gram-constraints}
G_{\lambda_a\lambda_b}\in\mathbb{N},\quad
G=G^T,\quad
G\succeq0,\quad
G=BB^T,
\end{equation}
where
\begin{equation}
B_{\lambda,\alpha}\in\mathbb{N}.
\end{equation}
The trivial $G$-irrep remains fixed by the unity condition in Eq.~\eqref{eq:Fusion-Unity-b}, namely $b_{\lambda,\alpha_0}=\delta_{\lambda,\lambda_0}$.}

{The numerical ramp data first restrict the admissible integer Gram matrices. Each surviving Gram matrix is then factorized over non-negative integers according to $G=BB^T$. When several factorizations exist, candidate branching matrices are enumerated rank by rank using the trial rank $r$ already introduced in Sec.~\ref{sec:Bootstrap-Algorithm}.}

{Each surviving branching matrix is subsequently passed to the algebraic bootstrap developed above. In particular, it must satisfy the monoidality condition
\begin{equation}
\label{eq:ramp-monoidality}
\sum_{\lambda_a,\lambda_b\in\hat{N}}
N_{\lambda_a\lambda_b}^{\lambda_c}
b_{\lambda_a,\alpha_i}
b_{\lambda_b,\alpha_j}=
\sum_{\alpha_k\in\hat{G}}
N_{\alpha_i\alpha_j}^{\alpha_k}
b_{\lambda_c,\alpha_k},
\end{equation}
together with dimensionality, non-negativity, unity, commutativity, associativity, and rigidity. The surviving solutions determine the admissible branching matrices, $G$-irrep dimensions, fusion rules, and character data.}

{The logical sequence of the ramp-only route is $K_{\lambda_a,\lambda_b}^{(N),{\rm c}}(t)\rightarrow G_{\lambda_a\lambda_b}\rightarrow B_{\lambda,\alpha}\rightarrow$ fusion bootstrap. Its domain of applicability is narrower than that of the plateau bootstrap because a universal RMT ramp requires chaotic dynamics, a common RMT universality class for the relevant hidden sectors, and a common time regime between the Thouless and Heisenberg scales. Within this regime, however, the ramp provides direct access to the unweighted branching Gram matrix and can serve as an independent numerical input for a limited symmetry bootstrap.}

\subsection{Examples}

\subsubsection{$S_3$-invariant Chain}
\begin{figure*}[t]
    \centering
    \includegraphics[width=0.9\linewidth]{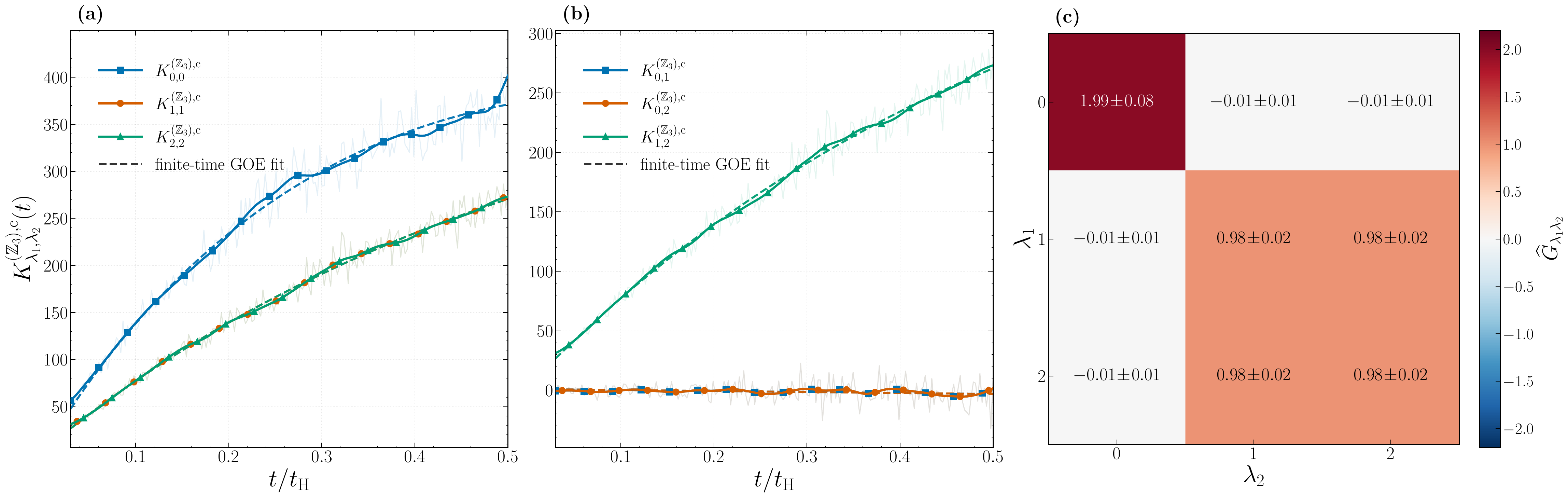}
    \caption{{
  Connected xSFF ramps for the $S_3$-invariant O'Brien--Fendley chain with $L=8$ sites at $\theta=0.28$, resolved into the manifest
  $N\cong\mathbb{Z}_3$ sectors and averaged over $300$ disorder realizations.
  The couplings are sampled independently from
  $J_j,h_j\sim\mathcal{N}(1,0.25^2)$.
  For each realization, all sectors are shifted by the same full-spectrum mean. A Gaussian spectral filter with $E_c=-0.472$ and
  $\sigma_E=1.233$ is then applied.
  The horizontal axis uses
  $t_{\mathrm H}=1196.79$, the median visible-sector Heisenberg time.
  Faint curves show the unsmoothed connected xSFF, solid curves with
  markers show Gaussian-smoothed data for visualization, and black dashed
  curves show fits to the finite-time GOE spectral form factor. The fits are performed on the unsmoothed data in physical time
  $t$, over the interval
  $0.03\leq t/t_{\mathrm H}\leq0.50$.
  \textbf{(a)} Diagonal components $K_{\lambda,\lambda}^{(\mathbb{Z}_3),\mathrm{c}}(t)$. \textbf{(b)} Off-diagonal components $K_{\lambda_1,\lambda_2}^{(\mathbb{Z}_3),\mathrm{c}}(t)$.
  \textbf{(c)} Fitted Gram matrix $\widehat{G}$; the quoted uncertainties
  are obtained from $400$ disorder-bootstrap resamples.
  The fit gives
  $\widehat{G}_{00}=1.989\pm0.077$,
  $\widehat{G}_{11}=\widehat{G}_{22}
  =\widehat{G}_{12}=0.984\pm0.016$, and $\widehat{G}_{01}=\widehat{G}_{02}=-0.0086\pm0.0061$. This result is consistent with the integer Gram matrix $G=BB^T$ associated with the $S_3$ branching structure.}}
    \label{fig:xSFF-s3_xsff_ramp}
\end{figure*}

{First, we consider the $S_3$-invariant O'Brien--Fendley chain [Eq.~\eqref{eq:OF-Hamiltonian}]. The numerical connected xSFF shown in Fig.~\ref{fig:xSFF-s3_xsff_ramp} follows the finite-time GOE form over the selected ramp window for both diagonal and off-diagonal components. The finite-time GOE fits shown in the figure give $\widehat{G}_{00}=1.989\pm0.077$, $\widehat{G}_{11}=\widehat{G}_{22}=\widehat{G}_{12}=0.984\pm0.016$, and $\widehat{G}_{01}=\widehat{G}_{02}=-0.0086\pm0.0061$, matching panel~(c). These values are consistent with the integer Gram matrix obtained from the $S_3$ branching structure,
\begin{equation}\label{app-eq:S3-Gram-Matrix}
G=BB^T=
\begin{pmatrix}
2&0&0\\
0&1&1\\
0&1&1
\end{pmatrix},
\end{equation}
where the branching matrix $B$ is given by Eq.~\eqref{eq:S3-Branching-Matrix}.
In particular, the ramp resolves two independent hidden irreps contributing to the $\lambda=0$ sector, a single common hidden irrep shared by the $\lambda=1$ and $\lambda=2$ sectors, and no common hidden irrep between $\lambda=0$ and either non-trivial $\mathbb{Z}_3$ sector. The agreement therefore verifies directly that the connected ramp measures the unweighted Gram matrix $G_{\lambda_a\lambda_b}$ predicted by Eq.~\eqref{eq:xSFF-ramp-Gram}.}

{The same data illustrate both bootstrap routes. For Route~I, the plateau bootstrap in Sec.~\ref{sec:S3-Inv-OF-Chain} gives the $S_3$ branching matrix in Eq.~\eqref{eq:S3-Branching-Matrix}, whose Gram matrix is precisely the integer matrix above. The fitted entries agree with this prediction and are inconsistent, within their quoted statistical uncertainties, with the doubled matrix required by an additional $\mathbb{Z}_2$ factor. This comparison assumes that both copies are active in the common universal spectral window, as specified in Eq.~\eqref{eq:Abelian-extension-Gram}. For Route~II, we estimate the numerical uncertainty of each fitted Gram-matrix element by non-parametric bootstrap resampling of the disorder realizations. For every bootstrap sample, the connected xSFF is recomputed and the full finite-time GOE form is refitted, yielding a bootstrap distribution of $\widehat G_{\lambda_a\lambda_b}$. We denote its standard deviation by $\sigma_{\lambda_a\lambda_b}$. To convert the continuous fit into discrete representation-theoretic candidates, we enumerate non-negative integer Gram matrices satisfying
\begin{equation}
\left|G_{\lambda_a\lambda_b}-\widehat G_{\lambda_a\lambda_b}\right|
\leq1.5\sigma_{\lambda_a\lambda_b}
\end{equation}
for every independent matrix element, together with symmetry, positive semi-definiteness, and the factorization condition $G=BB^T$. In this context, the $1.5\sigma$ interval serves as a predetermined numerical tolerance for candidate enumeration rather than a joint confidence interval. Enlarging this tolerance may yield multiple Gram matrices, resulting in possible multiple solutions at the minimal rank $r=r_*$. For the present data, the same integer Gram matrix is also selected using a $2\sigma$ tolerance, demonstrating that the result is insensitive to this choice. The surviving Gram matrix is Eq. \eqref{app-eq:S3-Gram-Matrix}, whose non-negative integer factorization yields Eq.~\eqref{eq:S3-Branching-Matrix}. Passing this branching matrix to the algebraic bootstrap recovers two one-dimensional irreps, one two-dimensional irrep, and the fusion rules of $S_3$. The example therefore demonstrates both roles of the ramp: it independently tests the completeness of a plateau-derived symmetry and, for a chaotic system with a resolved universal ramp, supplies sufficient branching data to drive a limited bootstrap without plateau input.}

\subsubsection{Self-dual KT chain: a hidden direct-product symmetry}
\label{sec:ramp-KT-direct-product}

{Consider the following self-dual spin-$1$ chain with open boundary conditions:
\begin{equation}
\begin{split}
H_{\rm sd}
={}&\sum_{i=1}^{L-1}J_i h_i^+
+\sum_{i=1}^{L}D_i(S_i^z)^2\\
&+\sum_{i=1}^{L-2}\kappa_i
\left(\{h_i^+,h_{i+1}^+\}+\{h_i^-,h_{i+1}^-\}\right),
\end{split}
\label{eq:KT-ramp-self-dual-H}
\end{equation}
where $\{A,B\}=AB+BA$ and
\begin{equation}
\begin{aligned}
h_i&=\mathbf S_i\cdot\mathbf S_{i+1},
&h_i^-&=h_i-h_i^+,\\
h_i^+&=S_i^yP_{i,i+1}S_{i+1}^y,
&P_{i,i+1}&=\frac{1+e^{{\rm i}\pi S_i^z}e^{{\rm i}\pi S_{i+1}^x}}{2}.
\end{aligned}
\label{eq:KT-ramp-bond-definitions}
\end{equation}
All couplings are independent, with $J_i,\kappa_i\sim\mathcal N(1,0.25^2)$ and $D_i\sim\mathcal N(0.5,0.25^2)$, removing translation and reflection symmetries. Every local term preserves the manifest symmetry
\begin{equation}
\begin{gathered}
N\cong\mathbb Z_2^x\times\mathbb Z_2^z,\\
U_\pi^x=\prod_j e^{{\rm i}\pi S_j^x},
\quad
U_\pi^z=\prod_j e^{{\rm i}\pi S_j^z}.
\end{gathered}
\label{eq:KT-ramp-manifest-symmetry}
\end{equation}
At $\kappa_i=0$, each $Q_i=(S_i^y)^2$ is also conserved. The $\{h_i^-,h_{i+1}^-\}$ interaction breaks these local conservation laws for generic nonzero $\kappa_i$.}

{The hidden symmetry follows from the parent Hamiltonian
\begin{equation}
\begin{split}
H_0={}&\sum_{i=1}^{L-1}J_i h_i
+\sum_{i=1}^{L}D_i(S_i^z)^2+\sum_{i=1}^{L-2}\kappa_i\{h_i,h_{i+1}\}
\end{split}
\label{eq:KT-ramp-parent-H}
\end{equation}
and the Kennedy--Tasaki transformation
\begin{equation}
U_{\rm KT}=\prod_{1\leq u<v\leq L}
e^{{\rm i}\pi S_u^zS_v^x},
\quad U_{\rm KT}^2=1.
\label{eq:KT-ramp-transformation}
\end{equation}
Since $U_{\rm KT}h_i^\pm U_{\rm KT}^{\dagger}=\pm h_i^\pm$, the interpolation
\begin{equation}
H(\eta)=(1-\eta)H_0+\eta U_{\rm KT}H_0U_{\rm KT}^{\dagger}
\label{eq:KT-ramp-interpolation}
\end{equation}
obeys $U_{\rm KT}H(\eta)U_{\rm KT}^{\dagger}=H(1-\eta)$ and gives Eq.~\eqref{eq:KT-ramp-self-dual-H} at $\eta=1/2$. In particular, dual averaging retains both the $++$ and $--$ anticommutators. Moreover, $U_{\rm KT}$ commutes with both manifest generators, so together they generate the direct-product symmetry
\begin{equation}
G\cong N\times\mathbb Z_2^{\rm KT}.
\label{eq:KT-ramp-direct-product}
\end{equation}}

{We apply the ramp bootstrap using only the manifest $N$-sector projectors. Each $\lambda\in\widehat N$ has two extensions $(\lambda,+)$ and $(\lambda,-)$ to $G$, distinguished by the KT eigenvalue. Their branching matrix is, up to a column permutation,
\begin{equation}
B=\begin{pmatrix}I_4&I_4\end{pmatrix},
\quad
G_{\lambda\mu}=(BB^T)_{\lambda\mu}=2\delta_{\lambda\mu}.
\label{eq:KT-ramp-product-Gram}
\end{equation}
For independent chaotic spectra in the fully resolved sectors, the hidden factor therefore doubles every diagonal ramp coefficient without producing off-diagonal overlap.}

{For $L=9$ and $300$ disorder realizations, we use the Gaussian filter $f_a(E)=\exp[-(E-E_{0,a})^2/(2\sigma_E^2)]$, with $\sigma_E=2.0392$ and $E_{0,a}=\overline E_a-0.435$, where $\overline E_a$ is the full-spectrum mean in realization $a$. The time-evolution phases also use the centered energies $E-E_{0,a}$. The common time unit is $t_{\rm H}=\operatorname{median}_\lambda[2\pi\rho_\lambda(E_0)]=1915.7$, where $\rho_\lambda(E_0)$ is the disorder-averaged Gaussian-filtered sector density.}

{To account for ramp curvature, we fit the finite-time GOE function in Eq.~\eqref{eq:GOE-form-factor-ramp} and extract its tangent:
\begin{equation}
\begin{aligned}
K_{\rm ref}^{\rm c}(t)&=c_{\rm ref}+A_{\rm ref}
k_{\rm GOE}(t/T_{\rm ref}),\\
\operatorname{Re}K_{\lambda\mu}^{(N),{\rm c}}(t)&=c_{\lambda\mu}+A_{\lambda\mu}
k_{\rm GOE}(t/T_*).
\end{aligned}
\label{eq:KT-ramp-full-GOE-fit}
\end{equation}
The reference consists of $400$ GOE spectra of dimension $4921$, rescaled to match the mean Gaussian-filtered physical density and evaluated with the same filter and time grid. All fits use unsmoothed data on $0.03\leq t/t_{\rm H}\leq0.50$. For each trial $T>0$, the diagonal offsets and amplitudes are obtained by linear least squares; minimizing their total residual determines the common $T_*$. This scale is then held fixed when fitting every sector pair. Neither an integer component count nor $T_*=T_{\rm ref}/2$ is imposed. Since $k'_{\rm GOE}(0)=2$, the fitted tangent ratios give
\begin{equation}
\widehat G_{\lambda\mu}
=\frac{A_{\lambda\mu}/T_*}{A_{\rm ref}/T_{\rm ref}}.
\label{eq:KT-ramp-tangent-ratio}
\end{equation}
Thus the Gram estimator uses the extrapolated GOE tangents, with no straight-line regression.}

\begin{figure*}[t]
    \centering
    \includegraphics[width=0.9\linewidth]{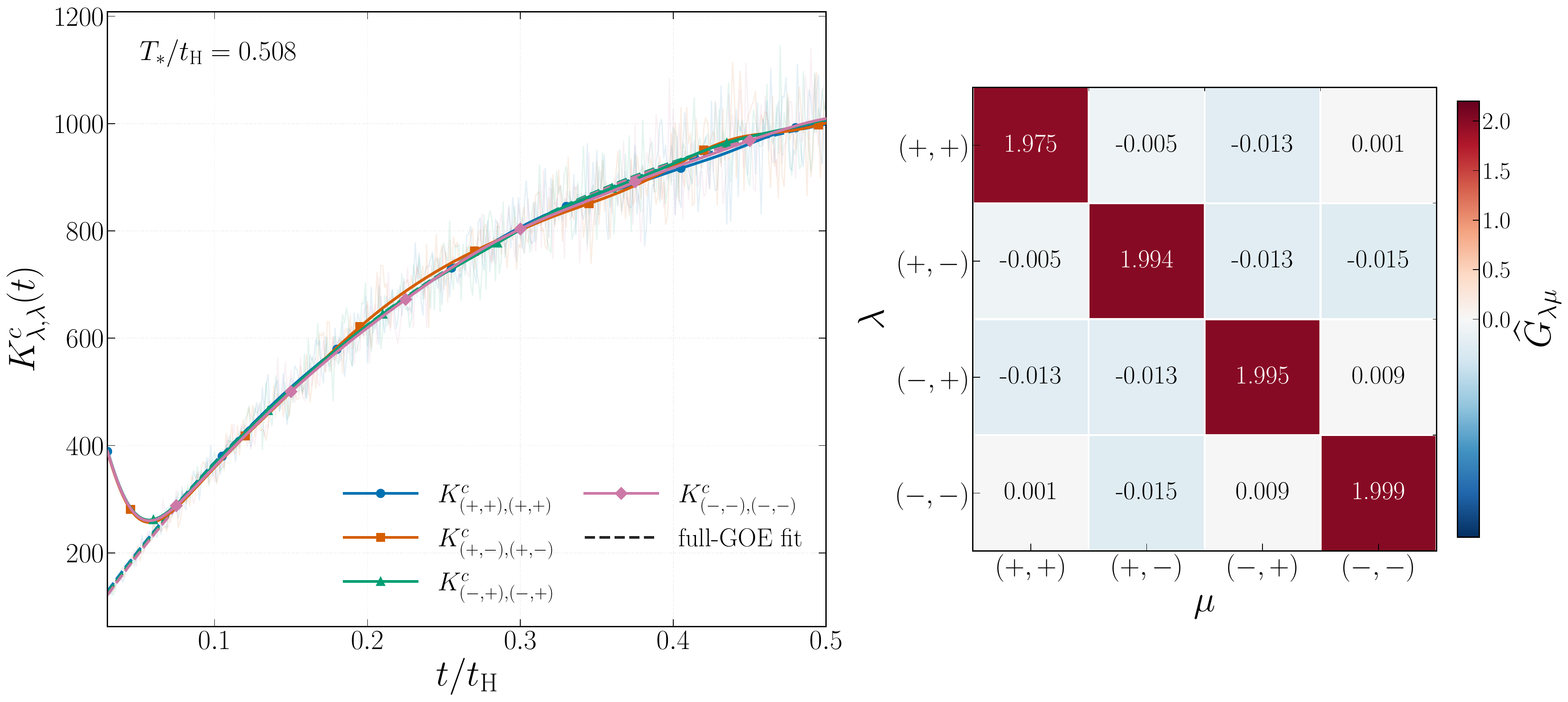}
    \caption{{Full-GOE extraction of the manifest-sector ramp Gram matrix for Eq.~\eqref{eq:KT-ramp-self-dual-H}, with $L=9$ and $300$ disorder realizations. \textbf{Left}: The four diagonal connected xSFFs. Faint curves show the unsmoothed data; colored curves with markers are smoothed for display. Black dashed curves are fits to the unsmoothed data over $0.03\leq t/t_{\rm H}\leq0.50$, with common scale $T_*/t_{\rm H}=0.508$. The earliest-time transient is not captured by the GOE fit; agreement improves along the subsequent rising ramp. \textbf{Right}: The Gram matrix extracted from the fitted tangent ratios, with diagonal entries close to two and off-diagonal entries close to zero, consistent with $G=2I_4$.}}
    \label{fig:KT-self-dual-full-GOE-ramp}
\end{figure*}

{The fit gives $T_{\rm ref}/t_{\rm H}=1.0058$, $T_*/t_{\rm H}=0.50841$, and $T_{\rm ref}/T_*=1.9782$. Independent diagonal fits give $T_\lambda/t_{\rm H}\in[0.49068,0.52913]$, supporting the common-scale approximation. Figure~\ref{fig:KT-self-dual-full-GOE-ramp} shows the fitted curves and the extracted matrix, which in the sector order $(++,+-,-+,--)$ is
\begin{equation}
\widehat G=
\begin{pmatrix}
1.9754&-0.0047&-0.0134&0.0010\\
-0.0047&1.9945&-0.0134&-0.0147\\
-0.0134&-0.0134&1.9951&0.0091\\
0.0010&-0.0147&0.0091&1.9992
\end{pmatrix}.
\label{eq:KT-ramp-fitted-Gram}
\end{equation}
Its mean diagonal is $1.99103$, and its largest off-diagonal magnitude is $0.01468$. The resulting standard errors are $0.0552$--$0.0572$ on the diagonal and $0.0101$--$0.0110$ off the diagonal, with a correlated diagonal mean $1.99103\pm0.05513$. A covariance-aware test of the ten independent matrix entries against $2I_4$ gives $\chi^2=8.212$ for $10$ degrees of freedom ($p=0.608$).}

{The measured matrix is therefore statistically consistent with $2I_4$ under the stated fit procedure. Figure~\ref{fig:KT-self-dual-full-GOE-ramp} also shows an early-time transient, so this consistency does not imply GOE behavior throughout the plotted interval. Rounding to this integer Gram matrix yields $B=(I_4\ I_4)$, up to column permutations and omission of zero columns, supplying branching data to the algebraic bootstrap.}

\subsection{Approximate Symmetry}
\label{sec:xSFF-approximate-symmetry}

{Finally, because approximate symmetries are known to affect the SFF ramp~\cite{PhysRevX.12.021009}, we investigate the xSFF ramp in the presence of approximate hidden symmetries. We assume the known subgroup $N$ remains exact, while additional transformations are only approximately compatible with the extended group $G \supset N$. This is captured by the controlled reference Hamiltonian
\begin{equation}
\begin{split}
H&=H_0+\epsilon V,
\end{split}
\label{eq:approx-reference-H}
\end{equation}
where $[H_0, U_g] = 0$ and $[V, U_n] = 0$ for all $g \in G$ and $n \in N$. Here, $U_g$ provides a representation of $g$, and the term $\epsilon V$ quantifies the deviation from exact $G$-invariance. Although the $N$-sector projectors remain exact, the branching multiplicities $b_{\lambda,\alpha}$ describe the unperturbed, exactly $G$-symmetric baseline ($H_0$), rather than exact conserved $G$ sectors of the full $H$ when $\epsilon \neq 0$. Notably, neither this formulation nor the associated branching analysis requires $N$ to be a normal subgroup of $G$.}

{When $\epsilon=0$, Eq.~\eqref{eq:filtered-trace-branching} holds as an exact identity. Applying Eq.~\eqref{eq:xSFF-ramp-full-RMT} to the $G$-symmetric reference Hamiltonian $H_0$ gives
\begin{multline}
\left.K_{\lambda_a,\lambda_b}^{(N),{\rm c}}(t)\right|_{\epsilon=0}
\approx
\sum_\alpha b_{\lambda_a,\alpha}b_{\lambda_b,\alpha}
\int\dd Ef(E)^2\varrho_\alpha(E)\\
\times k_{\rm GOE}\left(\frac{t}{2\pi\varrho_\alpha(E)}\right)
\approx
\frac{t}{\pi}\left[\int\dd Ef(E)^2\right]
G_{\lambda_a\lambda_b}.
\label{eq:approx-reference-ramp}
\end{multline}
The first approximation relies on the local RMT description. The second applies the expansion $k_{\rm GOE}(\tau)=2\tau+O(\tau^2)$ and strictly requires $t\ll2\pi\varrho_\alpha(E)$ for every contributing $\alpha$ wherever $f(E)^2$ is non-negligible. Consequently, the linear Gram regime requires both $t\gg t_{\rm Th}$ and $t\sigma_E\gg1$, while remaining well below all relevant Heisenberg scales. Here, the lower equilibration scale $t_{\rm Th}$ is determined by the internal dynamics of these $G$-symmetric reference blocks. Furthermore, the relaxation of the approximate symmetry can delay the onset of the pure RMT regime in the full $N$ sector~\cite{PhysRevX.12.021009}.}

{An approximate symmetry extension can preserve the early-time RMT behavior shown in the first line of Eq.~\eqref{eq:approx-reference-ramp} before deviations from exact $G$-invariance become spectrally resolved. This requires a clear temporal separation between intrablock equilibration and symmetry relaxation. We denote the relevant component-dependent crossover scales as $t_{\rm sym}^{\lambda_a\lambda_b}$. Once the correlations induced by the approximate symmetry subsequently relax, and assuming every exact $N$-sector $M_\lambda$ thermalizes into a single GOE block with no residual exact symmetries pairing different $N$ sectors, the late-time ramp instead satisfies
\begin{equation}
K_{\lambda_a,\lambda_b}^{(N),{\rm c}}(t)
\approx
\frac{t}{\pi}\left[\int\dd Ef(E)^2\right]
\delta_{\lambda_a,\lambda_b}.
\label{eq:approx-exact-N-ramp}
\end{equation}
This requires equilibration of the relevant slow modes before the
Heisenberg time. The two limiting coefficient matrices are therefore
$BB^T$ and $I$, rather than an integer Gram matrix valid throughout the
crossover. Diagonal enhancements $G_{\lambda\lambda}>1$ can relax toward one, whereas nonzero off-diagonal entries can relax toward zero. These statements apply also to simple $G$ and non-normal $N$.}

{The diagonal mechanism connects directly to the return-probability approximation introduced in Ref.~\cite{PhysRevX.12.021009}. For independently chaotic, multiplicity-free hidden blocks within a given $M_\lambda$, slow Markovian transitions weight the single-block leading ramp by the sum of the block return probabilities. Considering two such blocks, the symmetry-breaking perturbation induces an interblock transition rate $\gamma_\lambda$ representing the probability per unit time that a state leaks from one hidden block to the other. Assuming this rate is symmetric and remains approximately constant across the energy filter, the ramp evaluates to
\begin{equation}
\begin{split}
K_{\lambda,\lambda}^{(N),{\rm c}}(t)
&\approx\frac{t}{\pi}\int\dd Ef(E)^2
\tr\exp\left[
 t\begin{pmatrix}
 -\gamma_\lambda&\gamma_\lambda\\
 \gamma_\lambda&-\gamma_\lambda
 \end{pmatrix}\right]\\
&=\frac{t}{\pi}\int\dd Ef(E)^2
\left(1+e^{-2\gamma_\lambda t}\right).
\end{split}
\label{eq:approx-diagonal-return}
\end{equation}
Here, the diagonal entries of the matrix exponential correspond to the return probabilities. The transition rate matrix has eigenvalues $0$ and $-2\gamma_\lambda$, which establishes the symmetry-breaking crossover scale $t_{\rm sym}^{\lambda\lambda} = (2\gamma_\lambda)^{-1}$ for this two-block model. This reduction crucially assumes that intrablock equilibration occurs much faster than interblock leakage, and that the spectrum is sufficiently dense to support incoherent Markovian transitions. In contrast, systems with repeated identical branching copies require tracking coherence information alongside classical populations, and therefore cannot be described by this simple two-block rate equation.}

{In contrast, the off-diagonal mechanism for xSFF governs spectra residing in distinct, exact $N$ sectors. Let $E_n^{(\lambda)}$ denote the eigenvalues within $M_\lambda$, counting each symmetry-enforced $d_\lambda$-fold degeneracy only once. Following the normalization of Eq.~\eqref{eq:connected-xSFF-ramp}, we obtain the exact spectral expression:
\begin{equation}
\begin{split}
K_{\lambda_a,\lambda_b}^{(N),{\rm c}}(t)
=\operatorname{Re}\sum_{n,m}
\Big\langle&
\big[f(E_n^{(\lambda_a)})e^{-itE_n^{(\lambda_a)}}\big]\\
&\times
\big[f(E_m^{(\lambda_b)})e^{itE_m^{(\lambda_b)}}\big]
\Big\rangle_{\rm c},
\end{split}
\label{eq:approx-spectral-cumulant}
\end{equation}
where the connected average subtracts the product of the separate ensemble averages of the two bracketed terms. Consider a pair of states belonging to a common reference multiplet, such that their energies perfectly coincide at $\epsilon=0$. For non-zero $\epsilon$, we define their energy splitting as $\delta E=E_n^{(\lambda_a)}-E_m^{(\lambda_b)}$. Their contribution to the unsubtracted spectral sum includes the phase factor $e^{-it\delta E}$, the real part of which is $\cos(\delta E t)$. When averaged over a finite observation interval of duration $T$, this yields
\begin{equation}
\frac{1}{T}\int_0^T\dd t\cos(\delta E t)
=
\begin{cases}
\displaystyle\frac{\sin(\delta E T)}{\delta E T},&\delta E\neq0,\\[2mm]
1,&\delta E=0.
\end{cases}
\label{eq:approx-splitting-resolution}
\end{equation}
Consequently, a near-degenerate contribution remains unresolved for $T\vert{}\delta E\vert{}\ll1$ and completely dephases in the time average when $T\vert{}\delta E\vert{}\gg1$. While this identifies an inverse splitting resolution scale, it does not factorize the connected ramp: Eq.~\eqref{eq:approx-spectral-cumulant} also includes cross-terms from different reference levels as well as the connected subtraction. Therefore, a simple weighted cosine average of matched splittings does not yield a general formula for the normalized connected xSFF, and its inverse root-mean-square splitting is not necessarily equivalent to the ramp crossover time. Instead, the off-diagonal scales discussed below are extracted directly from the full measured correlator. Unlike the scalar autocorrelation studied in Ref.~\cite{PhysRevX.12.021009}, these off-diagonal entries independently identify which exact $N$ sectors retain approximately common multiplets, even if their individual diagonal spectra appear purely GOE-like. Finally, the decorrelated background in Eq.~\eqref{eq:approx-exact-N-ramp} reflects the standard universal absence of correlations between unrelated symmetry sectors~\cite{Braun2011}.}

{As a representative example, we consider the O'Brien--Fendley chain of
Eq.~\eqref{eq:OF-Hamiltonian} with
\begin{equation}
\begin{split}
&H_{\theta,\epsilon}
=H_\theta+\epsilon V,
\quad
V=i\sum_{j=1}^L h_j(X_j-X_j^\dagger),\\
&[H_{\theta,\epsilon},X]=0,
\quad
CH_{\theta,\epsilon}C=H_{\theta,-\epsilon}.
\end{split}
\label{eq:approx-S3-H}
\end{equation}
The known $N\cong\mathbb Z_3$ symmetry is exact for every $\epsilon$, whereas $C$ supplies the additional approximate $S_3$ structure. The exact anti-unitary symmetry $C\mathcal T$, with $\mathcal T$ the complex conjugation defined above, squares to one and acts within each
$\mathbb Z_3$ sector, so the comparison does not require a change of Dyson class. The reference branching matrix in Eq.~\eqref{eq:S3-Branching-Matrix} gives $G_{00}=2$ and $G_{11}=G_{22}=G_{12}=1$. Consequently, the diagonal $\lambda=0$ component probes mixing of the two reference $C$-parity blocks, whereas the $(1,2)$ component probes loss of the common doublet spectrum shared by $\lambda=1$ and $2$. In the latter case $\vec b_1=\vec b_2$ enforces equality of the two projected traces at $\epsilon=0$, and therefore
\begin{equation}
\left.
\frac{K_{1,2}^{(\mathbb Z_3),{\rm c}}(t)}
{\sqrt{K_{1,1}^{(\mathbb Z_3),{\rm c}}(t)
        K_{2,2}^{(\mathbb Z_3),{\rm c}}(t)}}
\right|_{\epsilon=0}=1
\label{eq:approx-S3-exact-coherence}
\end{equation}
whenever the denominator is nonzero. This equality is exact and does
not rely on the linear ramp approximation.}

\begin{figure*}[t]
    \centering
    \includegraphics[width=0.98\linewidth]{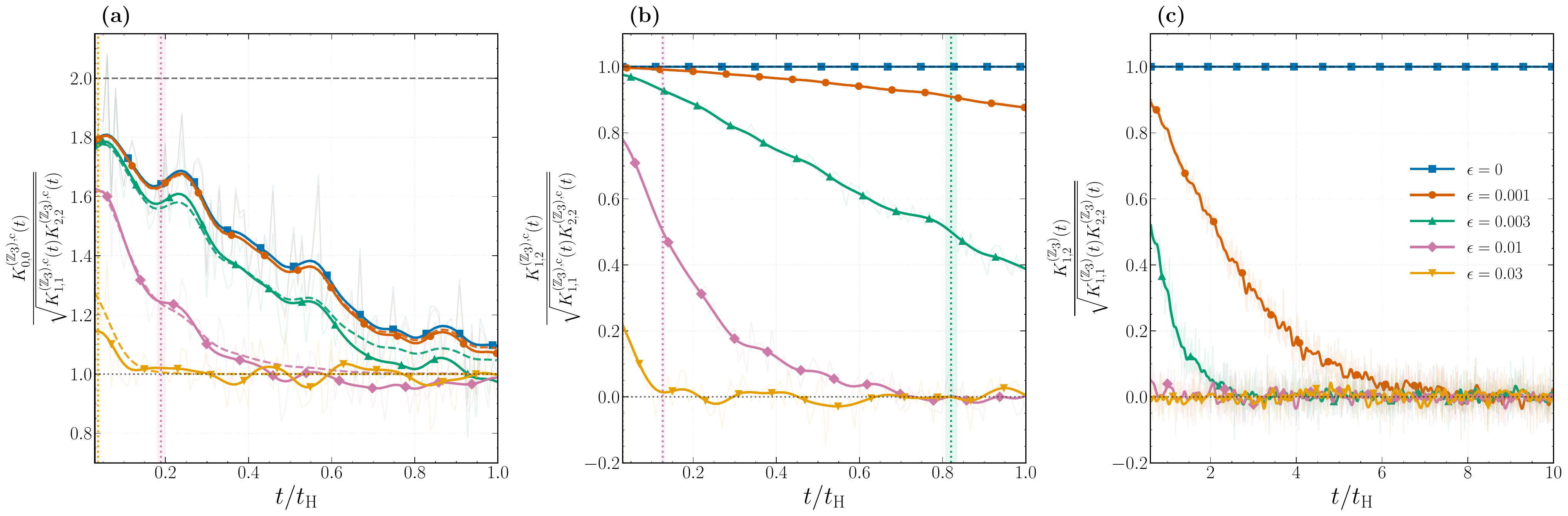}
    \caption{Approximate hidden $S_3$ symmetry in the O'Brien--Fendley
  chain with exact $N\cong\mathbb Z_3$ symmetry
  [Eq.~\eqref{eq:approx-S3-H}]. We use $L=8$, $\theta=0.28$, and
  $300$ matched disorder realizations with independent
  $J_j,h_j\sim\mathcal N(1,0.25^2)$ at
  $\epsilon=0,0.001,0.003,0.01,0.03$. The same disorder samples,
  reference full-spectrum shifts, and Gaussian filter
  ($E_c=-0.47410$, $\sigma_E=1.23343$) are used at every $\epsilon$,
  without unfolding. The time unit $t_{\rm H}=1196.80$ is $2\pi$ times
  the reference smooth density at $E_c$ in either of the identical
  $\lambda=1,2$ sectors. Faint curves show unsmoothed ensemble estimates;
  solid curves with markers show Gaussian smoothing with standard
  deviation three grid spacings, where $\Delta t/t_{\rm H}=0.00997$.
  \textbf{(a)}~The diagonal connected xSFF in the $\lambda=0$ sector,
  normalized by the geometric mean of the $\lambda=1,2$ diagonal
  components. The $\epsilon=0$ curve already decreases because of
  finite-time spectral curvature. Horizontal lines at two and one are
  leading-ramp benchmarks. Colored dashed curves are fits relative to
  this measured reference [Eq.~\eqref{eq:approx-S3-diagonal-fit}], giving
  $t_{\rm sym}^{00}/t_{\rm H}=0.189\pm0.010$ for $\epsilon=0.01$ and
  $0.0377\pm0.0049$ for $\epsilon=0.03$. The diagonal crossover is
  unresolved before $t_{\rm H}$ for $\epsilon=0.001,0.003$.
  \textbf{(b)}~The normalized off-diagonal connected xSFF in
  Eq.~\eqref{eq:approx-S3-offdiagonal-time}. It is exactly one at
  $\epsilon=0$. The first half-correlation crossings give
  $t_{\rm sym}^{12}/t_{\rm H}=0.820\pm0.014$ for $\epsilon=0.003$ and
  $0.127\pm0.005$ for $\epsilon=0.01$. For $\epsilon=0.001$, the ratio
  remains above $1/2$ throughout the panel. For $\epsilon=0.03$, it is
  already below $1/2$ at the lower cutoff, so only
  $t_{\rm sym}^{12}/t_{\rm H}\lesssim0.03$ can be reported.
  Vertical dotted lines mark the reported times, with bands showing
  one bootstrap standard deviation from $300$ matched-realization
  resamples. The orange line at the lower boundary of panel~(b) marks
  the cutoff bound rather than a measured crossing.
  \textbf{(c)}~The normalized ordinary, unfiltered xSFF
  $K_{1,2}^{(\mathbb Z_3)}(t)/
  \sqrt{K_{1,1}^{(\mathbb Z_3)}(t)K_{2,2}^{(\mathbb Z_3)}(t)}$
  over a longer time window. The $\epsilon=0$ ratio remains one,
  while every nonzero perturbation shown loses correlation, including
  $\epsilon=0.001$, whose decay occurs mainly beyond $t_{\rm H}$.}
    \label{fig:S3-approximate-tsym}
\end{figure*}

{Figure~\ref{fig:S3-approximate-tsym} separates the diagonal response [panel~(a)] from cross-sector decorrelation [panel~(b)]. In panel~(a), even the exactly symmetric $\epsilon=0$ curve decreases toward one because of finite-time spectral curvature; the value two is only a leading-ramp benchmark. This decrease alone is therefore not evidence of symmetry breaking. To isolate the additional relaxation at $\epsilon\neq0$, we fit relative to the measured $\epsilon=0$ baseline using the phenomenological interpolation
\begin{multline}
\frac{K_{0,0}^{(\mathbb Z_3),{\rm c}}(t)}
{\sqrt{K_{1,1}^{(\mathbb Z_3),{\rm c}}(t)
        K_{2,2}^{(\mathbb Z_3),{\rm c}}(t)}}-1\\
\approx e^{-t/t_{\rm sym}^{00}}
\left[
\left.
\frac{K_{0,0}^{(\mathbb Z_3),{\rm c}}(t)}
{\sqrt{K_{1,1}^{(\mathbb Z_3),{\rm c}}(t)
        K_{2,2}^{(\mathbb Z_3),{\rm c}}(t)}}
\right|_{\epsilon=0}-1\right].
\label{eq:approx-S3-diagonal-fit}
\end{multline}
The evaluated term on the right-hand side explicitly represents this exact $G$-symmetric reference ensemble, measured directly from the matched $\epsilon=0$ data. The baseline subtraction of $1$ approximates the ratio expected for single GOE blocks with comparable smooth spectral densities, though it is not an exact finite-time identity. We fit the unsmoothed ratios using unweighted least squares over the interval $0.05\leq t/t_{\rm H}\leq0.75$. While this procedure successfully preserves the measured spectral shape of the unperturbed reference system, the interpolation formula itself is not rigorously derived from the full finite-$\tau$ GOE crossover. Consequently, $t_{\rm sym}^{00}$ should be understood as an operational, single-mode fit scale rather than an independently measured physical transition rate.}

{For the off-diagonal channel, $t_{\rm sym}^{12}$ is instead defined operationally as the time of the first downward half-maximum crossing of the Gaussian-smoothed correlation ratio:
\begin{equation}
\left.
\frac{K_{1,2}^{(\mathbb Z_3),{\rm c}}(t)}
{\sqrt{K_{1,1}^{(\mathbb Z_3),{\rm c}}(t)
        K_{2,2}^{(\mathbb Z_3),{\rm c}}(t)}}
\right|_{t=t_{\rm sym}^{12}}=\frac12.
\label{eq:approx-S3-offdiagonal-time}
\end{equation}
Gaussian smoothing is applied to the complete ratio. If the ratio is already below $1/2$ at the lower cutoff $t_{\min}=0.03t_{\rm H}$, we report only the resolution bound $t_{\rm sym}^{12}\lesssim t_{\min}$. This half-correlation time differs from the diagonal exponential-fit scale $t_{\rm sym}^{00}$; the two values should not be interpreted as directly comparable decay rates. We estimate statistical uncertainties using $300$ matched-realization bootstrap resamples. For each resample, we recalculate the connected correlators and the unperturbed $G$-symmetric baseline (the $\epsilon=0$ reference) before repeating the fits and extracting the crossing times. Throughout this procedure, the energy filter, time grid, and smoothing prescription are held strictly fixed.}

{The two channels in Fig.~\ref{fig:S3-approximate-tsym}
have different resolution limits. At $\epsilon=0.003$, the diagonal
relaxation remains unresolved before $t_{\rm H}$, but the off-diagonal
ratio crosses $1/2$ at $t_{\rm sym}^{12}/t_{\rm H}=0.820\pm0.014$.
At $\epsilon=0.01$, both measures are resolved:
$t_{\rm sym}^{00}/t_{\rm H}=0.189\pm0.010$ and
$t_{\rm sym}^{12}/t_{\rm H}=0.127\pm0.005$.
At $\epsilon=0.03$, the diagonal fit gives
$t_{\rm sym}^{00}/t_{\rm H}=0.0377\pm0.0049$, while
the off-diagonal half crossing lies at or below the lower cutoff.
For $\epsilon=0.001$, neither measure is resolved before $t_{\rm H}$,
although panel~(c) shows decorrelation at longer times. A crossover
time below $t_{\rm H}$ need not lie within a universal ramp: it must
also exceed the intrablock equilibration time, which is particularly
relevant to the rapid relaxation at $\epsilon=0.03$.}

{Figure~\ref{fig:S3-approximate-tsym}(c) shows decay over a finite observation window. The infinite-time average follows from exact degeneracy counting and does not require RMT. For the ordinary, unfiltered xSFF at fixed finite size,
\begin{equation}
\lim_{T\to\infty}\frac1T\int_0^T\dd t
K_{\lambda_a,\lambda_b}^{(N)}(t)
=
\left\langle\sum_{n,m}
\delta_{E_n^{(\lambda_a)},E_m^{(\lambda_b)}}\right\rangle.
\label{eq:approx-exact-degeneracy-plateau}
\end{equation}
Thus, in the absence of residual exact cross-sector degeneracies, the infinite-time-averaged cross-plateau vanishes even if approximate multiplets produce substantial correlations at finite times. Figure~\ref{fig:S3-approximate-tsym}(c) is consistent with this conclusion: the exactly symmetric ratio remains one, whereas the perturbed ratios decay toward zero. The xSFF therefore tracks the loss of correlations between specific exact $N$ sectors, complementing the diagonal response.}

\section{Application of the xSFF: Exclusion of a Hidden Conserved-Order-Parameter Mechanism}
\label{sec:xSFF-order-parameter}

{In this appendix, we use the xSFF to perform a consistency check for the mechanism proposed in Ref.~\cite{PhysRevLett.133.176001} for continuous symmetry breaking in one dimension at $T=0$. We first summarize the construction of the mechanism and then use the xSFF to test its correctness.}

\subsection{Review of the frustration-free symmetry-breaking mechanism}
\label{sec:spin1-claim-OP}

{The Hohenberg--Mermin--Wagner theorem forbids continuous
symmetry breaking in $d\leq2$ at finite temperature, and such breaking is
also generically obstructed in one-dimensional quantum systems at
$T=0$~\cite{Coleman1973}. The ferromagnetic Heisenberg chain is a familiar
exception:
\begin{equation}
\begin{split}
 H^{\rm FM}&=\sum_i H_i^{\rm FM},\\
 H_i^{\rm FM}&=\frac{J}{2}\left[2s(2s+1)-
 (\mathbf s_i+\mathbf s_{i+1})^2\right],\quad J>0.
\end{split}
\label{eq:FM-Hamiltonian-OP}
\end{equation}
In the two-site spin-$j$ sector, $H_i^{\rm FM}$ has eigenvalue
$\frac{J}{2}[2s(2s+1)-j(j+1)]\geq0$. Its ground states form the maximal-spin
multiplet
\begin{equation}
\begin{split}
 &\ket{M}_{\rm FM}\propto
 (S^-)^{sL-M}\ket{\uparrow\uparrow\cdots\uparrow},
\end{split}
\end{equation}
Here, $M=-sL,\ldots,sL$ and the order parameter $S^z$ is conserved, $[H^{\rm FM},S^z]=0$, so the different magnetization sectors cannot mix.}

{Ref.~\cite{PhysRevLett.133.176001} instead considers a
periodic spin-$1$ chain. For the xSFF calculation, we allow the positive
coupling to vary from bond to bond in order to remove crystalline symmetries:
\begin{equation}
\begin{split}
H^{\rm chain}=\sum_{i=1}^{L}J_i\Big\{&
-2S_i^xS_{i+1}^x-2S_i^yS_{i+1}^y-S_i^zS_{i+1}^z
\\
&+\left[2-(S_i^z)^2\right]\times
\left[2-(S_{i+1}^z)^2\right]\Big\}.
\end{split}
\label{eq:spin1-random-chain-OP}
\end{equation}
The uniform limit $J_i=J$ is the $\Delta=1/2$ point of the frustration-free
deformation, while $\Delta=1$ gives the spin-$1$ Heisenberg ferromagnet. For
any $J_i>0$, there is one exact ground state in each sector
$M=-L,\ldots,L$:
\begin{equation}
 \ket{M}_{\rm chain}
 \propto
 \sum_{\substack{\{m_i\}\\ \sum_i m_i=M}}
 \ket{\{m_i\}}.
\label{eq:spin1-ground-manifold-OP}
\end{equation}
where $m_i\in\{-1,0,1\}$. For example, for $L=2$,
\begin{equation*}
\begin{aligned}
 \ket{2}&=\ket{1,1},\\
 \ket{1}&\propto\ket{1,0}+\ket{0,1},\\
 \ket{0}&\propto\ket{1,-1}+\ket{0,0}+\ket{-1,1},
\end{aligned}
\end{equation*}
and the negative-$M$ states follow by $m_i\mapsto-m_i$. The deformation
preserves this $(2L+1)$-dimensional
ground manifold but reduces the symmetry of the Hamiltonian to the
$\mathrm{U}(1)$ generated by $S^z=\sum_iS_i^z$. The proposed order parameter
is $S^x=\sum_iS_i^x$, which is not conserved:
\begin{equation}
 [H^{\rm chain},S^x]\neq0.
\label{eq:spin1-order-not-conserved-OP}
\end{equation}
Nevertheless, Ref.~\cite{PhysRevLett.133.176001} finds an extensive response
to an infinitesimal field:
\begin{equation}
\begin{split}
 m(0^+)&=\lim_{h\to0^+}\lim_{L\to\infty}
 \frac{1}{L}\bra{\Phi(h)}S^x\ket{\Phi(h)}\neq0,\\
 H(h)&=H^{\rm chain}-hS^x.
\end{split}
\label{eq:spin1-SSB-criterion-OP}
\end{equation}
The claim is therefore that the frustration-free ground manifold supports
$T=0$ symmetry breaking even though its physical order parameter is not
conserved. However, checking only one order parameter does not rule all the possibility that there may be other hidden conserved order parameters that are not manifest. In the following, we are going to use xSFF to perform the check over the spectrum to ensure there are no other conserved order parameters.}

\subsection{xSFF consistency check}
\label{sec:xSFF-hidden-OP}

{A conserved operator connecting different symmetry sectors
would produce exact cross-sector degeneracies and hence an off-diagonal xSFF.
We therefore resolve the known
$\mathrm{O}(2)\times\mathbb Z_2^\Theta$ symmetry formed by total spin $S^z$, phase rotation $e^{-i\pi S^x}$, time reversal symmetry $\Theta = e^{-i\pi S^y}K$ and compute the xSFF for
$L=8$ using the random-coupling Hamiltonian in
Eq.~\eqref{eq:spin1-random-chain-OP}. The positive random $J_i$ remove lattice
symmetries while leaving both the internal symmetry and the frustration-free
ground manifold unchanged.}

\begin{figure*}[t]
    \centering
    \includegraphics[width=0.8\linewidth]{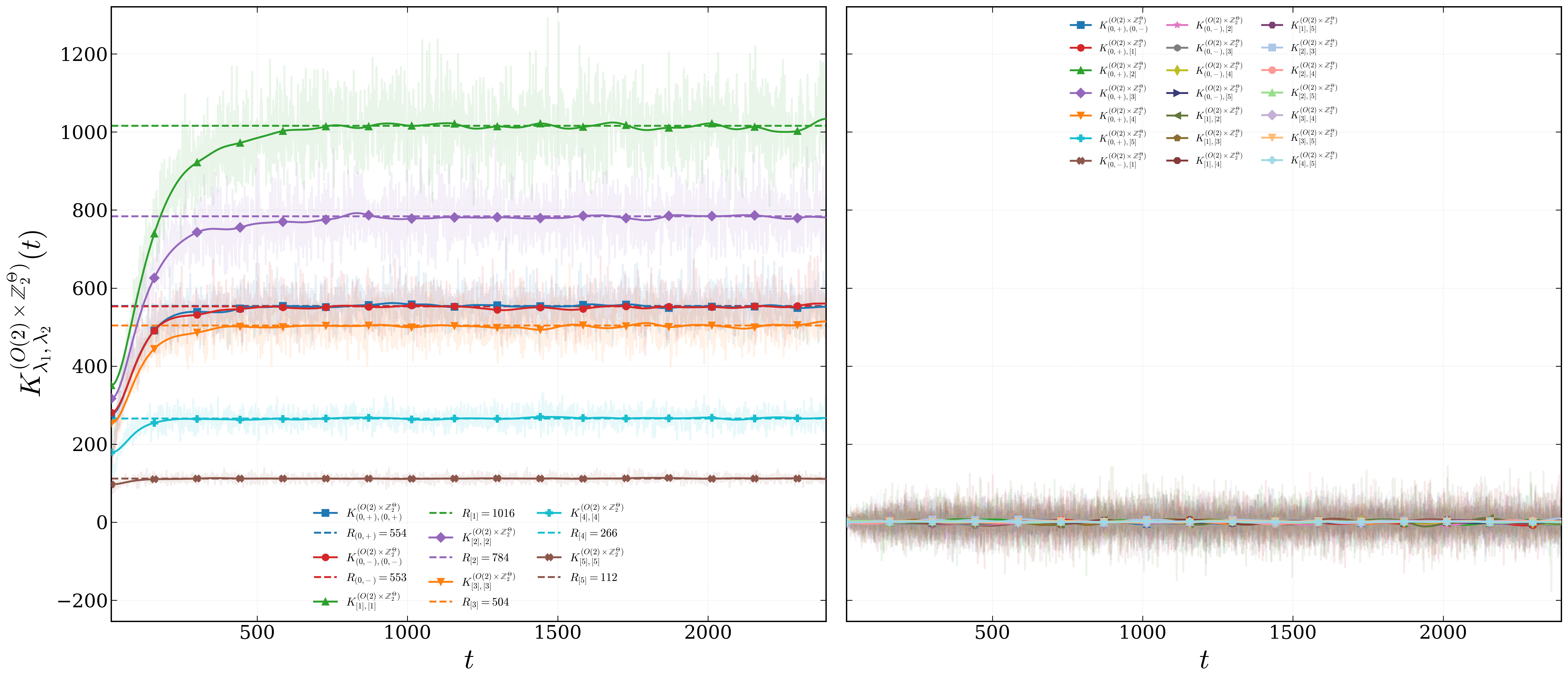}
    \caption{{xSFF for the $L=8$ spin-$1$ chain with positive random couplings $J_i\sim\mathcal N(1,0.25^2)$, averaged over $200$ realizations. The spectrum is resolved into $\mathrm{O}(2)\times\mathbb Z_2^\Theta$ sectors. \textbf{Left:} diagonal components. \textbf{Right:} off-diagonal components, for which no many-body-scale signal is observed.}}
    \label{fig:xSFF-spin1-OP}
\end{figure*}

{As shown in Fig.~\ref{fig:xSFF-spin1-OP}, the diagonal xSFFs
approach their expected plateaus, while no many-body-scale off-diagonal term
is observed. The small $O(1)$ residual is consistent with finite-size effects
and the known ground-state degeneracy. The xSFF therefore gives no evidence
for an additional conserved order parameter connecting different symmetry
sectors, supporting the mechanism proposed in
Ref.~\cite{PhysRevLett.133.176001}.}

\bibliography{reference.bib}
\end{document}